\documentclass[useAMS,usenatbib]{mn2e}

\usepackage{graphicx,color}
\usepackage{amssymb,amsmath}
\usepackage{amsmath} 
\usepackage{multicol}
\usepackage{latexsym}
\usepackage{amsfonts}
\usepackage{amssymb}
\usepackage{wasysym}
\usepackage{textcomp}
\usepackage{longtable,lscape}
\usepackage{rotating}
\usepackage{mathtools}
\usepackage{aas_macros}
\usepackage[dvipsnames]{xcolor}
\usepackage{morefloats}
\usepackage{afterpage}

\usepackage{natbib}
\usepackage{amssymb,amsmath}

\title[Enlightening the OCs dynamical evolution]{Enlightening the dynamical evolution of Galactic open clusters: an approach using \textit{Gaia} DR3 and analytical descriptions}
\author[M. S. Angelo et al.]{M. S. Angelo$^{1,2}$\thanks{E-mail:
mateusangelo@cefetmg.br}, J. F. C. Santos Jr.$^{2}$, F. F. S. Maia$^{3}$ and W. J. B. Corradi$^{4,2}$    \\ 
\noindent
$^1$Centro Federal de Educa\c{c}\~ao Tecnol\'ogica de Minas Gerais, Av. Monsenhor Luiz de Gonzaga, 103, 37250-000 Nepomuceno, MG, Brazil\\
$^2$Departamento de F\'isica, ICEx, Universidade Federal de Minas Gerais, Av. Ant\^onio Carlos 6627, 31270-901 Belo Horizonte, MG, Brazil\\
$^3$Universidade Federal do Rio de Janeiro, Instituto de F\'isica, 21941-972, Brazil\\
$^4$Laborat\'orio Nacional de Astrof\'isica, R. Estados Unidos 154, 37504-364 Itajub\'a, MG, Brazil}

\begin{document}

\date{Accepted XXX. Received XXX; in original form XXX}

\pagerange{\pageref{firstpage}--\pageref{lastpage}} \pubyear{2023}

\maketitle

\label{firstpage}

\begin{abstract}


Most stars in our Galaxy form in stellar aggregates, which can become long-lived structures called open clusters (OCs). Along their dynamical evolution, their gradual depletion leave some imprints on their structure. In this work, we employed astrometric, photometric and spectroscopic data from the \textit{Gaia} DR3 catalogue to uniformly characterize a sample of 60 OCs. Structural parameters (tidal, core and half-light radii, respectively, $r_t$, $r_c$ and $r_h$), age, mass ($M_{\textrm{clu}}$), distance, reddening, besides Jacobi radius ($R_J$) and half-light relaxation time ($t_{rh}$), are derived from radial density profiles and astrometrically decontaminated colour-magnitude diagrams. Ages and Galactocentric distances ($R_G$) range from 7.2$\,\lesssim\,$log($t.$yr$^{-1}$)$\,\lesssim\,$9.8 and 6$\,\lesssim\,R_G$(kpc)$\,\lesssim\,$12. Analytical expressions derived from $N$-body simulations, taken from the literature, are also employed to estimate the OC initial mass ($M_{\textrm{ini}}$) and mass loss due to exclusively dynamical effects. Both $r_c$ and the tidal filling ratio, $r_h/R_J$, tend to decrease with the dynamical age (=$t/t_{rh}$), indicating the shrinking of the OCs' internal structure as consequence of internal dynamical relaxation. This dependence seems differentially affected by the external tidal field, since OCs at smaller $R_G$ tend to be dynamically older and have smaller $M_{\textrm{clu}}/M_{\textrm{ini}}$ ratios. In this sense, for $R_G\lesssim8\,$kpc, the $r_h/R_J$ ratio presents a slight positive correlation with $R_G$. Beyond this limit, there is a dichotomy in which more massive OCs tend to be more compact and therefore less subject to tidal stripping in comparison to those less massive and looser OCs at similar $R_G$. Besides, the $r_t/R_J$ ratio also tends to correlate positively with $R_G$.





\end{abstract}


\section{Introduction}

It is well-established that stars form in clustered environments rather than in isolation (e.g., \citeauthor{Lada:2003}\,\,\citeyear{Lada:2003}), from the gravitational collapse and fragmentation of a progenitor molecular cloud. The gravitationally bound groups resulting from this process are called stellar clusters. In the Milky Way (MW), those self-gravitating stellar systems classified as open clusters (OCs) comprise wide ranges in age, mass and Galactocentric distance (e.g., \citeauthor{Lynga:1982}\,\,\citeyear{Lynga:1982}; \citeauthor{Bica:2019}\,\,\citeyear{Bica:2019}; \citeauthor{Dias:2021a}\,\,\citeyear{Dias:2021a}; \citeauthor{Cantat-Gaudin:2020}\,\,\citeyear{Cantat-Gaudin:2020}) and are therefore essential tools that can help constraining theories of molecular cloud fragmentation, star formation and stellar evolution (e.g., \citeauthor{Krumholz:2019}\,\,\citeyear{Krumholz:2019}; \citeauthor{Dalessandro:2021}\,\,\citeyear{Dalessandro:2021}; \citeauthor{Darma:2021}\,\,\citeyear{Darma:2021}; \citeauthor{Camargo:2009}\,\,\citeyear{Camargo:2009}), besides probing Galactic chemodynamical properties. They are also excellent laboratories to study stellar dynamics \citep{Friel:1995}. 


In this context, the description of the intricate interplay between physical processes that lead bound clusters to dissolution and the signature of such processes on the clusters' morphology is a debated topic. A number of studies have been dedicated to improve this comprehension from a theoretical point of view.

\cite{de-La-Fuente-Marcos:1997} performed a set of $N$-body simulations to study the influence of different mass spectra on the dynamical evolution of OCs, truncated at their tidal radius. The employed models include mass loss due to stellar evolution and primordial binaries. It was found that the evolution of the modeled clusters is highly dependent on their initial number of stars and initial mass function. \cite{Trenti:2013}, in turn, employed $N$-body simulations with a variety of realistic initial mass functions and initial conditions to investigate the energy equipartition effect over many two-body relaxation times. Previously, the seminal work of \cite{Vishniac:1978} established the conditions for reaching dynamical equilibrium in stellar systems with a continuous distribution of masses.  

\cite{Terlevich:1987} studied the evolution of clusters with $N\sim1000$ stars with initial masses following a power-law (slope $\alpha=-2.75$) initial mass function; mass loss by stellar evolution and binaries were included. For the external interactions, a smooth gravitational field and the effects of tidal heating due to encounters with giant molecular clouds were also considered. It was found evidence of mass segregation due to the preferential escape of low luminosity stars. It was also concluded that the external perturbations effectively affect the clusters velocity distribution and shape their outer structure. In this framework, \cite{Engle:1999} established that initially Roche volume underfilling OCs are less subject to tidal stresses and tend to survive longer.   

\cite{Portegies-Zwart:2001} performed detailed $N$-body simulations using initial conditions representative of young Galactic OCs; they employed realistic mass functions, primordial binaries and also considered the external potential of the MW. Evidence of mass segregation is verified after one cluster relaxation time. They found agreement between the luminosity functions predicted by their models with the observed ones for four well-known OCs (namely, Pleiades, Praesepe, Hyades and NGC\,3680) at their respective ages. They also found that stars tend to escape along the line connecting the cluster to the Galactic centre, through the first and second Lagrangian points, which may cause flattening of the cluster stellar distribution. Previously, \cite{Fukushige:2000} demonstrated that it is important to consider the finite time it takes for a star to find one of the Lagrangian points (where the escape energy is lowest; \citeauthor{Gieles:2008}\,\,\citeyear{Gieles:2008}, hereafter GB08; \citeauthor{Baumgardt:2003}\,\,\citeyear{Baumgardt:2003}, hereafter BM03) and escape through it.          

BM03 employed $N$-body simulations, which incorporated mass loss through stellar evolution, to investigate the process of disruption of clusters subject to an external tidal field. From the outcomes of their simulations (see also \citeauthor{Gieles:2004}\,\,\citeyear{Gieles:2004}), they proposed a scaling law between the disruption time ($t_{\textrm{dis}}$) and the cluster initial mass ($M_{\textrm{ini}}$) under the form $t_{\textrm{dis}}=t_0(M_{\textrm{ini}}/M_{\odot})^\gamma$, where $\gamma=0.62$. The $\gamma$ exponent is in agreement with \cite{Boutloukos:2003}, who derived this value from mass and age histograms of cluster samples in four different galaxies (M\,51, M\,33, the Small Magellanic Cloud and the MW); $t_0$ is a constant whose value depends on the galactic environment (\citeauthor{Lamers:2005a}\,\,2005a, hereafter LGPZ05).

In a subsequent paper, \citeauthor{Lamers:2005b}\,\,(2005b, hereafter LGB05) presented approximated analytical expressions from the outcomes of BM03, combined with GALEV evolutionary models (\citeauthor{Schulz:2002}\,\,\citeyear{Schulz:2002}; \citeauthor{Anders:2003}\,\,\citeyear{Anders:2003}), in order to describe the cluster mass loss due to stellar evolution and dynamical processes. Their analysis proved to be consistent with the outcomes from detailed $N$-body simulations. They demonstrated that $t_{\textrm{dis}}$ is not only correlated with the cluster initial mass, but at all times it scales with the present mass under the form proposed by BM03.

\cite{Miholics:2014} investigated how changes in the external tidal field affect the cluster dynamics by modeling the accretion process of a stellar cluster originally located in a dwarf galaxy and then falling on the MW. They found that, after $\sim$2 relaxation times, the stellar distribution readjusts itself to the new potential and the cluster size becomes indistinguishable from another one always living on the MW at compatible Galactocentric distance ($R_G$). Subsequently, \cite{Miholics:2016} performed an analogous procedure, this time employing time-dependant external potentials and found that the dynamical evolution of a star cluster is determined by whichever galaxy has the strongest tidal field at its position; as before, its structure quickly mimics that of a cluster born in the MW on the same orbit.

Other works worth to mention have: ($i$) modeled single mass clusters and analysed their evolution along common sequences on dynamical \textit{Luminosity} versus \textit{Temperature} diagram \citep{Kupper:2008}, ($ii$) self-consistently studied the formation and co-evolution of stellar clusters and their host galaxies through cosmic time (like in the E-MOSAICS simulations; \citeauthor{Reina-Campos:2019}\,\,\citeyear{Reina-Campos:2019}; \citeauthor{Pfeffer:2018}\,\,\citeyear{Pfeffer:2018}), ($iii$) simulated young star clusters with different degrees of initial substructure, providing insights into the formation process and subsequent dynamical evolution of star clusters (\citeauthor{Darma:2021}\,\,\citeyear{Darma:2021}; \citeauthor{Dalessandro:2021}\,\,\citeyear{Dalessandro:2021}), ($iv$) investigated the impact of different gas explusion regimes (impulsive approximation or adiabatic process), during the early evolutionary stages, on the clusters properties along their subsequent evolution (\citeauthor{Pang:2021}\,\,\citeyear{Pang:2021}; see also \citeauthor{Hills:1980}\,\,\citeyear{Hills:1980}, \citeauthor{Kroupa:2001a}\,\,\citeyear{Kroupa:2001a} and \citeauthor{Leveque:2022}\,\,\citeyear{Leveque:2022}), ($v$) employed advanced $N$-body models (via, e.g., the {\fontfamily{ptm}\selectfont{NBODY6}} code; \citeauthor{Aarseth:1999}\,\,\citeyear{Aarseth:1999}, \citeyear{Aarseth:2003}) to explore the most important physical mechanisms that mold the size scale of star clusters subject to the Galactic tidal field (\citeauthor{Madrid:2012}\,\,\citeyear{Madrid:2012}, hereafter MHS12; \citeauthor{Renaud:2011}\,\,\citeyear{Renaud:2011}; \citeauthor{Webb:2013}\,\,\citeyear{Webb:2013}; \citeauthor{Moreno:2014}\,\,\citeyear{Moreno:2014}; \citeauthor{Cai:2016}\,\,\citeyear{Cai:2016}; \citeauthor{Madrid:2017}\,\,\citeyear{Madrid:2017}), ($vi$) proposed general descriptions for the life cycle of clusters based on relations between half-mass density, cluster mass and galactocentric radius \citep{Gieles:2011a}, ($vii$) derived the response of a cluster to tidal perturbations due to collisions with giant molecular clouds (\citeauthor{Gieles:2016}\,\,\citeyear{Gieles:2016}; \citeauthor{Theuns:1991}\,\,\citeyear{Theuns:1991}; \citeauthor{Spitzer:1958}\,\,\citeyear{Spitzer:1958}; \citeauthor{van-den-Bergh:1980}\,\,\citeyear{van-den-Bergh:1980}; \citeauthor{Portegies-Zwart:2010}\,\,\citeyear{Portegies-Zwart:2010}) and provided expressions for the disruption time from the fractional mass loss scaled with the fractional energy gain during these collisions \citep{Gieles:2006}, ($viii$) investigated the impact of passages through the MW disc (\citeauthor{Ostriker:1972}\,\,\citeyear{Ostriker:1972}; \citeauthor{Gieles:2007}\,\,\citeyear{Gieles:2007}; \citeauthor{Gnedin:1999}\,\,\citeyear{Gnedin:1999}; \citeauthor{Martinez-Medina:2017}\,\,\citeyear{Martinez-Medina:2017}), which can enhance the mass-loss process due to tidal heating and shocks \citep{Webb:2014}.

From the observational side, an increasingly large number of studies have benefited from the spatial coverage and high precision data provided by the most recent releases of the \textit{Gaia} catalogue (DR2: \citeauthor{Gaia-Collaboration:2018}\,\,\citeyear{Gaia-Collaboration:2018}; EDR3: \citeauthor{Gaia-Collaboration:2021}\,\,\citeyear{Gaia-Collaboration:2021}; DR3, \citeauthor{Gaia-Collaboration:2022}\,\,\citeyear{Gaia-Collaboration:2022}) and focused on the determination of the OCs structural parameters. Recently, \cite{Tarricq:2022} employed data from the \textit{Gaia} EDR3 catalogue and performed \citeauthor{King:1962}'s\,\,(\citeyear{King:1962}) profile fitting to radial density profiles (RDP) of a sample of 389 Galactic OCs, in order to derive their core and tidal radii. They also systematically searched for the presence of extended external structures, like outer haloes and tidal tails. Some possible evolutionary connections between cluster structure, age, number of members and degree of mass segregation were also investigated. 

Using \textit{Gaia} EDR3 data, \cite{Zhong:2022} proposed a modified model with the aim of providing a better description of the clusters density profile, by incorporating two main structural components: the \textit{core} component, described by the \cite{King:1962} model, and the \textit{outer halo} component, described by a logarithmic Gaussian function. Their model is parametrized in terms of 4 characteristic radii: core and tidal radii, as inferred from the King profile; $r_0$ and $r_e$, which account for, respectively, the mean and boundary radius of outer halo members.

Also recently, \cite{Pang:2021} benefited from \textit{Gaia} EDR3 data for OCs located in the solar neighborhood to analyze their three-dimensional morphology. The spatial distribution of member candidate stars within the tidal radius was fit using ellipsoid models. This procedure allowed to identify elongated filament-like substructures in the younger clusters of their sample, while tidal tails have been detected for the older ones. It was also investigated, from $N$-body models, how the different regimes of gas expulsion, during the early evolutionary stages, could have affected the clusters properties at their ages. The elongated structure of Galactic OCs was also discussed by \cite{Chen:2004}, who employed data from the 2MASS catalogue \citep{Skrutskie:2006} to analyze clusters' morphological parameters by means of projected stellar density distributions.

Other previous noteworthy observational works, focused on the OCs' morphology and on empirical evidences related to their dynamical evolution, have: ($i$) searched for relations of cluster parameters with cluster age and $R_G$ \citep{Bonatto:2005a}, ($ii$) characterized low-contrast OCs close to the Galactic plane, by means of colour-magnitude diagrams (CMDs) and RDPs built after dedicated field-star decontamination and colour-magnitude filtered photometry (\citeauthor{Camargo:2009}\,\,\citeyear{Camargo:2009}; \citeauthor{Bica:2011}\,\,\citeyear{Bica:2011}), ($iii$) used the integrated form of the \citeauthor{King:1962}'s\,\,(\citeyear{King:1962}) model aiming to the determination of structural parameters even in the case of poorly populated clusters, thus allowing uniformity in the fitting procedure (\citeauthor{Piskunov:2007}\,\,\citeyear{Piskunov:2007}; \citeauthor{Kharchenko:2013}\,\,\citeyear{Kharchenko:2013}), ($iv$) employed density versus radius diagram to investigate the universality of young cluster sequences \citep{Pfalzner:2009}, ($v$) searched for correlations involving OCs parameters to provide observational constraints to the disc properties in the solar neighborhood \citep{Tadross:2002}. Detailed reviews on cluster evolution can be found in \cite{Vesperini:2010}, \cite{Portegies-Zwart:2010}, \cite{Renaud:2018}, and \cite{Krumholz:2019}.

In the present paper, besides astrometric and photometric information, we have also incorporated spectroscopic data from the most updated version (DR3) of the \textit{Gaia} catalogue in order to refine the characterization of the investigated sample. We have searched the catalogues of \citeauthor{Dias:2021a}\,\,(\citeyear{Dias:2021a}, hereafter DMML21) and \cite{Cantat-Gaudin:2020} looking for non-embedded OCs containing significant number of catalogued members ($N\gtrsim100$), presenting reasonable contrasts with the local field population (central to background stellar density typically larger than $\sim4$; see Section~\ref{sec:struct_analysis}), defining unambigous evolutionary sequences on astrometrically decontaminated CMDs and notable concentrations on the astrometric space (Section~\ref{sec:results}). This cluster selection allowed us to properly derive structural parameters from proper motion filtered RDPs and to determine fundamental astrophysical parameters from well-constrained member star lists. The selected clusters span moderately wide ranges in age, Galactocentric distance and dynamical evolutionary stages (see Section~\ref{sec:discussion}).  



The present work is part of an ongoing project whose main goal is to characterize the clusters dynamical state from an observational perspective and to provide some insights regarding the possible imprints that the internal dynamics, regulated by the external tidal field, may have produced on the OCs' structure. In what follows, our discussions are enlighted by the outcomes from previous numerical works, like some of those mentioned above.

This paper is organized as follows: in Section~\ref{sec:sample_and_data} we present our sample and the collected data; the structural analysis is shown in Section~\ref{sec:struct_analysis}; in Sections~\ref{sec:membership} and \ref{sec:results}, we present our membership assignment procedure, combined the astrometric, photometric and spectroscopic information to obtain member stars lists and the clusters' astrophysical parameters; in Section~\ref{sec:analysis}, we present the analytical formulas employed to analyse the OCs' properties; our results are discussed in Section~\ref{sec:discussion}; in Section~\ref{sec:conclusions}, we present our main conclusions.

\section{Sample and data}
\label{sec:sample_and_data}

\subsection{Data source}

Astrometric, photometric and radial velocity ($V_{\textrm{rad}}$) data from the \textit{Gaia} DR3 main table ({\fontfamily{ptm}\selectfont\textit{gaiadr3.gaia\_source}}) were downloaded from the \textit{Gaia} Archive\footnote[1]{https://gea.esac.esa.int/archive/} for each investigated OC in circular regions centred on the coordinates informed in DMML21 catalogue and with extraction radius ($r_{\textrm{extr}}$) larger than $\sim5\times$ the listed diameters (typically, $r_{\textrm{extr}}\gtrsim2^{\circ}$). Also from the same table, we extracted, in the same regions, atmospheric parameters (effective temperature, $T_{\textrm{eff}}$; surface gravity, log\,$g$; metallicity, $[Fe/H]$) estimated from low-resolution BP/RP spectra via the {\fontfamily{ptm}\selectfont GSP-Phot} module within the \textit{Gaia}'s astrophysical parameters inference system (Apsis; \citeauthor{Andrae:2022}\,\,\citeyear{Andrae:2022}). Additionally, $T_{\textrm{eff}}$, log\,$g$ and metallic abundances ($[Fe/M]$ and $[M/H]$) derived from the {\fontfamily{ptm}\selectfont GSP-Spec} module, based on mid-resolution ($R\approx11\,500$) RVS spectra \citep{Recio-Blanco:2022}, were also extracted. To complement our database, $T_{\textrm{eff}}$ and log\,$g$ values were also taken from the ESP-HS\footnote[2]{This \textit{hot stars} specialized module assumes a solar chemical composition, and therefore no corresponding metallicity value is saved in the catalogue; see figure 7 of \cite{Creevey:2022} for information about the parameter spaces covered by the different stellar modules in Apsis.} module (table {\fontfamily{ptm}\selectfont\textit{gaiadr3.astrophysical\_parameters}}), an internal Apsis algorithm dedicated to the fitting of BP/RP+RVS spectra for hot stars ($T_{\textrm{eff}}>7500\,$K; \citeauthor{Creevey:2022}\,\,\citeyear{Creevey:2022}; \citeauthor{Fouesneau:2022}\,\,\citeyear{Fouesneau:2022}). Data from different tables were then cross-matched via the {\fontfamily{ptm}\selectfont\textit{source\_id}} unique catalogue identifier and then a master table was created for each investigated OC.

\subsection{Data filtering and corrections}
\label{sec:data_filtering_corrections}

Following the prescriptions informed in the \textit{Gaia} documentation\footnote[3]{https://gea.esac.esa.int/archive/documentation/GDR3/ (online documentation)}, we applied the available recipes\footnote[4]{https://www.cosmos.esa.int/web/gaia/dr3-software-tools} accompanying the third data release in order to correct the original data for parallax zero-point \citep{Lindegren:2021} and to provide a corrected version of the photometric flux excess factor ($E(BP/RP)$; \citeauthor{Riello:2021}\,\,\citeyear{Riello:2021}). No corrections in flux or $G$-band magnitudes have been applied, since they are already implemented in DR3 \citep{Gaia-Collaboration:2022}. 

After that, in order to ensure the best possible quality of our data and to remove spurious astrometric and/or photometric solutions, we applied quality filters limiting our database to those sources consistent with the following conditions

\begin{flalign}
    & \vert C^{*}\vert < 5\,\sigma_{C^*}  & \\
    & RUWE<1.4, &
\end{flalign}

\noindent where $C^*$ is the corrected value of $E(BP/RP)$, $\sigma_{C^*}$ is taken from equation\,18 of \cite{Riello:2021} and $RUWE$ is the \textit{renormalised unit weight error} \citep{Lindegren:2021a}. We also restricted our sample to stars brighter than $G=19\,$mag, which is the nominal magnitude limit to ensure a minimum of 5 transits for a source ({\fontfamily{ptm}\selectfont\textit{matched\_transits}}$\,\ge\,$5) and 9 visibility periods ({\fontfamily{ptm}\selectfont\textit{visibility\_periods\_used}}$\,\ge\,$9), thus allowing astrometric and photometric completeness \citep{Fabricius:2021}.

Additionally, the surface gravity and metallicity\footnote[5]{In \textit{Gaia} DR3, the neutral iron abundance $[Fe/H]$ is obtained from the outcomes of the {\fontfamily{ptm}\selectfont{GSP-Spec}} module via the relation $[Fe/H]=[Fe/M]+[M/H]$; in our database, the uncertainty in $[Fe/H]$ consists in the sum in quadrature of the individual uncertainties in $[Fe/M]$ and $[M/H]$.} values obtained from the {\fontfamily{ptm}\selectfont GSP-Spec} module have been recalibrated by applying low-degree polynomials, following the recommendations stated by \citeauthor{Recio-Blanco:2022}\,\,(\citeyear{Recio-Blanco:2022}; their equations 1 and 3). The proper coefficients are informed in their tables 3 and 4, with the uncertainties available in their table E.1. The uncertainties of these coefficients and also of the original log\,$g$ and $[Fe/H]$ catalogued values have been propagated into the final values for both parameters within our database. The $[Fe/H]$ values obtained from low-resolution BP/RP spectra (i.e., via the {\fontfamily{ptm}\selectfont GSP-Phot} module) were not recalibrated (although a recipe is provided\footnote[6]{https://github.com/mpi-astronomy/gdr3apcal}) since, as stated by \cite{Andrae:2022}, they are only useful at a qualitative level. These metallicities are employed here only for comparison purposes (see Section~\ref{sec:results}).       

Finally, the $V_{\textrm{rad}}$ values have been recalibrated according to equation 1 of \cite{Blomme:2022}, in the case of hot stars ({\fontfamily{ptm}\selectfont\textit{rv\_template\_teff}}$\,\ge\,$8500\,K), and equation 5 of \cite{Katz:2022}, for cooler ones ({\fontfamily{ptm}\selectfont\textit{rv\_template\_teff}}\,$<$\,8500\,K). Again, the proper calibration errors have been propagated into the final $V_{\textrm{rad}}$ uncertainties.

\subsection{Projected cartesian coordinates}


The $\alpha, \delta$ coordinates for each star were projected on the plane of the sky along the line-of-sight vector through the literature cluster centre ($\alpha_c, \delta_c,$ taken from DMML21), as suggested by \cite{van-de-Ven:2006}. This way, the equatorial coordinates have been transformed into cartesian ones ($X,Y$) through the relations:

\begin{flalign}   
   & X = r_0\,\textrm{cos}\,\delta\,.\,\textrm{sin}(\alpha - \alpha_c)  \label{eq:project_alpha_delta1} &  \\ 
   & Y = r_0\,[\textrm{sin}\,\delta\,.\,\textrm{cos}\,\delta_c\,-\,\textrm{sin}\,\delta_c\,.\,\textrm{cos}\,\delta\,.\,\textrm{cos}(\alpha - \alpha_c)],  \label{eq:project_alpha_delta2} &  
\end{flalign}

\noindent where the scaling factor $r_0=180/\pi$ is applied to result $X$ and $Y$ in units of degrees. With the above transformations, the cluster centre is automatically at the origin, that is, (X,Y)=(0,0) for ($\alpha, \delta$) = ($\alpha_c, \delta_c$) and the radial distance of each star to the cluster centre is simply $r = \sqrt{X^2 + Y^2}$.

When obtaining the angular distance $r$ on the celestial sphere for two sources with a small separation, the above equations reduce to the approximate formula $r\cong\sqrt{(\Delta\,\alpha)^2 + (\Delta\,\delta)^2}$, where $\Delta\,\alpha=(\alpha-\alpha_c)\,\textrm{cos}\,\delta$ and $\Delta\,\delta=\delta-\delta_c$. 


\subsection{Proper motions filtering}
\label{sec:prop_motions_filtering}

\begin{figure*}
\begin{center}

\parbox[c]{1.00\textwidth}
  {
   \begin{center}
       \includegraphics[width=0.331\textwidth]{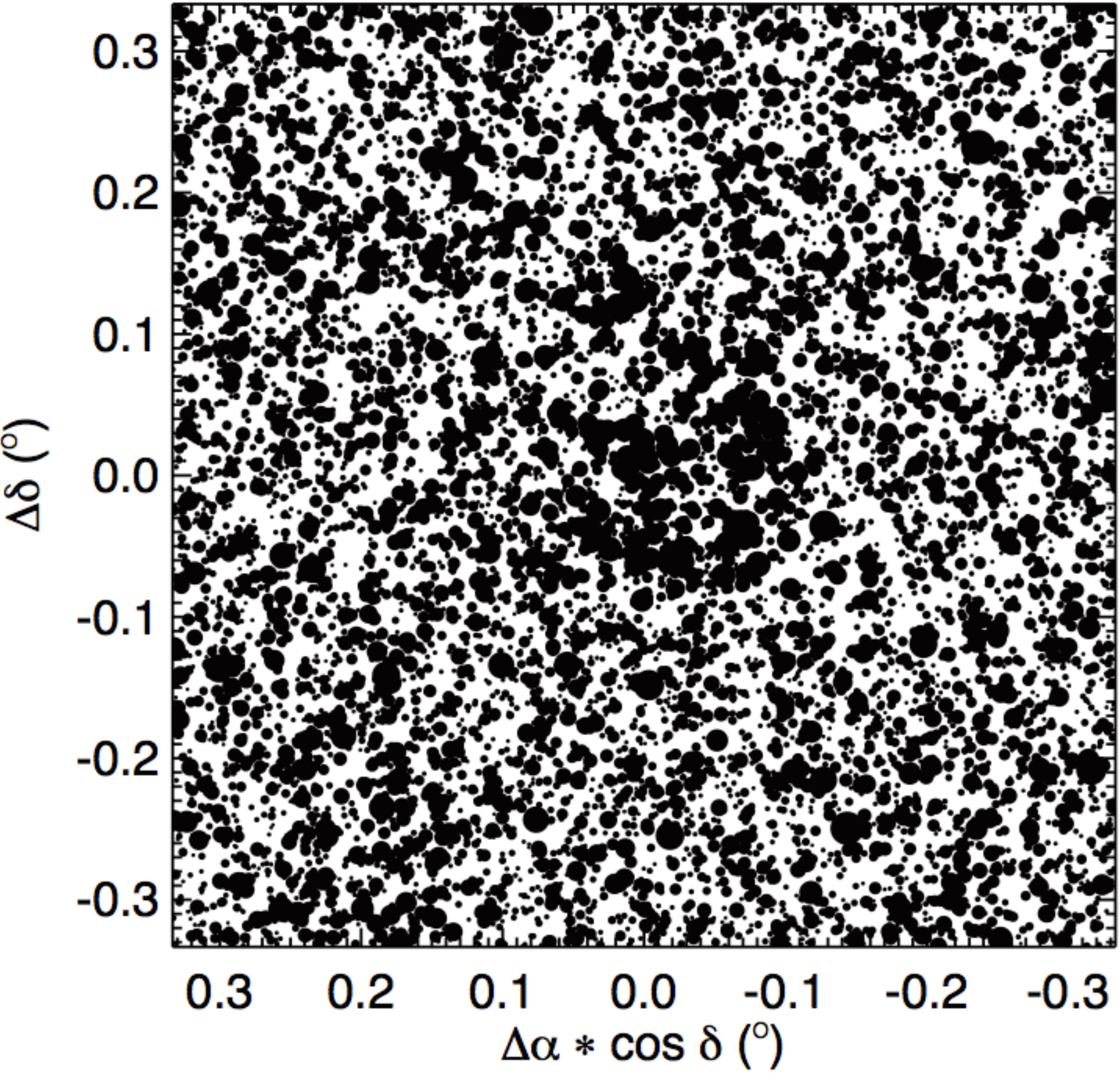}   
       \includegraphics[width=0.328\textwidth]{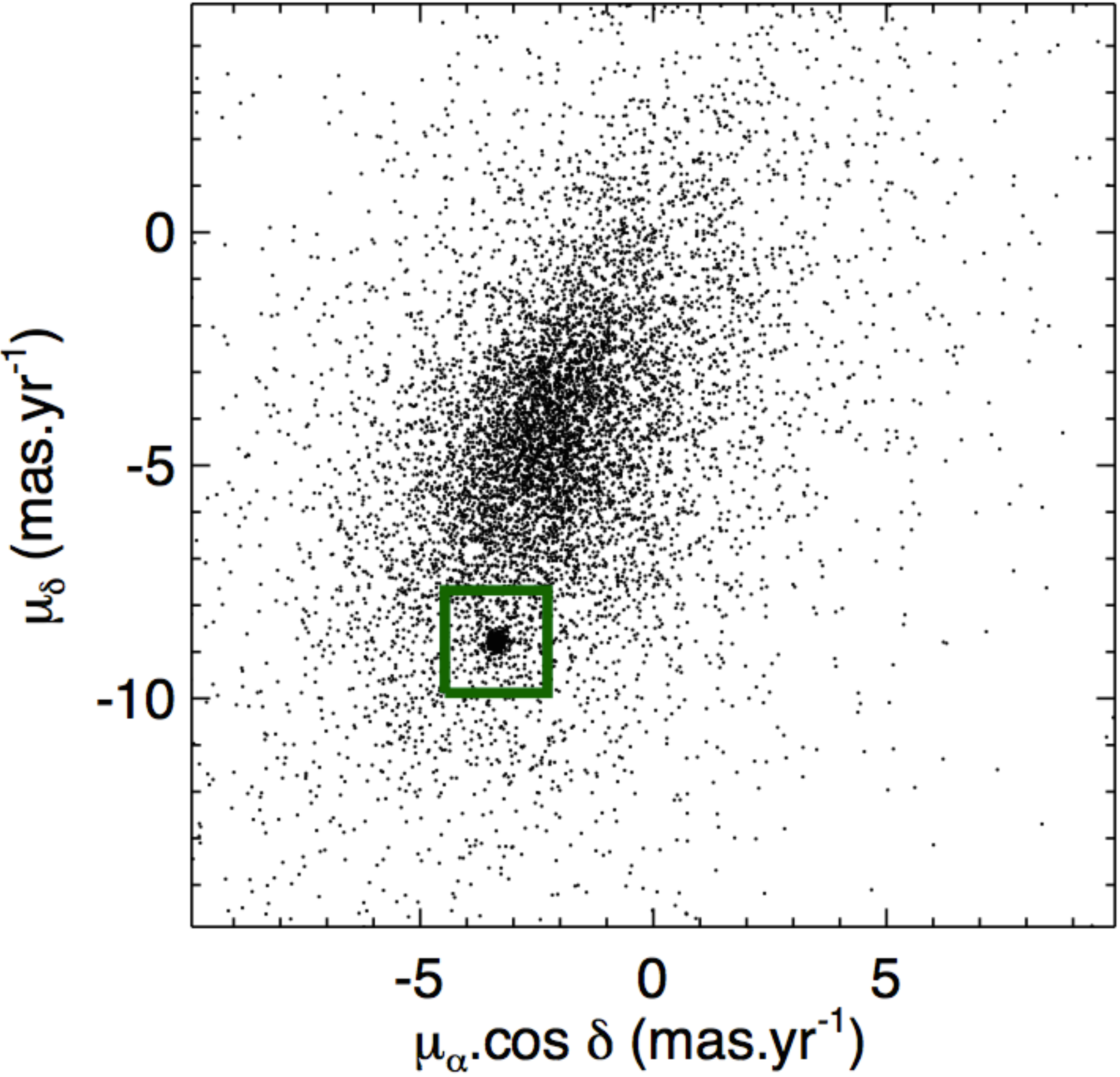}  
       \includegraphics[width=0.331\textwidth]{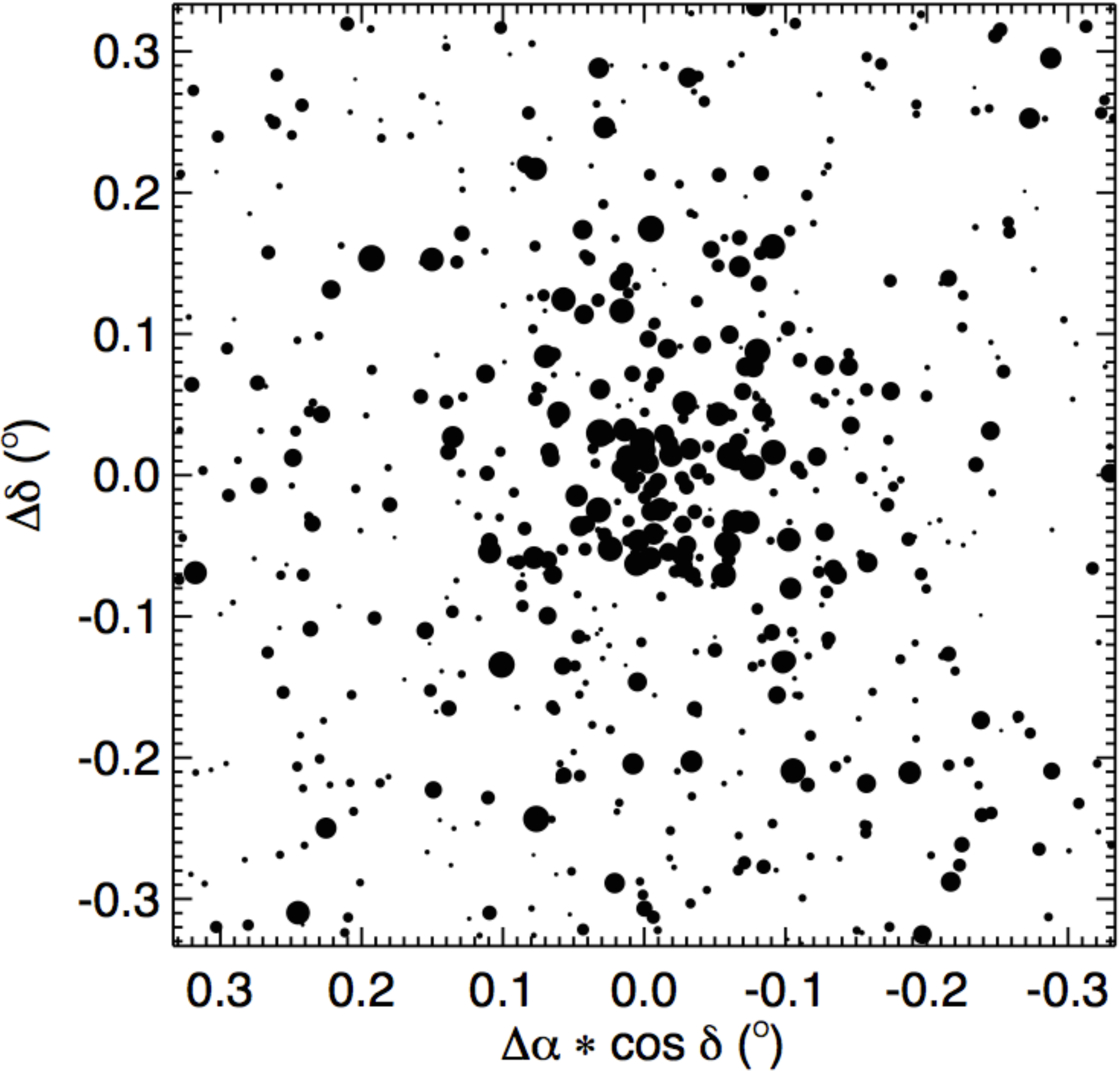}   
    \end{center}    
  }
\caption{ \textit{Left panel}: skymap for stars in a field of $40'\times40'$ centered on the OC NGC\,6811. Equatorial coordinates have been projected on the cartesian plane tangent to the cluster centre on the celestial sphere (equations \ref{eq:project_alpha_delta1} and \ref{eq:project_alpha_delta2}). The stars sizes are proportional to their brightness in the $G$-band. \textit{Middle panel}: VPD for the sample of stars in the left panel; the green box shows the proper motions filter. \textit{Right panel}: reconstructed skymap for NGC\,6811, after applying the proper motions filter. It is noticeable the improvement in the contrast cluster-field. }

\label{fig:ilustra_preanalise_NGC6811}
\end{center}
\end{figure*}

The signature of each cluster on the vector-point diagram (VPD) was identified by looking for detached concentration of stars (defined almost by members of a given OC). Restricting the analysis to those stars located close to these ``clumpy" groups (Figure~\ref{fig:ilustra_preanalise_NGC6811}) allows us to filter out most of the contamination by Galactic field stars projected in the cluster area, therefore optimizing the contrast cluster-field (see, e.g., \citeauthor{Bonatto:2007}\,\,\citeyear{Bonatto:2007}, who employed an analogous procedure by means of colour filters applied on CMDs).

This filtering procedure is illustrated in Figure~\ref{fig:ilustra_preanalise_NGC6811} for the OC NGC\,6811: (i) the left panel shows the whole set of stars (after applying the data quality filters, as explained in Section~\ref{sec:data_filtering_corrections}) located in a square area of $40\times40\,$arcmin$^2$, centered on the coordinates informed in DMML21; (ii) the middle panel exhibits the VPD for this sample of stars, with the green box indicating the proper motions filter; its size (for our sample, this filter is at least one order of magnitude larger than the intrinsic dispersions in $\mu_{\alpha}\,\textrm{cos}\,\delta$ and $\mu_{\delta}$, as inferred after setting memberships; see Section~\ref{sec:membership} and Table~\ref{tab:investig_sample}) is large enough to encompass the cluster member stars, but small enough to eliminate most of the contamination by the field population; (iii) the right panel shows the proper motions filtered skymap of the OC NGC\,6811, from which the subsequent steps (see Section~\ref{sec:struct_analysis}) are performed.

\begin{table*}
\centering
\tiny
\rotcaption{ Central coordinates, Galactocentric distances, structural and fundamental parameters, mean proper motion components and half-light relaxation times ($t_{rh}$) for the studied sample. }
\begin{sideways}
\begin{minipage}{240mm}

\begin{tabular}{lcccrcccrrccrrrr}

 Cluster             & $RA$                            & $DEC$                           & $\ell$             & $b$               & R$_G^{(*)}$        & $r_c$              & $r_{h}$            & $r_t$                & $(m-M)_0$            & $E(B-V)$             & log\,$t$            & $[Fe/H]^{(**)}$        & $\langle\mu_{\alpha}\,\textrm{cos}\,\delta\rangle^{\dag}$    & $\langle\mu_{\delta}\rangle^{\dag}$     & $t_{rh}^{\dag\dag}$  \\    
                     &($h$:$m$:$s$)                    & ($^{\circ}$:$'$:$''$)           & $(^\circ)$         & $(^\circ)$        &  (kpc)             &  (pc)              &     (pc)           &  (pc)                & (mag)                & (mag)                & (dex)               & (dex)                  & (mas yr$^{-1}$)                                              & (mas yr$^{-1}$)                         & (Myr)                \\     
                                                                                                                                                                                                                                                                                                                                                                                                                                                                        
\hline                                                                                                                                                                                                                                                                                                                                                                                                                                                                  
NGC\,129             & 00:30:32                        &  60:13:32                       & 120.32             &  -2.55            & 9.1\,$\pm$\,0.5    & 3.23\,$\pm$\,0.79  &4.11\,$\pm$\,0.55   &11.43\,$\pm$\,1.42    &11.30\,$\pm$\,0.25    &0.62\,$\pm$\,0.07     &8.05\,$\pm$\,0.30    & 0.22\,$\pm$\,0.21      & -2.59\,$\pm$\,0.07                                           & -1.18\,$\pm$\,0.08                   & 239\,$\pm$\,49       \\ 
NGC\,188             & 00:47:18                        &  85:13:59                       & 122.84             &  22.37            & 9.0\,$\pm$\,0.5    & 2.72\,$\pm$\,0.51  &4.52\,$\pm$\,0.60   &15.61\,$\pm$\,2.68    &11.15\,$\pm$\,0.20    &0.11\,$\pm$\,0.05     &9.80\,$\pm$\,0.15    & 0.05\,$\pm$\,0.20      & -2.32\,$\pm$\,0.08                                           & -1.01\,$\pm$\,0.11                   & 480\,$\pm$\,98       \\ 
NGC\,559             & 01:29:30                        &  63:18:29                       & 127.20             &   0.75            & 9.3\,$\pm$\,0.5    & 1.65\,$\pm$\,0.19  &3.11\,$\pm$\,0.49   &11.97\,$\pm$\,3.19    &11.45\,$\pm$\,0.30    &0.70\,$\pm$\,0.01     &8.90\,$\pm$\,0.10    &-0.13\,$\pm$\,0.23      & -4.27\,$\pm$\,0.06                                           &  0.18\,$\pm$\,0.03                   & 201\,$\pm$\,49       \\ 
NGC\,654             & 01:44:04                        &  61:52:55                       & 129.09             &  -0.36            & 9.5\,$\pm$\,0.6    & 0.93\,$\pm$\,0.12  &2.22\,$\pm$\,0.45   &10.51\,$\pm$\,3.70    &11.65\,$\pm$\,0.40    &1.03\,$\pm$\,0.10     &7.30\,$\pm$\,0.20    &-0.13\,$\pm$\,0.31      & -1.16\,$\pm$\,0.14                                           & -0.32\,$\pm$\,0.12                   & 100\,$\pm$\,31       \\ 
NGC\,752             & 01:57:05                        &  37:49:51                       & 136.96             & -23.29            & 8.3\,$\pm$\,0.5    & 3.28\,$\pm$\,0.46  &3.59\,$\pm$\,0.29   & 8.89\,$\pm$\,0.69    & 8.22\,$\pm$\,0.15    &0.06\,$\pm$\,0.03     &9.20\,$\pm$\,0.05    & 0.05\,$\pm$\,0.15      &  9.76\,$\pm$\,0.19                                           &-11.82\,$\pm$\,0.21                   & 140\,$\pm$\,23       \\ 
NGC\,1027            & 02:42:36                        &  61:36:30                       & 135.75             &   1.55            & 8.7\,$\pm$\,0.5    & 2.59\,$\pm$\,0.40  &4.14\,$\pm$\,0.83   &13.90\,$\pm$\,4.65    & 9.95\,$\pm$\,0.25    &0.47\,$\pm$\,0.08     &7.95\,$\pm$\,0.20    & 0.05\,$\pm$\,0.20      & -1.78\,$\pm$\,0.13                                           &  2.03\,$\pm$\,0.14                   & 222\,$\pm$\,67       \\ 
NGC\,1647            & 04:46:08                        &  19:02:55                       & 180.37             & -16.79            & 8.5\,$\pm$\,0.5    & 3.98\,$\pm$\,0.95  &4.19\,$\pm$\,0.66   &10.08\,$\pm$\,2.05    & 8.50\,$\pm$\,0.30    &0.45\,$\pm$\,0.08     &8.35\,$\pm$\,0.25    & 0.10\,$\pm$\,0.23      & -1.08\,$\pm$\,0.15                                           & -1.54\,$\pm$\,0.17                   & 200\,$\pm$\,48       \\ 
NGC\,1817            & 05:12:38                        &  16:41:38                       & 186.19             & -13.03            & 9.5\,$\pm$\,0.5    & 5.35\,$\pm$\,1.14  &6.40\,$\pm$\,1.04   &17.00\,$\pm$\,3.89    &10.95\,$\pm$\,0.20    &0.23\,$\pm$\,0.05     &9.05\,$\pm$\,0.10    &-0.06\,$\pm$\,0.13      &  0.43\,$\pm$\,0.03                                           & -0.93\,$\pm$\,0.07                   & 528\,$\pm$\,131      \\ 
M\,37                & 05:52:20                        &  32:32:38                       & 177.64             &   3.09            & 9.2\,$\pm$\,0.5    & 1.92\,$\pm$\,0.14  &3.80\,$\pm$\,0.49   &15.28\,$\pm$\,3.32    &10.39\,$\pm$\,0.30    &0.26\,$\pm$\,0.05     &8.85\,$\pm$\,0.10    & 0.00\,$\pm$\,0.23      &  1.88\,$\pm$\,0.14                                           & -5.62\,$\pm$\,0.14                   & 326\,$\pm$\,64       \\ 
NGC\,2141            & 06:02:54                        &  10:27:08                       & 198.04             &  -5.80            &11.7\,$\pm$\,0.8    & 2.83\,$\pm$\,0.28  &6.42\,$\pm$\,1.18   &29.08\,$\pm$\,9.44    &12.95\,$\pm$\,0.35    &0.29\,$\pm$\,0.07     &9.40\,$\pm$\,0.20    & 0.22\,$\pm$\,0.17      & -0.09\,$\pm$\,0.13                                           & -0.75\,$\pm$\,0.11                   & 942\,$\pm$\,262      \\ 
NGC\,2168            & 06:09:16                        &  24:20:10                       & 186.61             &   2.23            & 8.8\,$\pm$\,0.5    & 3.04\,$\pm$\,0.38  &4.70\,$\pm$\,0.70   &15.35\,$\pm$\,3.67    & 9.54\,$\pm$\,0.30    &0.28\,$\pm$\,0.07     &8.20\,$\pm$\,0.15    & 0.05\,$\pm$\,0.15      &  2.27\,$\pm$\,0.15                                           & -2.90\,$\pm$\,0.15                   & 315\,$\pm$\,71       \\ 
NGC\,2204            & 06:15:35                        & -18:39:09                       & 226.02             & -16.11            &11.1\,$\pm$\,0.5    & 4.81\,$\pm$\,0.82  &9.81\,$\pm$\,1.92   &40.39\,$\pm$\,13.08   &13.03\,$\pm$\,0.15    &0.09\,$\pm$\,0.05     &9.30\,$\pm$\,0.05    &-0.22\,$\pm$\,0.19      & -0.57\,$\pm$\,0.06                                           &  1.95\,$\pm$\,0.05                   &1420\,$\pm$\,420      \\ 
NGC\,2243            & 06:29:32                        & -31:17:06                       & 239.48             & -18.01            &10.4\,$\pm$\,0.5    & 1.88\,$\pm$\,0.31  &5.38\,$\pm$\,0.93   &30.19\,$\pm$\,8.47    &12.90\,$\pm$\,0.20    &0.10\,$\pm$\,0.05     &9.55\,$\pm$\,0.10    &-0.47\,$\pm$\,0.33      & -1.26\,$\pm$\,0.03                                           &  5.50\,$\pm$\,0.04                   & 581\,$\pm$\,153      \\ 
Collinder\,110       & 06:38:50                        &  02:03:55                       & 209.64             &  -1.89            & 9.8\,$\pm$\,0.6    & 5.43\,$\pm$\,0.82  &8.14\,$\pm$\,1.14   &25.88\,$\pm$\,5.43    &11.47\,$\pm$\,0.30    &0.52\,$\pm$\,0.10     &9.25\,$\pm$\,0.10    &-0.13\,$\pm$\,0.23      & -1.08\,$\pm$\,0.07                                           & -2.03\,$\pm$\,0.07                   &1206\,$\pm$\,257      \\ 
NGC\,2287            & 06:45:54                        & -20:43:25                       & 230.98             & -10.43            & 8.4\,$\pm$\,0.5    & 2.40\,$\pm$\,0.21  &4.79\,$\pm$\,1.11   &19.35\,$\pm$\,8.08    & 9.10\,$\pm$\,0.30    &0.06\,$\pm$\,0.07     &8.35\,$\pm$\,0.15    & 0.00\,$\pm$\,0.23      & -4.37\,$\pm$\,0.15                                           & -1.36\,$\pm$\,0.15                   & 234\,$\pm$\,82       \\ 
NGC\,2323            & 07:02:51                        & -08:22:08                       & 221.64             &  -1.29            & 8.7\,$\pm$\,0.5    & 1.61\,$\pm$\,0.14  &2.81\,$\pm$\,0.58   &10.13\,$\pm$\,3.73    & 9.65\,$\pm$\,0.30    &0.23\,$\pm$\,0.08     &8.25\,$\pm$\,0.20    & 0.00\,$\pm$\,0.23      & -0.72\,$\pm$\,0.23                                           & -0.61\,$\pm$\,0.16                   & 131\,$\pm$\,41       \\ 
NGC\,2353            & 07:14:38                        & -10:16:34                       & 224.68             &   0.40            & 8.8\,$\pm$\,0.5    & 3.44\,$\pm$\,0.73  &3.61\,$\pm$\,0.56   & 8.69\,$\pm$\,1.86    &10.18\,$\pm$\,0.30    &0.17\,$\pm$\,0.07     &8.15\,$\pm$\,0.30    & 0.00\,$\pm$\,0.23      & -1.10\,$\pm$\,0.08                                           &  0.79\,$\pm$\,0.07                   & 131\,$\pm$\,32       \\ 
Berkeley\,36         & 07:16:24                        & -13:11:07                       & 227.50             &  -0.56            &11.6\,$\pm$\,0.7    & 2.39\,$\pm$\,0.42  &5.59\,$\pm$\,1.31   &25.93\,$\pm$\,10.49   &13.30\,$\pm$\,0.40    &0.63\,$\pm$\,0.01     &9.65\,$\pm$\,0.15    &-0.32\,$\pm$\,0.36      & -1.72\,$\pm$\,0.10                                           &  0.95\,$\pm$\,0.09                   & 705\,$\pm$\,252      \\ 
NGC\,2360            & 07:17:51                        & -15:37:46                       & 229.80             &  -1.41            & 8.6\,$\pm$\,0.5    & 1.61\,$\pm$\,0.18  &3.11\,$\pm$\,0.59   &12.20\,$\pm$\,4.04    & 9.90\,$\pm$\,0.20    &0.12\,$\pm$\,0.05     &9.15\,$\pm$\,0.10    &-0.22\,$\pm$\,0.19      &  0.39\,$\pm$\,0.13                                           &  5.63\,$\pm$\,0.12                   & 168\,$\pm$\,49       \\ 
Haffner\,11          & 07:35:22                        & -27:42:00                       & 242.39             &  -3.54            &11.1\,$\pm$\,0.6    & 3.76\,$\pm$\,0.63  &4.29\,$\pm$\,0.52   &11.00\,$\pm$\,1.78    &13.40\,$\pm$\,0.30    &0.59\,$\pm$\,0.05     &8.95\,$\pm$\,0.05    &-0.06\,$\pm$\,0.20      & -1.50\,$\pm$\,0.05                                           &  3.16\,$\pm$\,0.09                   & 302\,$\pm$\,61       \\ 
NGC\,2422            & 07:36:28                        & -14:29:29                       & 230.96             &   3.13            & 8.3\,$\pm$\,0.5    & 1.42\,$\pm$\,0.26  &2.53\,$\pm$\,0.56   & 9.26\,$\pm$\,3.46    & 8.18\,$\pm$\,0.25    &0.09\,$\pm$\,0.08     &8.05\,$\pm$\,0.25    & 0.10\,$\pm$\,0.18      & -7.03\,$\pm$\,0.23                                           &  1.02\,$\pm$\,0.20                   &  74\,$\pm$\,25       \\ 
Melotte\,71          & 07:37:35                        & -12:04:00                       & 228.95             &   4.51            & 9.4\,$\pm$\,0.5    & 2.07\,$\pm$\,0.34  &4.35\,$\pm$\,0.75   &18.40\,$\pm$\,5.12    &11.42\,$\pm$\,0.25    &0.18\,$\pm$\,0.05     &9.10\,$\pm$\,0.05    &-0.13\,$\pm$\,0.16      & -2.38\,$\pm$\,0.05                                           &  4.21\,$\pm$\,0.06                   & 338\,$\pm$\,88       \\ 
NGC\,2432            & 07:40:54                        & -19:05:06                       & 235.47             &   1.78            & 9.0\,$\pm$\,0.5    & 2.02\,$\pm$\,0.51  &1.89\,$\pm$\,0.24   & 4.18\,$\pm$\,0.43    &10.99\,$\pm$\,0.25    &0.23\,$\pm$\,0.07     &8.95\,$\pm$\,0.10    & 0.00\,$\pm$\,0.23      & -0.69\,$\pm$\,0.07                                           &  1.62\,$\pm$\,0.04                   &  54\,$\pm$\,11       \\ 
NGC\,2477            & 07:52:11                        & -38:33:38                       & 253.57             &  -5.84            & 8.5\,$\pm$\,0.5    & 2.63\,$\pm$\,0.37  &4.83\,$\pm$\,0.69   &18.13\,$\pm$\,3.97    &10.59\,$\pm$\,0.30    &0.40\,$\pm$\,0.10     &9.05\,$\pm$\,0.10    &-0.13\,$\pm$\,0.23      & -2.43\,$\pm$\,0.15                                           &  0.91\,$\pm$\,0.15                   & 605\,$\pm$\,129      \\ 
NGC\,2516            & 07:57:48                        & -60:50:08                       & 273.86             & -15.87            & 8.0\,$\pm$\,0.5    & 2.77\,$\pm$\,0.39  &3.87\,$\pm$\,0.55   &11.61\,$\pm$\,2.51    & 7.90\,$\pm$\,0.30    &0.12\,$\pm$\,0.10     &8.20\,$\pm$\,0.15    & 0.05\,$\pm$\,0.20      & -4.64\,$\pm$\,0.47                                           & 11.22\,$\pm$\,0.38                   & 231\,$\pm$\,49       \\ 
NGC\,2539            & 08:10:42                        & -12:50:40                       & 233.72             &  11.11            & 8.7\,$\pm$\,0.5    & 2.52\,$\pm$\,0.38  &4.68\,$\pm$\,0.94   &17.72\,$\pm$\,6.03    &10.23\,$\pm$\,0.30    &0.08\,$\pm$\,0.01     &8.85\,$\pm$\,0.15    &-0.06\,$\pm$\,0.26      & -2.33\,$\pm$\,0.07                                           & -0.54\,$\pm$\,0.07                   & 276\,$\pm$\,85       \\ 
Haffner\,22          & 08:12:24                        & -27:55:12                       & 246.78             &   3.37            & 9.4\,$\pm$\,0.5    & 2.91\,$\pm$\,0.35  &6.09\,$\pm$\,1.30   &25.64\,$\pm$\,9.71    &12.16\,$\pm$\,0.40    &0.21\,$\pm$\,0.07     &9.35\,$\pm$\,0.10    &-0.13\,$\pm$\,0.23      & -1.63\,$\pm$\,0.08                                           &  2.90\,$\pm$\,0.07                   & 579\,$\pm$\,188      \\ 
NGC\,2660            & 08:42:40                        & -47:12:25                       & 265.93             &  -3.01            & 8.5\,$\pm$\,0.5    & 0.86\,$\pm$\,0.21  &2.04\,$\pm$\,0.35   & 9.61\,$\pm$\,2.13    &11.96\,$\pm$\,0.25    &0.48\,$\pm$\,0.04     &9.15\,$\pm$\,0.05    &-0.22\,$\pm$\,0.28      & -2.73\,$\pm$\,0.06                                           &  5.21\,$\pm$\,0.03                   & 106\,$\pm$\,28       \\ 
M\,67                & 08:51:29                        &  11:50:14                       & 215.69             &  31.92            & 8.5\,$\pm$\,0.5    & 1.57\,$\pm$\,0.20  &2.73\,$\pm$\,0.43   & 9.81\,$\pm$\,2.53    & 9.35\,$\pm$\,0.20    &0.05\,$\pm$\,0.05     &9.65\,$\pm$\,0.10    & 0.00\,$\pm$\,0.11      &-10.96\,$\pm$\,0.17                                           & -2.91\,$\pm$\,0.18                   & 184\,$\pm$\,44       \\ 
NGC\,3114            & 10:02:00                        & -60:01:59                       & 283.25             &  -3.81            & 7.8\,$\pm$\,0.5    & 4.70\,$\pm$\,0.88  &5.80\,$\pm$\,1.04   &15.77\,$\pm$\,4.35    & 9.85\,$\pm$\,0.25    &0.11\,$\pm$\,0.06     &8.10\,$\pm$\,0.15    & 0.10\,$\pm$\,0.18      & -7.37\,$\pm$\,0.14                                           &  3.82\,$\pm$\,0.16                   & 443\,$\pm$\,120      \\ 
IC\,2714             & 11:17:24                        & -62:44:49                       & 292.40             &  -1.78            & 7.6\,$\pm$\,0.5    & 2.36\,$\pm$\,0.25  &3.32\,$\pm$\,0.48   &10.01\,$\pm$\,2.38    &10.29\,$\pm$\,0.30    &0.40\,$\pm$\,0.05     &8.80\,$\pm$\,0.15    &-0.13\,$\pm$\,0.23      & -7.59\,$\pm$\,0.08                                           &  2.69\,$\pm$\,0.12                   & 199\,$\pm$\,44       \\ 
Melotte\,105         & 11:19:39                        & -63:28:55                       & 292.90             &  -2.42            & 7.5\,$\pm$\,0.5    & 1.13\,$\pm$\,0.11  &2.04\,$\pm$\,0.38   & 7.57\,$\pm$\,2.46    &11.33\,$\pm$\,0.30    &0.49\,$\pm$\,0.10     &8.60\,$\pm$\,0.15    &-0.13\,$\pm$\,0.23      & -6.77\,$\pm$\,0.09                                           &  2.18\,$\pm$\,0.09                   &  98\,$\pm$\,28       \\ 
NGC\,3766            & 11:36:15                        & -61:37:01                       & 294.12             &  -0.04            & 7.5\,$\pm$\,0.5    & 2.37\,$\pm$\,0.24  &3.66\,$\pm$\,0.45   &11.92\,$\pm$\,2.30    &11.10\,$\pm$\,0.30    &0.26\,$\pm$\,0.10     &7.55\,$\pm$\,0.20    & 0.10\,$\pm$\,0.18      & -6.72\,$\pm$\,0.09                                           &  0.99\,$\pm$\,0.10                   & 254\,$\pm$\,47       \\ 
NGC\,3960            & 11:50:35                        & -55:40:10                       & 294.37             &   6.18            & 7.4\,$\pm$\,0.5    & 2.24\,$\pm$\,0.47  &3.55\,$\pm$\,0.59   &11.79\,$\pm$\,2.79    &11.53\,$\pm$\,0.25    &0.33\,$\pm$\,0.06     &9.00\,$\pm$\,0.10    & 0.00\,$\pm$\,0.17      & -6.52\,$\pm$\,0.07                                           &  1.88\,$\pm$\,0.07                   & 218\,$\pm$\,55       \\ 
Juchert\,13          & 12:01:35                        & -64:05:39                       & 297.53             &  -1.75            & 7.2\,$\pm$\,0.5    & 1.84\,$\pm$\,0.14  &3.15\,$\pm$\,0.65   &11.14\,$\pm$\,4.06    &12.20\,$\pm$\,0.40    &0.83\,$\pm$\,0.10     &9.15\,$\pm$\,0.15    &-0.06\,$\pm$\,0.26      & -8.16\,$\pm$\,0.05                                           &  0.29\,$\pm$\,0.06                   & 207\,$\pm$\,65       \\ 
NGC\,4052            & 12:01:53                        & -63:13:34                       & 297.38             &  -0.90            & 7.2\,$\pm$\,0.5    & 3.31\,$\pm$\,0.52  &3.20\,$\pm$\,0.31   & 7.23\,$\pm$\,0.78    &11.83\,$\pm$\,0.30    &0.33\,$\pm$\,0.07     &8.65\,$\pm$\,0.15    & 0.00\,$\pm$\,0.23      & -6.81\,$\pm$\,0.06                                           &  0.11\,$\pm$\,0.06                   & 144\,$\pm$\,24       \\ 
Collinder\,261       & 12:38:03                        & -68:22:40                       & 301.70             &  -5.54            & 7.0\,$\pm$\,0.5    & 3.03\,$\pm$\,0.56  &4.85\,$\pm$\,0.59   &16.22\,$\pm$\,2.33    &12.08\,$\pm$\,0.30    &0.36\,$\pm$\,0.07     &9.75\,$\pm$\,0.15    &-0.06\,$\pm$\,0.20      & -6.37\,$\pm$\,0.11                                           & -2.68\,$\pm$\,0.11                   & 863\,$\pm$\,159      \\ 
NGC\,4815            & 12:57:56                        & -64:57:19                       & 303.63             &  -2.10            & 7.0\,$\pm$\,0.5    & 1.59\,$\pm$\,0.26  &2.47\,$\pm$\,0.42   & 8.08\,$\pm$\,2.16    &11.70\,$\pm$\,0.40    &0.75\,$\pm$\,0.10     &8.80\,$\pm$\,0.15    &-0.06\,$\pm$\,0.26      & -5.83\,$\pm$\,0.15                                           & -0.89\,$\pm$\,0.16                   & 191\,$\pm$\,49       \\ 
NGC\,5316            & 13:54:02                        & -61:50:38                       & 310.24             &   0.10            & 7.3\,$\pm$\,0.5    & 1.99\,$\pm$\,0.37  &1.90\,$\pm$\,0.25   & 4.26\,$\pm$\,0.72    &10.44\,$\pm$\,0.30    &0.35\,$\pm$\,0.10     &8.20\,$\pm$\,0.10    & 0.05\,$\pm$\,0.15      & -6.33\,$\pm$\,0.10                                           & -1.50\,$\pm$\,0.09                   &  54\,$\pm$\,11       \\ 
NGC\,5715            & 14:43:25                        & -57:35:07                       & 317.52             &   2.09            & 7.0\,$\pm$\,0.5    & 0.99\,$\pm$\,0.11  &1.52\,$\pm$\,0.33   & 4.95\,$\pm$\,1.87    &10.85\,$\pm$\,0.30    &0.57\,$\pm$\,0.10     &8.90\,$\pm$\,0.10    & 0.05\,$\pm$\,0.20      & -3.45\,$\pm$\,0.08                                           & -2.31\,$\pm$\,0.08                   &  46\,$\pm$\,16       \\ 
NGC\,6124            & 16:25:16                        & -40:38:40                       & 340.73             &   6.01            & 7.5\,$\pm$\,0.5    & 1.33\,$\pm$\,0.20  &2.08\,$\pm$\,0.33   & 6.82\,$\pm$\,1.72    & 8.75\,$\pm$\,0.30    &0.82\,$\pm$\,0.12     &8.10\,$\pm$\,0.15    & 0.10\,$\pm$\,0.18      & -0.22\,$\pm$\,0.27                                           & -2.11\,$\pm$\,0.28                   &  89\,$\pm$\,21       \\ 
NGC\,6134            & 16:27:47                        & -49:08:54                       & 334.92             &  -0.21            & 7.2\,$\pm$\,0.5    & 0.99\,$\pm$\,0.20  &2.47\,$\pm$\,0.59   &12.22\,$\pm$\,5.02    & 9.87\,$\pm$\,0.20    &0.40\,$\pm$\,0.04     &9.15\,$\pm$\,0.05    &-0.06\,$\pm$\,0.13      &  2.15\,$\pm$\,0.14                                           & -4.44\,$\pm$\,0.14                   & 128\,$\pm$\,46       \\ 
NGC\,6192            & 16:40:17                        & -43:22:04                       & 340.65             &   2.14            & 6.6\,$\pm$\,0.5    & 1.54\,$\pm$\,0.28  &3.25\,$\pm$\,0.61   &13.78\,$\pm$\,4.18    &10.90\,$\pm$\,0.30    &0.74\,$\pm$\,0.10     &8.20\,$\pm$\,0.20    & 0.05\,$\pm$\,0.25      &  1.63\,$\pm$\,0.10                                           & -0.21\,$\pm$\,0.10                   & 184\,$\pm$\,53       \\ 
NGC\,6242            & 16:55:36                        & -39:28:43                       & 345.45             &   2.46            & 6.9\,$\pm$\,0.5    & 1.86\,$\pm$\,0.38  &2.48\,$\pm$\,0.47   & 7.15\,$\pm$\,2.12    &10.25\,$\pm$\,0.25    &0.49\,$\pm$\,0.07     &7.70\,$\pm$\,0.15    & 0.00\,$\pm$\,0.23      &  1.11\,$\pm$\,0.12                                           & -0.88\,$\pm$\,0.15                   &  98\,$\pm$\,29       \\ 
NGC\,6253            & 16:59:04                        & -52:41:53                       & 335.46             &  -6.26            & 6.7\,$\pm$\,0.5    & 1.40\,$\pm$\,0.21  &1.95\,$\pm$\,0.29   & 5.86\,$\pm$\,1.36    &10.82\,$\pm$\,0.30    &0.27\,$\pm$\,0.07     &9.55\,$\pm$\,0.10    & 0.22\,$\pm$\,0.17      & -4.56\,$\pm$\,0.11                                           & -5.29\,$\pm$\,0.11                   & 119\,$\pm$\,28       \\ 
IC\,4651             & 17:24:54                        & -49:57:42                       & 340.10             &  -7.90            & 7.2\,$\pm$\,0.5    & 1.93\,$\pm$\,0.23  &2.61\,$\pm$\,0.49   & 7.62\,$\pm$\,2.43    & 9.62\,$\pm$\,0.25    &0.13\,$\pm$\,0.06     &9.30\,$\pm$\,0.10    & 0.05\,$\pm$\,0.20      & -2.43\,$\pm$\,0.16                                           & -5.05\,$\pm$\,0.15                   & 145\,$\pm$\,42       \\ 
Dias\,6              & 18:30:29                        & -12:19:34                       &  19.59             &  -1.03            & 6.2\,$\pm$\,0.6    & 0.76\,$\pm$\,0.12  &1.36\,$\pm$\,0.27   & 5.05\,$\pm$\,1.65    &11.50\,$\pm$\,0.30    &0.99\,$\pm$\,0.10     &8.90\,$\pm$\,0.10    &-0.06\,$\pm$\,0.26      &  0.52\,$\pm$\,0.10                                           & -0.58\,$\pm$\,0.08                   &  50\,$\pm$\,16       \\ 
Ruprecht\,171        & 18:32:03                        & -16:02:45                       &  16.45             &  -3.09            & 6.7\,$\pm$\,0.5    & 1.85\,$\pm$\,0.40  &3.86\,$\pm$\,0.55   &16.26\,$\pm$\,2.70    &10.70\,$\pm$\,0.20    &0.30\,$\pm$\,0.04     &9.50\,$\pm$\,0.05    & 0.00\,$\pm$\,0.17      &  7.71\,$\pm$\,0.09                                           &  1.08\,$\pm$\,0.09                   & 310\,$\pm$\,68       \\ 
Czernik\,38          & 18:49:46                        &  04:57:33                       &  37.17             &   2.62            & 6.8\,$\pm$\,0.5    & 1.08\,$\pm$\,0.10  &2.07\,$\pm$\,0.44   & 8.07\,$\pm$\,3.06    &11.05\,$\pm$\,0.40    &1.74\,$\pm$\,0.10     &8.45\,$\pm$\,0.15    & 0.10\,$\pm$\,0.18      & -1.80\,$\pm$\,0.08                                           & -5.06\,$\pm$\,0.07                   &  94\,$\pm$\,31       \\ 
Berkeley\,81         & 19:01:39                        &  00:26:59                       &  33.70             &  -2.48            & 6.3\,$\pm$\,0.5    & 0.93\,$\pm$\,0.11  &1.67\,$\pm$\,0.23   & 6.16\,$\pm$\,1.38    &11.65\,$\pm$\,0.30    &1.08\,$\pm$\,0.06     &9.10\,$\pm$\,0.10    &-0.22\,$\pm$\,0.28      & -1.13\,$\pm$\,0.12                                           & -1.94\,$\pm$\,0.11                   &  66\,$\pm$\,15       \\ 
NGC\,6791            & 19:20:52                        &  37:46:47                       &  69.96             &  10.91            & 7.7\,$\pm$\,0.5    & 3.74\,$\pm$\,0.60  &6.35\,$\pm$\,1.06   &22.33\,$\pm$\,5.94    &13.09\,$\pm$\,0.20    &0.13\,$\pm$\,0.04     &9.90\,$\pm$\,0.15    & 0.41\,$\pm$\,0.11      & -0.42\,$\pm$\,0.06                                           & -2.28\,$\pm$\,0.04                   &1521\,$\pm$\,384      \\ 
NGC\,6802            & 19:30:37                        &  20:15:14                       &  55.33             &   0.91            & 7.1\,$\pm$\,0.5    & 1.09\,$\pm$\,0.18  &3.03\,$\pm$\,0.73   &16.61\,$\pm$\,7.11    &11.25\,$\pm$\,0.30    &0.88\,$\pm$\,0.06     &9.00\,$\pm$\,0.10    &-0.06\,$\pm$\,0.20      & -2.84\,$\pm$\,0.03                                           & -6.43\,$\pm$\,0.11                   & 231\,$\pm$\,84       \\ 
NGC\,6811            & 19:37:23                        &  46:22:23                       &  79.21             &  12.00            & 7.9\,$\pm$\,0.5    & 1.57\,$\pm$\,0.23  &2.54\,$\pm$\,0.40   & 8.55\,$\pm$\,2.15    &10.00\,$\pm$\,0.20    &0.06\,$\pm$\,0.05     &9.05\,$\pm$\,0.10    & 0.00\,$\pm$\,0.11      & -3.35\,$\pm$\,0.08                                           & -8.79\,$\pm$\,0.08                   &  91\,$\pm$\,23       \\ 
NGC\,6819            & 19:41:21                        &  40:12:27                       &  73.98             &   8.48            & 7.7\,$\pm$\,0.5    & 2.44\,$\pm$\,0.42  &4.51\,$\pm$\,0.73   &17.03\,$\pm$\,4.25    &11.90\,$\pm$\,0.20    &0.16\,$\pm$\,0.05     &9.40\,$\pm$\,0.10    & 0.10\,$\pm$\,0.18      & -2.89\,$\pm$\,0.08                                           & -3.87\,$\pm$\,0.09                   & 555\,$\pm$\,136      \\ 
NGC\,6866$^a$        & 20:03:54                        &  44:08:59                       &  79.56             &   6.84            & 7.9\,$\pm$\,0.5    & 0.90\,$\pm$\,0.21  &2.02\,$\pm$\,0.52   & 9.05\,$\pm$\,3.94    &10.45\,$\pm$\,0.30    &0.16\,$\pm$\,0.07     &8.90\,$\pm$\,0.10    & 0.05\,$\pm$\,0.20      & -1.38\,$\pm$\,0.07                                           & -5.77\,$\pm$\,0.06                   &  61\,$\pm$\,24       \\ 
Berkeley\,89         & 20:24:29                        &  46:02:12                       &  83.14             &   4.84            & 8.2\,$\pm$\,0.5    & 2.27\,$\pm$\,0.41  &3.67\,$\pm$\,0.69   &12.42\,$\pm$\,3.75    &12.30\,$\pm$\,0.30    &0.70\,$\pm$\,0.05     &9.30\,$\pm$\,0.10    & 0.00\,$\pm$\,0.23      & -1.99\,$\pm$\,0.04                                           & -2.32\,$\pm$\,0.03                   & 251\,$\pm$\,73       \\ 
NGC\,7044            & 21:13:08                        &  42:30:14                       &  85.89             &  -4.15            & 8.3\,$\pm$\,0.5    & 1.46\,$\pm$\,0.28  &3.88\,$\pm$\,0.82   &20.35\,$\pm$\,7.12    &12.35\,$\pm$\,0.30    &0.76\,$\pm$\,0.06     &9.20\,$\pm$\,0.10    & 0.00\,$\pm$\,0.23      & -4.94\,$\pm$\,0.07                                           & -5.57\,$\pm$\,0.07                   & 403\,$\pm$\,128      \\ 
NGC\,7142            & 21:45:08                        &  65:46:19                       & 105.35             &   9.48            & 8.7\,$\pm$\,0.5    & 1.65\,$\pm$\,0.21  &3.83\,$\pm$\,0.90   &17.75\,$\pm$\,7.49    &11.45\,$\pm$\,0.30    &0.41\,$\pm$\,0.05     &9.55\,$\pm$\,0.15    & 0.05\,$\pm$\,0.15      & -2.68\,$\pm$\,0.06                                           & -1.35\,$\pm$\,0.06                   & 302\,$\pm$\,108      \\ 
NGC\,7654            & 23:24:44                        &  61:34:48                       & 112.82             &   0.43            & 8.6\,$\pm$\,0.5    & 2.02\,$\pm$\,0.58  &2.29\,$\pm$\,0.34   & 5.83\,$\pm$\,0.80    &10.63\,$\pm$\,0.30    &0.74\,$\pm$\,0.08     &7.80\,$\pm$\,0.15    & 0.00\,$\pm$\,0.23      & -1.90\,$\pm$\,0.12                                           & -1.19\,$\pm$\,0.13                   & 118\,$\pm$\,27       \\ 
NGC\,7789            & 23:57:33                        &  56:43:27                       & 115.53             &  -5.37            & 9.0\,$\pm$\,0.5    & 3.63\,$\pm$\,0.65  &7.53\,$\pm$\,1.32   &31.47\,$\pm$\,8.67    &11.35\,$\pm$\,0.30    &0.30\,$\pm$\,0.07     &9.20\,$\pm$\,0.10    & 0.00\,$\pm$\,0.17      & -0.92\,$\pm$\,0.12                                           & -1.95\,$\pm$\,0.13                   &1436\,$\pm$\,378      \\ 
\hline

\end{tabular}                                                                                                                                                                                                                                                                                                                                                                                                                                                                     
                                                                                                                                                                                                                                                                                                                                                                                                                                                                                  
$^{(*)}$  For the Sun, it is assumed $R_{G,\odot}$\,=\,8.0\,$\pm$\,0.5\,kpc \citep{Reid:1993} \\                                                                                                                                                                                                                                                                                                                                                                                                                  
$^{(**)}$ Obtained from isochrone fits, using as initial guesses the mean spectroscopic $[Fe/H]$ values for member stars (See Sections \ref{sec:astrometric_spec_params} and \ref{sec:isoc_fit} for details). \\                                                                                                                                                                                                                                                                                                                                                                                                                 
$^{\dag}$ The numbers after the ``$\pm$"\,signal correspond to the intrinsic (i.e., corrected for measurement uncertainties) dispersions derived from the member stars data. \\ 
$^{\dag\dag}$ Equation~\ref{eq:trh} (Section~\ref{sec:analysis}). \\ 
$^{a}$ For NGC\,6866, we have assumed $r_t\,\sim\,R_{\textrm{lim}}$, due to large density fluctuations in this cluster external structure (see Section~\ref{sec:struct_analysis} and Figure\,B5 of the online Supplementary Material). \\

\end{minipage}
\end{sideways}
\label{tab:investig_sample}
\end{table*}

\begin{table*}
\centering
\small
\caption{ Total masses and number of stars, Jacobi radii, initial mass estimates, contrast parameters and stellar densities for the studied sample. }
\begin{minipage}{160mm}

\begin{tabular}{lrrrrrrr}

 Cluster             & $M_{\textrm{clu}}$           & $N_{\textrm{clu}}$        & $R_J^{\dag}$        & $M_{\textrm{ini}}^{\dag}$        &      $\delta_c$         & $\sigma_0$                     & $\sigma_{\textrm{bg}}$  \\    
                     & ($\times\,10^3\,M_{\odot}$)  &    ($\times10^2$)         & (pc)                & ($\times\,10^3\,M_{\odot}$)      &                         & (stars/$\textrm{arcmin}^2$)    & (stars/$\textrm{arcmin}^2$)  \\     
                                                                                                                                                                                     
\hline                                                                                                                                                                       
NGC\,129             & 2.2 \,$\pm$\,0.1             & 50\,$\pm$\, 2             &12.4\,$\pm$\, 0.9    & 2.9\,$\pm$\, 0.2                &   4.18\,$\pm$\,0.70     &   2.42\,$\pm$\,0.51            &    0.76\,$\pm$\,0.05    \\ 
NGC\,188             & 5.5 \,$\pm$\,0.2             &158\,$\pm$\, 6             &24.6\,$\pm$\, 1.8    &25.6\,$\pm$\, 4.7                &  44.71\,$\pm$\,11.23    &   6.57\,$\pm$\,1.17            &    0.15\,$\pm$\,0.03    \\ 
NGC\,559             & 3.2 \,$\pm$\,0.1             & 81\,$\pm$\, 3             &13.6\,$\pm$\, 2.7    & 7.7\,$\pm$\, 0.6                &  25.64\,$\pm$\,2.51     &  12.92\,$\pm$\,1.04            &    0.52\,$\pm$\,0.03    \\ 
NGC\,654             & 3.3 \,$\pm$\,0.2             & 67\,$\pm$\, 5             &14.5\,$\pm$\, 1.1    & 3.6\,$\pm$\, 0.1                &  13.72\,$\pm$\,2.20     &  15.35\,$\pm$\,2.30            &    1.21\,$\pm$\,0.10    \\ 
NGC\,752             & 0.62\,$\pm$\,0.05            & 16\,$\pm$\, 2             & 8.3\,$\pm$\, 0.8    & 5.3\,$\pm$\, 0.5                &   9.37\,$\pm$\,2.34     &   0.11\,$\pm$\,0.03            &   0.013\,$\pm$\,0.001   \\ 
NGC\,1027            & 1.32\,$\pm$\,0.04            & 33\,$\pm$\, 1             & 9.9\,$\pm$\, 1.0    & 1.8\,$\pm$\, 0.1                &   3.77\,$\pm$\,0.41     &   1.27\,$\pm$\,0.19            &    0.46\,$\pm$\,0.01    \\ 
NGC\,1647            & 1.08\,$\pm$\,0.03            & 26\,$\pm$\, 1             & 9.8\,$\pm$\, 1.1    & 1.9\,$\pm$\, 0.2                &   5.62\,$\pm$\,1.09     &   0.31\,$\pm$\,0.07            &    0.07\,$\pm$\,0.01    \\ 
NGC\,1817            & 2.3 \,$\pm$\,0.1             & 59\,$\pm$\, 2             &16.2\,$\pm$\, 2.9    & 5.7\,$\pm$\, 0.5                &   4.95\,$\pm$\,0.46     &   1.31\,$\pm$\,0.15            &    0.33\,$\pm$\,0.01    \\ 
M\,37                & 5.3 \,$\pm$\,0.1             &135\,$\pm$\, 3             &16.5\,$\pm$\, 2.1    &10.3\,$\pm$\, 0.6                &  10.93\,$\pm$\,3.20     &   8.76\,$\pm$\,1.14            &    0.44\,$\pm$\,0.04    \\ 
NGC\,2141            & 9.2 \,$\pm$\,0.2             &250\,$\pm$\, 8             &31.0\,$\pm$\, 3.3    &19.5\,$\pm$\, 2.3                &  14.49\,$\pm$\,1.70     &  27.56\,$\pm$\,3.07            &    2.04\,$\pm$\,0.12    \\ 
NGC\,2168            & 2.5 \,$\pm$\,0.1             & 58\,$\pm$\, 2             &12.4\,$\pm$\, 1.1    & 3.5\,$\pm$\, 0.1                &   8.10\,$\pm$\,0.85     &   1.36\,$\pm$\,0.14            &    0.19\,$\pm$\,0.01    \\ 
NGC\,2204            & 5.1 \,$\pm$\,0.1             &139\,$\pm$\, 5             &28.2\,$\pm$\, 3.2    &10.1\,$\pm$\, 0.3                &  15.62\,$\pm$\,1.80     &   6.26\,$\pm$\,0.73            &    0.43\,$\pm$\,0.02    \\ 
NGC\,2243            & 4.6 \,$\pm$\,0.1             &132\,$\pm$\, 4             &25.9\,$\pm$\, 5.6    &12.9\,$\pm$\, 1.3                & 161.44\,$\pm$\,26.6     &  27.74\,$\pm$\,2.38            &    0.17\,$\pm$\,0.02    \\ 
Collinder\,110       & 7.0 \,$\pm$\,0.2             &190\,$\pm$\, 6             &19.0\,$\pm$\, 1.5    &19.0\,$\pm$\, 1.6                &   5.61\,$\pm$\,0.62     &   3.80\,$\pm$\,0.47            &    0.82\,$\pm$\,0.04    \\ 
NGC\,2287            & 1.00\,$\pm$\,0.03            & 24\,$\pm$\, 1             & 9.4\,$\pm$\, 0.7    & 1.7\,$\pm$\, 0.1                &  24.77\,$\pm$\,5.95     &   0.65\,$\pm$\,0.13            &   0.027\,$\pm$\,0.004   \\ 
NGC\,2323            & 1.70\,$\pm$\,0.05            & 41\,$\pm$\, 1             &10.7\,$\pm$\, 0.9    & 2.6\,$\pm$\, 0.2                &   2.89\,$\pm$\,0.39     &   2.00\,$\pm$\,0.40            &    1.06\,$\pm$\,0.05    \\ 
NGC\,2353            & 0.58\,$\pm$\,0.03            & 14\,$\pm$\, 1             & 7.6\,$\pm$\, 0.6    & 1.0\,$\pm$\, 0.1                &   2.67\,$\pm$\,0.37     &   1.01\,$\pm$\,0.21            &    0.60\,$\pm$\,0.04    \\ 
Berkeley\,36         & 6.5 \,$\pm$\, 0.3            &188\,$\pm$\,10             &21.1\,$\pm$\, 1.9    &31.3\,$\pm$\, 6.3                &  10.62\,$\pm$\,1.50     &  18.74\,$\pm$\,2.57            &    1.95\,$\pm$\,0.14    \\ 
NGC\,2360            & 1.8 \,$\pm$\, 0.1            & 48\,$\pm$\, 2             &11.0\,$\pm$\, 0.8    & 8.2\,$\pm$\, 1.0                &  49.21\,$\pm$\,12.4     &   3.74\,$\pm$\,0.74            &    0.08\,$\pm$\,0.01    \\ 
Haffner\,11          & 2.4 \,$\pm$\, 0.1            & 65\,$\pm$\, 5             &17.8\,$\pm$\, 2.1    & 4.9\,$\pm$\, 0.3                &   5.74\,$\pm$\,1.33     &   9.81\,$\pm$\,2.60            &    2.07\,$\pm$\,0.20    \\ 
NGC\,2422            & 0.69\,$\pm$\,0.03            & 15\,$\pm$\, 0             & 7.8\,$\pm$\, 0.6    & 1.1\,$\pm$\, 0.1                &  16.22\,$\pm$\,3.97     &   0.62\,$\pm$\,0.11            &    0.04\,$\pm$\,0.01    \\ 
Melotte\,71          & 3.1 \,$\pm$\, 0.1            & 82\,$\pm$\, 2             &14.9\,$\pm$\, 1.8    & 8.8\,$\pm$\, 0.4                &  28.42\,$\pm$\,4.76     &   9.04\,$\pm$\,1.38            &    0.33\,$\pm$\,0.03    \\ 
NGC\,2432            & 0.69\,$\pm$\,0.04            & 18\,$\pm$\, 1             & 8.2\,$\pm$\, 0.8    & 3.3\,$\pm$\, 0.4                &   4.27\,$\pm$\,0.56     &   4.20\,$\pm$\,0.67            &    1.29\,$\pm$\,0.08    \\ 
NGC\,2477            & 9.8 \,$\pm$\, 0.2            &254\,$\pm$\, 5             &20.0\,$\pm$\, 2.8    &20.0\,$\pm$\, 1.2                &   8.17\,$\pm$\,0.84     &  10.31\,$\pm$\,1.11            &    1.44\,$\pm$\,0.07    \\ 
NGC\,2516            & 2.13\,$\pm$\,0.04            & 52\,$\pm$\, 1             &11.5\,$\pm$\, 0.9    & 3.1\,$\pm$\, 0.1                &   6.01\,$\pm$\,0.59     &   0.55\,$\pm$\,0.06            &    0.11\,$\pm$\,0.01    \\ 
NGC\,2539            & 1.4 \,$\pm$\, 0.1            & 36\,$\pm$\, 2             &11.7\,$\pm$\, 1.1    & 3.6\,$\pm$\, 0.4                &  11.48\,$\pm$\,2.41     &   1.74\,$\pm$\,0.39            &    0.17\,$\pm$\,0.01    \\ 
Haffner\,22          & 3.2 \,$\pm$\, 0.1            & 87\,$\pm$\, 4             &15.3\,$\pm$\, 1.6    &13.5\,$\pm$\, 1.6                &   3.27\,$\pm$\,0.51     &   4.02\,$\pm$\,0.91            &    1.78\,$\pm$\,0.05    \\ 
NGC\,2660            & 2.9 \,$\pm$\, 0.1            & 77\,$\pm$\, 3             &13.7\,$\pm$\, 1.1    & 9.7\,$\pm$\, 0.6                &  18.83\,$\pm$\,2.16     &  30.67\,$\pm$\,2.57            &    1.72\,$\pm$\,0.15    \\ 
M\,67                & 3.3 \,$\pm$\, 0.1            & 94\,$\pm$\, 4             &17.8\,$\pm$\, 2.0    &20.0\,$\pm$\, 2.9                &  32.23\,$\pm$\,11.5     &   3.16\,$\pm$\,0.82            &    0.10\,$\pm$\,0.03    \\ 
NGC\,3114            & 2.9 \,$\pm$\, 0.1            & 66\,$\pm$\, 2             &12.4\,$\pm$\, 1.0    & 3.9\,$\pm$\, 0.1                &   2.82\,$\pm$\,0.46     &   0.87\,$\pm$\,0.22            &    0.48\,$\pm$\,0.02    \\ 
IC\,2714             & 2.4 \,$\pm$\, 0.1            & 61\,$\pm$\, 2             &11.3\,$\pm$\, 0.9    & 5.8\,$\pm$\, 0.7                &   3.43\,$\pm$\,0.31     &   3.14\,$\pm$\,0.39            &    1.29\,$\pm$\,0.03    \\ 
Melotte\,105         & 2.6 \,$\pm$\, 0.1            & 65\,$\pm$\, 2             &11.5\,$\pm$\, 1.0    & 4.9\,$\pm$\, 0.4                &  11.14\,$\pm$\,1.27     &  19.77\,$\pm$\,1.52            &    1.95\,$\pm$\,0.19    \\ 
NGC\,3766            & 4.7 \,$\pm$\, 0.1            &101\,$\pm$\, 3             &14.0\,$\pm$\, 1.1    & 5.5\,$\pm$\, 0.1                &   5.43\,$\pm$\,0.6      &   7.46\,$\pm$\,0.98            &    1.68\,$\pm$\,0.08    \\ 
NGC\,3960            & 2.3 \,$\pm$\, 0.1            & 59\,$\pm$\, 2             &12.5\,$\pm$\, 1.2    & 6.5\,$\pm$\, 0.6                &   6.75\,$\pm$\,1.35     &   7.32\,$\pm$\,1.66            &    1.27\,$\pm$\,0.08    \\ 
Juchert\,13          & 3.1 \,$\pm$\, 0.1            & 82\,$\pm$\, 4             &11.9\,$\pm$\, 1.4    &12.4\,$\pm$\, 2.2                &   4.63\,$\pm$\,0.71     &  11.23\,$\pm$\,2.14            &    3.10\,$\pm$\,0.15    \\ 
NGC\,4052            & 1.3 \,$\pm$\, 0.1            & 32\,$\pm$\, 2             & 8.8\,$\pm$\, 0.8    & 3.3\,$\pm$\, 0.4                &   3.38\,$\pm$\,0.59     &   5.02\,$\pm$\,1.21            &    2.11\,$\pm$\,0.11    \\ 
Collinder\,261       &18.4 \,$\pm$\, 0.3            &533\,$\pm$\,11             &25.5\,$\pm$\, 2.8    &76.3\,$\pm$\,13.5                &  21.34\,$\pm$\,3.00     &  32.36\,$\pm$\,4.38            &    1.59\,$\pm$\,0.09    \\ 
NGC\,4815            & 6.6 \,$\pm$\, 0.1            &173\,$\pm$\, 5             &15.4\,$\pm$\, 1.5    &12.5\,$\pm$\, 1.0                &   3.25\,$\pm$\,0.19     &  20.51\,$\pm$\,1.57            &    9.09\,$\pm$\,0.35    \\ 
NGC\,5316            & 0.86\,$\pm$\,0.04            & 20\,$\pm$\, 1             & 7.8\,$\pm$\, 0.7    & 1.5\,$\pm$\, 0.1                &   1.65\,$\pm$\,0.30     &   0.97\,$\pm$\,0.45            &    1.49\,$\pm$\,0.05    \\ 
NGC\,5715            & 1.1 \,$\pm$\, 0.1            & 28\,$\pm$\, 2             & 8.3\,$\pm$\, 0.8    & 4.5\,$\pm$\, 0.6                &   6.18\,$\pm$\,1.84     &   5.69\,$\pm$\,1.96            &    1.10\,$\pm$\,0.09    \\ 
NGC\,6124            & 2.4 \,$\pm$\, 0.1            & 54\,$\pm$\, 2             &11.1\,$\pm$\, 0.9    & 3.3\,$\pm$\, 0.1                &   3.65\,$\pm$\,0.56     &   1.77\,$\pm$\,0.36            &    0.67\,$\pm$\,0.05    \\ 
NGC\,6134            & 2.3 \,$\pm$\, 0.1            & 59\,$\pm$\, 2             &10.5\,$\pm$\, 0.9    &10.8\,$\pm$\, 0.7                &  28.26\,$\pm$\,4.34     &   6.14\,$\pm$\,0.71            &    0.23\,$\pm$\,0.02    \\ 
NGC\,6192            & 2.7 \,$\pm$\, 0.1            & 62\,$\pm$\, 2             &10.5\,$\pm$\, 1.4    & 3.9\,$\pm$\, 0.3                &  23.10\,$\pm$\,3.05     &   8.30\,$\pm$\,0.83            &    0.38\,$\pm$\,0.04    \\ 
NGC\,6242            & 1.7 \,$\pm$\, 0.1            & 37\,$\pm$\, 2             & 9.4\,$\pm$\, 0.9    & 2.1\,$\pm$\, 0.1                &   8.10\,$\pm$\,1.93     &   5.30\,$\pm$\,1.39            &    0.75\,$\pm$\,0.05    \\ 
NGC\,6253            & 4.3 \,$\pm$\, 0.2            &117\,$\pm$\, 5             &13.5\,$\pm$\, 2.0    &34.0\,$\pm$\, 5.4                &   4.18\,$\pm$\,0.48     &  13.48\,$\pm$\,1.91            &    4.24\,$\pm$\,0.23    \\ 
IC\,4651             & 2.4 \,$\pm$\, 0.1            & 65\,$\pm$\, 3             &11.4\,$\pm$\, 1.0    &14.0\,$\pm$\, 2.0                &   2.92\,$\pm$\,0.46     &   2.14\,$\pm$\,0.51            &    1.12\,$\pm$\,0.03    \\ 
Dias\,6              & 2.2 \,$\pm$\, 0.1            & 56\,$\pm$\, 4             & 9.6\,$\pm$\, 1.2    & 7.3\,$\pm$\, 0.8                &   8.16\,$\pm$\,2.38     &  13.41\,$\pm$\,4.41            &    1.87\,$\pm$\,0.09    \\ 
Ruprecht\,171        & 3.5 \,$\pm$\, 0.1            & 98\,$\pm$\, 4             &11.6\,$\pm$\, 2.6    &35.1\,$\pm$\, 3.2                & 210.32\,$\pm$\,49.62    &   5.77\,$\pm$\,0.91            &    0.03\,$\pm$\,0.01    \\ 
Czernik\,38          & 2.5 \,$\pm$\, 0.1            & 60\,$\pm$\, 4             &10.6\,$\pm$\, 1.7    & 4.5\,$\pm$\, 0.3                &  16.71\,$\pm$\,3.23     &  10.17\,$\pm$\,1.86            &    0.65\,$\pm$\,0.06    \\ 
Berkeley\,81         & 1.9 \,$\pm$\, 0.1            & 49\,$\pm$\, 4             & 9.5\,$\pm$\, 1.2    & 9.3\,$\pm$\, 1.6                &   6.99\,$\pm$\,1.26     &  15.52\,$\pm$\,3.20            &    2.59\,$\pm$\,0.10    \\ 
NGC\,6791            &27.3 \,$\pm$\, 0.5            &795\,$\pm$\,16             &38.1\,$\pm$\,13.2    &86.5\,$\pm$\,13.1                &  55.18\,$\pm$\,5.60     &  48.21\,$\pm$\,3.16            &    0.89\,$\pm$\,0.07    \\ 
NGC\,6802            & 4.7 \,$\pm$\, 0.1            &124\,$\pm$\, 4             &13.3\,$\pm$\, 1.5    &12.6\,$\pm$\, 1.1                &   5.76\,$\pm$\,0.48     &  18.36\,$\pm$\,1.78            &    3.86\,$\pm$\,0.11    \\ 
NGC\,6811            & 0.82\,$\pm$\,0.04            & 21\,$\pm$\, 1             & 9.0\,$\pm$\, 1.3    & 4.3\,$\pm$\, 0.6                &  12.83\,$\pm$\,2.82     &   1.36\,$\pm$\,0.27            &    0.12\,$\pm$\,0.01    \\ 
NGC\,6819            & 9.0 \,$\pm$\, 0.2            &247\,$\pm$\, 6             &23.2\,$\pm$\, 2.1    &24.0\,$\pm$\, 2.0                &  18.48\,$\pm$\,1.62     &  20.74\,$\pm$\,1.76            &    1.19\,$\pm$\,0.04    \\ 
NGC\,6866            &0.73 \,$\pm$\,0.03            & 18\,$\pm$\, 1             & 8.1\,$\pm$\, 1.1    & 3.2\,$\pm$\, 0.4                &  10.84\,$\pm$\,2.40     &   2.45\,$\pm$\,0.53            &    0.25\,$\pm$\,0.03    \\ 
Berkeley\,89         & 2.6 \,$\pm$\, 0.1            & 71\,$\pm$\, 4             &14.0\,$\pm$\, 2.9    &11.8\,$\pm$\, 1.5                &  13.80\,$\pm$\,1.62     &   8.78\,$\pm$\,0.87            &    0.69\,$\pm$\,0.05    \\ 
NGC\,7044            & 7.6 \,$\pm$\, 0.2            &202\,$\pm$\, 7             &19.6\,$\pm$\, 3.5    &18.9\,$\pm$\, 1.6                &  32.91\,$\pm$\,3.01     &  34.57\,$\pm$\,2.70            &    1.08\,$\pm$\,0.06    \\ 
NGC\,7142            & 3.3 \,$\pm$\, 0.1            & 93\,$\pm$\, 5             &17.2\,$\pm$\, 2.1    &17.4\,$\pm$\, 3.5                &   6.87\,$\pm$\,1.09     &   7.83\,$\pm$\,1.37            &    1.33\,$\pm$\,0.08    \\ 
NGC\,7654            & 3.5 \,$\pm$\, 0.1            & 79\,$\pm$\, 2             &13.8\,$\pm$\, 1.0    & 4.3\,$\pm$\, 0.1                &   6.22\,$\pm$\,0.78     &   8.19\,$\pm$\,1.18            &    1.57\,$\pm$\,0.07    \\ 
NGC\,7789            &15.4 \,$\pm$\, 0.2            &408\,$\pm$\, 7             &25.1\,$\pm$\, 3.7    &31.3\,$\pm$\, 1.8                &  13.20\,$\pm$\,1.23     &  10.88\,$\pm$\,0.97            &    0.89\,$\pm$\,0.04    \\ 
\hline

\end{tabular}


$^{\dag}$ See Section~\ref{sec:analysis}.   \\

\end{minipage}
\label{tab:investig_sample_compl}
\end{table*}

\subsection{Sample and parameters compilation}

Our investigated sample consists of 60 OCs, as shown in Table~\ref{tab:investig_sample}. Most of these OCs present density contrast parameter, defined as $\delta_c=(\sigma_0/\sigma_{\textrm{bg}})+1$ (analogously to \citeauthor{Bica:2005a}\,\,\citeyear{Bica:2005a}; $\sigma_0$ is the background-subtracted stellar density at the innermost radial bins and $\sigma_{\textrm{bg}}$ is the mean background density; see Section~\ref{sec:struct_analysis}), larger than $\sim4$. Although there are some lower contrast OCs added to our sample, these ones present deep main sequences ($\gtrsim\,4$\,mag below the turnoff) on decontaminated CMDs and recognizable evolutionary sequences (see Sections~\ref{sec:membership}, \ref{sec:results} and Figure~\ref{fig:CMD_3clusters_maintext}). The log\,(age.yr$^{-1}$) and $R_G$ values span the intervals $\sim$7.2 - 9.8 and $\sim6$ - 12\,kpc, respectively (Table~\ref{tab:investig_sample}).


The clusters central coordinates and structural parameters (tidal, core and half-light radii, respectively: $r_t$, $r_c$, $r_h$), have been derived according to the procedures outlined in the next section. Table~\ref{tab:investig_sample} also contains fundamental astrophysical parameters (true distance modulus, metallicity, age and interstellar reddening), Galactocentric distances, besides half-light relaxation times ($t_{rh}$) and mean proper motion components. In turn, Table~\ref{tab:investig_sample_compl} contains the OCs' mass ($M_{\textrm{clu}}$), Jacobi radius ($R_J$) and contrast parameter ($\delta_c$), together with the central ($\sigma_0$) and mean background ($\sigma_{\textrm{bg}}$) stellar densities. Initial mass ($M_{\textrm{ini}}$) estimates, obtained based on analytical expressions taken from the literature, are also informed in the same table. These additional parameters are derived in Sections \ref{sec:membership}, \ref{sec:results} and \ref{sec:analysis}.

\section{Centre redetermination and structural parameters}
\label{sec:struct_analysis}

\begin{figure*}
\begin{center}

\parbox[c]{1.00\textwidth}
  {
   \begin{center}
       \includegraphics[width=0.95\textwidth]{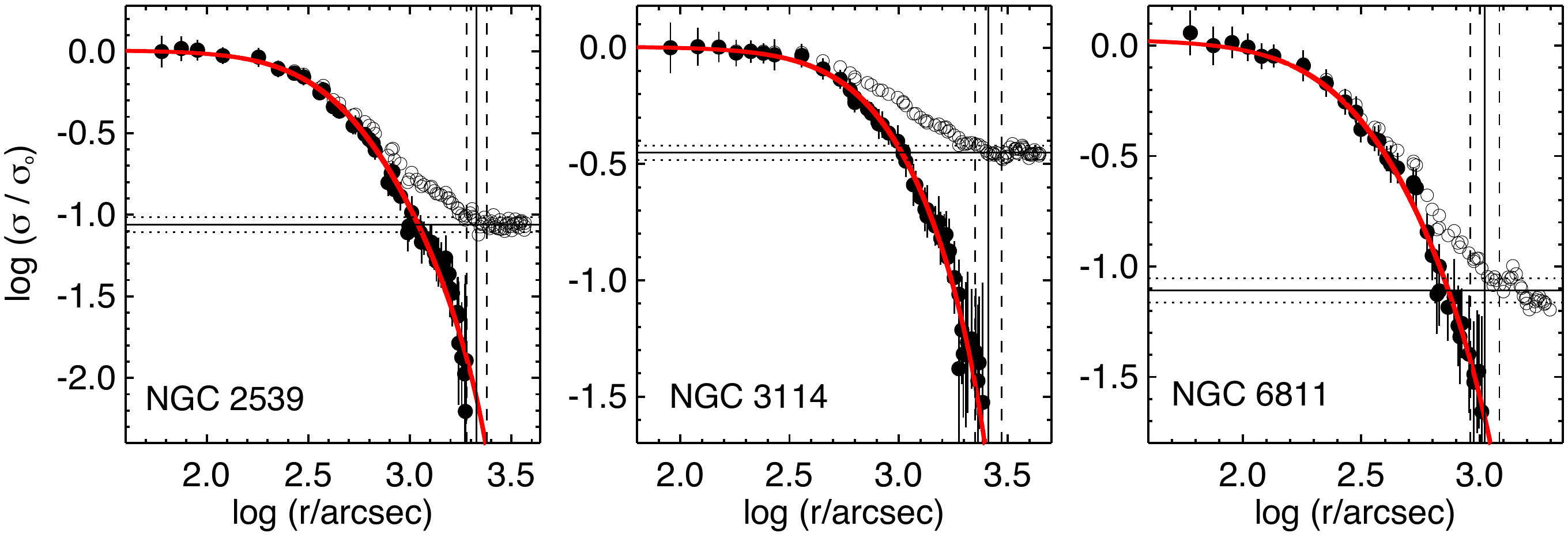}   
    \end{center}    
  }
\caption{ RDPs for the OCs NGC\,2539, NGC\,3114 and NGC\,6811 (see the online supplementary material for other RDPs). The filled (open) symbols represent the normalized and (non-)background subtracted profiles. Error bars come from Poisson statistics. The continous red line is the fit of the \citeauthor{King:1962}\,\,(\citeyear{King:1962}) profile to the data. The horizontal continous line is the mean background density ($\sigma_{\textrm{bg}}$), with the corresponding uncertainty (dotted lines). The limiting radius ($R_{\textrm{lim}}$) is indicated by a vertical line (dashed lines for its uncertainty). }

\label{fig:RDP_3clusters_maintext}
\end{center}
\end{figure*}

In this step, we employed the proper motions filtered skymap (right panel of Figure~\ref{fig:ilustra_preanalise_NGC6811}) of each OC and built a regular grid of ($X,Y$) coordinates (equations~\ref{eq:project_alpha_delta1} and \ref{eq:project_alpha_delta2}) surrounding the literature centre. Typically, 20\,$\times$\,20 central coordinates, with equal spacing of $\sim0.5{\arcmin}-1.0{\arcmin}$, have been employed. Then we devised an algorithm that runs through the whole points in this grid and, for each position, a RDP is built by counting the number of stars in annular concentric rings of different widths and dividing this number by the ring area, that is: $\sigma\,(\bar{r}_k)=N_k/[\pi(r_{k}^{2}-r_{k-1}^2)]_{k=1,2,3...}$, where $\bar{r}_k=(r_{k}+r_{k-1})/2$. A background subtracted RDP is then built by performing $\sigma_{\textrm{sub}}\,(\bar{r}_k)=\sigma\,(\bar{r}_k) - \sigma_{bg}$; $\sigma_{bg}$ is the mean background density, obtained by simply averaging the set of density values in the interval $r\,\ge\,R_{\textrm{lim}}$ (cluster's limiting radius), where the density values fluctuate around a nearly constant value (Figure~\ref{fig:RDP_3clusters_maintext}). 

We then fit the set of $\sigma_{\textrm{sub}}$ values employing \citeauthor{King:1962}'s\,\,(\citeyear{King:1962}) profile, defined as

\begin{equation}
   \sigma(r)\,\propto\,\left[\frac{1}{\sqrt{1+(r/r_c)^2}} - \frac{1}{\sqrt{1+(r_t/r_c)^2}}\right]^2
\end{equation}

\noindent by means of $\chi^2$ minimization. The core radius ($r_c$) provides a length scale of the cluster's inner structure, while its overall empirical size (corresponding to the truncation radius of the King profile) is determined by the tidal radius ($r_t$). This latter parameter should not be confused with the Jacobi radius ($R_J$; see Section~\ref{sec:analysis}), which is the limit beyond which a star is more subject to the external tidal field than to the cluster's gravitational pull (e.g., \citeauthor{Renaud:2011}\,\,\citeyear{Renaud:2011}; \citeauthor{Portegies-Zwart:2010}\,\,\citeyear{Portegies-Zwart:2010}; \citeauthor{von-Hoerner:1957}\,\,\citeyear{von-Hoerner:1957}). Its value depends on the cluster mass and its location within the host galaxy.


The redetermined central coordinates are those from which we obtained the highest density in the innermost region with minimum residuals, thus resulting in a smooth stellar RDP (see, e.g., \citeauthor{Bica:2011}\,\,\citeyear{Bica:2011}). The final outcomes of this procedure is shown in Figure~\ref{fig:RDP_3clusters_maintext} for 3 investigated OCs, taken as illustrative examples. Other RDPs are available in the online supplementary material. From now on, the same procedure will be employed for other figures. 

We have estimated the projected half-light radius ($r_{hp}$) from King model parameters using the calibration proposed by \citeauthor{Santos:2020}\,\,(\citeyear{Santos:2020}, their equation 9), which was then converted to the three-dimensional value under the assumption that mass follows light and by assuming $r_h=f_h\,r_{hp}$, where $f_h$ is the conversion ratio. Here we employed $f_h=1.33$ (\citeauthor{Baumgardt:2010}\,\,\citeyear{Baumgardt:2010}, hereafter BPG10). In order to allow for possible variations in this factor (which can range from $1.31-1.38$; \citeauthor{King:1966}\,\,\citeyear{King:1966}; \citeauthor{Wilson:1975}\,\,\citeyear{Wilson:1975}), an uncertainty of $\sim0.04$ in $f_h$ has been propagated into our final uncertainty estimates for $r_h$.

\section{Membership determination}
\label{sec:membership}

The unique precision of the astrometric and photometric information in the \textit{Gaia} catalogue allows to identify groups of stars with coherent proper motions and parallaxes and to statistically differentiate them from representative samples of field stars. This strategy has been employed in a number of previous works in order to identify member candidate stars of OCs (e.g., \citeauthor{Cantat-Gaudin:2020}\,\,\citeyear{Cantat-Gaudin:2020}; \citeauthor{Bisht:2021}\,\,\citeyear{Bisht:2021}; \citeauthor{Ferreira:2021}\,\,\citeyear{Ferreira:2021}), therefore improving the determination of astrophysical parameters via astrometrically decontaminated CMDs. 

In this paper, we employed the method proposed by \citeauthor{Angelo:2019a}\,\,(\citeyear{Angelo:2019a}, hereafter ASCM19) in order to assign membership probabilities ($P$) to stars within the tidal radius of each investigated OC. After that, by restricting the sample of stars to those with high $P$ values, we can identify unambigous evolutionary sequences on the CMDs, thus providing useful constraints for isochrone fitting (see Section~\ref{sec:isoc_fit}). 

We are well aware that some of the investigated OCs (e.g., M\,67, NGC\,2516, NGC\,752) present external structures (elongated tidal tails and extended haloes) beyond the derived tidal radius (Section~\ref{sec:struct_analysis}), as demonstrated in previous works (e.g., \citeauthor{Tarricq:2022}\,\,\citeyear{Tarricq:2022}; \citeauthor{Carrera:2019}\,\,\citeyear{Carrera:2019}; see also \citeauthor{Roser:2019}\,\,\citeyear{Roser:2019}; \citeauthor{Meingast:2019}\,\,\citeyear{Meingast:2019}). However, extending the search of member stars to regions considerably larger than the cluster's $r_t$ may result in a non-negligible number of false-positives due to a progressively smaller contrast with the general Galactic field population, specially in the case of OCs projected against dense fields (e.g., \citeauthor{Zhong:2022}\,\,\citeyear{Zhong:2022}; \citeauthor{Pang:2021}\,\,\citeyear{Pang:2021}; \citeauthor{Krone-Martins:2014}\,\,\citeyear{Krone-Martins:2014}). This way, to optimize our decontamination method performance, warrant uniformity in our treatment and to identify candidate members gravitationally bound to each cluster, we have restricted our search to the more contrasting region $r\leq r_t$.     

In what follows, we briefly describe the main steps of our method.

\begin{itemize}

  \item After applying the proper motions filter (Section~\ref{sec:prop_motions_filtering}) to our database, we took those stars within the cluster tidal radius ($r\leq r_t$) and also selected stars in an annular comparison field, centered on the redetermined cluster coordinates (Table~\ref{tab:investig_sample}). 
  
  
  \item The astrometric space ($\varpi, \mu_{\alpha}\,\textrm{cos}\,\delta$, $\mu_{\delta}$) defined by cluster and field stars is divided in a regular grid of cells (typical widths of $\Delta\varpi\sim0.1\,$mas, $\Delta\mu_{\alpha}\,\textrm{cos}\,\delta\,\sim\,\Delta\mu_{\delta}\sim0.5\,$mas.yr$^{-1}$) and then membership likelihoods ($\ell_{\textrm{star}}$) are assigned to cluster and field stars using multivariate gaussians (equation 1 of ASCM19); 
  
  \item For each cell, the set of $\ell_{\textrm{star}}$ values are then inserted into entropy-like functions ($S$, equation 3 of ASCM19) and those cells for which $S_{\textrm{cluster}}<S_{\textrm{field}}$ are flagged. After that, an exponential factor is evaluated (equation 4 of ASCM19), which considers the overdensity of cluster stars within the cell in relation to the average of counts across the overall grid;
  
  
  \item The dependence on the initial grid configuration is alleviated by varying their sizes by 1/3 in each dimension and the complete procedure is repeated. After all iterations, the final membership probabilities ($P$) are derived.  

\end{itemize}

This procedure allows to identify concentrated groups of cluster stars in the astrometric space, statistically contrastant with the distribution of field star samples. Then we restricted each cluster sample to the high membership stars ($P\gtrsim0.7$) and combined the spectroscopic and photometric data in order to derive the OCs' astrophysical parameters, as shown in Section~\ref{sec:results} below.

\section{Results}
\label{sec:results}

\subsection{Astrometric and spectroscopic parameters}
\label{sec:astrometric_spec_params}

\begin{figure*}
\begin{center}

\parbox[c]{1.00\textwidth}
  {
   \begin{center}
       \includegraphics[width=0.95\textwidth]{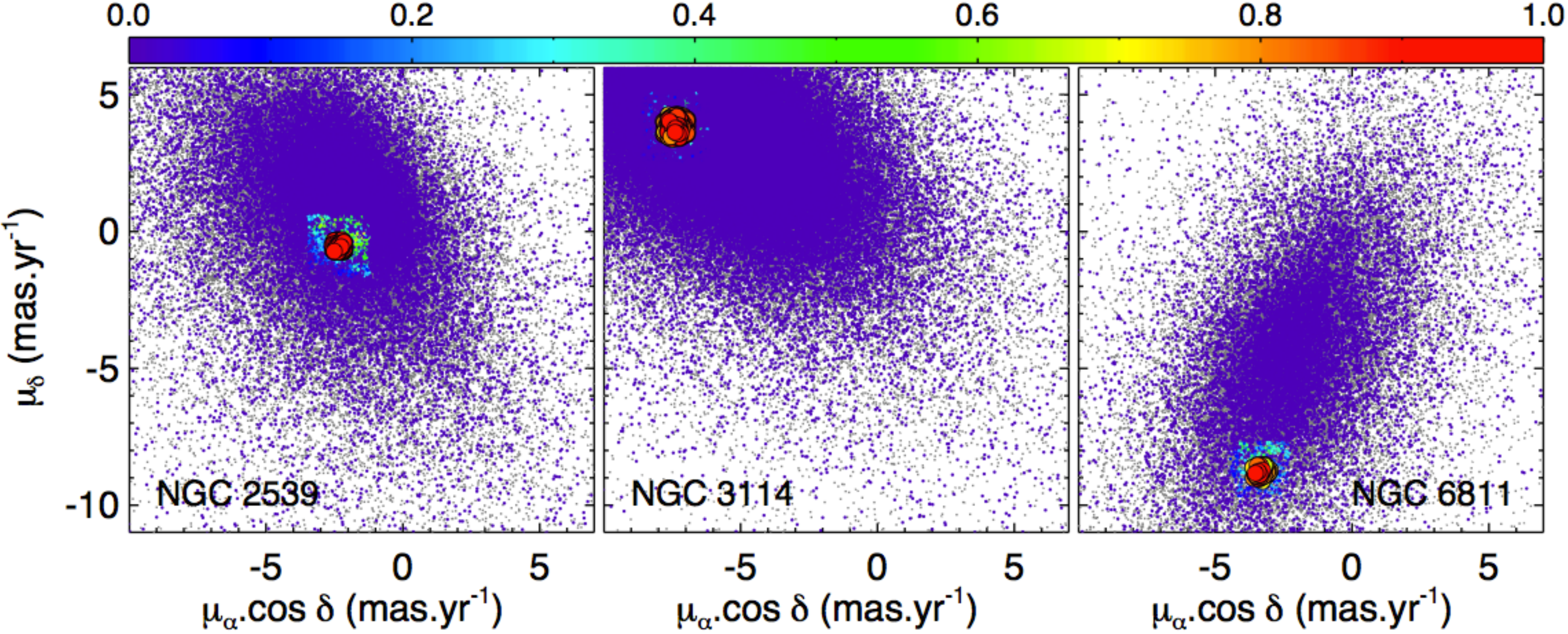}   
    \end{center}    
  }
\caption{ VPDs for the OCs NGC\,2539, NGC\,3114 and NGC\,6811 (see the online supplementary material for other VPDs). The symbol colours are assigned according to the membership probability ($P$, Section~\ref{sec:membership}), as indicated by the colourbar. Larger filled circles are stars with $P\gtrsim0.7$ and small grey dots are stars in a comparison field. }

\label{fig:VPD_3clusters_maintext}
\end{center}
\end{figure*}

\begin{figure*}
\begin{center}

\parbox[c]{1.00\textwidth}
  {
   \begin{center}
       \includegraphics[width=0.95\textwidth]{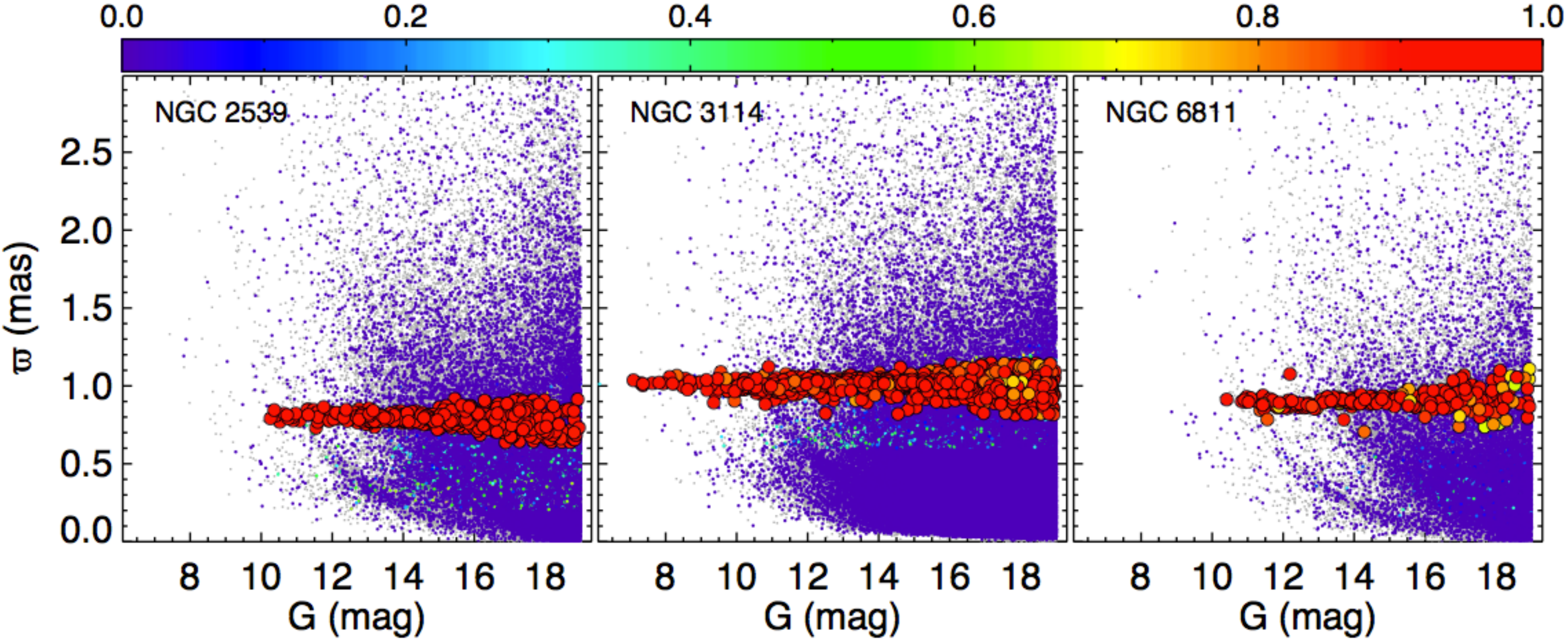}   
    \end{center}    
  }
\caption{ Parallax versus $G$ magnitude for the OCs NGC\,2539, NGC\,3114 and NGC\,6811 (see the online supplementary material for other $\varpi\,\times\,G$ plots). Symbol convention is the same as Figure~\ref{fig:VPD_3clusters_maintext}. }

\label{fig:plx_Gmag_3clusters_maintext}
\end{center}
\end{figure*}

Figure~\ref{fig:VPD_3clusters_maintext} shows the vector-point diagram (VPD) for 3 investigated OCs: NGC\,2539, NGC\,3114 and NGC\,6811. The symbol colours are representative of the membership probabilities (indicated by the colourbar) assigned to stars within the cluster's tidal radius (that is, $r\leq r_t$), according to the procedure outlined in Section~\ref{sec:membership}. The larger filled circles represent stars with $P\gtrsim0.7$. Stars within the respective annular comparison field (Section~\ref{sec:membership}) are plotted with small grey dots in each panel. Figure~\ref{fig:plx_Gmag_3clusters_maintext}, in turn, allows to verify the dispersion of parallax ($\varpi$) as function of $G$ magnitude. The same symbol convention of Figure~\ref{fig:VPD_3clusters_maintext} is employed. Again, we can verify a concentration of the $\varpi$ values defined by the high-membership stars, with an increasing dispersion for fainter $G$ magnitudes, due to the progressively larger uncertainties (typically, $\Delta\varpi\sim0.05\,$mas for $G\sim16\,$mag; $\Delta\varpi\sim0.1\,$mas for $G\sim18\,$mag; $\Delta\varpi\sim0.2\,$mas for $G\sim19\,$mag).


\begin{figure*}
\begin{center}

\parbox[c]{1.00\textwidth}
  {
   \begin{center}
       \includegraphics[width=0.95\textwidth]{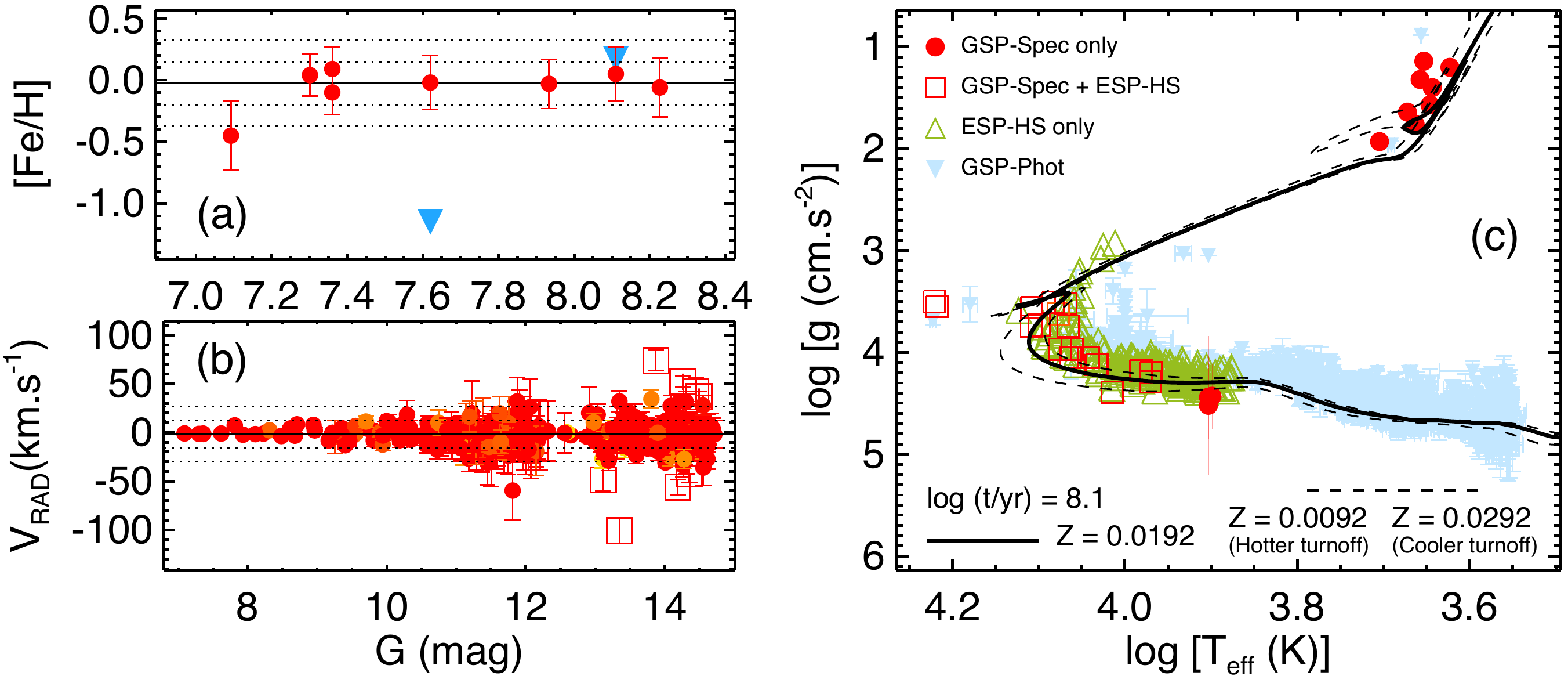}   
    \end{center}    
  }
\caption{   Plots with spectroscopic data for high-membership stars of NGC\,3114 (see the online supplementary material for other plots). \textit{Panel (a)}: Metallicity $[Fe/H]$ versus $G$-band magnitude. Filled circles are stars with atmospheric parameters obtained by the {\fontfamily{ptm}\selectfont GSP-Spec} module. The blue triangles represent values obtained from {\fontfamily{ptm}\selectfont GSP-Phot}. Continuous and dotted horizontal lines are the median of the $[Fe/H]$ values (as inferred from {\fontfamily{ptm}\selectfont GSP-Spec}) and the mean absolute deviations at the 1 and 2\,$\sigma$ levels. \textit{Panel (b)}: $V_{\textrm{rad}}$ versus $G$-band magnitude. Open squares are those stars that deviate by more than 2$\sigma$ from the median value. \textit{Panel (c)}: spectroscopic \textit{Hertzprung-Russel diagram} for NGC\,3114 (only stars with $P\gtrsim0.7$ are shown). Filled circles represent stars that had their atmospheric parameters derived by {\fontfamily{ptm}\selectfont GSP-Spec}. Stars represented by open green triangles had their parameters estimated only by the especialized ESP-HS modulus, while the light blue points are the set of parameters obtained from {\fontfamily{ptm}\selectfont GSP-Phot} for the whole sample of members. The open squares represent those stars analysed by both the {\fontfamily{ptm}\selectfont GSP-Spec} \textit{and} also the {\fontfamily{ptm}\selectfont ESP-HS} modulus (see Figure~\ref{fig:compara_Teff_logg_GSPSPEC_ESPHS}). The continuous and the 2 dashed lines (showing hotter and cooler turnoff points) are, respectively, PARSEC \citep{Bressan:2012} isochrones (log\,($t$.yr$^{-1}$)=8.2) with Z=0.0192 ($[Fe/H]\simeq0.10$; the same employed in this cluster CMD, see Figure~\ref{fig:CMD_3clusters_maintext}), Z=0.0092 ($[Fe/H]\simeq$\,$-$0.22) and Z=0.0292 ($[Fe/H]\simeq$\,0.28). The two open squares corresponding to log\,$T_{\textrm{eff}}\,($K$)\simeq4.2$ represent stars with \textit{source\_ID}  5256597008888640640 and 5256713454040430336 and both are marked, for illustration purposes, with cyan contoured symbols in Figure~\ref{fig:CMD_3clusters_maintext} (middle panel; see below).   }

\label{fig:HRD_NGC3114}
\end{center}
\end{figure*}

Panel (a) of Figure~\ref{fig:HRD_NGC3114} shows the metallicity $[Fe/H]$ for member stars of NGC\,3114 as derived from the RVS spectra, via the {\fontfamily{ptm}\selectfont GSP-Spec} module (filled circles; see Section~\ref{sec:sample_and_data}). For a qualitative comparison, we have overplotted the $[Fe/H]$ as derived from  {\fontfamily{ptm}\selectfont GSP-Phot} (blue triangles; only 2 stars in the present case), which employs low-resolution BP/RP spectra. It is noticeable that one of the {\fontfamily{ptm}\selectfont GSP-Phot} stars present very discrepant metallicity. In this sense, it is important to stress that the same given star can have two or more sets of parameters derived within the Apsis procedure (Creevey et al. 2022) and different moduli can result in significantly different estimates. In the present example, the most discrepant star has $[Fe/H]_{\textrm{GSP-Phot}}\simeq-1.15\,$dex and $[Fe/H]_{\textrm{GSP-Spec}}\simeq-0.02\,$dex.

The continuous horizontal line is the median obtained from the dispersion of the {\fontfamily{ptm}\selectfont GSP-Spec} metallicities and the dotted lines represent 1 and 2\,$\sigma$ median absolute deviations. Panel (b) shows the dispersion of $V_{\textrm{rad}}$ for member stars (filled symbols) as function of $G$ magnitude. As in the previous panel, the continuous and dotted horizontal lines are, respectively, the median and the median absolute deviation for this subsample. The open squares are stars whose $V_{\textrm{rad}}$ value deviates from the median by more than $2\,\sigma$, considering uncertainties. These stars were considered less-probable members, but have not been excluded from the analysis, due to possible binarity. Although the {\fontfamily{ptm}\selectfont DSC} (Discrete Source Classifier) module within the \textit{Gaia} catalogue informs the probability of a source being a physical binary (field {\fontfamily{ptm}\selectfont\textit{classprob\_dsc\_combmod\_binarystar}} within the {\fontfamily{ptm}\selectfont\textit{astrophysical\_parameters}} table), it is currently adviced against the use of this value, since some improvements are needed in the global class priors, as explained in \cite{Babusiaux:2022} and in the DR3 online documentation.


Panel (c) is the spectroscopic \textit{Hertzprung-Russel diagram} (HRD) for stars of NGC\,3114 with $P\gtrsim0.7$. Filled circles are stars with atmospheric parameters obtained from {\fontfamily{ptm}\selectfont GSP-Spec}. The open light green triangles are stars with $T_{\textrm{eff}}$ and log\,$g$ derived exclusively from the {\fontfamily{ptm}\selectfont ESP-HS} module. For comparison, we have also overplotted in the HRD, for the whole sample of high-membership stars, the $T_{\textrm{eff}}$ and log\,$g$ (when available) as obtained from the {\fontfamily{ptm}\selectfont GSP-Phot} algorithm (light blue symbols).

The open squares in panel (c) are 19 stars for which the atmospheric parameters were determined from {\fontfamily{ptm}\selectfont GSP-Spec} \textit{and}, more accurately, from the specialized {\fontfamily{ptm}\selectfont ESP-HS} module (for this group of stars, the $T_{\textrm{eff}}$ and log\,$g$ values shown in the HRD are those obtained from this latter module). In some cases, the specialized moduli within Apsis provide better estimates for the atmospheric parameters than the general parametrizers \citep{Fouesneau:2022}, since they deal with specific spectral types. 

\begin{figure}
\begin{center}

\parbox[c]{0.48\textwidth}
  {
   \begin{center}
       \includegraphics[width=0.48\textwidth]{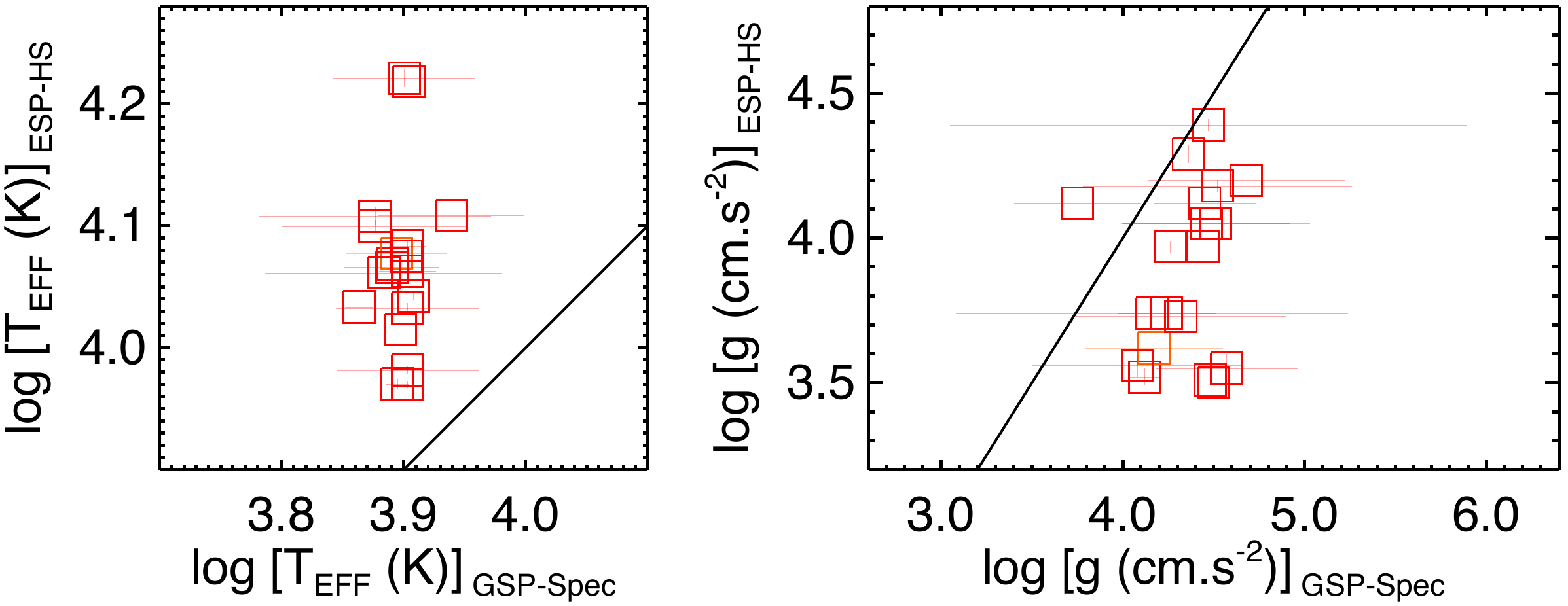}   
    \end{center}    
  }
\caption{ Comparison between the atmospheric parameters derived by two different Apsis moduli ({\fontfamily{ptm}\selectfont GSP-Spec} and {\fontfamily{ptm}\selectfont ESP-HS}) for 19 member stars of NGC\,3114 (open squares in Figure~\ref{fig:HRD_NGC3114}). The black continuous line is the identity relation. }

\label{fig:compara_Teff_logg_GSPSPEC_ESPHS}
\end{center}
\end{figure}

Differences between parameters derived by different moduli are illustrated in Figure~\ref{fig:compara_Teff_logg_GSPSPEC_ESPHS}. For the 19 member stars highlighted in the HRD of NGC\,3114 (open squares; spectral types A and B: $9300\,\lesssim\,T_{\textrm{eff}}(K)\,\lesssim\,16500$), it is noticeable that the specialized module provides considerably higher effective temperatures and, in most cases, smaller surface gravities compared to the general parametrizer. In this plot, uncertainties for the {\fontfamily{ptm}\selectfont GSP-Spec} data are much larger, since the corresponding parameters have been recalibrated (as explained in Section~\ref{sec:data_filtering_corrections}) and the uncertainties in the transformation equations have been propagated into the final values. In the case of {\fontfamily{ptm}\selectfont ESP-HS} data, the set of parameters were extracted directly from the catalogue and no transformations have been applied. In general, uncertainties in the spectroscopic data within the \textit{Gaia} DR3 catalogue seem underestimated (see, e.g., section 2.7 of \citeauthor{Andrae:2022}\,\,\citeyear{Andrae:2022} and Appendix D of \citeauthor{Recio-Blanco:2022}\,\,\citeyear{Recio-Blanco:2022}).


\begin{figure*}
\begin{center}

\parbox[c]{1.00\textwidth}
  {
   \begin{center}
       \includegraphics[width=0.95\textwidth]{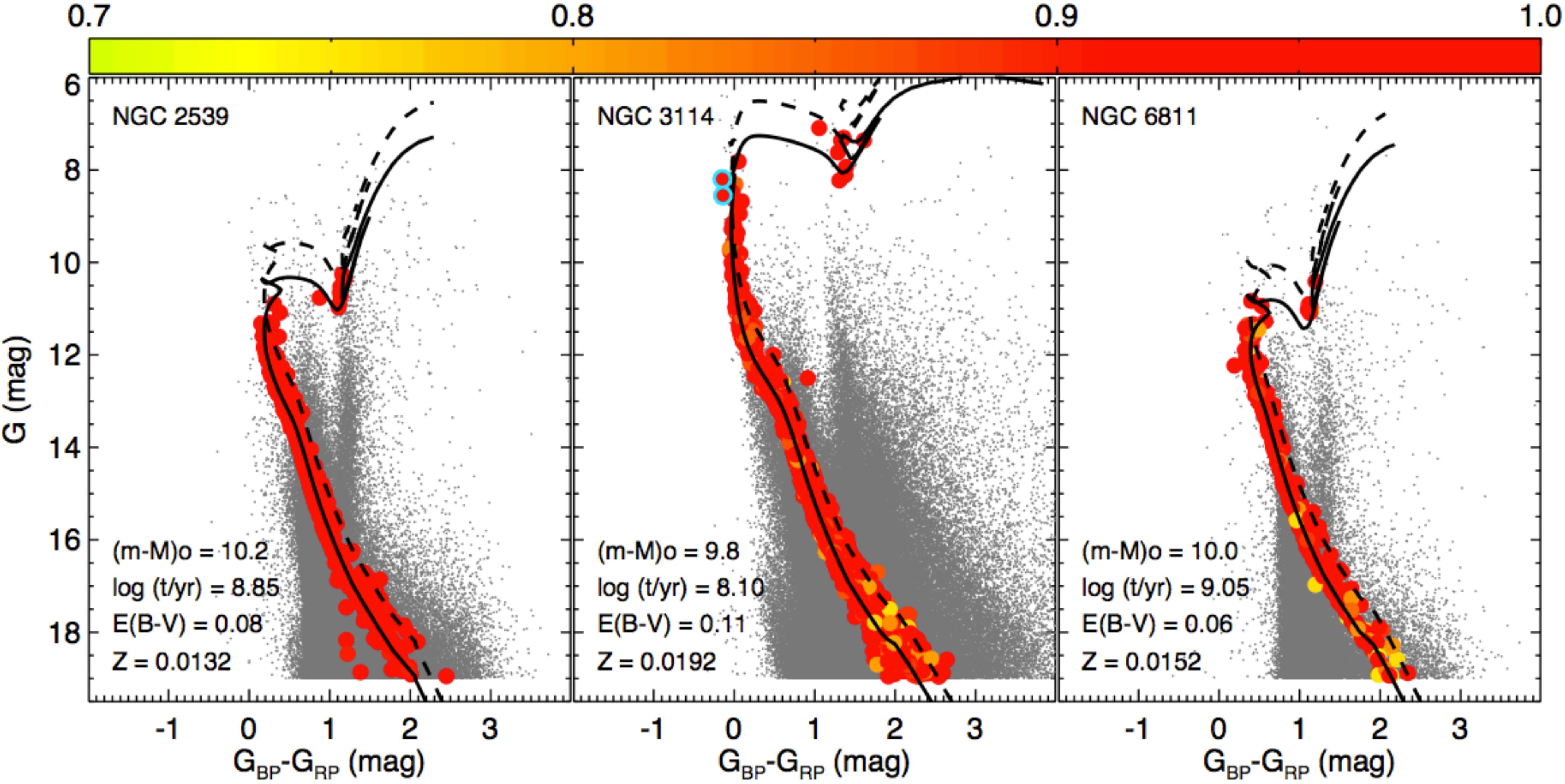}   
    \end{center}    
  }
\caption{ Decontaminated CMDs $G\,\times\,(G_{\textrm{BP}}-G_{\textrm{RP}})$ for the OCs NGC\,2539, NGC\,3114 and NGC\,6811 (CMDs for other OCs are found in the online Supplementary Material). Only stars with $P\gtrsim0.7$ are shown (Section~\ref{sec:membership}). Small grey dots are stars in a comparison field. The continous lines are PARSEC isochrones, properly shifted according to the clusters fundamental parameters, as indicated. The dashed lines represent the same isochrones, but shifted vertically by $-0.75\,$mag, representing the locus of unresolved binaries of equal mass components. The two circled dots in cyan in NGC\,3114 CMD identify stars with \textit{source\_ID} 5256597008888640640 and 5256713454040430336, both with effective temperature log\,$T_{\textrm{eff}} (K)\simeq4.2$ (see Figure~\ref{fig:HRD_NGC3114}, panel $c$). }

\label{fig:CMD_3clusters_maintext}
\end{center}
\end{figure*}

The plots in Figure~\ref{fig:HRD_NGC3114}, constructed for each OC, can be used to perform initial guesses for the cluster metallicity (from the mean $[Fe/H]$ value in panel $a$) and age (from panel $c$, in which the turnoff, the red giant branch and the red clump, when present, provide useful constraints); at this stage, it is important to mention that no calculations are performed based directly on these diagrams (see Sections~\ref{sec:mass_functions} and \ref{sec:analysis}). The spectroscopic HRD (panel $c$) is particularly useful, as the intrinsic position of stars do not depend on distance or interstellar reddening. We have overplotted a log\,$t$\,=\,8.1 PARSEC isochrone \citep{Bressan:2012} with overall metallicity $Z=0.0192$ ($[Fe/H]\simeq\,$log\,$(Z/Z_{\odot});\,Z_{\odot}=0.0152$; \citeauthor{Bonfanti:2016}\,\,\citeyear{Bonfanti:2016}), which is the best fitted isochrone superimposed to the data in the cluster CMD (see Section~\ref{sec:isoc_fit} and Figure~\ref{fig:CMD_3clusters_maintext}). To evaluate the effect of the metallicity in this diagram, we have also shown, illustratively, two other isochrones (dashed lines) with this same age, but representing a chemically poorer stellar group ($Z=0.0092$; $[Fe/H]\simeq-0.22$) and a richer one ($Z=0.0292$; $[Fe/H]\simeq+0.28$), as indicated in the legend.

\subsection{Isochrone fit and fundamental parameters}
\label{sec:isoc_fit}

The initial estimates for the overall metallicity $Z$ and log\,$t$ allowed to alleviate the degeneracy of the isochrone fit solutions in Figure~\ref{fig:CMD_3clusters_maintext}, which shows the astrometrically decontaminated CMDs for 3 investigated OCs (namely, NGC\,2539, NGC\,3114 and NGC\,6811). Initial guess for the colour excess, $E(B-V)$, was taken from DMML21. In the case of the true distance modulus, $(m-M)_0$, an initial guess was obtained by simply inverting the mean parallax of the high-membership stars ($P\gtrsim0.7$).

Then we built a grid of parameters allowing for variations, in relation to the initial estimates, at maximum levels of $\sim0.01\,$dex and $\sim0.2\,$dex in $Z$ and log\,$t$, respectively, and maximum variations of about $\sim0.5\,$mag and $\sim0.2\,$mag in $(m-M)_0$ and $E(B-V)$, respectively. The step sizes in each parameter are: $\sim0.001\,$ dex for $Z$, 0.05\,dex for log\,$t$, 0.05\,mag for $(m-M)_0$ and 0.01\,mag for $E(B-V)$. For each grid point, we evaluated the distance of each star in the $G\times(G_{BP}-G_{RP})$ to the closest isochrone point, considering the whole list of high-membership stars, and the adopted solution corresponds to the set of parameters that provided minimal residuals.

For each OC, the isochrone fit solution was carefully inspected in order to ensure a proper match of the clusters' key evolutionary stages (the main sequence, the turnoff point, the subgiant and red giant branches and the red clump, if present). Uncertainties in the fundamental parameters $(m-M)_0$, log\,$t$, $E(B-V)$ and $Z$ have been determined by successively shifting the isochrone around the optimal solution until the evolutionary sequences no longer produce a proper match to the loci defined by the high-membership stars along the cluster CMD. The final results are informed in Table~\ref{tab:investig_sample}.

\subsection{Mass functions}
\label{sec:mass_functions}

\begin{figure}
\begin{center}

\parbox[c]{0.48\textwidth}
  {
   \begin{center}
       \includegraphics[width=0.48\textwidth]{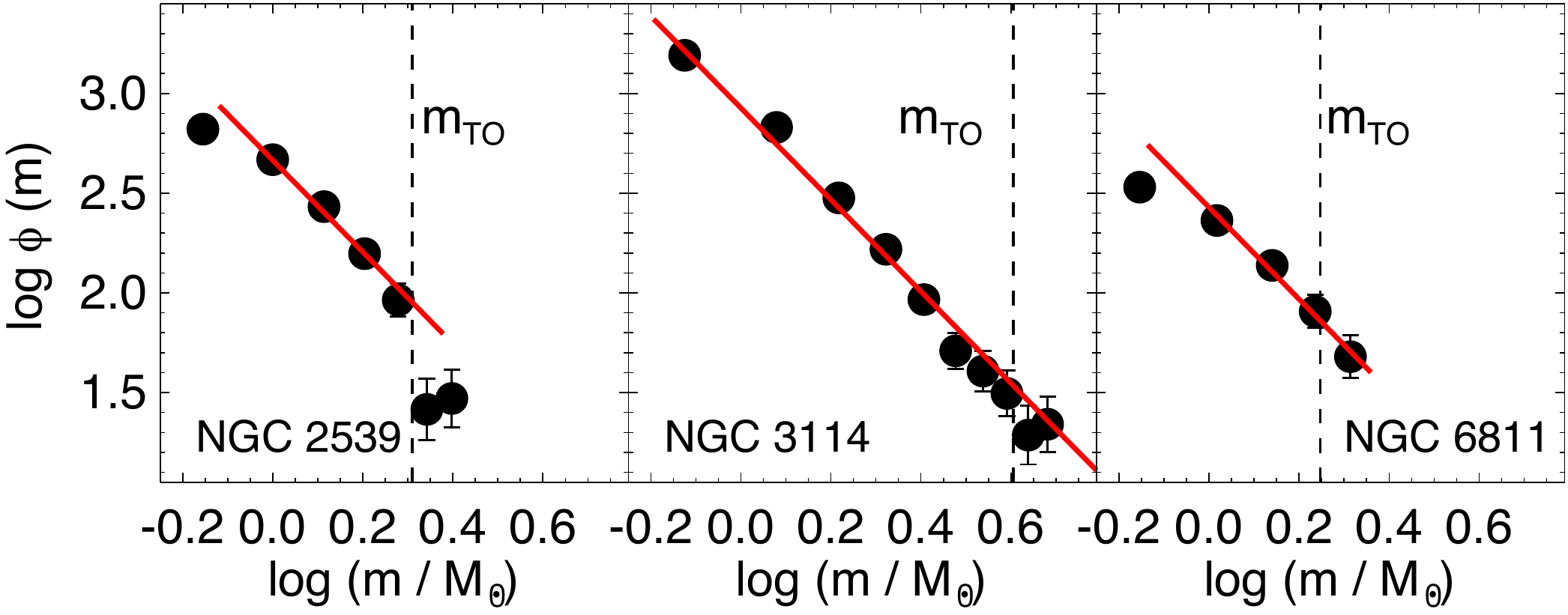}   
    \end{center}    
  }
\caption{ Mass function of the OCs NGC\,2539, NGC\,3114 and NGC\,6811 (see Appendix~\ref{sec:suppl_MFs} of the online supplementary material for other OCs). The red line is the normalized initial mass function of \citeauthor{Kroupa:2001}\,\,(\citeyear{Kroupa:2001}) overplotted to the data. The vertical line indicates the mass at the turnoff ($m_{TO}$). }

\label{fig:MF_3clusters_maintext}
\end{center}
\end{figure}

We employed the decontaminated CMD $G\times(G_{\textrm{BP}}-G_{\textrm{RP}})$ of each OC in our sample (Figure~\ref{fig:CMD_3clusters_maintext}) and estimated individual masses from interpolation of the $G$ magnitude for each member star across the best fitted isochrone, properly shifted according to the cluster distance modulus and reddening (Section~\ref{sec:isoc_fit} and Table~\ref{tab:investig_sample}). After that, by counting the number of stars within linear bins of mass, the cluster mass function (MF; $\phi\,(m)=dN/dm$) was constructed. Errorbars come from applying Poisson statistics. The MFs for 3 investigated OCs are shown in Figure~\ref{fig:MF_3clusters_maintext}. In the mass ranges $\sim[1.0\,,\,1.9]\,\textrm{M}_{\odot}$ for NGC\,2539, $\sim[0.8\,,\,4.8]\,\textrm{M}_{\odot}$ for NGC\,3114 and $\sim[1.0\,,\,2.1]\,\textrm{M}_{\odot}$ for NGC\,6811, the observed MFs can be expressed as power laws under the form $\phi(m)=A\,m^{-(1+\chi)}$, where $(1+\chi)$ is the MF slope and $A$ is a normalization constant. The derived $\chi$ values for these 3 OCs resulted $\chi_{N2539}=1.4\,\pm\,0.3$, $\chi_{N3114}=1.3\,\pm\,0.1$ and $\chi_{N6811}=1.2\,\pm\,0.3$, values that are compatible with the initial mass function (IMF) of \cite{Kroupa:2001}, considering uncertainties.

We determined the observed cluster mass contained in the above mass intervals by adding up the contribution of each individual bin; the Kroupa's IMF was then normalized according to this value (an analogous procedure was employed for all OCs; see Appendix~\ref{sec:suppl_MFs}) and then overplotted on the observed mass function in Figure~\ref{fig:MF_3clusters_maintext}. The OCs total mass ($M_{\textrm{clu}}$) and number of stars ($N_{\textrm{clu}}$), Table~\ref{tab:investig_sample_compl}, have been derived by integration of the normalized IMF until the theoretical lower mass limit of $\sim0.1M_{\odot}$. Uncertainties in $M_{\textrm{clu}}$ come from error propagation. Some OCs present signals of lower mass stars depletion (e.g., NGC\,2539 and NGC\,6811), since the observed MFs depart from Kroupa's law towards lower stellar masses, possibly due to their preferential evaporation (e.g., BM03; \citeauthor{de-La-Fuente-Marcos:1997}\,\,\citeyear{de-La-Fuente-Marcos:1997}). In the case of NGC\,2539, the two higher mass bins also depart from the IMF, which may be due to a combination of low-number statistics, stochasticity (e.g., \citeauthor{Santos:1997}\,\,\citeyear{Santos:1997}) and/or shorter evolutionary time-scales (e.g., \citeauthor{Valegard:2021}\,\,\citeyear{Valegard:2021}). The mass contained in these bins has been incorporated into $M_{\textrm{clu}}$. Other OCs (e.g, NGC\,3114) have their MF bins compatible with the IMF along the complete observed mass domain.

Estimates of upper limit of mass in dark stellar remnants (white dwarfs, neutron stars and stellar black holes) have also been added to the final $M_{\textrm{clu}}$ values (Table~\ref{tab:investig_sample_compl}), for which we assumed no natal kicks and the Kroupa's IMF along with the zero-points in Fig.~\ref{fig:MF_3clusters_maintext}. However, as shown in Appendix~\ref{sec:mass_dark_remnants}, the estimated mass fraction in dark remannts is small, which makes their contribution to the total mass also small.

\section{Analysis}
\label{sec:analysis}

\subsection{Jacobi radius}
\label{sec:rJ}

The Jacobi radius ($R_J$) establishes the limit of the cluster gravitational influence on a star, taking into account the external Galactic tidal field. It can be assumed as the distance between the cluster centre and the Lagrangian point $L_1$. From the the linearized equations of motion for a star submitted to the gravitational potential of both the cluster and the host galaxy, \cite{Renaud:2011} present a derivation for $R_J$ based on the tidal tensor of the total potential (their equation 10):

\begin{equation}
  R_J = \left(\frac{G\,M_{\textrm{clu}}}{\lambda_{e,1}}\right)^{1/3},
  \label{eq:R_Jacobi}
\end{equation}

\noindent where $G$ is the gravitational constant, $M_{\textrm{clu}}$ is the cluster mass (Table~\ref{tab:investig_sample_compl}) and $\lambda_{e,1}$ is the largest eigenvalue of the tidal tensor, given by the expression (see section 2 of \citeauthor{Renaud:2011}\,\,\citeyear{Renaud:2011}):

\begin{equation}
   \lambda_{e,1} = -\left(\frac{\partial^2\phi_G}{\partial x'^2}\right)_{R_G} - \left(-\frac{\partial^2\phi_G}{\partial z'^2}\right)_{R_G}.
   \label{eq:eigenvalue_lambda1}
\end{equation}

\noindent The above partial derivatives are determined after expressing the total potential $\phi_G$ (see below) in terms of a right-handed coordinates system ($x', y', z'$) centered on the cluster and with the $x'$-axis oriented along the Galactic centre $-$ cluster direction. 

In the present paper, the MW gravitational potential ($\phi_G$) is considered a superposition of three components (e.g., \citeauthor{Haghi:2015}\,\,\citeyear{Haghi:2015}; \citeauthor{Darma:2021}\,\,\citeyear{Darma:2021}): a bulge ($\phi_B$), a disc ($\phi_D$) and a dark matter halo ($\phi_H$; this way, $\phi_G=\phi_B\,+\,\phi_D\,+\,\phi_H$), which we adopt from \cite{Hernquist:1990}, \cite{Miyamoto:1975} and \citeauthor{Sanderson:2017}\,\,(\citeyear{Sanderson:2017}; see also \citeauthor{Navarro:1996}\,\,\citeyear{Navarro:1996}), respectively. The bulge, disc and halo potentials can be modelled as

\begin{flalign}
    & \phi_B = -\frac{G\,M_B}{r+r_B}   \label{eq:phi_B}  & \\
    & \phi_D = -\frac{G\,M_D}{  \sqrt{\rho^2 + (a + \sqrt{z^2 + b^2})^2}    }   \label{eq:phi_D} &   \\
    & \phi_H = -\frac{G\,M_s}{(\textrm{ln}\,2 - 1/2)}\frac{\textrm{ln}(1+r/r_s)}{r}\,\,\,.   \label{eq:phi_H}   &      
\end{flalign}

\noindent In the above expressions, $r$ is the Galactocentric distance, $\rho$ and $z$ are the polar Galactic coordinates. The parameters $M_B=2.5\times10^{10}\,$M$_{\odot}$, $r_B=0.5\,$kpc, $M_D=7.5\times10^{10}\,$M$_{\odot}$, $a=5.4\,$kpc and $b=0.3\,$kpc were obtained from \cite{Haghi:2015}; in turn, the values for the scale radius $r_s=15.19\,$kpc and mass $M_s=1.87\times10^{11}\,$M$_{\odot}$ come from \cite{Sanderson:2017}.

Although equation~\ref{eq:eigenvalue_lambda1} is appliable to an arbitrary galactic potential, its analytical simplicity is restricted to circular orbits. This way, we applied the correction proposed by \cite{Webb:2013}, in order to take into account the effect of the orbital eccentricity ($\epsilon$) on the derived $R_J$ for the investigated OCs. The set of $\epsilon$ values were taken from \cite{Tarricq:2021} and are typically smaller than 0.1 (for the present sample, $\langle\epsilon\rangle$=0.08$\pm0.05$), which indicates nearly circular orbits.

In order to estimate the uncertainty in $R_J$, for each cluster we ran ten thousand redrawings allowing for variations in each parameter $\theta_i$ (with respective uncertainty $\Delta\,\theta_i$, where $\theta_1=M_{\textrm{clu}}$; $\theta_2=R_G$; $\theta_3=\epsilon$,...) entering in the above formulation, over the interval $\theta_i\pm\Delta\theta_i$. The $\Delta\,R_J$ value corresponds to twice the dispersion obtained after the whole redrawings procedure.

\subsection{Half-light relaxation time}

We have also derived the cluster half-light relaxation time, expressed as \citep{Spitzer:1971}:

\begin{equation}
   t_{rh}=(8.9\times10^5\,\textrm{yr})\,\frac{M_{\textrm{clu}}^{1/2}\,r_h^{3/2}}{\langle m\rangle\,\textrm{log}_{10}(0.4M_{\textrm{clu}}/\langle m\rangle)},
   \label{eq:trh}
\end{equation}

\noindent  where $\langle m\rangle=M_{\textrm{clu}}/N_{\textrm{clu}}$ (Table~\ref{tab:investig_sample_compl}). The half-light relaxation time can be interpreted as a dynamical timescale during which the stellar system tends to dynamical equilibrium, continuously (re)populating the high-velocity tail of its velocity distribution and, consequently, losing a given fraction of its stellar content to the field (e.g., \citeauthor{Portegies-Zwart:2010}\,\,\citeyear{Portegies-Zwart:2010}).

The derived values of $t_{rh}$ and $R_J$ for our sample, together with the corresponding uncertainties, are informed in Tables~\ref{tab:investig_sample} and \ref{tab:investig_sample_compl}, respectively.

\subsection{Initial mass estimates}
\label{sec:Mini_t95}

In order to investigate how stellar evolution, tidal forces and the internal relaxation have driven the stellar mass loss process and shaped the OCs structure, we have employed analytical expressions for the disruption of star clusters presented by LGPZ05 and LGB05. These formulas showed consistency with the outcomes of a large set of $N$-body simulations (BM03), which include stellar evolution and internal interactions in multimass star clusters, under the influence of an external tidal field. 

Based on the GALEV models (\citeauthor{Schulz:2002}\,\,\citeyear{Schulz:2002}; \citeauthor{Anders:2003}\,\,\citeyear{Anders:2003}) for simple stellar populations at different metallicities, LGB05 showed that the fraction of the initial cluster mass ($M_{\textrm{ini}}$) that is lost by stellar evolution is a function of time under the form

\begin{equation}
   \textrm{log}\,q_{\textrm{ev}} (t) = (\textrm{log}\,t - a_{\textrm{ev}})^{b_{\textrm{ev}}} + c_{\textrm{ev}}\,\,\,(t\,>\,12.5\,\textrm{Myr}),
   \label{eq:q_ev}
\end{equation}

\noindent where $q_{\textrm{ev}}=\Delta M_{\textrm{ev}}/M_{\textrm{ini}}$ and $\Delta M_{\textrm{ev}}$ is the mass lost by stellar evolution only. The $a_{\textrm{ev}}$, $b_{\textrm{ev}}$ and $c_{\textrm{ev}}$ coefficients depend only slightly on the cluster overall metallicity $Z$, as shown in table 1 of LGB05. The $Z$ values for our investigated OCs (obtained from the $[Fe/H]$ values in Table~\ref{tab:investig_sample}) were interpolated across these tabulated values and the proper coefficients were obtained. 


Describing the cluster mass loss rate due to both stellar evolution and dynamical effects under the form $(dM/dt) = (dM/dt)_{\textrm{ev}}+(dM/dt)_{\textrm{dyn}}$ and assuming that the disruption timescale is proportional to $M_{\textrm{ini}}^{\gamma}$, LGB05 showed that the mass decrease of a cluster can be well described by the following formula

\begin{equation}
   \mu(t;M_{\textrm{ini}})\equiv\frac{M_{\textrm{clu}}}{M_{\textrm{ini}}}\simeq\left\{  [\mu_{\textrm{ev}} (t)]^\gamma - \frac{\gamma}{M_{\textrm{ini}}^\gamma}\,\frac{t}{t_0} \right\}^{1/\gamma}.
   \label{eq:Mclu_Mini}
\end{equation}

\noindent In this expression, $\mu_{\textrm{ev}}\,(t)=1-q_{\textrm{ev}} (t)$, and $\gamma=0.62$ (LGPZ05; \citeauthor{Boutloukos:2003}\,\,\citeyear{Boutloukos:2003}; \citeauthor{de-Grijs:2005}\,\,\citeyear{de-Grijs:2005}); both $M_{\textrm{clu}}$ and $M_{\textrm{ini}}$ are expressed in $M_{\odot}$. This relation can easily be inverted to express $M_{\textrm{ini}}$ in terms of the present-day cluster mass, $M_{\textrm{clu}}$ (see equation 7 of LGB05). The constant $t_0$ depends on the tidal field of the particular galaxy in which the cluster moves and on the eccentricity ($\epsilon$) of its orbit. It can be expressed as (see section 2 of LGPZ05)  

\begin{equation}
    t_0 \simeq C_{\textrm{env,0}}\,(1-\epsilon)\,10^{-4\gamma}\,(\rho_{\textrm{amb}}/M_{\odot}\,\textrm{pc}^{-3})^{-0.5},
    \label{eq:t0}
\end{equation}

\noindent where $C_{\textrm{env,0}}\simeq810\,$Myr for clusters moving in the Galactic potential field and $\rho_{\textrm{amb}}$ is the ambient density evaluated at the apogalactic radius (LGPZ05), which can be found by applying Poisson's

\begin{equation}
    \rho_{\textrm{amb}}=\frac{1}{4\pi G}\nabla^2 [\phi_B(r)\,+\,\phi_D(\rho,z)\,+\,\phi_H(r)]\,\,, 
    \label{eq:rho_amb}
\end{equation}

\noindent where $\phi_B$, $\phi_D$ and $\phi_H$ are taken from equations~\ref{eq:phi_B}, \ref{eq:phi_D} and \ref{eq:phi_H}.

In order to obtain the uncertainty in $M_{\textrm{ini}}$ (equation~\ref{eq:Mclu_Mini}), a procedure analogous to that of $R_J$ (Section~\ref{sec:rJ})  was employed: $\Delta\,M_i$ corresponds to twice the dispersion obtained after a ten thousand random redrawings applied to each independ variable ($\theta_i$), over the corresponding associated error ($\Delta\,\theta_i$). Finally, the fraction of mass lost due to exclusively dynamical effects ($\Delta M_{\textrm{dyn}}/M_{\textrm{ini}}$) can be estimated from the expression

\begin{equation}
  \Delta M_{\textrm{dyn}}/M_{\textrm{ini}} = 1 - q_{\textrm{ev}} - (M_{\textrm{clu}}/M_{\textrm{ini}}).  
\end{equation}

\section{Discussion}
\label{sec:discussion}

\begin{table}
 \small
\begin{minipage}{85mm}
  \caption{ Symbol convention and colours used in Section~\ref{sec:discussion}. }
  \label{tab:symbols_convention}
 \begin{tabular}{ccccc}

  \hline
  \multicolumn{5}{c}{$R_G$ intervals (in kpc)}                                                                      \\ 
  \hline
  \hline	
  6.0$-$7.0         & 7.0$-$9.0               &  9.0$-$11.0               & 11.0$-$12.0              &              \\ 
  \hline
  \Huge{$\bullet$}  & \Large{$\blacksquare$}  & \Large{$\blacktriangle$}  & \Large{$\blacklozenge$}  &              \\                                                                                                                                                
  \hline
  \hline
  \multicolumn{5}{c}{ Colours (see Figure~\ref{fig:rh_rJ__logttrh0__Rgal}, panel \textit{b}) }                      \\ 
  \hline
  \textcolor{orange}{orange}   &  \multicolumn{4}{c}{\textcolor{orange}{$R_G\leq8.0\,$kpc}}                         \\  
                               &  \multicolumn{4}{c}{   }                                                           \\                                                          
  \textcolor{green}{green}     &  \multicolumn{4}{c}{\textcolor{green}{$8<R_G\,$(kpc)$\leq10$}}                     \\  
                               &  \multicolumn{4}{c}{\textcolor{green}{$r_h/R_J>0.35$ (\textit{looser group})}}     \\
                               &  \multicolumn{4}{c}{   }                                                           \\
  \textcolor{blue}{blue}       &  \multicolumn{4}{c}{\textcolor{blue}{$8<R_G\,$(kpc)$\leq10$}}                      \\ 
                               & \multicolumn{4}{c}{\textcolor{blue}{$r_h/R_J\leq0.35$ (\textit{compact group})}}   \\ 
                               &  \multicolumn{4}{c}{   }                                                           \\                                          
  \textcolor{purple}{purple}   &  \multicolumn{4}{c}{\textcolor{purple}{$R_G>10\,$kpc}}                             \\

\hline

\end{tabular}

\end{minipage}
\end{table}

\begin{figure*}
\begin{center}

\parbox[c]{1.0\textwidth}
  {
   \begin{center}
    \includegraphics[width=1.0\textwidth]{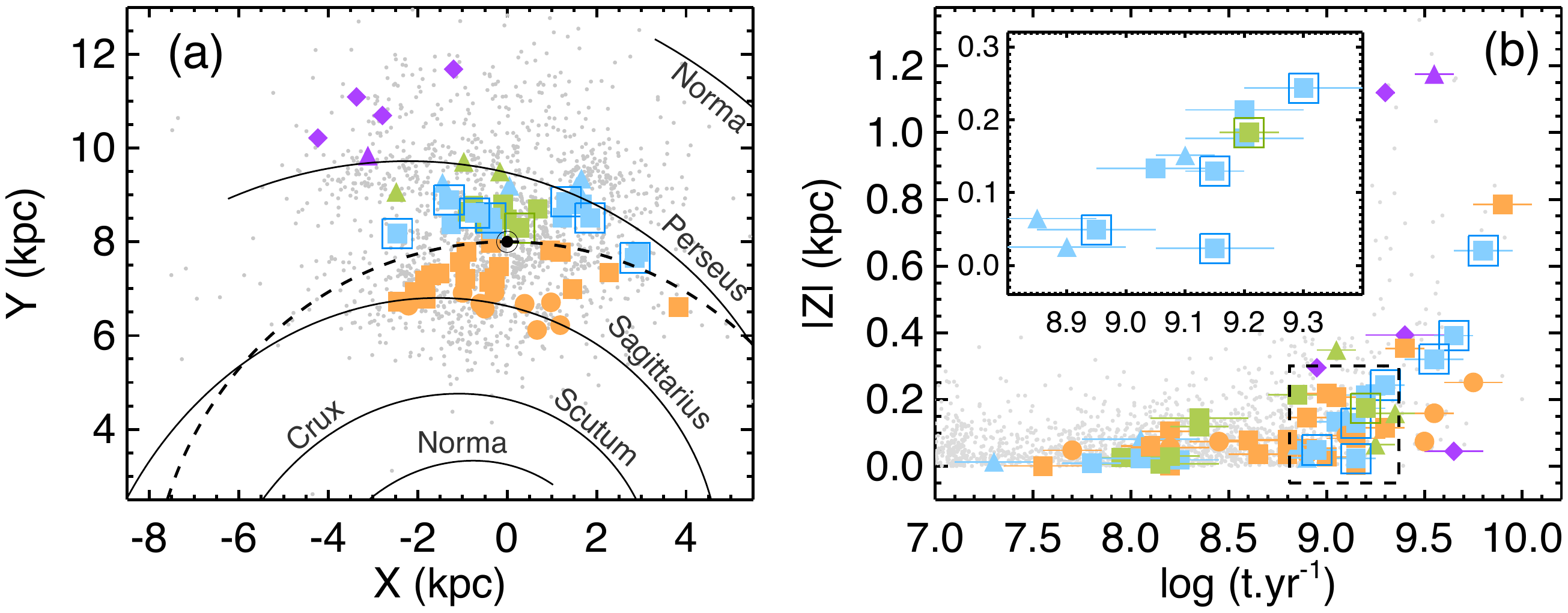}  
    \end{center}    
  }
\caption{  \textit{Panel (a)}: Disposal of the 60 investigated OCs along the Galactic plane. The position of the Sun and the solar circle are identified. The spiral pattern is also shown (Vall\'ee\,\citeyear{Vallee:1995}, \citeyear{Vallee:2008}). \textit{Panel (b):} Distance to the Galactic plane ($|Z|$) as function of log\,$t$. The inset highlights part of our sample containing nearly coeval OCs (dashed rectangle; see text for details). Symbol convention and colours for both panels are specified in Table~\ref{tab:symbols_convention}. The contoured blue squares refer to a subsample of dynamically evolved OCs (see details in the text following Figure~\ref{fig:rh_rJ__logttrh0__Rgal}). }

\label{fig:Galactic_plane_and_Zgal_age}
\end{center}
\end{figure*}

In the present section, we intend to investigate the evolutionary stages of the present sample by exploring possible connections among the structural parameters ($r_t$, $r_c$, $r_{h}$), relaxation time ($t_{rh}$), $R_J$ and $R_G$. Evolution-related parameters (ages, stellar masses) and estimates based on analytical description of the disruption of star clusters (Section~\ref{sec:Mini_t95}) are also employed. In the figures of this section, the investigated OCs have been categorized according to their $R_G$ and $r_h/R_J$ intervals, following the the symbol convention described in Table~\ref{tab:symbols_convention}, except when otherwise indicated.

Figure~\ref{fig:Galactic_plane_and_Zgal_age} shows the disposal of our sample along the Galactic plane (panel \textit{a}) and perpendicular to it (panel \textit{b}). The position of the spiral arms were taken from Vall\'ee\,(\citeyear{Vallee:1995}, \citeyear{Vallee:2008}). The inset in panel (b) highlights part of the investigated OCs (see Figure~\ref{fig:rh_rJ__logttrh0__Rgal} and the discussions following it). The 60 OCs are distributed along the four Galactic quadrants and those more distant from the Galactic disc tend to be older, following the overall trend of literature OCs (DMML21; small grey dots).

\begin{figure}
\begin{center}

\parbox[c]{0.45\textwidth}
  {
   \begin{center}
    \includegraphics[width=0.45\textwidth]{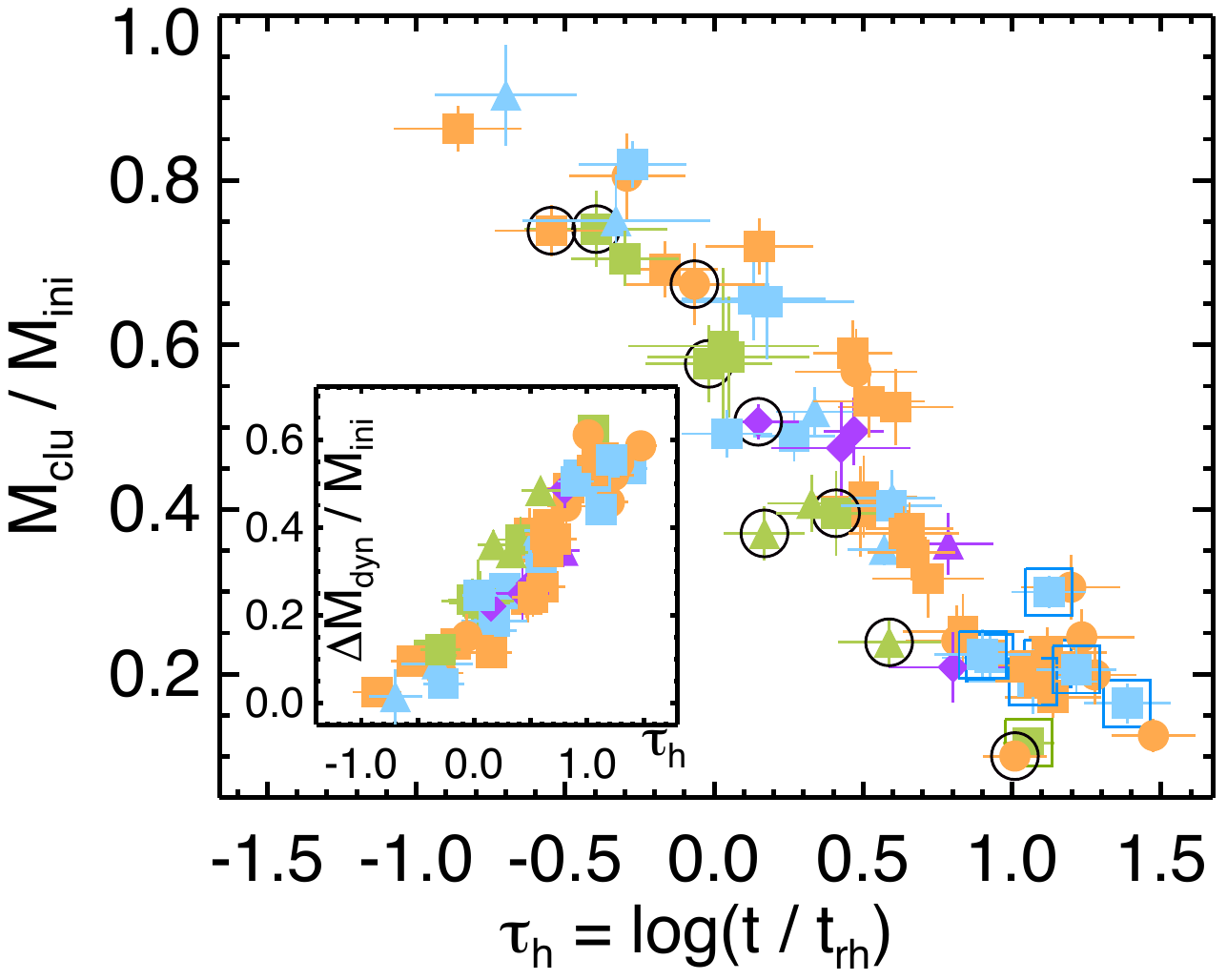}  
    \end{center}    
 }
\caption{ Present-day cluster mass fraction ($M_{\textrm{clu}}/M_{\textrm{ini}}$) as function of $\tau_h$ = log\,(age/$t_{rh}$). Symbols and colours identify $R_G$ and $r_h/R_J$ intervals, as detailed in Table~\ref{tab:symbols_convention} (see also Figure~\ref{fig:rh_rJ__logttrh0__Rgal}). The circled symbols represent tidally overfilled clusters with $r_t/R_J\gtrsim1.25$, as will be explained in Figure~\ref{fig:rt_Rj_versus_Rgal} and in the text following it. The inset shows the fraction of mass lost by purelly dynamical effects ($\Delta M_{\textrm{dyn}}/M_{\textrm{ini}}$) as function of $\tau_h$. }

\label{fig:Mclu_Mini_versus_logt_trh}
\end{center}
\end{figure}

Figure~\ref{fig:Mclu_Mini_versus_logt_trh} shows an anticorrelation between $M_{\textrm{clu}}/M_{\textrm{ini}}$ and the dynamical age, expressed as $\tau_h=\textrm{log}\,($age/$t_{rh}$), which can be assumed as a measure of how dynamically evolved a cluster is. In fact, since a fraction of the cluster mass is lost at each relaxation time \citep{Spitzer:1987}, the more its age surpasses $t_{rh}$, the greater the expected fraction of mass loss. Taking, for example, the more dynamically evolved OCs in our sample ($\tau_h\gtrsim0.3$, that is, age\,$\gtrsim2\,\times\,t_{rh}$), $\sim90\%$ of them present $M_{\textrm{clu}}/M_{\textrm{ini}}\lesssim0.5$, that is, significantly mass-depleted. In our case, all OCs older than $\sim170\,$Myr (log\,$t\gtrsim8.23$) are dynamically evolved (i.e., age/$t_{rh}\gtrsim1$). The inset in the same figure shows a positive correlation, as expected, between the fraction ($f_{\textrm{dyn}}$=$\Delta M_{\textrm{dyn}}/M_{\textrm{ini}}$) of mass lost exclusively due to dynamical effects and $\tau_h$. OCs with log\,$t\lesssim8$ in our sample (namely: NGC\,654, NGC\,1027, NGC\,3766, NGC\,6242 and NGC7654) are dynamically unevolved ($\tau_h < 0$) and present $f_{\textrm{dyn}}$ below $\sim0.1$. In this same age range, the expected mass loss by stellar evolution only ($q_{\textrm{ev}}$; equation~\ref{eq:q_ev}) is larger, varying from $\sim$0.1 to $\sim0.15$.

\begin{figure}
\begin{center}

\parbox[c]{0.40\textwidth}
  {
   \begin{center}
       \includegraphics[width=0.40\textwidth]{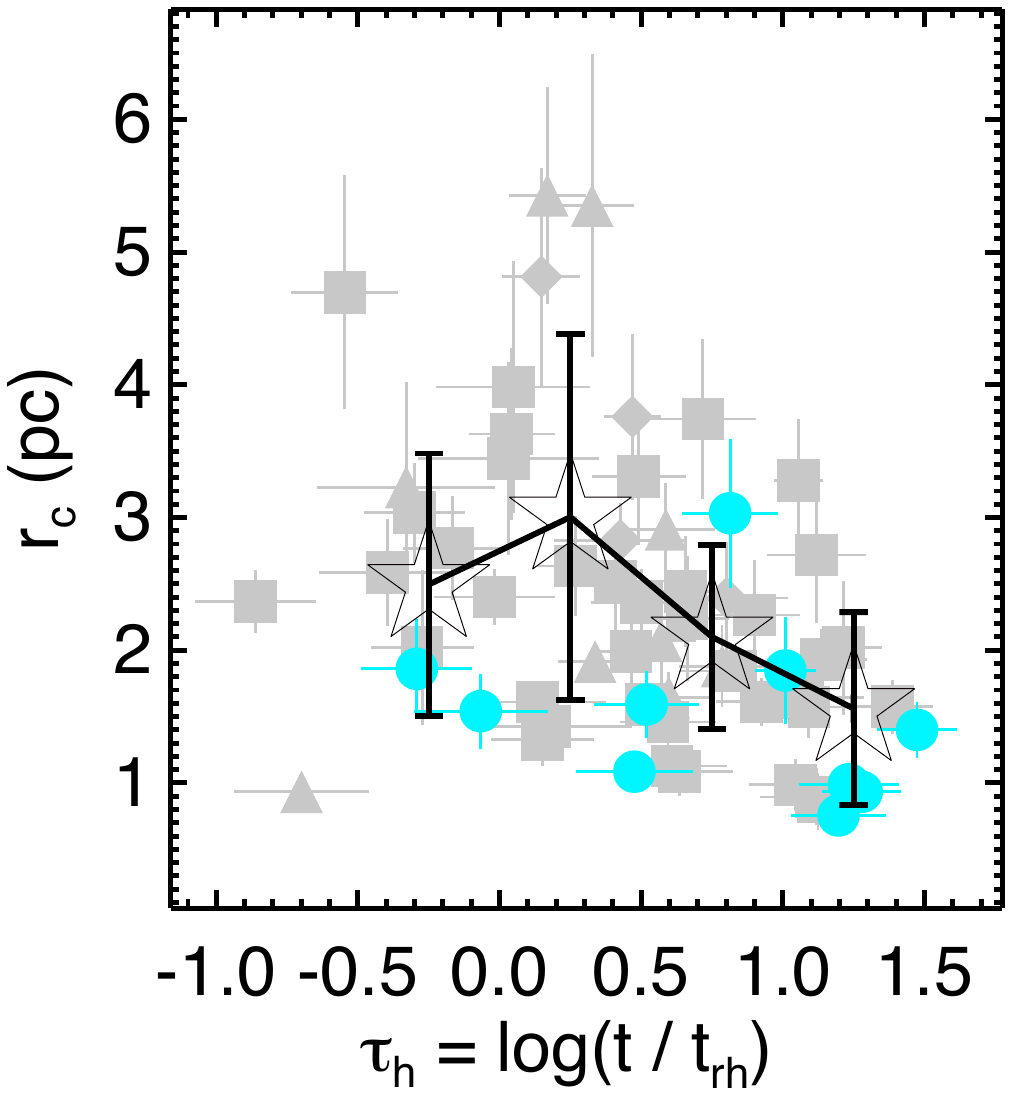}  
    \end{center}    
  }

\caption{ $r_c$ versus $\tau_{h}$(=log\,($t/t_{rh}$)) for the investigaged sample. The cyan symbols highlight the OCs located at $R_G\lesssim7\,$kpc; other OCs are plotted in grey colour. The open stars and the associated error bars represent, respectively, the mean and dispersion of the $r_c$ values for OCs within 4 bins: $\tau_h<0.0$, $\tau_h$ between 0.0$-$0.5, $\tau_h$ between 0.5$-$1.0 and $\tau_h>1.0$. }


\label{fig:rc_versus_tauh}
\end{center}
\end{figure}

Figure~\ref{fig:rc_versus_tauh} shows the plot $r_c$ as function of the dynamical age. The set of investigated clusters have been grouped in 4 bins of $\tau_h$: $\tau_h<0.0$, $0.0\le\tau_h<0.5$, $0.5\le\tau_h<1.0$ and $\tau_h\ge1.0$). Within each bin, the mean and dispersion of the $r_c$ values have been determined and indicated in the figure by, respectively, the vertical position of the large open stars and the corresponding error bars. For those clusters presenting signals of dynamical evolution (i.e., $\tau_h\gtrsim0$), we can note a general trend in which both the mean $r_c$ values and the associated dispersion tend to decrease slightly with $\tau_h$.

This way, $r_c$ seems to shrink along the cluster dynamical evolution, which suggests a progressively larger degree of compactness of the cluster central mass distribution. This result may be interpreted as a consequence of the migration of higher mass stars to the cluster's core due to two-body interactions (\citeauthor{Heggie:2003}\,\,\citeyear{Heggie:2003}; \citeauthor{Portegies-Zwart:2010}\,\,\citeyear{Portegies-Zwart:2010}), making the central parts denser. As stated by \cite{Chen:2004}, this effect also increases the core's sphericity.

In this same sense, \cite{Tarricq:2022} verified a systematic decrease of both $r_c$ and its dispersion as function of cluster age for a sample of 389 local (distance $\lesssim$ 500\,pc) OCs. Their figure 6 demonstrates that, even though there are young OCs that can have very concentrated cores, this feature is more common for evolved ones. In this context, the role of the external Galactic potential can not be ruled out given the results shown in Figure~\ref{fig:rc_versus_tauh}. Those OCs located at $R_G\leq7\,$kpc (coloured filled circles) tend to be concentrated in the bottom part of the plot and 8 out of 10 present $\tau_h\gtrsim0.0$. Apparently, their greater proximity to the Galactic centre may have accelerated their dynamical evolution (GB08; \citeauthor{Piatti:2019a}\,\,\citeyear{Piatti:2019a}; \citeauthor{Vesperini:2010}\,\,\citeyear{Vesperini:2010}; see also Figure~\ref{fig:rh_rJ__logttrh0__Rgal} below), since more compact central structures tend to result in smaller dynamical timescales (e.g., $t_{rh}$), thus speeding up the escape rate of stars and the process of mass segregation (MHS12; \citeauthor{Spitzer:1969}\,\,\citeyear{Spitzer:1969}).

\begin{figure}
\begin{center}

\parbox[c]{0.49\textwidth}
  {
   \begin{center}
    \includegraphics[width=0.49\textwidth]{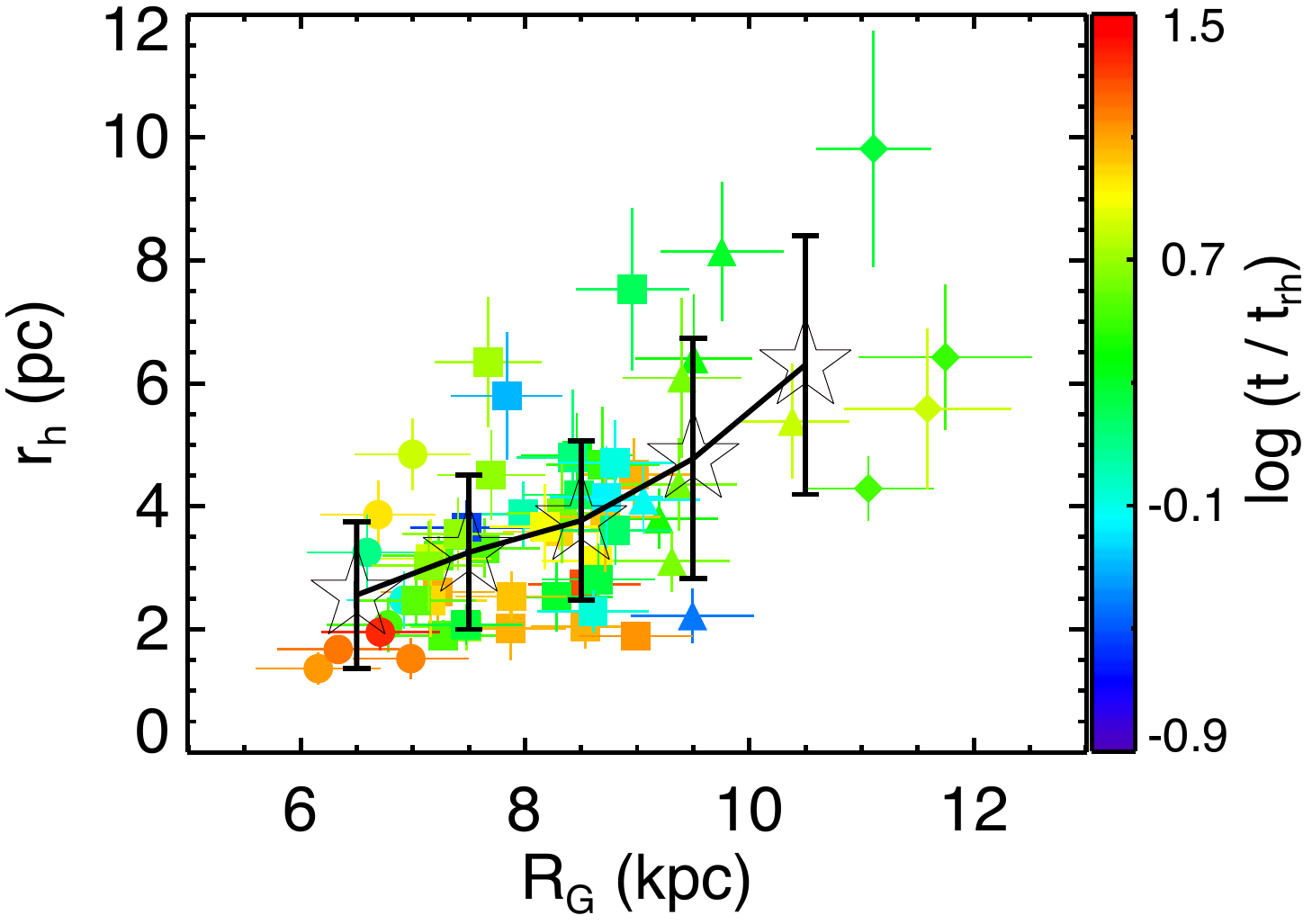}  
    \end{center}    
  }

\caption{ Half-light radius ($r_h$) as function of the Galactocentric distance; the symbol colours indicate the degree of dynamical evolution (as inferred from the log\,($t/t_{rh}$) ratio). }


\label{fig:rh_versus_Rgal}
\end{center}
\end{figure}


Figure~\ref{fig:rh_versus_Rgal} exhibits a positive correlation, although with some dispersion, between $r_h$ and $R_G$. This trend suggests that OCs located at larger $R_G$ are allowed to relax their internal mass distribution across larger dimensions, within the allowed tidal volume (as given by the Jacobi radius; equation~\ref{eq:R_Jacobi}), without being tidally disrupted. This result is consistent with the outcomes from $N$-body simulations performed by \citeauthor{Miholics:2014}\,\,(\citeyear{Miholics:2014}; see, e.g., their figure 1), who showed that, at a given age, simulated clusters submitted to a weaker external potential present larger $r_h$. Additionally, in Figure~\ref{fig:rh_versus_Rgal} we can note a preferential concentration of the more dynamically evolved clusters at smaller $R_G$. It is expected that the greater proximity to the Galactic centre increases the evaporation rate (GB08) and thus the fraction of mass loss due to dynamical effects (see also the inset in Figure~\ref{fig:Mclu_Mini_versus_logt_trh}).

\begin{figure*}
\begin{center}

\parbox[c]{01.00\textwidth}
  {
   \begin{center}
    \includegraphics[width=1.00\textwidth]{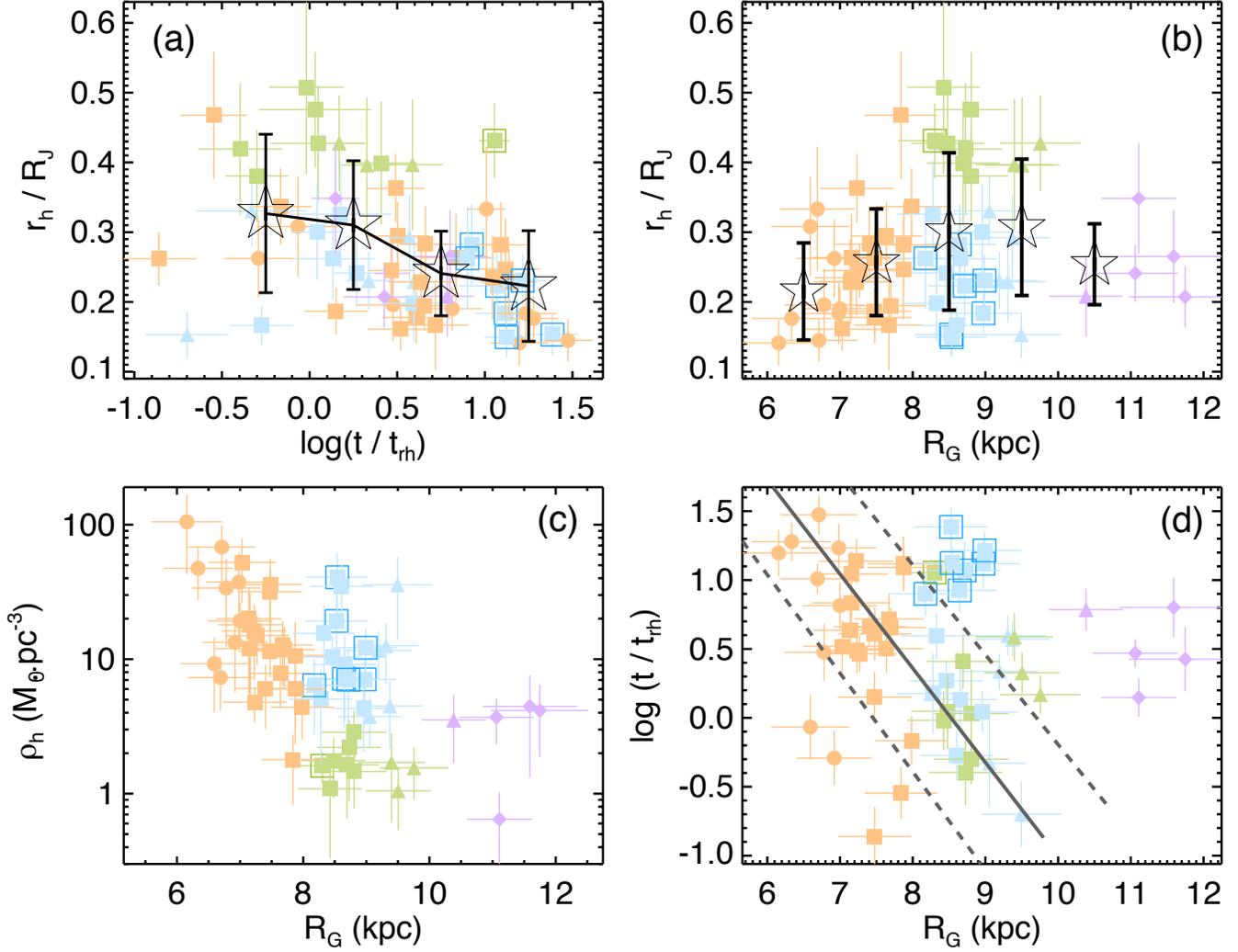}  
    \end{center}    
  }
\caption{ \textit{Panel (a)}: $r_h/R_J$ ratio as function of the dynamical age. The open stars and the associated error bars represent, respectively, the mean and dispersion of the $r_h/R_J$ values for OCs within the same $\tau_h$ bins as those of Figure~\ref{fig:rc_versus_tauh}. The symbol colours and the contoured ones are explained in the other panels. \textit{Panel (b)}: $r_h/R_J$ ratio as function of $R_G$. The investigated sample was separated in 4 subsamples, identified with different colours (see text for details). The open stars with error bars represent the mean and dispersion of the $r_h/R_J$ values within 5 $R_G$ bins: $R_G\leq7.0\,$kpc, $7.0<R_{G}\,(\textrm{kpc})\leq8$, $8<R_{G}\,(\textrm{kpc})\leq9$, $9<R_{G}\,(\textrm{kpc})\leq10$ and $R_G>10\,$kpc. \textit{Panel (c)}: half-light density ($\rho_h$) as function of $R_G$. In the range of $R_G$ between $\sim6-10\,$kpc, OCs plotted with orange and green colours follow a decreasing trend. \textit{Panel (d)}: Dynamical age versus $R_G$. Symbol colours are the same of panel (b). The contoured symbols identify dinamically evolved OCs with $\tau_h>0.7$ and located at $R_G$ in the range $8-10\,$kpc. The continuous line was plotted just to guide the eye and it represents a linear fit to the data of clusters located at $R_G\lesssim10\,$kpc (the dashed lines represent the uncertainties).}

\label{fig:rh_rJ__logttrh0__Rgal}
\end{center}
\end{figure*}


The Roche volume filling factor $r_{h}/R_J$ provides some insights regarding the dynamical state of a cluster subject to an external potential, since it indicates how tidally filling a cluster is (e.g., \citeauthor{Santos:2020}\,\,\citeyear{Santos:2020}). Once a cluster is adjusted to the given tidal conditions, the mass loss process is driven by stellar evolution and internal relaxation, regulated by the external potential, which sets the limits for the cluster expansion at each position within the Galaxy. It is therefore useful to check some possible connections between the $r_{h}/R_J$ ratio as function of both age/$t_{rh}$ and $R_G$.

Despite the large dispersion of the data, panel (a) of Figure~\ref{fig:rh_rJ__logttrh0__Rgal} shows an apparent overall decrease of $r_{h}/R_J$ with the dynamical age (as suggested by the open stars, which highlight the mean $r_h/R_J$ values in different $\tau_h$ bins), that is, progressively more compact internal structure as the system becomes dynamically older, analogously to what was verified for the $r_c$ shrinking (Figure~\ref{fig:rc_versus_tauh}). The data becomes slightly less dispersed for the dynamically older sample ($\tau_h\gtrsim0.5$) in comparison to the dynamically younger OCs ($\tau_h\lesssim0.5$). Panel (b), in turn, shows the $r_h/R_J$ ratio as function of $R_G$. The loci of observed data suggests that the Galactic tidal field may have impacted the OCs dynamical evolution. A positive correlation between the plotted quantities is verified for OCs in the range $R_G\lesssim8\,$kpc (orange symbols) and the set of $r_h/R_J$ values become more dispersed for larger $R_G$.   



For smaller $R_G$, the stronger external gravitational field may have been more effective in shaping the OCs' mass distribution and therefore accelerating their mass loss process, since those located at inner orbits in the Galaxy tend to be dynamically older (larger age/$t_{rh}$ ratio, panel $d$) and therefore to present smaller $M_{\textrm{clu}}/M_{\textrm{ini}}$ (following the general trend of Figure~\ref{fig:Mclu_Mini_versus_logt_trh}) compared to their larger $R_{G}$ counterparts. On the other side, clusters subject to less intense external potential can occupy larger fractions of the allowed tidal volume. Our investigated sample presents maximum $r_h/R_J$ value of about $\sim$0.5; this upper limit is consistent with BPG10's results, who verified that globular clusters with large $r_h/R_J$ ($\gtrsim0.5$) are basically absent in their sample, since such clusters would be subject to strong tidal forces and have small dissolution times. Despite this, cases of even more tidally influenced clusters (presenting $r_h/R_J\gtrsim0.5$) are reported in the literature, e.g., in the Small Magellanic Cloud (figure 16 of \citeauthor{Santos:2020}\,\,\citeyear{Santos:2020}).



In the range $R_G\sim8-10\,$kpc, we see a dichotomy in the distribution of $r_h/R_J$ (panel (b) of Figure~\ref{fig:rh_rJ__logttrh0__Rgal}), which suggests two regimes of dynamical evolution. Interestingly, an analogous separation was verified by BPG10 for globulars, although with different $r_h/R_J$ (in their case, a group of clusters with $r_h/R_J<0.05$ and other one with $0.07 < r_h/R_J < 0.3$; their figure 2). Clusters represented by light green points\footnote[8]{Namely, in ascending order of $\rmn{RA}$ (Table~\ref{tab:investig_sample}): NGC\,752, NGC\,1027, NGC\,1647, NGC\,1817, NGC\,2168, Collinder\,110, NGC\,2287, NGC\,2353, NGC\,2539 and Haffner\,22.} (\textit{looser group}; see Table~\ref{tab:symbols_convention}) in Figure~\ref{fig:rh_rJ__logttrh0__Rgal} present $r_{h}/R_J\gtrsim0.35$ and are therefore more tidally influenced compared to the more \textit{compact group} ($r_{h}/R_J\lesssim0.35$; see Table~\ref{tab:symbols_convention}), identified with light blue symbols\footnote[9]{Namely, in ascending order of $\rmn{RA}$ (Table~\ref{tab:investig_sample}): NGC\,129, NGC\,188, NGC\,559, NGC\,654, M\,37, NGC\,2323, NGC\,2360, NGC\,2422, Melotte\,71, NGC\,2432, NGC\,2477, NGC\,2660, M\,67, Berkeley\,89, NGC\,7044, NGC\,7142, NGC\,7654, NGC\,7789.}. This difference between both groups may be, at least partially, attributed to different clusters' masses, since the median of the present-day masses for the more compact group is $M_{\textrm{clu}}\sim3200\,M_{\odot}$, which is $\sim$2.5 times larger than the median of $M_{\textrm{clu}}$ for the \textit{looser group}. Since these both subsamples are at comparable $R_G$, those OCs with larger masses (and therefore larger half-light density, for comparable $r_h$ values) may have their evolution more importantly driven by the internal relaxation and seem to be more stable against tidal disruption. 

As shown in panel (c) of Figure~\ref{fig:rh_rJ__logttrh0__Rgal}, the OCs located at $R_G\le8\,$kpc (orange symbols), together with those ones in the \textit{looser group}, present an overall decrease of their half-light densities ($\rho_h$) as function of $R_G$. Due to their typically smaller $\rho_h$, clusters in the \textit{looser group} may be more affected by tidal stresses. Indeed, as stated by GB08, the larger the $r_h/R_J$, the larger the expected fraction of stars lost by evaporation at each $t_{rh}$ for clusters in the tidal regime ($r_h/R_J\gtrsim0.05$, as is the case of all investigated OCs in our sample). The difference in $r_h/R_J$ between the \textit{more compact} and the \textit{looser} groups of OCs, even in the case of clusters at comparable dynamical stage (as inferred from their age/$t_{rh}$ ratio; panels $a$ and $d$) and compatible Galactocentric distances, suggests that the clusters' initial formation conditions may also play a role. This statement is particularly true in the case of NGC\,654 and NGC\,7654, both dynamically unevolved systems (log\,(age/$t_{rh}$)$\lesssim-0.3$) with relatively low $r_h/R_J$ ($\lesssim0.13$) ratio, which means that both may have been compact at birth. 

Panel (d) of Figure~\ref{fig:rh_rJ__logttrh0__Rgal} shows that, in the range $R_G\lesssim10\,$kpc, clusters located at smaller $R_G$ tend to be in a more advanced dynamical stage, following the general trend indicated by the continuous grey line (linear fit to these data, with the corresponding uncertainties also shown). As stated before, it seems that the stronger external potential possibly made their dynamical evolution differentially faster. OCs represented by light blue symbols (the \textit{compact group} in panel $b$) can be divided in two subgroups: 7 dynamically older OCs ($\tau_h\,>\,0.7$; $M_{\textrm{clu}}/M_{\textrm{ini}}$ between $\sim0.16 - 0.30$; see Figure~\ref{fig:Mclu_Mini_versus_logt_trh}) and 11 OCs with $\tau_h\,<\,0.7$ (for which the $M_{\textrm{clu}}/M_{\textrm{ini}}$ ratios are between $\sim 0.35 - 0.90$). Since these two subgroups are located at similar $R_G$, they are exposed to similar external tidal conditions. From the inset in Figure~\ref{fig:Galactic_plane_and_Zgal_age} (panel $b$), it is noticeable that part of these two subgroups comprise similar age and $|Z|$ ranges, therefore suggesting that differences in the dynamical stage among these two subgroups may be traced back to their formation conditions. Analogous statements may be drawn for the OC NGC\,752 (contoured light green symbol), the only member of the \textit{looser group} (panel $b$) among the more dynamically evolved OCs (log$(\textrm{age}/t_{rh})_{\textrm{NGC\,752}}\simeq1.1$).


The purple symbols represent OCs located at $R_G\,\gtrsim\,10\,$kpc (namely, NGC2141 NGC\,2204, NGC\,2243, Berkeley\,36 and Haffner\,11); due to the low number of objects in this range, no general statements can be drawn, except that they present signals of dynamical evolution, since all are older ($t\gtrsim890\,$Myr) than their respective $t_{rh}$ and seem to have lost more than $\sim50\%$ of their initial mass. Panels (a) and (b) of Figure~\ref{fig:rh_rJ__logttrh0__Rgal} suggest that they can accomodate their stellar content across different percentages of their Roche volume, without being severely shaped by the (weaker) external potential. This statement is particularly true in the case of NGC\,2204, which is relatively massive ($M_{\textrm{clu}}\approx5100\,M_{\odot}$, among the 25\% more massive clusters of our sample) and presents an extended structure ($r_{h}=9.8\,$pc, the largest in our sample; see Figure~\ref{fig:rh_versus_Rgal}), being the less dense ($\rho_{h}$\,$\simeq$\,0.65\,$M_{\odot}/$pc$^3$; see Figure~\ref{fig:rh_rJ__logttrh0__Rgal}, panel $c$) among the investigated OCs.

\begin{figure}
\begin{center}

\parbox[c]{0.45\textwidth}
  {
   \begin{center}
    \includegraphics[width=0.45\textwidth]{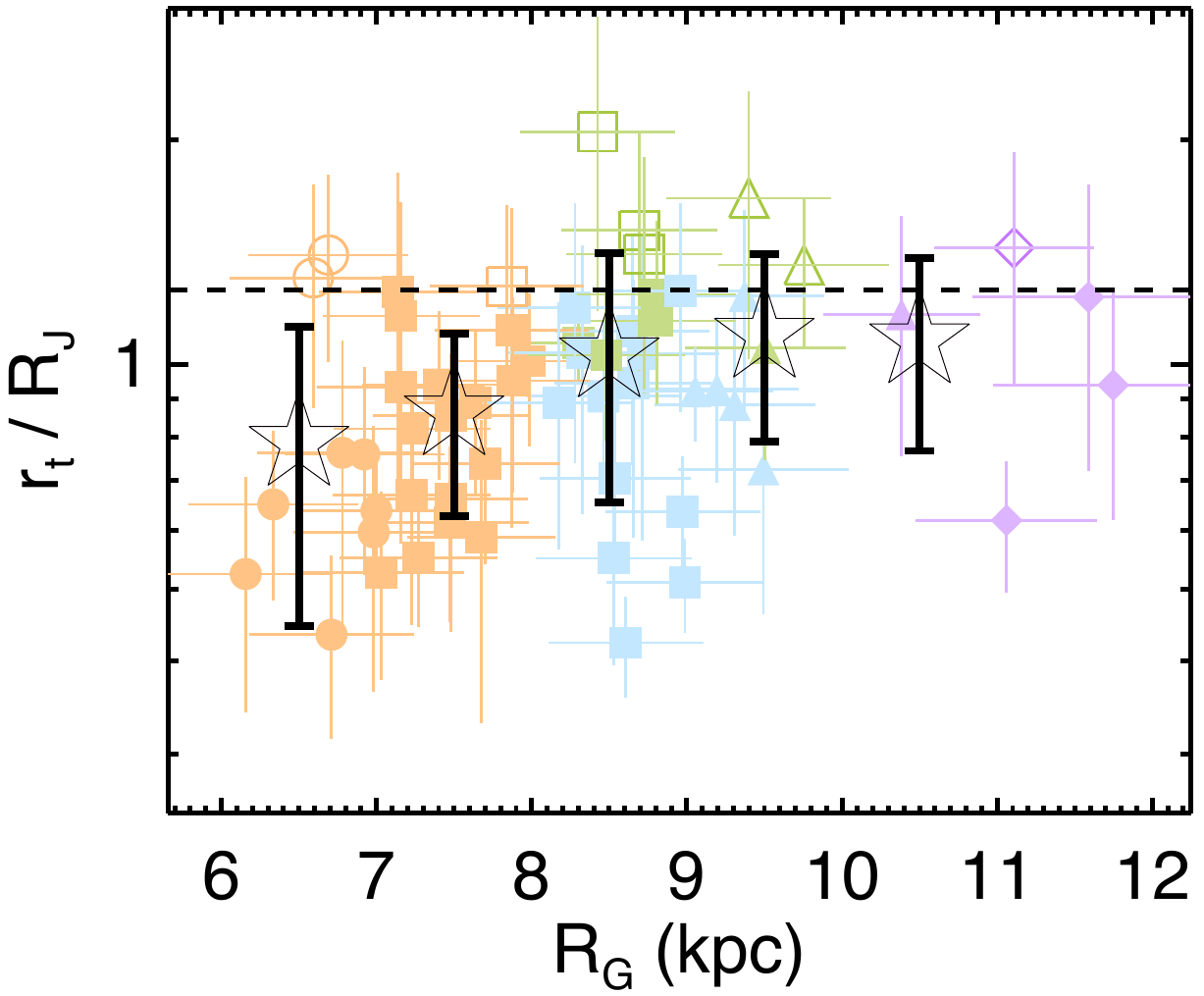}  
    \end{center}    
 }
\caption{ Roche volume filling factor, expressed as $r_t/R_J$, as function of $R_G$. The open symbols indicate OCs with $r_t/R_J\gtrsim1.25$ (dashed line; see also the circled points in Figure~\ref{fig:Mclu_Mini_versus_logt_trh}). The open stars represent the mean values of $r_t/R_J$ for OCs in the same $R_G$ bins as those of panel (b) in Figure~\ref{fig:rh_rJ__logttrh0__Rgal}; the associated dispersions are indicated by the error bars. }

\label{fig:rt_Rj_versus_Rgal}
\end{center}
\end{figure}

Figure~\ref{fig:rt_Rj_versus_Rgal} allows to verify how the clusters' external structure is affected by variations in the external Galactic potential. To accomplish this verification, we have plotted the Roche volume filling factor, expressed as the tidal to Jacobi radius ratio ($r_t/R_J$), as function of the Galactocentric distance. Symbols and colours are the same of previous panels, except for those OCs with significant extra-tidal components (in our case, $r_t/R_J\gtrsim1.25$; see below), represented by open symbols (namely, in ascending order of $R_G$: NGC\,1027, NGC\,2204, Collinder\,110, NGC\,2287, NGC\,2539, Haffner\,22, NGC\,3114, NGC\,6192 and Ruprecht\,171).



The mean values of $r_t/R_J$ (as indicated by the open stars) suggest an overall positive correlation between both plotted quantities in Figure~\ref{fig:rt_Rj_versus_Rgal} for $R_G\lesssim9\,$kpc. In this range, clusters at inner orbits are more subject to the truncating effects imposed by the stronger Galactic tidal field and their more compact structure favour their survival against tidal stripping. In the range of $R_G$ between $\sim6-9\,$kpc, OCs can occupy a progressively larger fraction of their Roche lobe, as they are submitted to a weaker external tidal field, up to the point of being tidally filled/overfilled.

Interestingly, the circled symbols in Figure~\ref{fig:Mclu_Mini_versus_logt_trh} (which represent tidaly overfilled OCs with $r_t/R_J\gtrsim1.25$) with $\tau_h\gtrsim0.0$ tend to occupy the lower envelope of points in the plot, that is, for a given $\tau_h$, the circled symbols tend to be slightly displaced towards smaller $M_{\textrm{clu}}/M_\textrm{ini}$ ratios compared to the corresponding OCs at compatible dynamical stage. This result may be justified from the fact that clusters with $r_t/r_J\gtrsim1$ are more susceptible to tidal effects leading to mass-loss (e.g., \citeauthor{Heggie:2003}\,\,\citeyear{Heggie:2003}; \citeauthor{Ernst:2015}\,\,\citeyear{Ernst:2015}). Besides, from the outcomes of $N$-body simulations, GB08 show that, for a given set of initial conditions, tidally filled OCs are expected to survive for a lesser number of initial relaxation times than lobe Roche underfilling ones due to increased mass loss. It is important to note that the presence of extra-tidal components may be due to energetic stars changing their status from bound to unbound or due to stars being recapturated by the cluster (see \citeauthor{Fukushige:2000}\,\,\citeyear{Fukushige:2000}, who pointed out that potential escapers can remain gravitationally bound due to the presence of, e.g., temporary periodic orbits close to the cluster outskirts; their section 3.2).

Based on the outcomes presented in this section, it becomes clear that the clusters' structural parameters can not be considered as single functions of, e.g., time. It is the interplay between formation conditions, stellar mass loss, internal relaxation and the influence of the external tidal field that determine the state of a cluster at a given age. These different aspects should be taken into account when searching for evolutionary connections between the clusters structure, position within the Galaxy and the dynamical timescales.

\section{Summary and concluding remarks}
\label{sec:conclusions}

In the present work, we characterized the dynamical state of a set of 60 Galactic OCs covering moderately large ranges in age (7.2$\,\lesssim\,$log($t.$yr$^{-1}$)$\,\lesssim\,$9.8) and Galactocentric distance (6$\,\lesssim\,R_G$(kpc)$\,\lesssim\,$12). We benefited from the high-precision astrometric and photometric data extracted from the most updated version of the \textit{Gaia} catalogue (Data Release 3), supplemented with the now available spectroscopic information for stars in the areas of the investigated clusters. This set of data, together with a decontamination algorithm that assigns membership probabilities for cluster stars, allowed us to establish optimized member star lists, thus improving the determination of the OCs astrophysical parameters. The set of results obtained in this way were complemented with parameters taken from analytical expressions that describe the disruption of star clusters in tidal fields, which are based on the outcomes of $N$-body simulations. This strategy allowed us to estimate initial masses ($M_{\textrm{ini}}$) and the fraction of mass loss due to dynamical interactions ($\Delta\,M_{\textrm{dyn}}/M_{\textrm{ini}}$). 


We pursued a comprehensive view on the set of structural parameters associated with the internal evolution (two-body relaxation) and also with the tidal conditions in which a stellar system is immersed. The analysis of the dispersion of the core radii revealed a shrinking of the $r_c$ values as function of the dynamical age (=$t/t_{rh}$), which may be interpreted as a consequence of the migration of higher mass stars to the cluster’s core due to two-body interactions. During this process, it
was shown that the external tidal field plays a fundamental role, since OCs located at smaller $R_G$ tend to present
smaller $r_c$ values and larger dynamical ages.

Analogously to the $r_c$, the tidal filling ratio, expressed as $r_h/R_J$, also presents an apparent anticorrelation with the dynamical age, which means that the clusters' central structure tend to be denser as they become dynamically older. Regarding its dependence with the external tidal field, we found some distinct groups that suggest apparent differing evolutionary regimes: ($i$) in the range of $R_G$ between $\sim6\,$-$\,8\,$kpc, there is a slight positive correlation between $r_h/R_J$ and $R_G$, thus suggesting that more compact internal structures favour the OCs survival against the more intense external tidal stresses; ($ii$) in the range of $R_G$ between $\sim8$\,-\,10\,kpc, there is a dichotomy in the distribution of $r_h/R_J$ values, in the sense that more massive OCs are also denser and thus less subject to tidal stripping; consequently, their evolution seem more importantly driven by the internal relaxation process compared to looser OCs at similar $R_G$. In general, the Galactic potential seems to have crucial importance to the clusters dynamical evolution, since those at inner orbits tend to present larger dynamical ages and, consequently, larger fractions of mass lost due to dynamical effects (i.e., larger $\Delta M_{\textrm{dyn}}/M_{\textrm{ini}}$).

For $R_G\gtrsim10\,$kpc, the low number of investigated OCs precludes conclusive statements; despite this, the dispersion in $r_h/R_J$ and $\rho_h$ at this more external regions suggests that such clusters can relax their internal mass content across considerable fractions of their Roche volumes, without being tidally disrupted. Given the weaker tidal field these clusters experience, those with more compact structures may have formed quite compact and/or their internal mass distribution may have been shaped by their internal evolution. 

In turn, the external clusters' structure seems noticeably affected by the Galactic tidal field. There is an apparent positive correlation between the Roche volume filling factor ($r_t/R_J$) and $R_G$ in the range $R_G\lesssim9\,$kpc, where OCs seem more subject to the truncating effects imposed by the Galactic tidal field. No trends are verified for larger $R_G$, at least for the presently analysed sample. We identified 9 significantly tidally overfilled clusters, for which $r_t/R_J\gtrsim1.25$. Interestingly, these stellar groups tend to present smaller $M_{\textrm{clu}}/M_{\textrm{ini}}$ ratios compared to the their counterpars at similar dynamical ages. This trend may be consequence of increased mass loss process due to tidal stripping. 


Investigations based on observed data can indeed benefit from the outcomes of theoretical works (e.g., based on $N$-body simulations) devoted to a detailed description of the physical processes that drive clusters evolution and, at the same time, provide empirical constraints to such models. The increase in the number of OCs characterized in detail and with uniform analysis procedures, together with the progressive improvements in the precision of astrometric, photometric and spectroscopic data (either from the \textit{Gaia} mission or from ground-based telescopes), will shed more light on the debated topic of clusters dissolution and contribute to the general knowledge of the Milky Way dynamical and chemical evolution.


\section{Acknowledgments}

We thank the anonymous referee for a detailed critical review, which helped to improve the quality and clarity of the paper. The authors acknowledge financial support from Conselho Nacional de Desenvolvimento Científico e Tecnológico - CNPq (proc. 404482/2021-0). F.F.S.M. acknowledges financial support from FAPERJ (proc. E-26/201.386/2022 and E-26/211.475/2021) This research has made use of the VizieR catalogue access tool, CDS, Strasbourg, France. This research has made use of the SIMBAD database, operated at CDS, Strasbourg, France. This work has made use of data from the European Space Agency (ESA) mission \textit{Gaia} (https://www.cosmos.esa.int/gaia), processed by the \textit{Gaia} Data Processing and Analysis Consortium (DPAC, https://www.cosmos.esa.int/web/gaia/dpac/consortium). Funding for the DPAC has been provided by national institutions, in particular the institutions participating in the \textit{Gaia} Multilateral Agreement. This research has made use of \textit{Aladin sky atlas} developed at CDS, Strasbourg Observatory, France.

\subsection*{Data availability}
\textit{The data underlying this article are available in the article and in its online supplementary material.}

\bibliographystyle{mn2e}
\bibliography{referencias}

\newpage
\newpage
\newpage

\appendix

\section{Contribution of dark stellar remnants to the total cluster mass}
\label{sec:mass_dark_remnants}

The total masses ($M_{\textrm{clu}}$) informed in Table~\ref{tab:investig_sample_compl} incorporate estimates of upper limit of mass possibly retained in the cluster under the form of white dwarfs (WDs), neutron stars (NSs) and stellar black holes (BHs), whose contribution to $M_{\textrm{clu}}$ depends on the cluster age.

For each cluster age and metallicity, we determined the PARSEC model with maximum initial mass still present in the corresponding isochrone ($m_0^{\textrm{isoc}}$). Then we integrated the normalized initial mass function of Kroupa (see Section~\ref{sec:mass_functions}) in the range $m_0^{\textrm{isoc}}\,\lesssim\,m_0\,(\textrm{M}_{\odot})\lesssim100\,\textrm{M}_{\odot}$. That is, stars whose initial mass $m_0$ are higher than $m_0^{\textrm{isoc}}$ evolved to a final state characterized by stellar remannts, namely: ($i$) WDs, for $m_0 < 8\,\textrm{M}_{\odot}$, ($ii$) NSs, for $8\,\leq\,m_0(\textrm{M}_{\odot})\,<\,30$, ($iii$) BHs, for $m_0\,\ge\,30\,M_{\odot}$.

For these estimates, we employed the relations of \cite{Kruijssen:2009}, who provided analytic expressions that link the star’s initial mass ($m_0$) with the correspondent remnant mass:

\begin{flalign}
  &  m_{WD} = 0.109\,m_0 + 0.394   & \\
  &  m_{NS} = 0.03636\,(m_0 - 8) + 1.02  &  \\
  &  m_{BH} = 0.06\,(m_0 - 30) + 8.3   &      
\end{flalign}

\begin{figure}
\begin{center}

\parbox[c]{0.45\textwidth}
  {
   \begin{center}
    \includegraphics[width=0.45\textwidth]{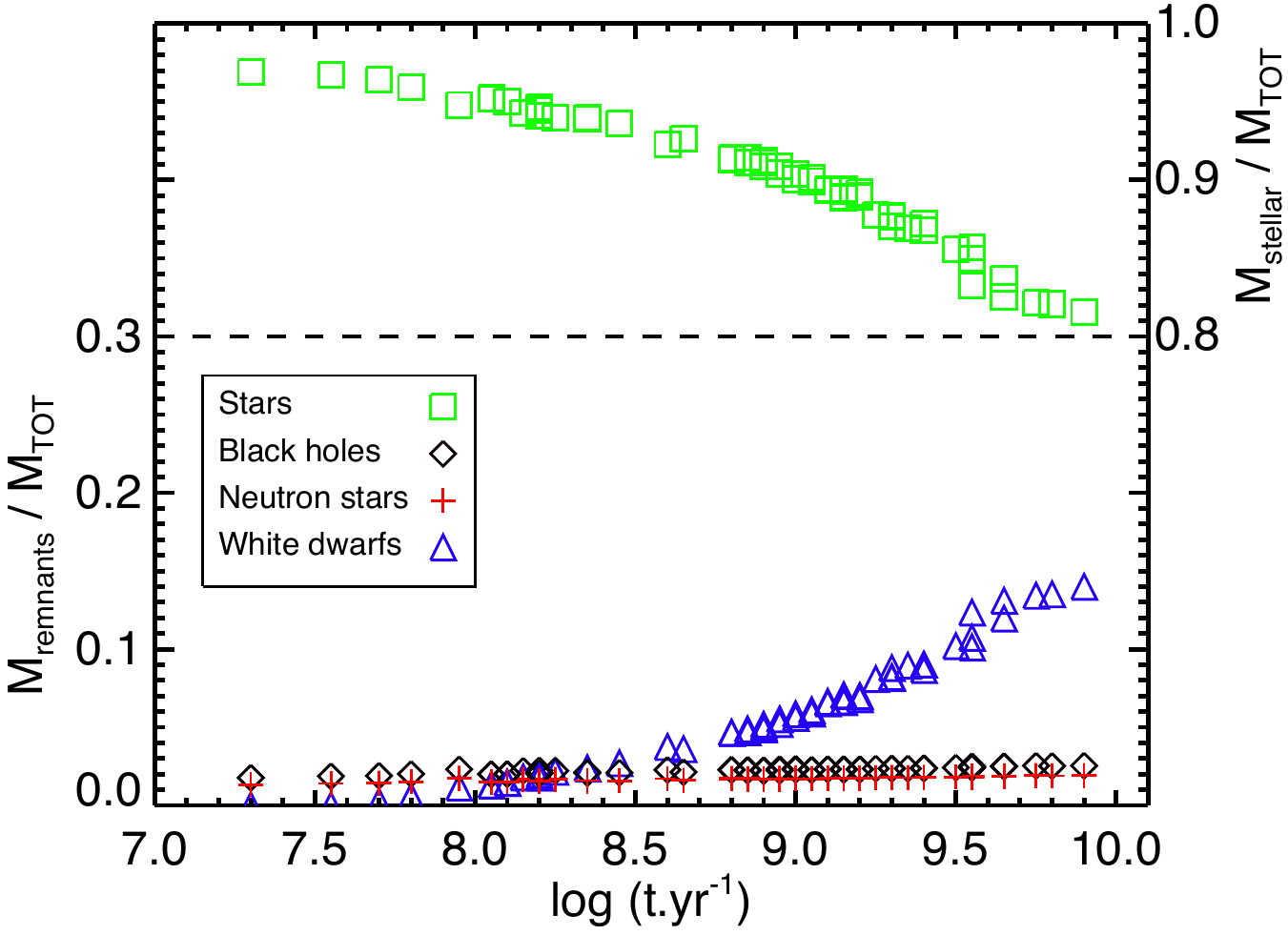}  
    \end{center}    
 }
\caption{ Mass contribution of stellar remnants as function of age for the 60 investigated OCs. The horizontal dashed line indicates a change in the scale of the vertical axis. }

\label{fig:mass_remnants}
\end{center}
\end{figure}

\noindent In Figure~\ref{fig:mass_remnants} we show the  contribution of the different stellar remnants to the total mass, for each investigated OC. The stellar component is also shown in the upper part of the figure. As expected, it is noticeable an increase in the $M_{\textrm{remnants}}/M_{\textrm{tot}}$ ratio as function of age, specially for WDs. For most ages, this ratio is smaller than $\sim10\%$.

\section{Supplementary figures - Radial density profiles}
This Appendix shows the RDPs for 57 investigated OCs (Figures\,B1 to B5) not shown in the manuscript. 

\begin{figure*}
\begin{center}

\parbox[c]{1.00\textwidth}
  {
   \begin{center}
    \includegraphics[width=1.00\textwidth]{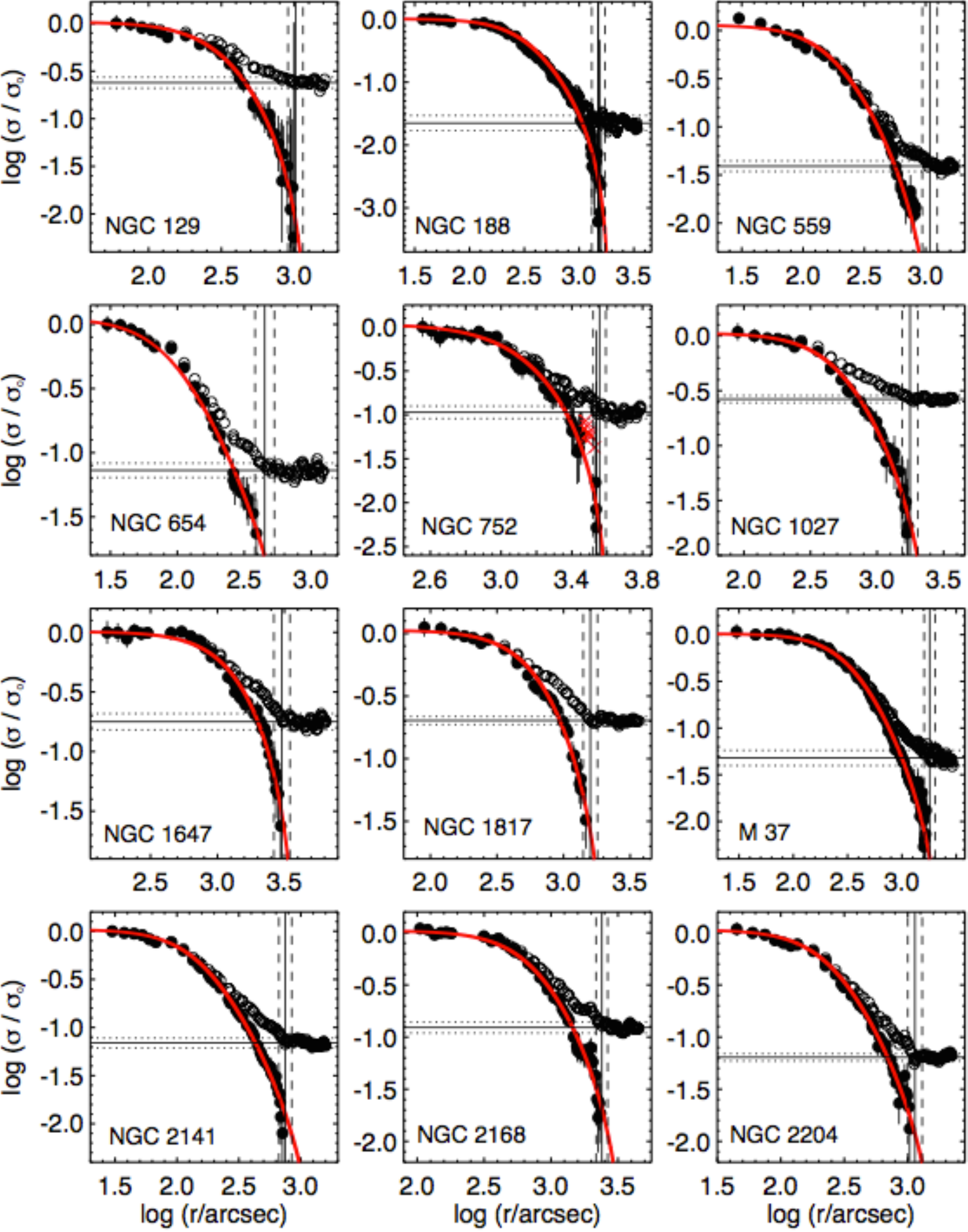}  
    \end{center}    
  }
\caption{ Same as Figure\,2 of the manuscript, but for the OCs indicated in each panel.  NGC\,752's RDP present a considerable fluctuation at $r\sim50\,$arcmin (log\,($r$.arcsec$^{-1}$)\,$\simeq$\,3.48) and the corresponding points (red crosses) have been excluded from the fit. }

\label{fig:RDP_SupplMater1}
\end{center}
\end{figure*}

\begin{figure*}
\begin{center}

\parbox[c]{1.00\textwidth}
  {
   \begin{center}
    \includegraphics[width=1.00\textwidth]{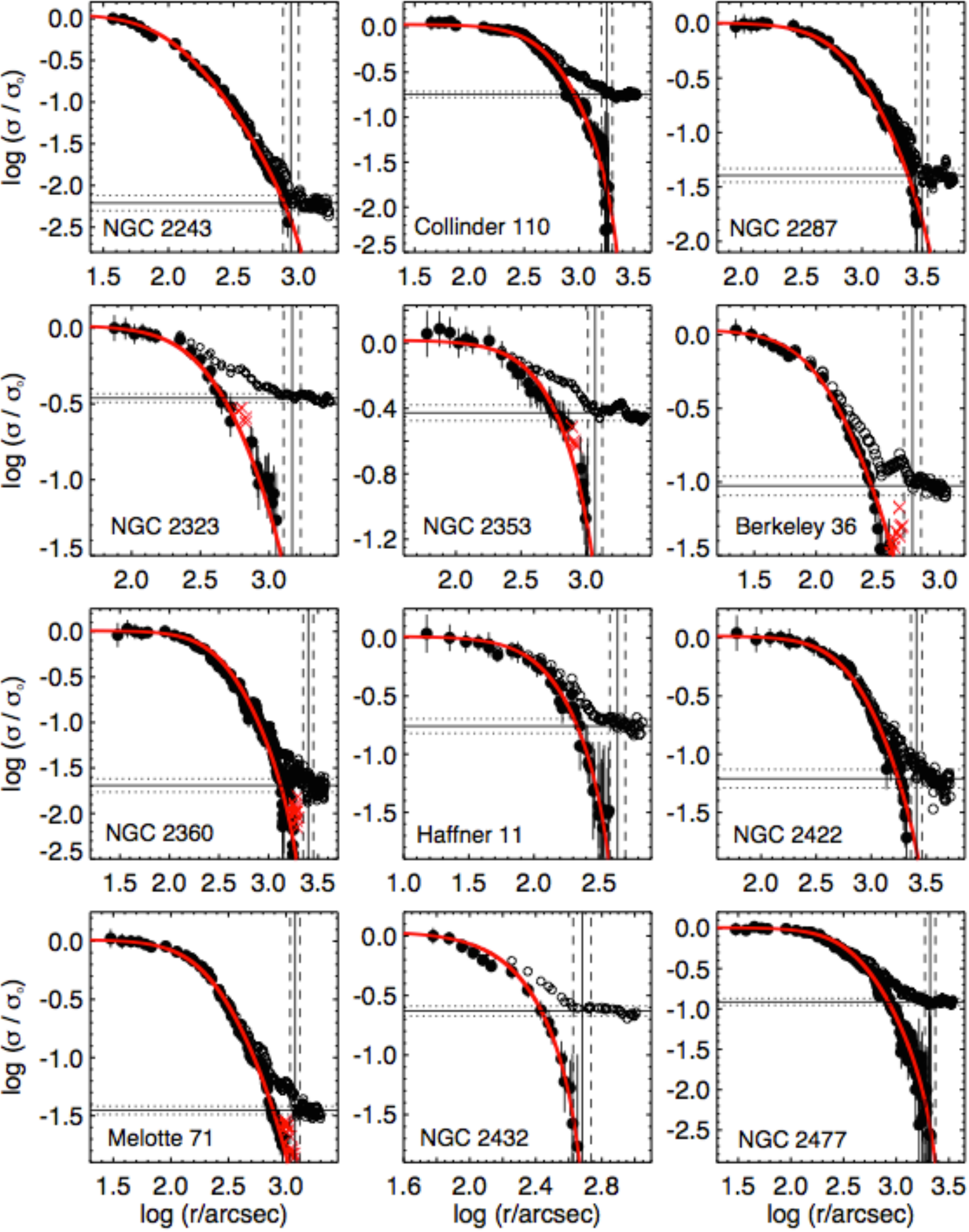}  
    \end{center}    
  }
\caption{ Same as Figure\,2 of the manuscript, but for the OCs indicated in each panel.  The red crosses in some panels represent excluded points from the fit. }

\label{fig:RDP_SupplMater2}
\end{center}
\end{figure*}

\begin{figure*}
\begin{center}

\parbox[c]{1.00\textwidth}
  {
   \begin{center}
    \includegraphics[width=1.00\textwidth]{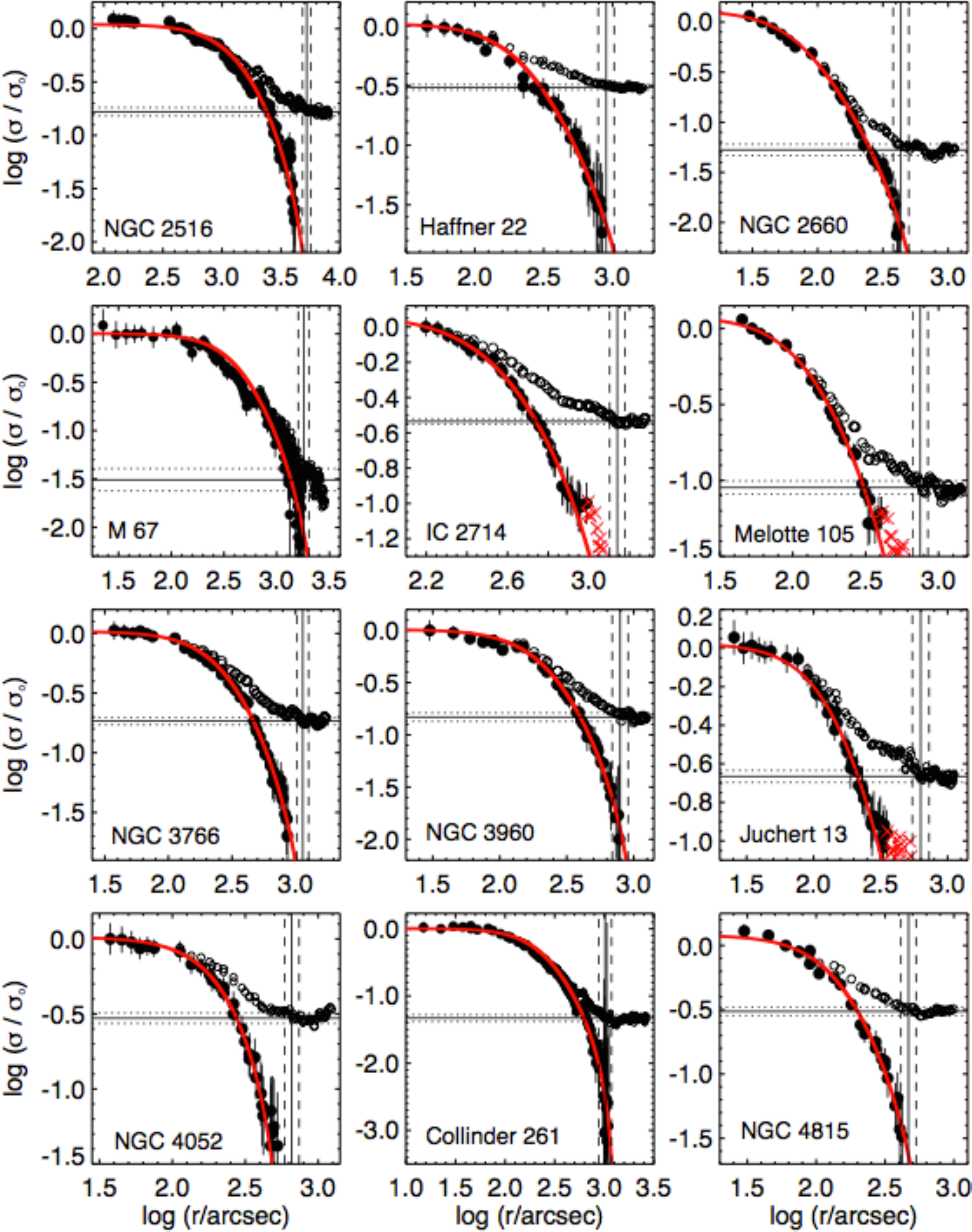}  
    \end{center}    
  }
\caption{ Same as Figure\,2 of the manuscript, but for the OCs indicated in each panel.  The red crosses in some panels represent excluded points from the fit, due to large fluctuactions of density in the cluster outskirts; in these cases (IC\,2714, Melotte\,105 and Juchert\,13), the fits have been truncated at inner radial bins.  }

\label{fig:RDP_SupplMater3}
\end{center}
\end{figure*}

\begin{figure*}
\begin{center}

\parbox[c]{1.00\textwidth}
  {
   \begin{center}
    \includegraphics[width=1.00\textwidth]{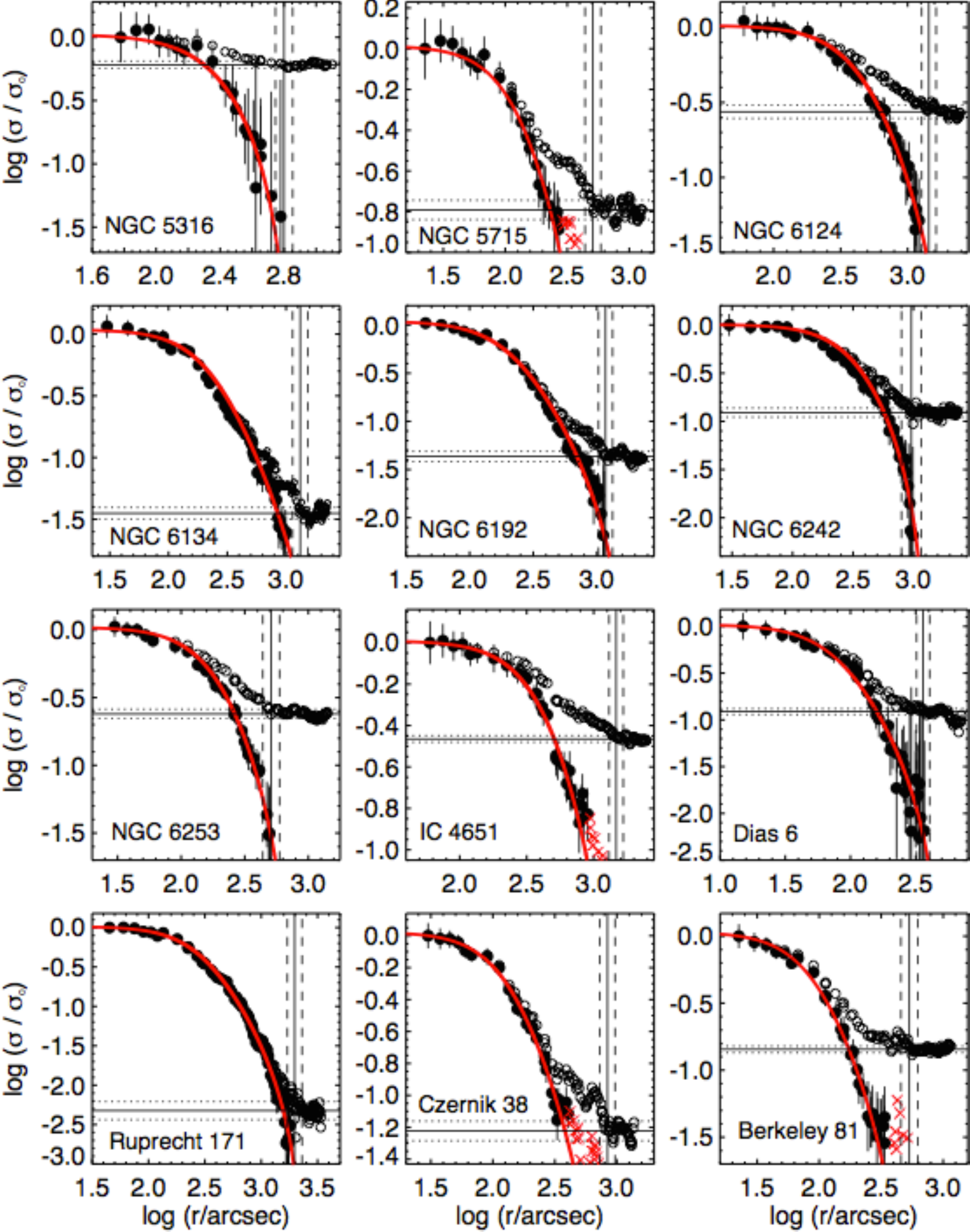}  
    \end{center}    
  }
\caption{ Same as Figure\,2 of the manuscript, but for the OCs indicated in each panel.  The red crosses in some panels represent excluded points from the fit, due to large fluctuactions of density in the cluster outskirts; in these cases (NGC\,5715, IC\,4651, Czernik\,38 and Berkeley\,81), the fits have been truncated at inner radial bins.  }

\label{fig:RDP_SupplMater4}
\end{center}
\end{figure*}

\begin{figure*}
\begin{center}

\parbox[c]{1.00\textwidth}
  {
   \begin{center}
    \includegraphics[width=1.00\textwidth]{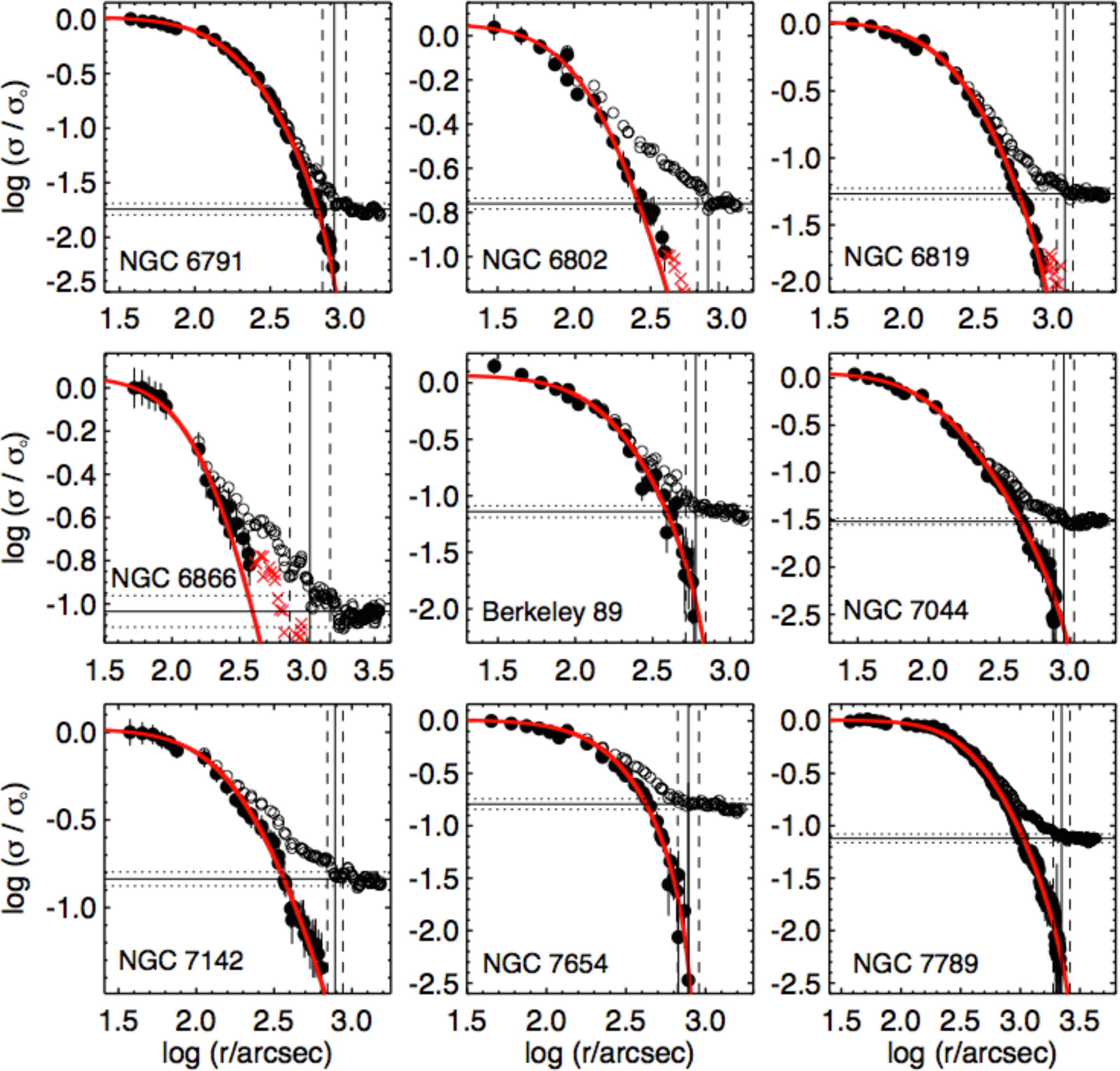}  
    \end{center}    
  }
\caption{ Same as Figure\,2 of the manuscript, but for the OCs indicated in each panel.  The red crosses in some panels represent excluded points from the fit, due to large fluctuactions of density in the cluster outskirts; in these cases (NGC\,6802, NGC\,6819 and NGC\,6866), the fits have been truncated at inner radial bins. In the case of NGC\,6866, the severe fluctuations precluded a proper fit of $r_t$ and therefore we assumed $r_t\sim\,R_{\textrm{lim}}$. }

\label{fig:RDP_SupplMater5}
\end{center}
\end{figure*}

\clearpage

\section{Supplementary figures - Colour-magnitude diagrams}
This Appendix shows the CMDs for 57 investigated OCs (Figures\,C1 to C7) not shown in the manuscript.

\begin{figure*}
\begin{center}

\parbox[c]{1.00\textwidth}
  {
   \begin{center}
    \includegraphics[width=1.00\textwidth]{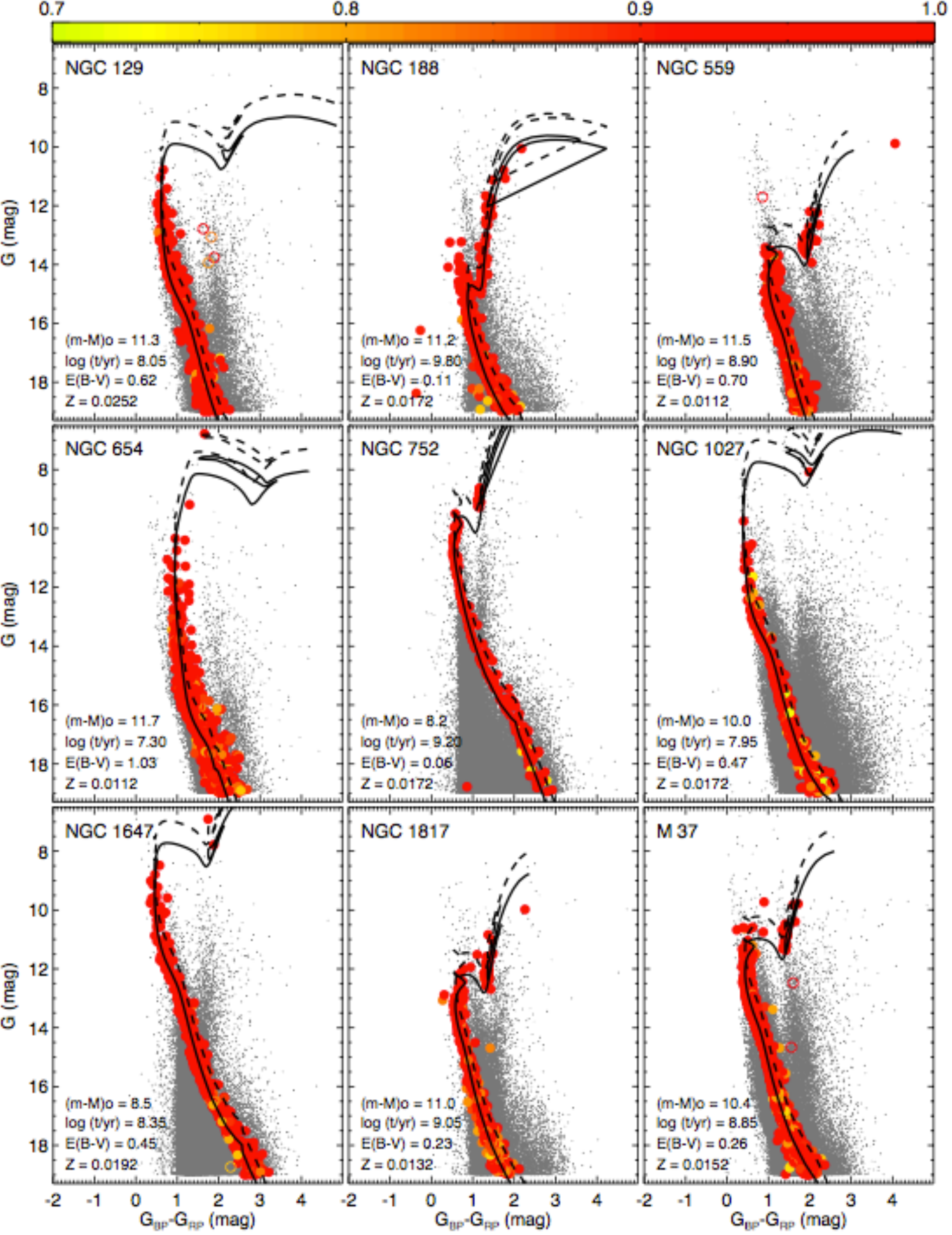}  
    \end{center}    
  }
\caption{ Same as Figure\,7 of the manuscript, but for the OCs indicated in each panel. }

\label{fig:CMD_SupplMater1}
\end{center}
\end{figure*}

\begin{figure*}
\begin{center}

\parbox[c]{1.00\textwidth}
  {
   \begin{center}
    \includegraphics[width=1.00\textwidth]{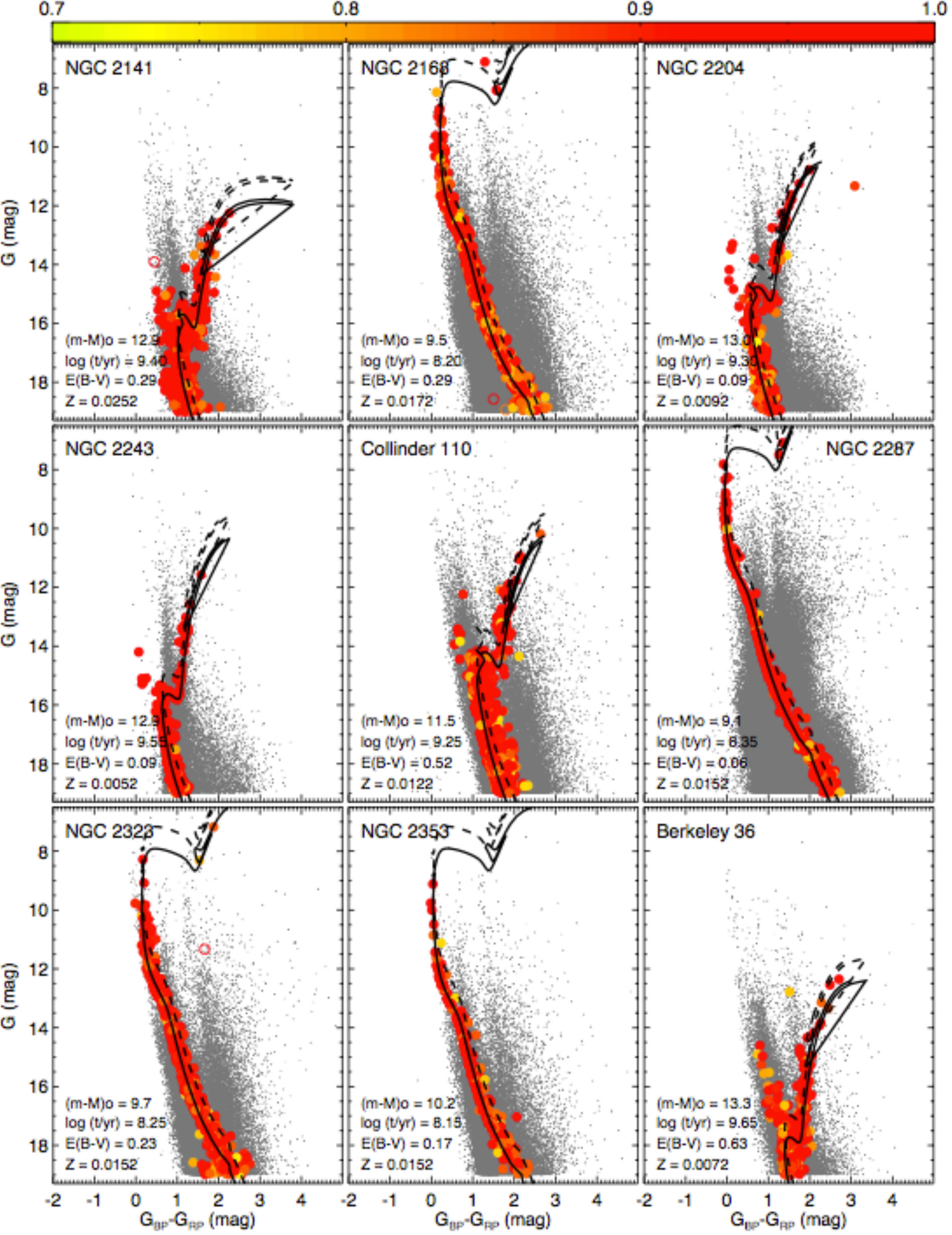}  
    \end{center}    
  }
\caption{ Same as Figure\,7 of the manuscript, but for the OCs indicated in each panel. }

\label{fig:CMD_SupplMater2}
\end{center}
\end{figure*}

\begin{figure*}
\begin{center}

\parbox[c]{1.00\textwidth}
  {
   \begin{center}
    \includegraphics[width=1.00\textwidth]{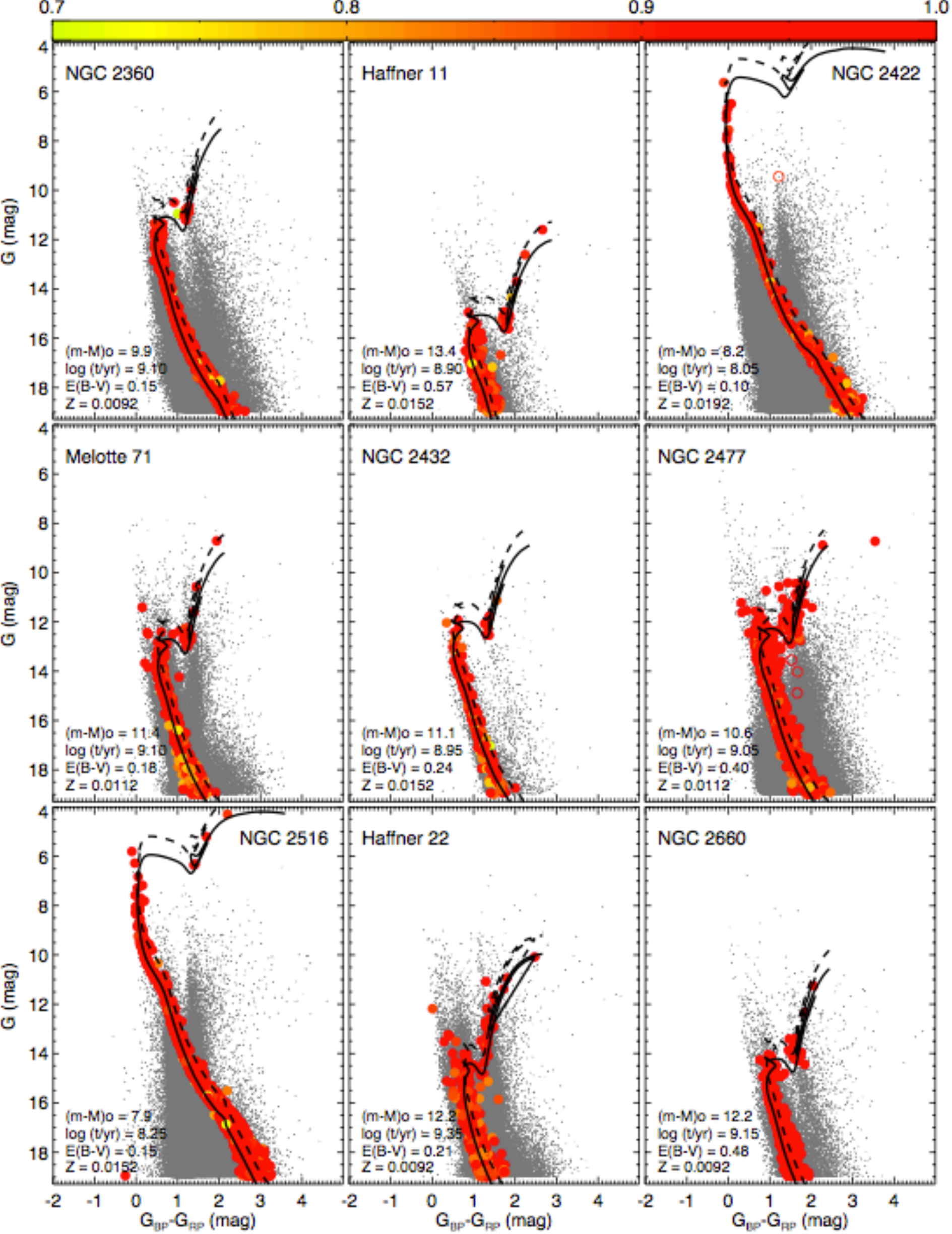}  
    \end{center}    
  }
\caption{ Same as Figure\,7 of the manuscript, but for the OCs indicated in each panel. }

\label{fig:CMD_SupplMater3}
\end{center}
\end{figure*}

\begin{figure*}
\begin{center}

\parbox[c]{1.00\textwidth}
  {
   \begin{center}
    \includegraphics[width=1.00\textwidth]{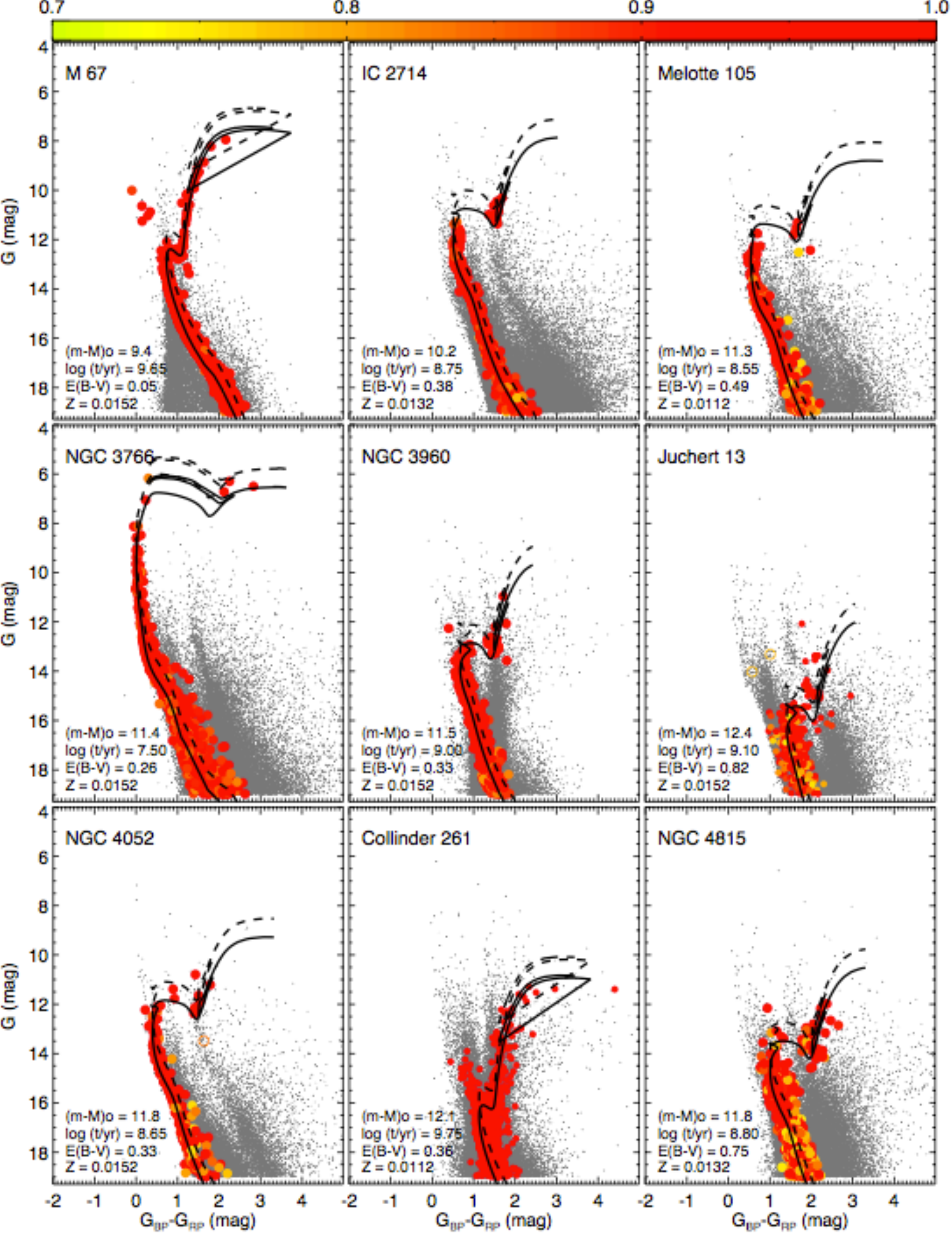}  
    \end{center}    
  }
\caption{ Same as Figure\,7 of the manuscript, but for the OCs indicated in each panel. }

\label{fig:CMD_SupplMater4}
\end{center}
\end{figure*}

\begin{figure*}
\begin{center}

\parbox[c]{1.00\textwidth}
  {
   \begin{center}
    \includegraphics[width=1.00\textwidth]{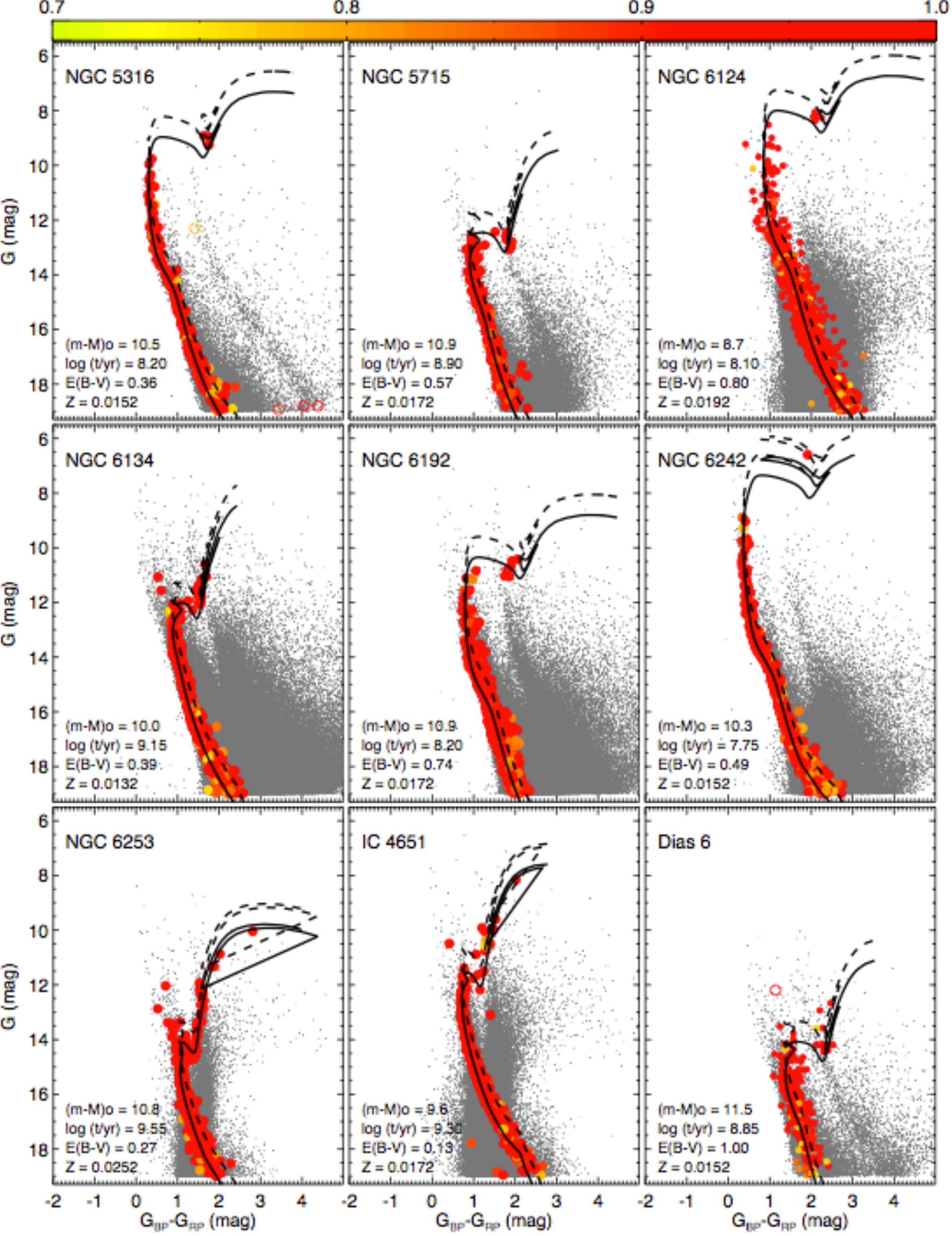}  
    \end{center}    
  }
\caption{ Same as Figure\,7 of the manuscript, but for the OCs indicated in each panel. }

\label{fig:CMD_SupplMater5}
\end{center}
\end{figure*}

\begin{figure*}
\begin{center}

\parbox[c]{1.00\textwidth}
  {
   \begin{center}
    \includegraphics[width=1.00\textwidth]{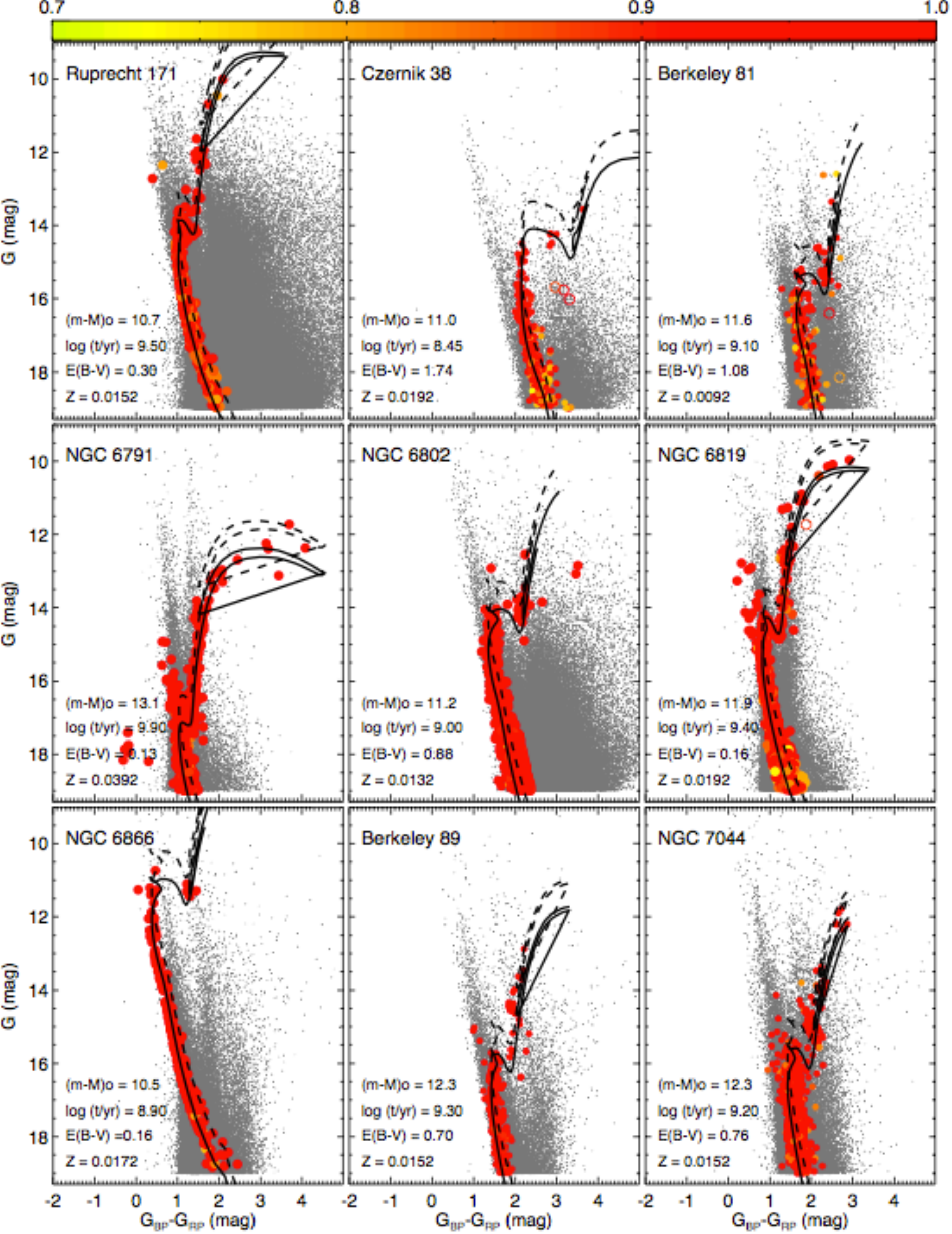}  
    \end{center}    
  }
\caption{ Same as Figure\,7 of the manuscript, but for the OCs indicated in each panel. }

\label{fig:CMD_SupplMater6}
\end{center}
\end{figure*}

\begin{figure*}
\begin{center}

\parbox[c]{1.00\textwidth}
  {
   \begin{center}
    \includegraphics[width=1.00\textwidth]{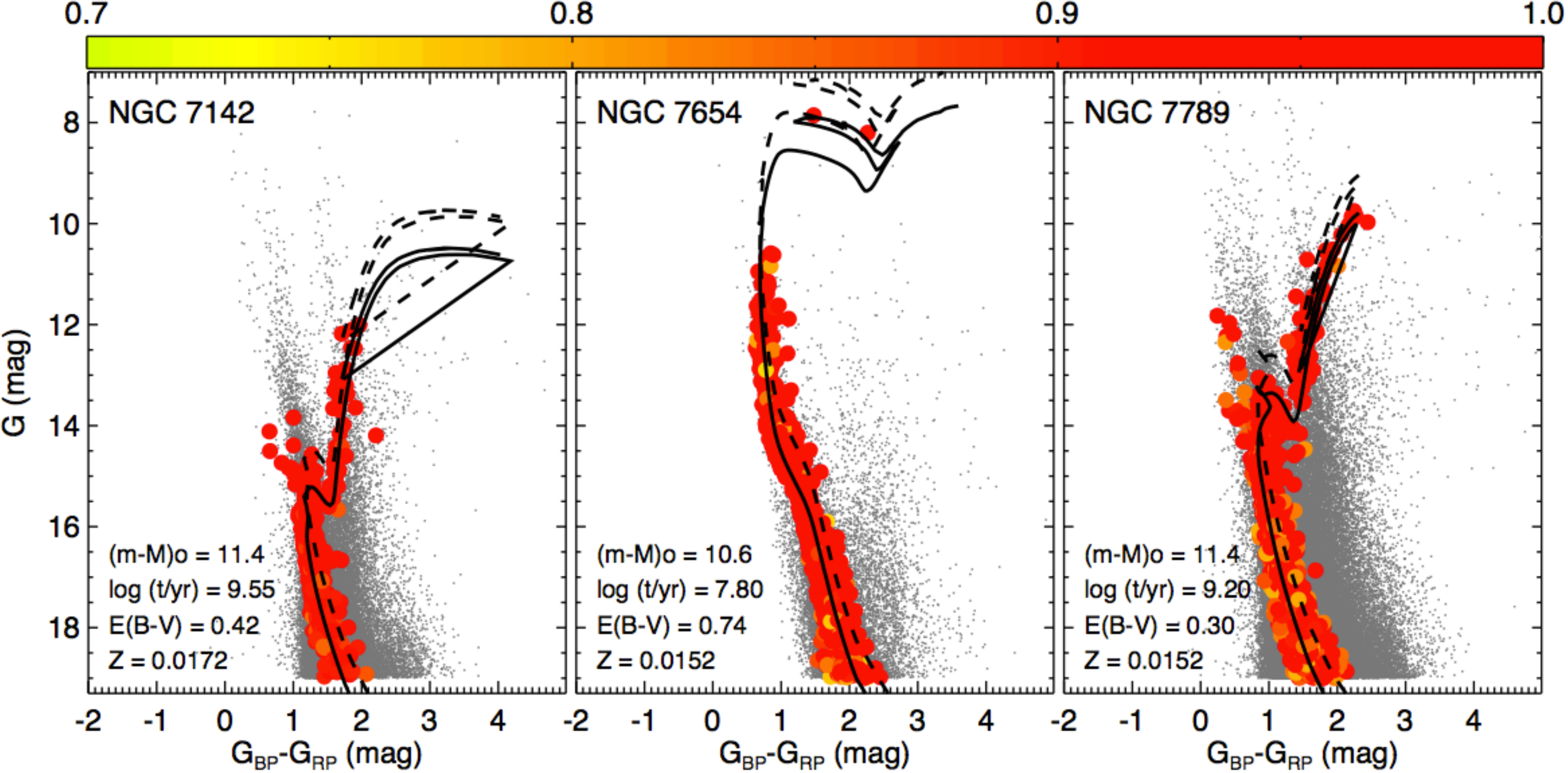}  
    \end{center}    
  }
\caption{ Same as Figure\,7 of the manuscript, but for the OCs indicated in each panel. }

\label{fig:CMD_SupplMater7}
\end{center}
\end{figure*}

\clearpage

\section{Supplementary figures - Vector-point diagrams}
This Appendix shows the VPDs for 57 investigated OCs (Figures\,D1 to D5) not shown in the manuscript.

\begin{figure*}
\begin{center}

\parbox[c]{1.00\textwidth}
  {
   \begin{center}
    \includegraphics[width=1.00\textwidth]{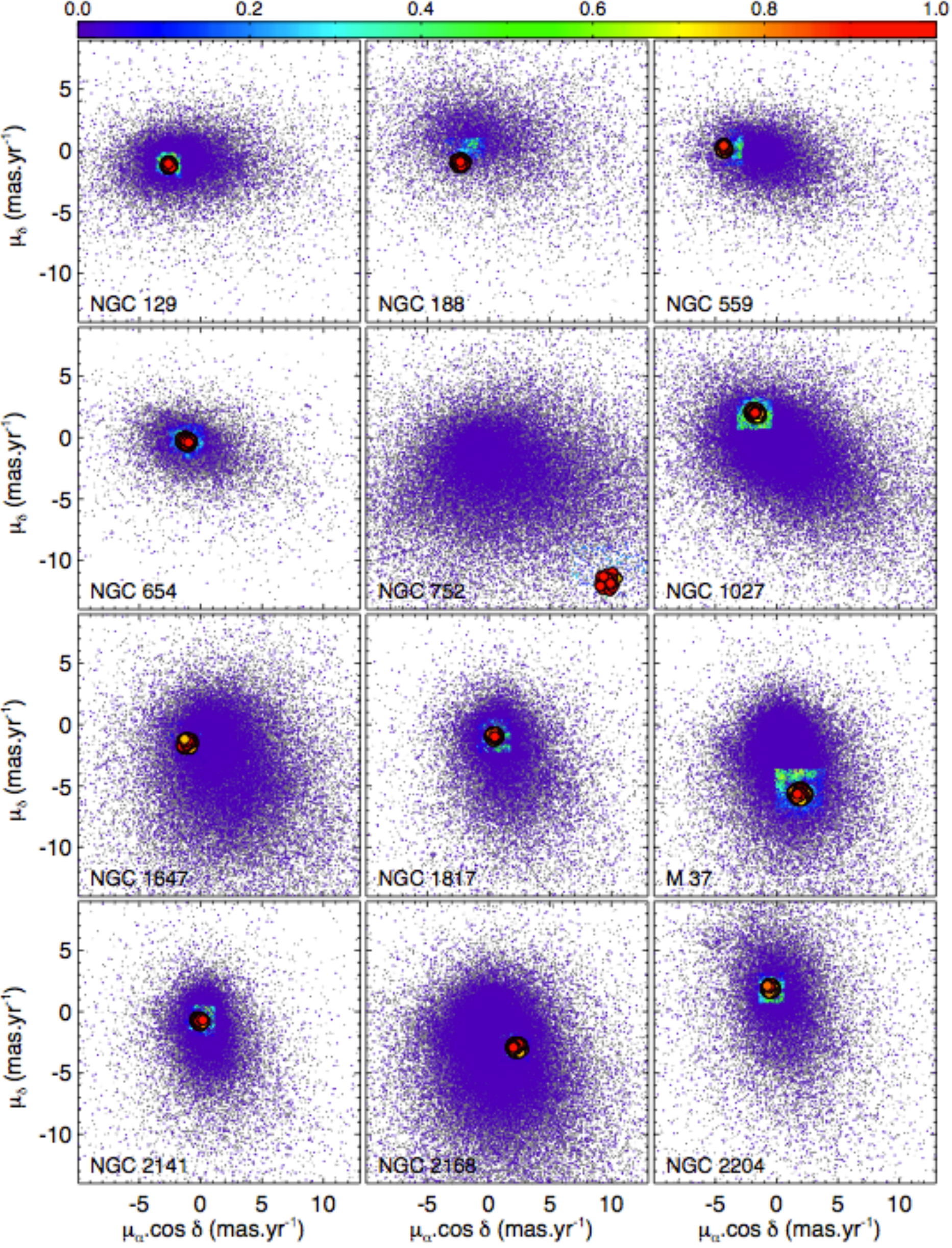}  
    \end{center}    
  }
\caption{ Same as Figure\,3 of the manuscript, but for the OCs indicated in each panel. }

\label{fig:VPD_SupplMater1}
\end{center}
\end{figure*}

\begin{figure*}
\begin{center}

\parbox[c]{1.00\textwidth}
  {
   \begin{center}
    \includegraphics[width=1.00\textwidth]{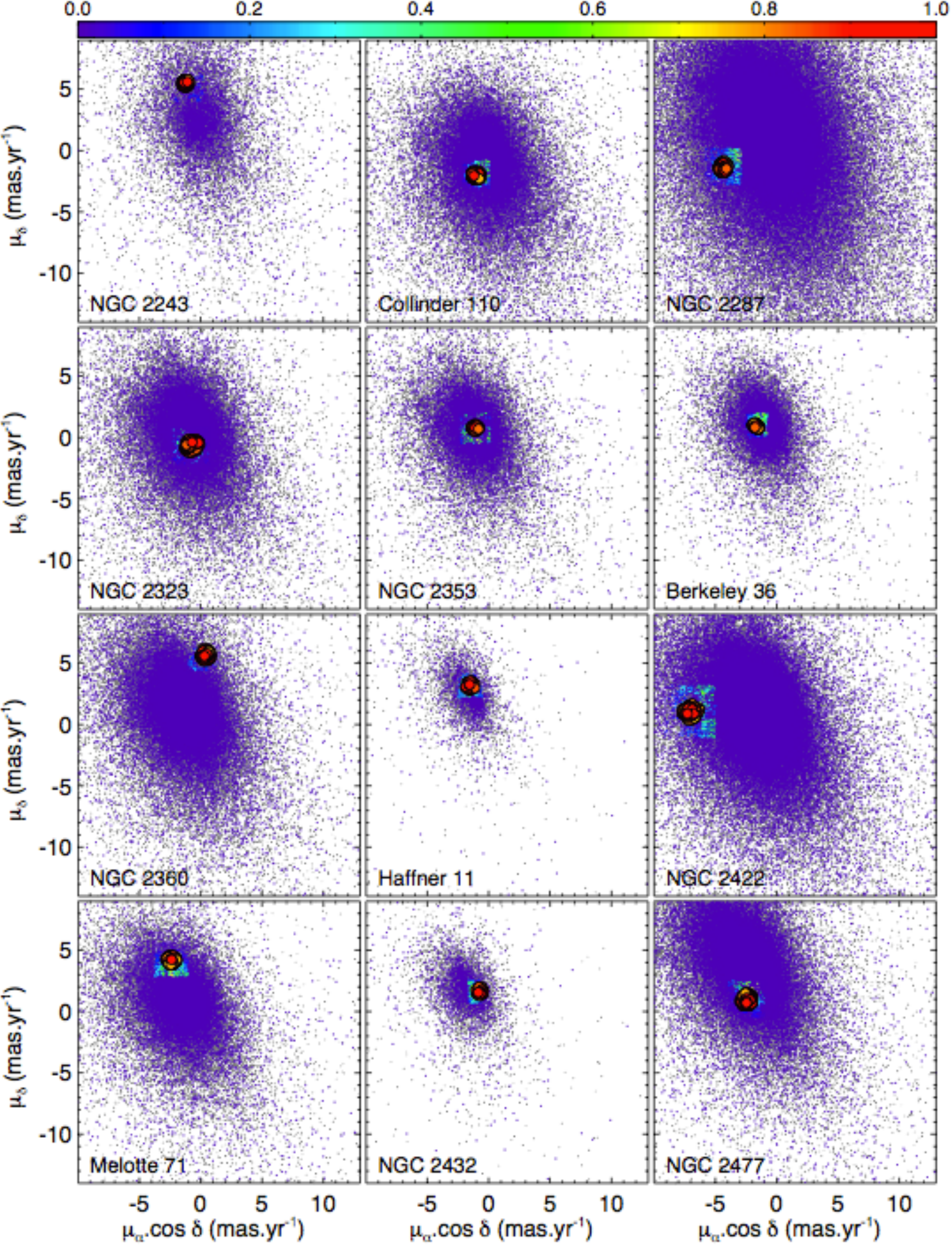}  
    \end{center}    
  }
\caption{ Same as Figure\,3 of the manuscript, but for the OCs indicated in each panel. }

\label{fig:VPD_SupplMater2}
\end{center}
\end{figure*}

\begin{figure*}
\begin{center}

\parbox[c]{1.00\textwidth}
  {
   \begin{center}
    \includegraphics[width=1.00\textwidth]{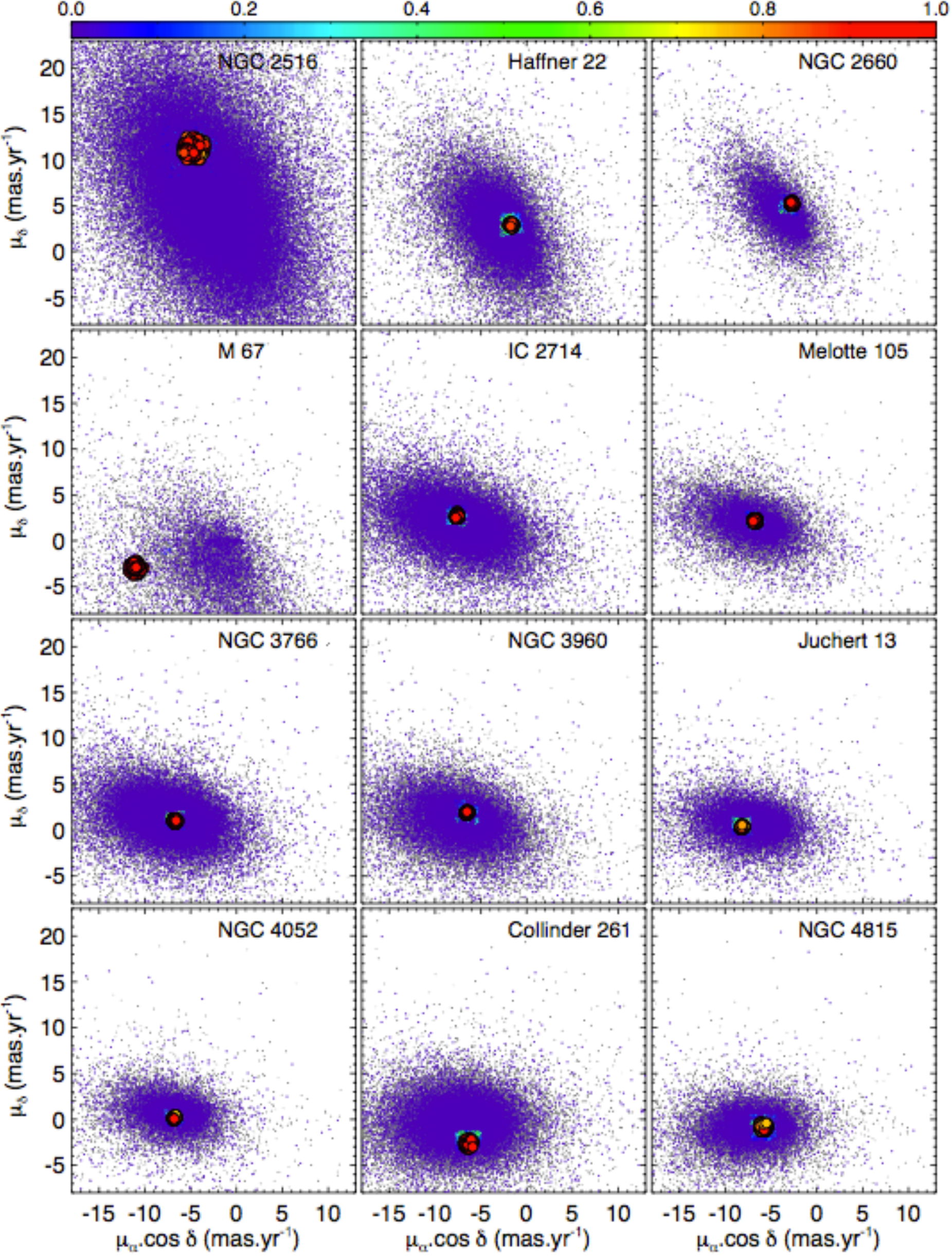}  
    \end{center}    
  }
\caption{ Same as Figure\,3 of the manuscript, but for the OCs indicated in each panel. }

\label{fig:VPD_SupplMater3}
\end{center}
\end{figure*}

\begin{figure*}
\begin{center}

\parbox[c]{1.00\textwidth}
  {
   \begin{center}
    \includegraphics[width=1.00\textwidth]{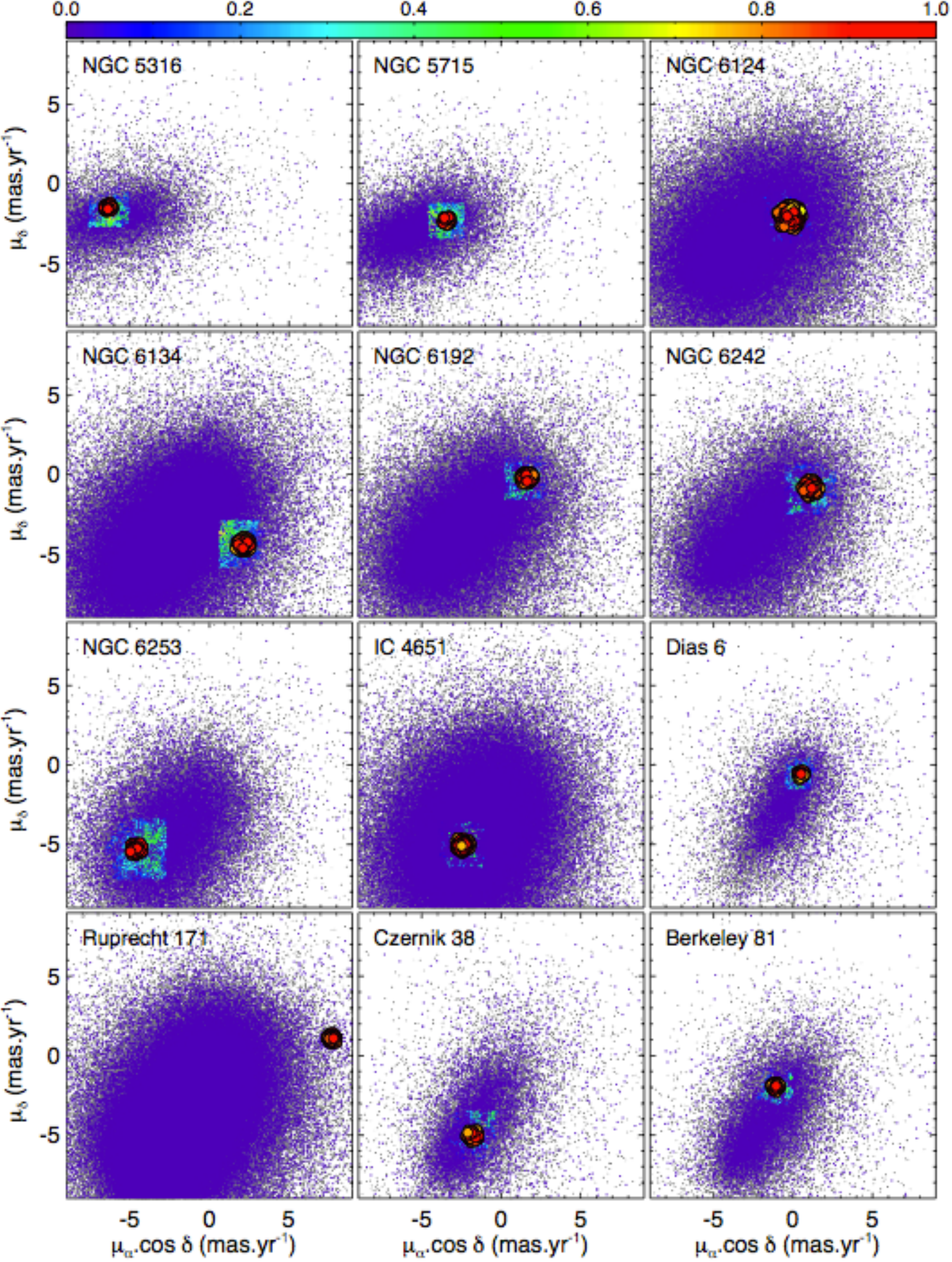}  
    \end{center}    
  }
\caption{ Same as Figure\,3 of the manuscript, but for the OCs indicated in each panel. }

\label{fig:VPD_SupplMater4}
\end{center}
\end{figure*}

\begin{figure*}
\begin{center}

\parbox[c]{1.00\textwidth}
  {
   \begin{center}
    \includegraphics[width=1.00\textwidth]{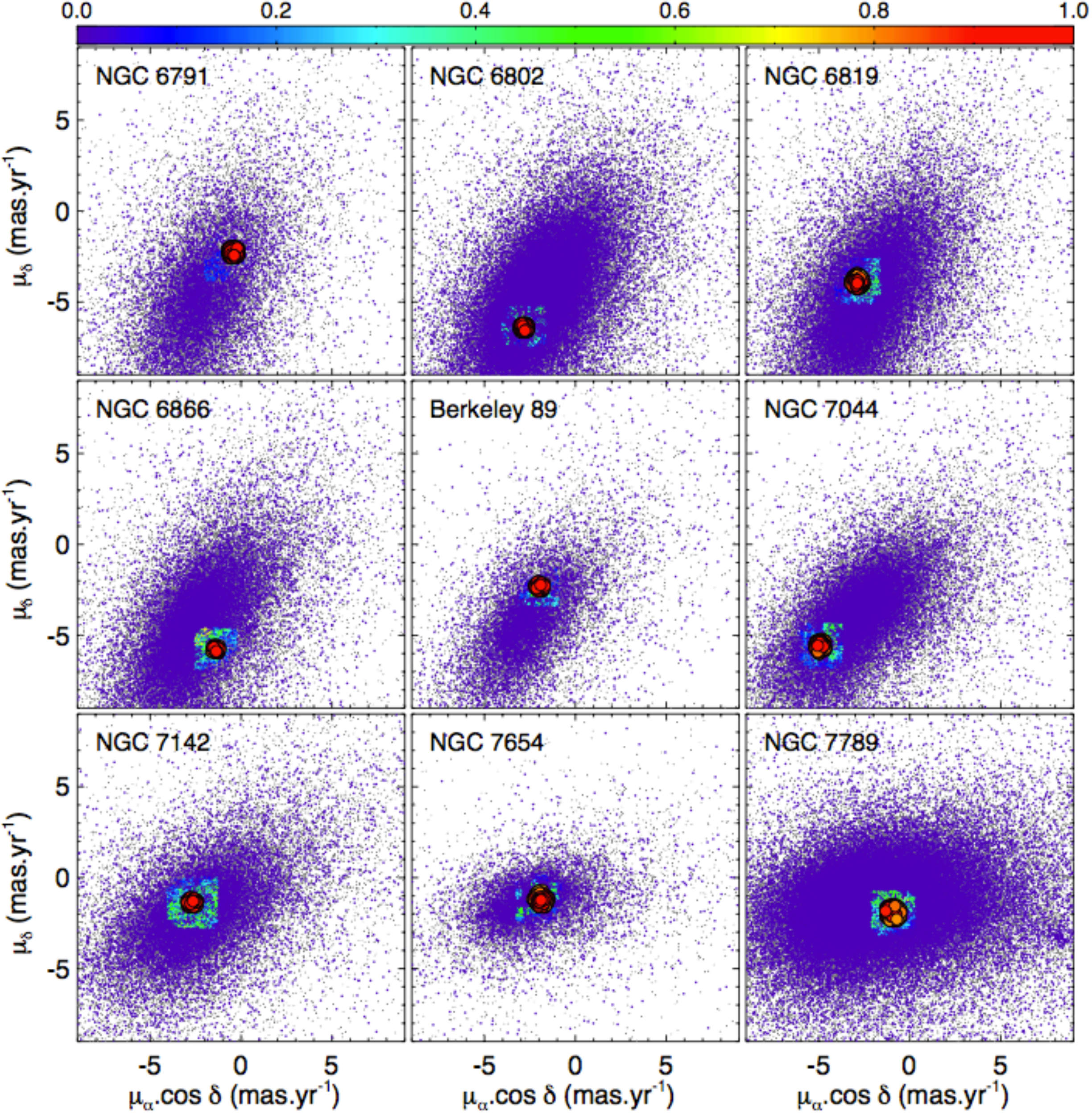}  
    \end{center}    
  }
\caption{ Same as Figure\,3 of the manuscript, but for the OCs indicated in each panel. }

\label{fig:VPD_SupplMater5}
\end{center}
\end{figure*}

\clearpage

\section{Supplementary figures - Parallax versus $G$ magnitude plots}
This Appendix shows the parallax\,$\times$\,$G\,$magnitude\, plots for 57 investigated OCs (Figures\,E1 to E5) not shown in the manuscript.

\begin{figure*}
\begin{center}

\parbox[c]{1.00\textwidth}
  {
   \begin{center}
    \includegraphics[width=1.00\textwidth]{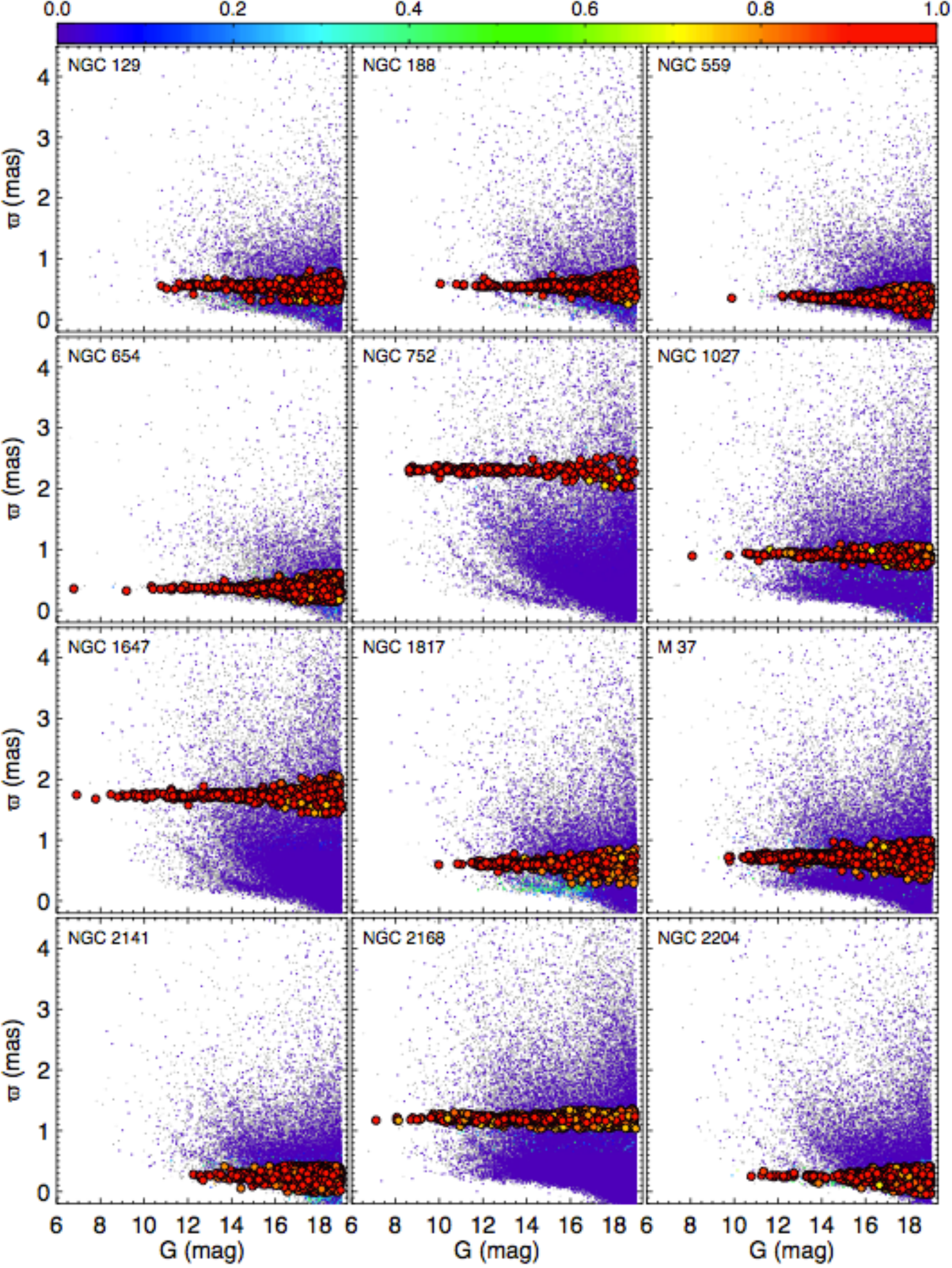}  
    \end{center}    
  }
\caption{ Same as Figure\,4 of the manuscript, but for the OCs indicated in each panel. }

\label{fig:plxvsG_SupplMater1}
\end{center}
\end{figure*}

\begin{figure*}
\begin{center}

\parbox[c]{1.00\textwidth}
  {
   \begin{center}
    \includegraphics[width=1.00\textwidth]{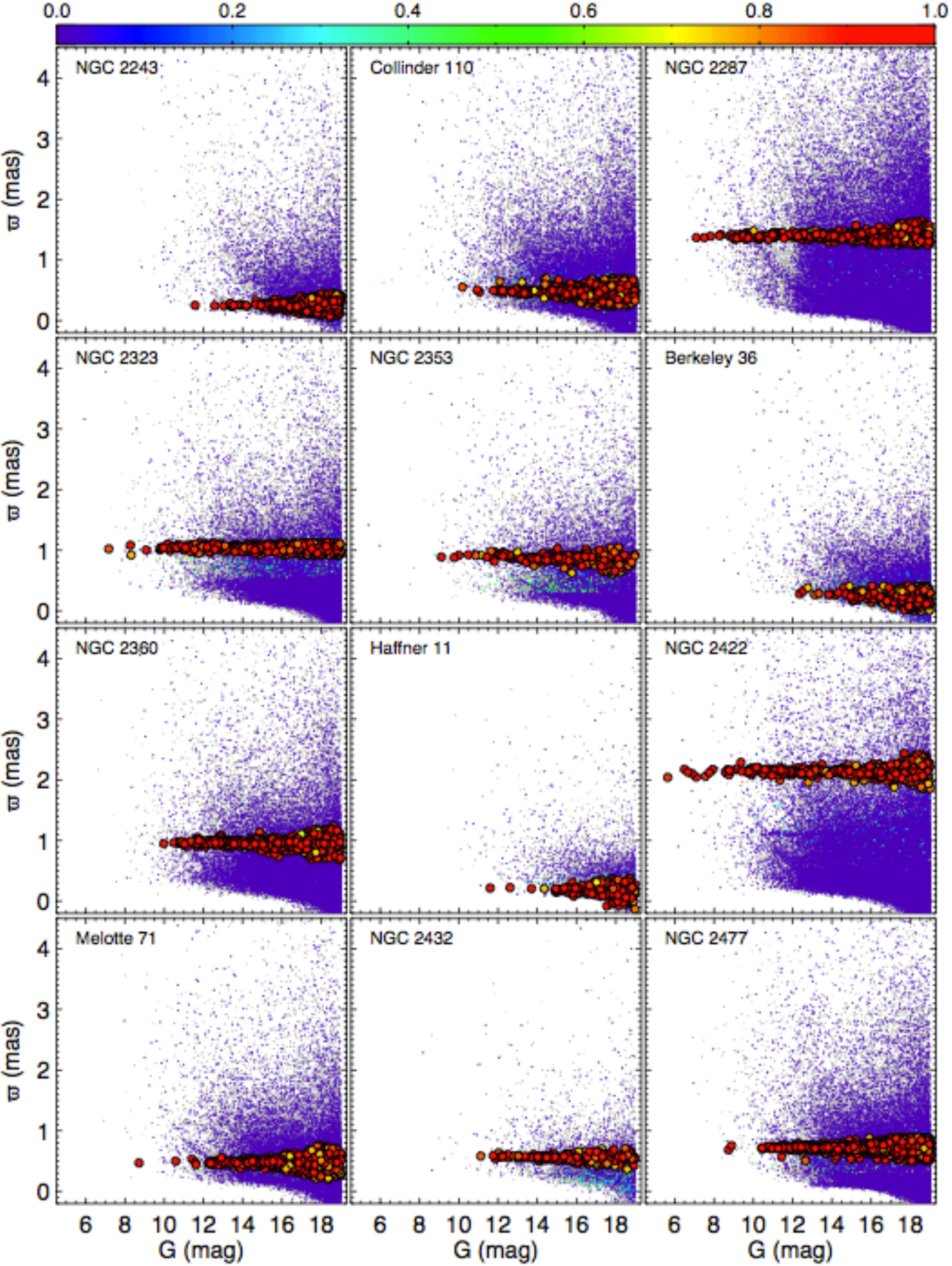}  
    \end{center}    
  }
\caption{ Same as Figure\,4 of the manuscript, but for the OCs indicated in each panel. }

\label{fig:plxvsG_SupplMater2}
\end{center}
\end{figure*}

\begin{figure*}
\begin{center}

\parbox[c]{1.00\textwidth}
  {
   \begin{center}
    \includegraphics[width=1.00\textwidth]{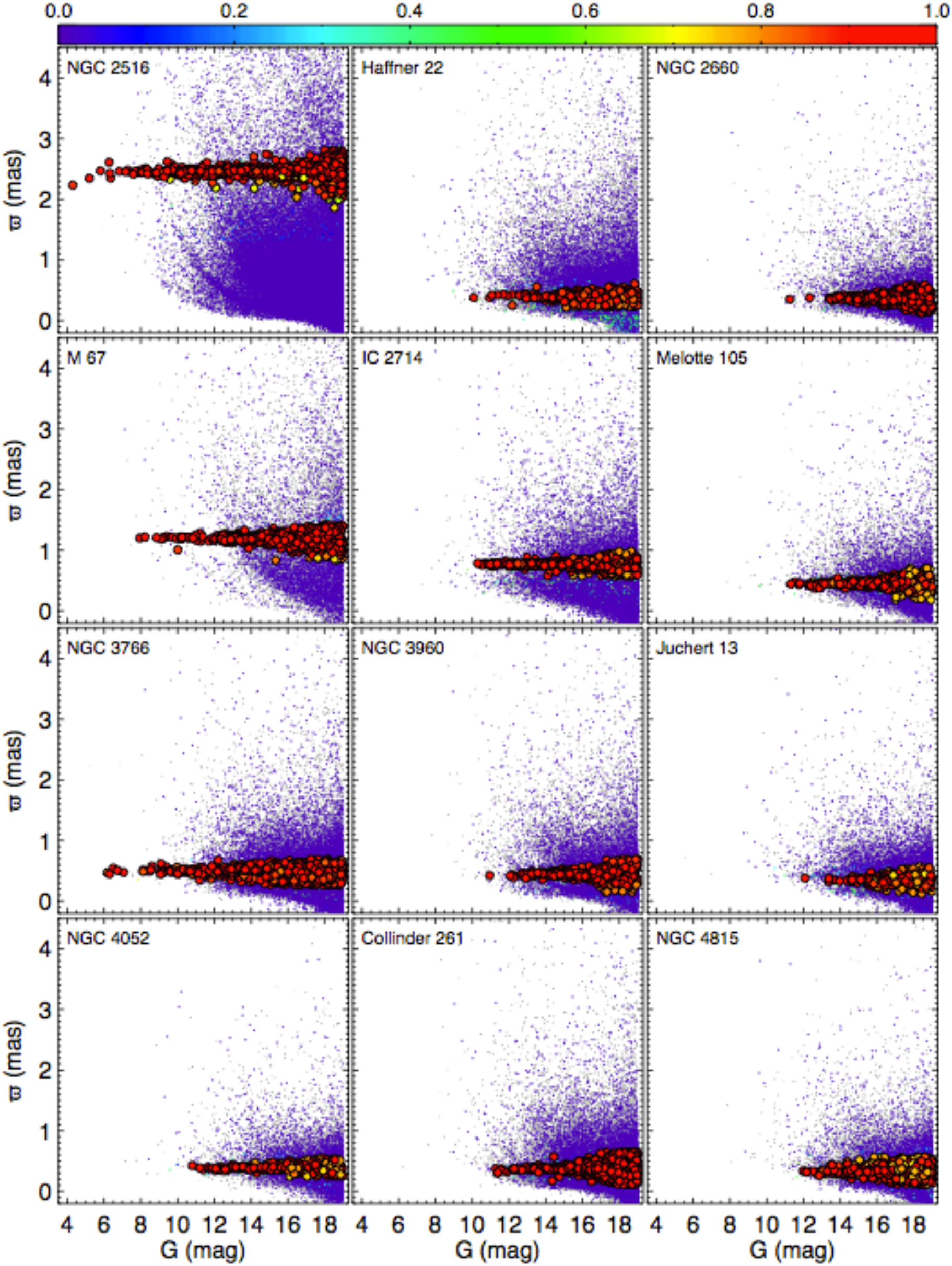}  
    \end{center}    
  }
\caption{ Same as Figure\,4 of the manuscript, but for the OCs indicated in each panel. }

\label{fig:plxvsG_SupplMater3}
\end{center}
\end{figure*}

\begin{figure*}
\begin{center}

\parbox[c]{1.00\textwidth}
  {
   \begin{center}
    \includegraphics[width=1.00\textwidth]{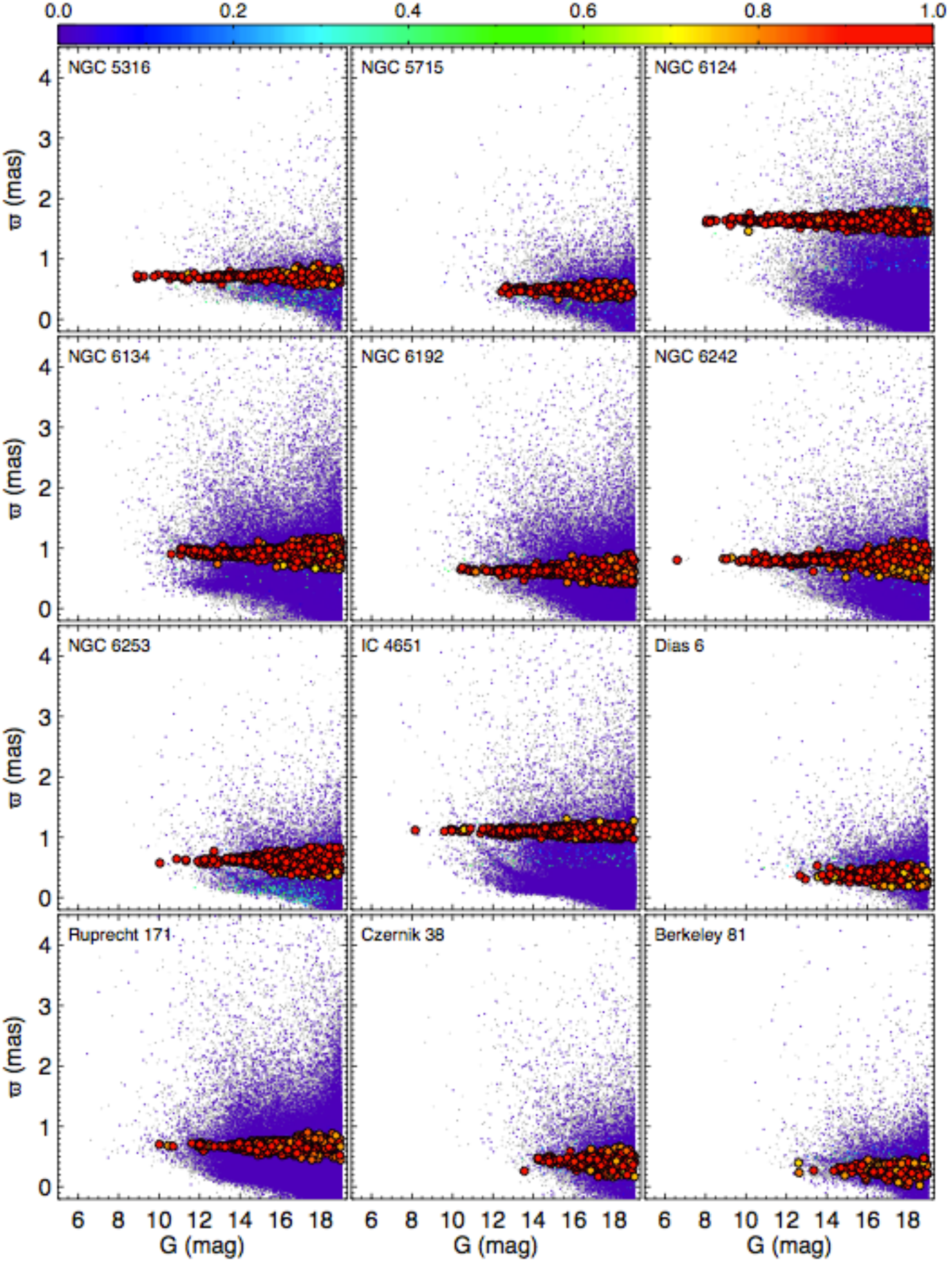}  
    \end{center}    
  }
\caption{ Same as Figure\,4 of the manuscript, but for the OCs indicated in each panel. }

\label{fig:plxvsG_SupplMater4}
\end{center}
\end{figure*}

\begin{figure*}
\begin{center}

\parbox[c]{1.00\textwidth}
  {
   \begin{center}
    \includegraphics[width=1.00\textwidth]{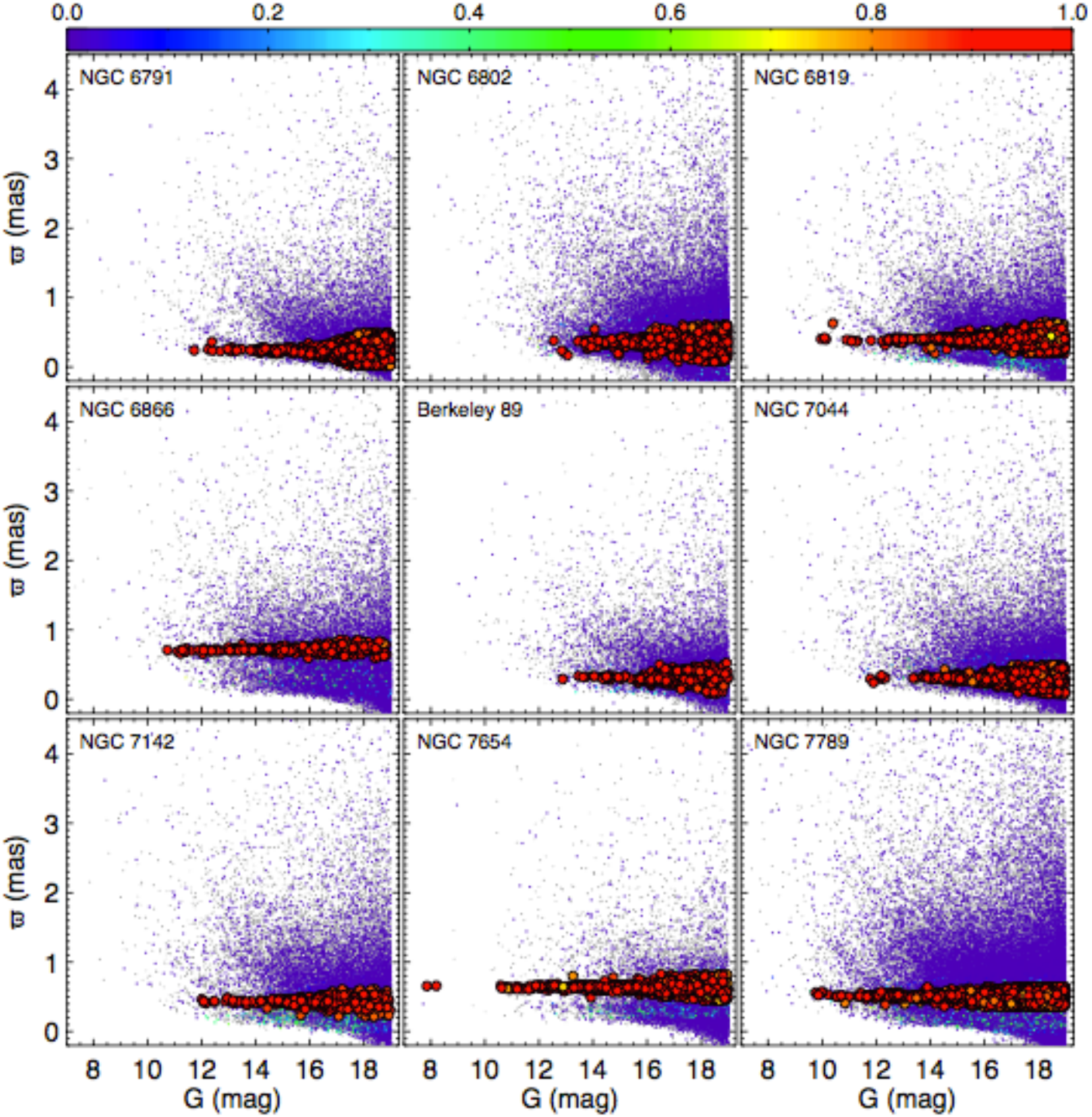}  
    \end{center}    
  }
\caption{ Same as Figure\,4 of the manuscript, but for the OCs indicated in each panel. }

\label{fig:plxvsG_SupplMater5}
\end{center}
\end{figure*}

\clearpage

\section{Supplementary figures - Mass functions}
\label{sec:suppl_MFs}
This Appendix shows the mass functions for 57 investigated OCs (Figures\,F1 to F5) not shown in the manuscript.

\begin{figure*}
\begin{center}

\parbox[c]{0.90\textwidth}
  {
   \begin{center}
    \includegraphics[width=0.90\textwidth]{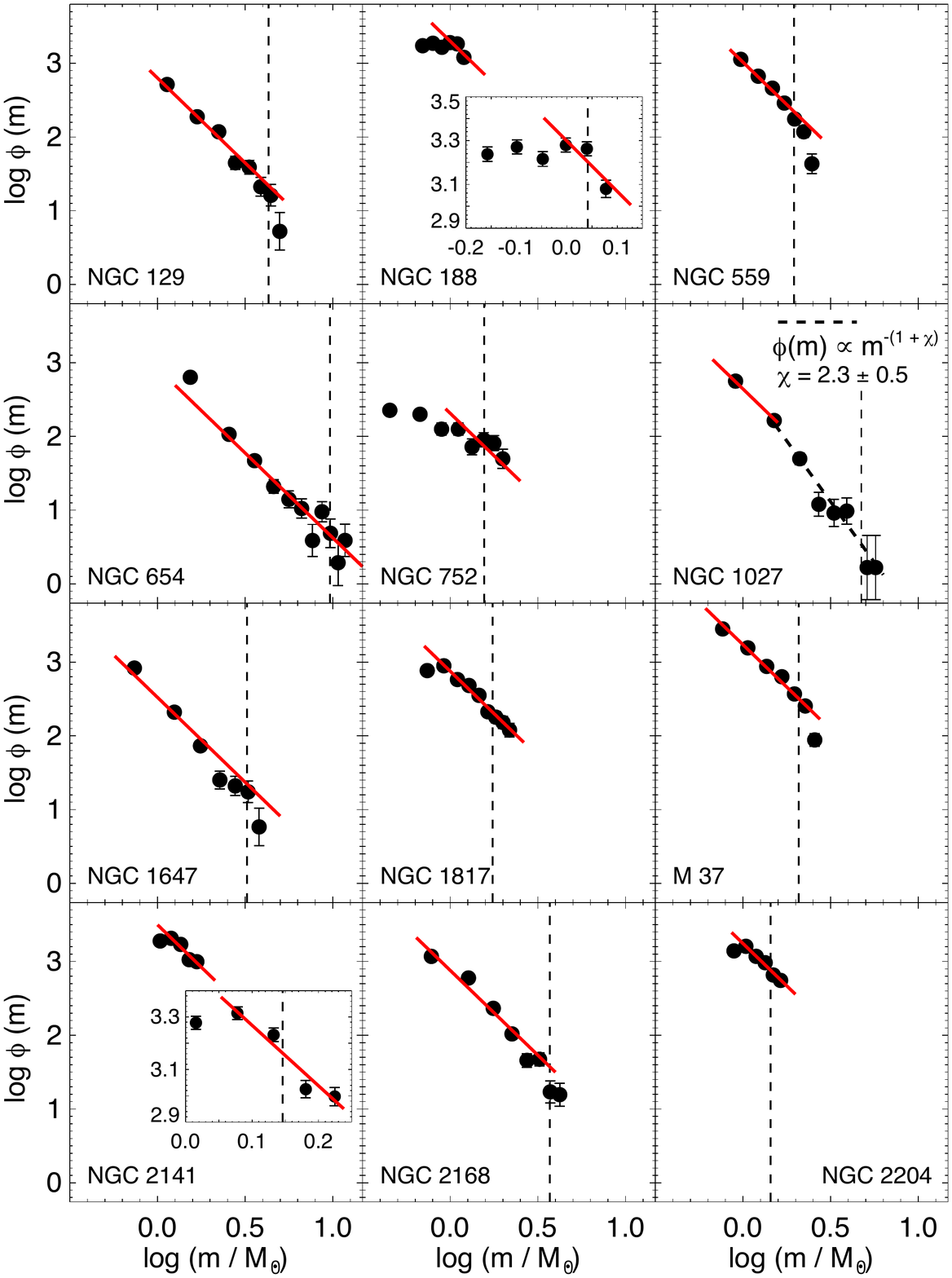}  
    \end{center}    
  }
\caption{ Same as Figure\,8 of the manuscript, but for the OCs indicated in each panel. In the case of NGC\,188 and NGC\,2141, the inset exhibits more clearly the corresponding MF. In the case of NGC\,1027, the higher mass bins present a steeper relation in comparison to Kroupa's IMF and a linear fit (in log-log scale) has been performed. The mass function slope is indicated. The same procedure has been employed in the next figures. }

\label{fig:MFs_SupplMater1}
\end{center}
\end{figure*}

\begin{figure*}
\begin{center}

\parbox[c]{0.90\textwidth}
  {
   \begin{center}
    \includegraphics[width=0.90\textwidth]{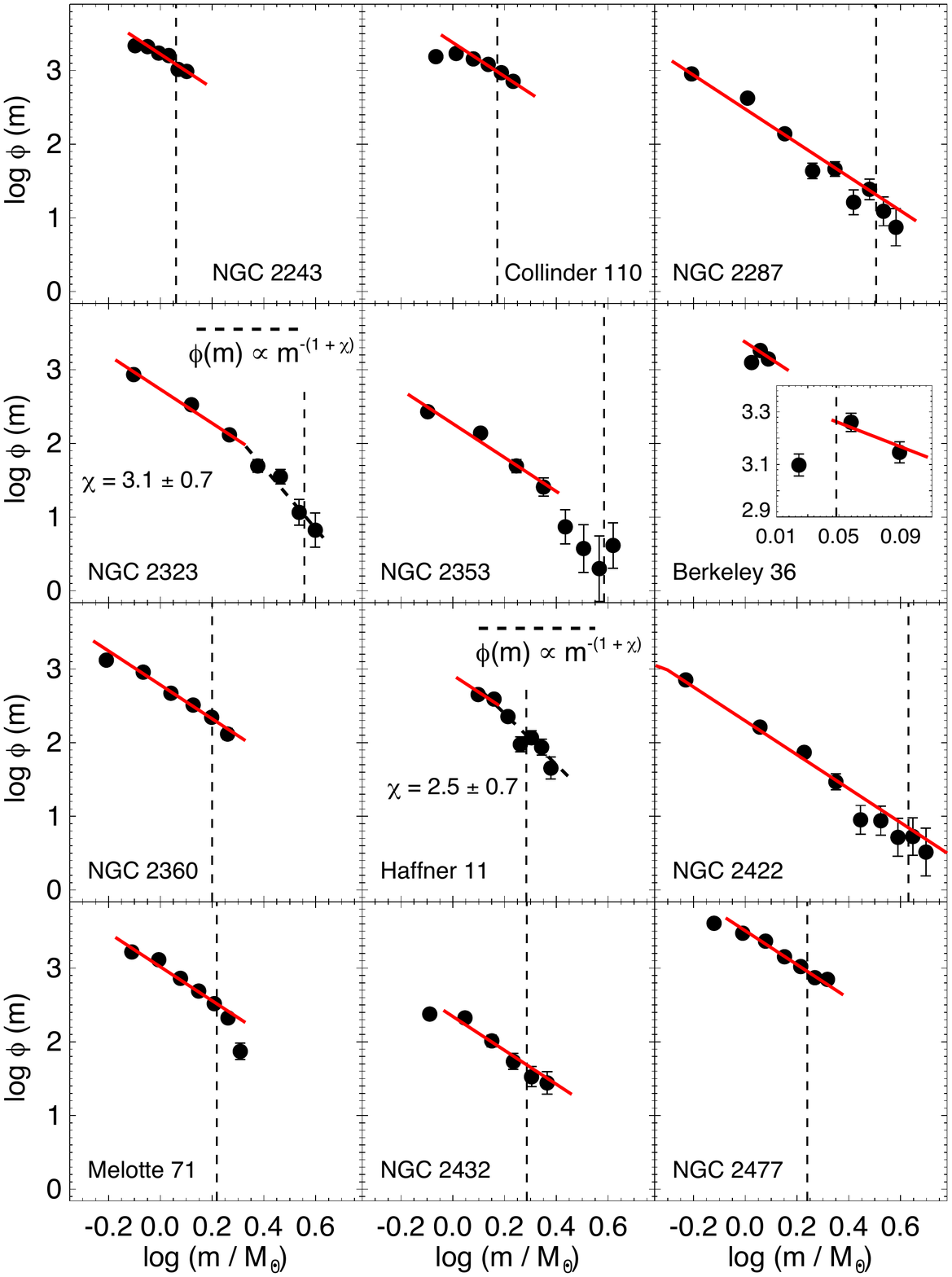}  
    \end{center}    
  }
\caption{ Same as Same as Figure\,8 of the manuscript, but for the OCs indicated in each panel. }

\label{fig:MFs_SupplMater2}
\end{center}
\end{figure*}

\begin{figure*}
\begin{center}

\parbox[c]{0.90\textwidth}
  {
   \begin{center}
    \includegraphics[width=0.90\textwidth]{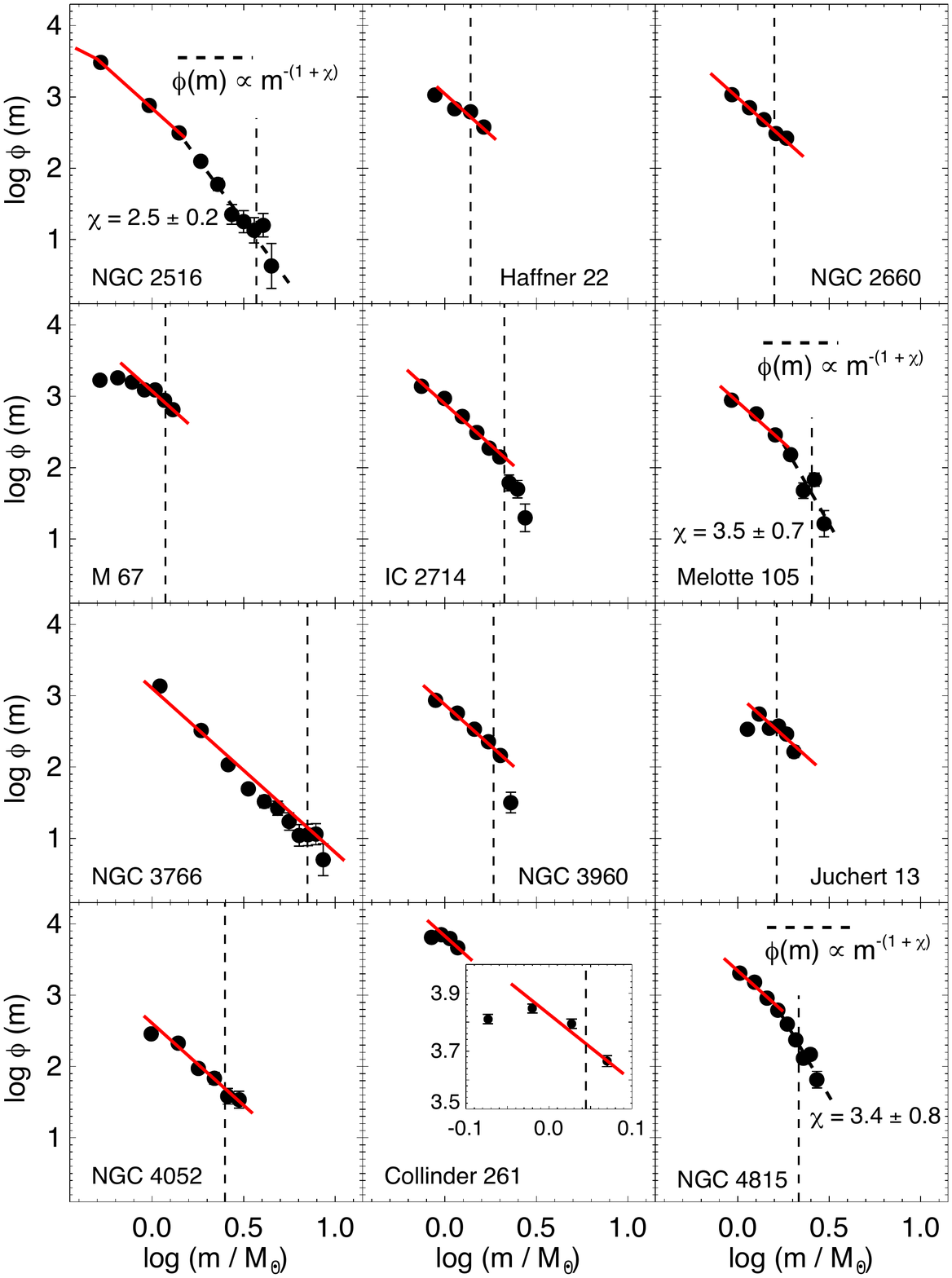}  
    \end{center}    
  }
\caption{ Same as Same as Figure\,8 of the manuscript, but for the OCs indicated in each panel. }

\label{fig:MFs_SupplMater3}
\end{center}
\end{figure*}

\begin{figure*}
\begin{center}

\parbox[c]{0.90\textwidth}
  {
   \begin{center}
    \includegraphics[width=0.90\textwidth]{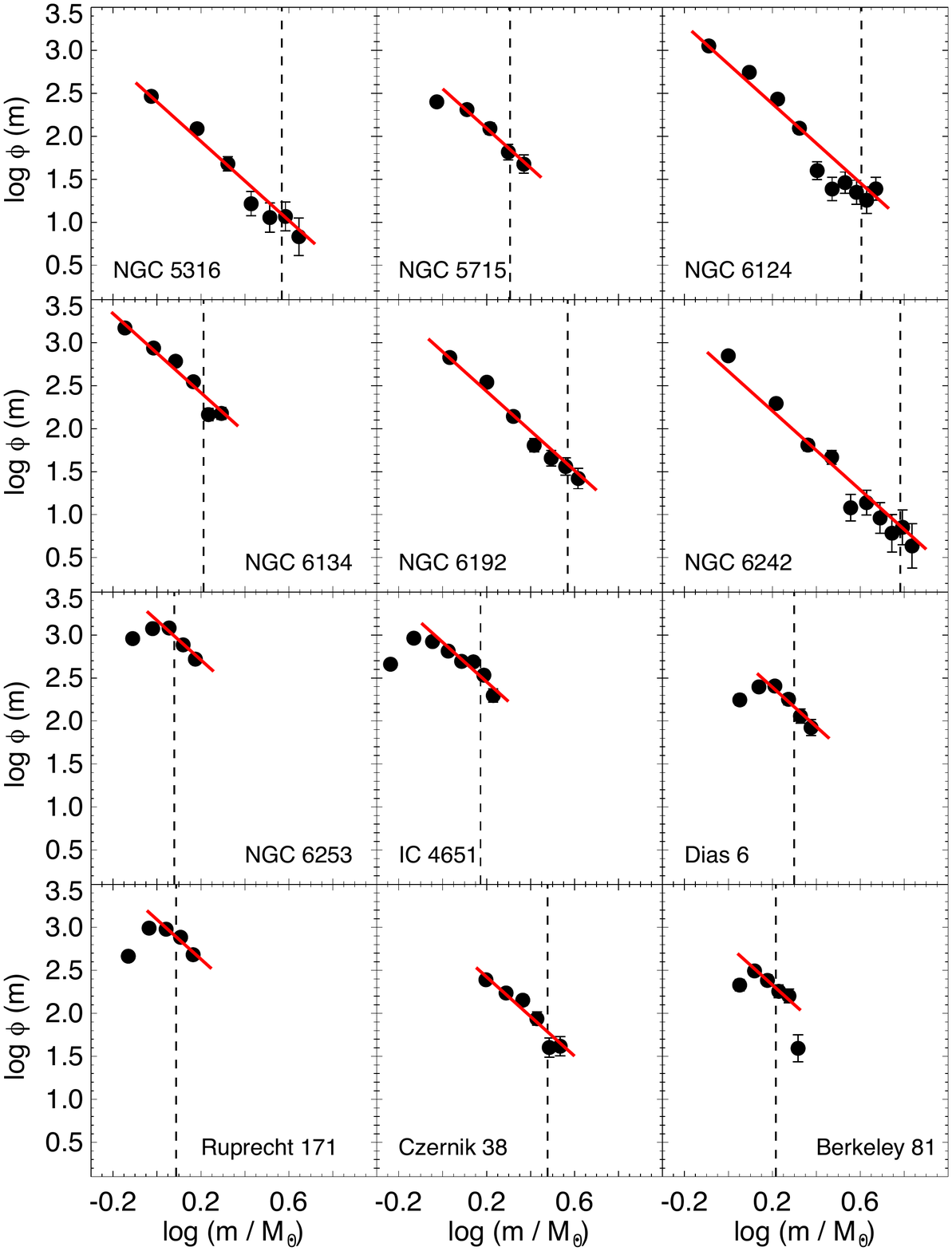}  
    \end{center}    
  }
\caption{ Same as Same as Figure\,8 of the manuscript, but for the OCs indicated in each panel. }

\label{fig:MFs_SupplMater4}
\end{center}
\end{figure*}

\begin{figure*}
\begin{center}

\parbox[c]{0.90\textwidth}
  {
   \begin{center}
    \includegraphics[width=0.90\textwidth]{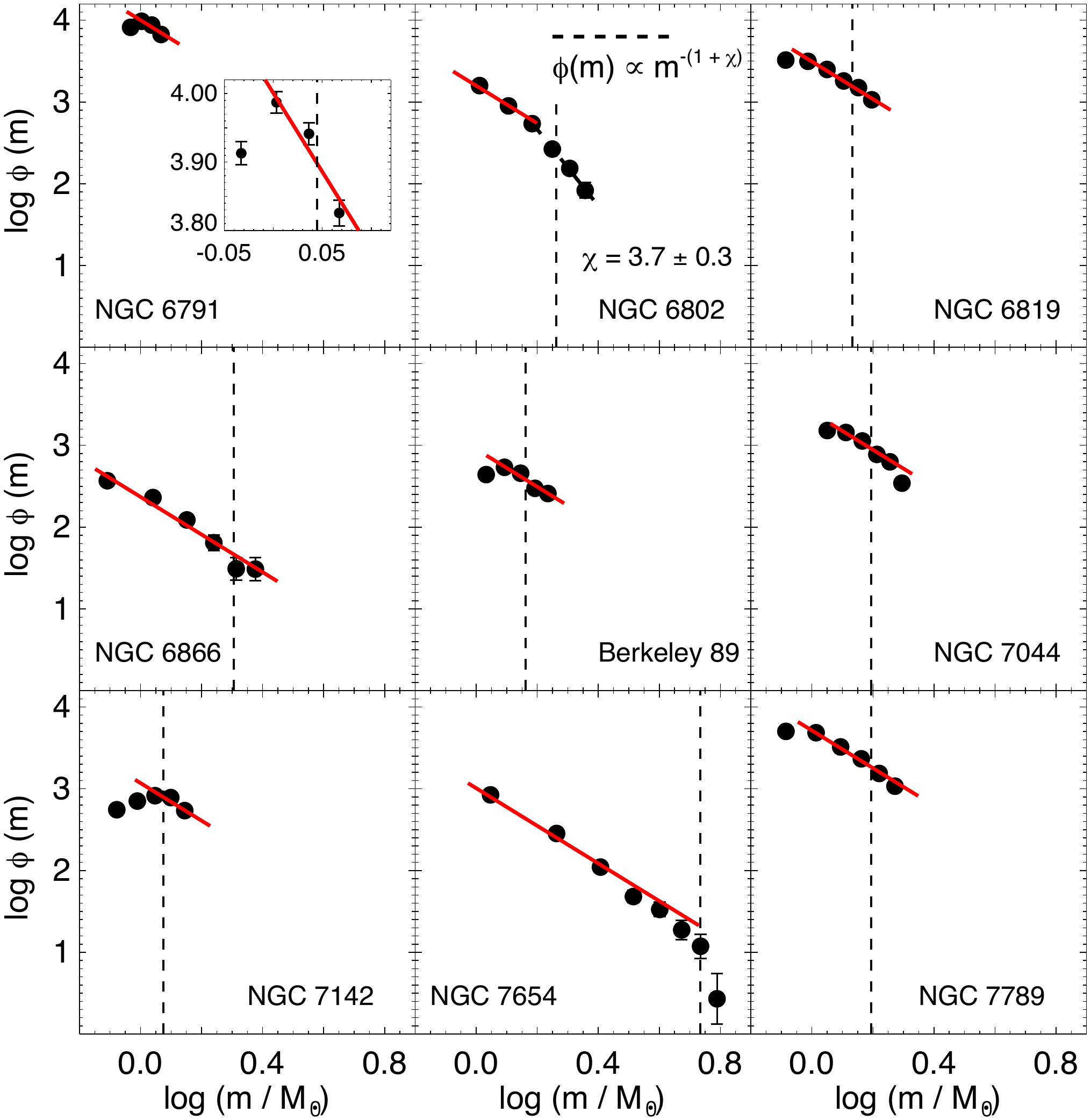}  
    \end{center}    
  }
\caption{ Same as Same as Figure\,8 of the manuscript, but for the OCs indicated in each panel. }

\label{fig:MFs_SupplMater5}
\end{center}
\end{figure*}

\clearpage

\section{Supplementary figures - Spectroscopic data}
This Appendix shows the set of plots containing spectroscopic data for 59 investigated OCs (Figures\,G1 to G59) not shown in the manuscript.

\begin{figure*}
\begin{center}

\parbox[c]{0.70\textwidth}
  {
   \begin{center}
    \includegraphics[width=0.70\textwidth]{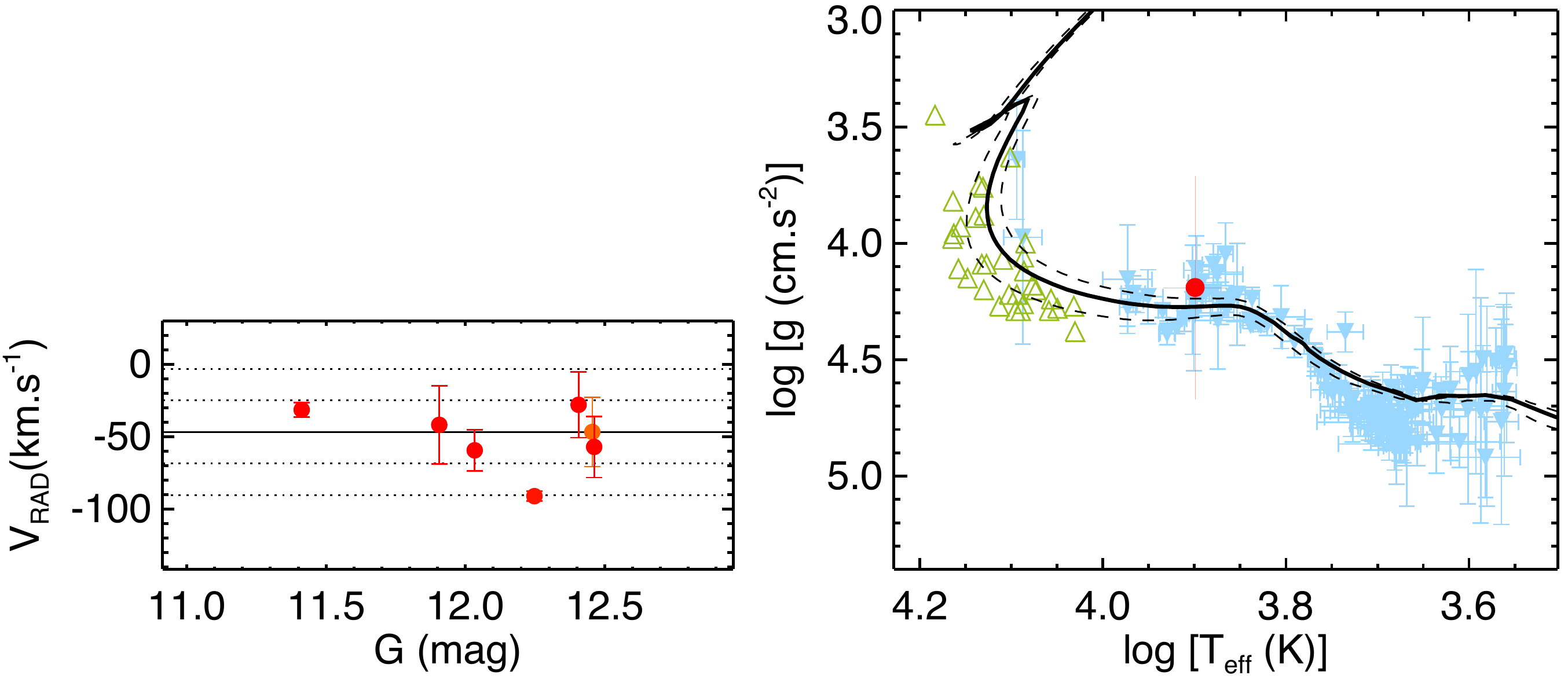}  
    \end{center}    
  }
\caption{ Same as Figure\,5 of the manuscript, but for the OC NGC\,129. No $[Fe/H]$ values available for the set of member stars. }

\label{fig:HRD_NGC129}
\end{center}
\end{figure*}

\begin{figure*}
\begin{center}

\parbox[c]{0.70\textwidth}
  {
   \begin{center}
    \includegraphics[width=0.70\textwidth]{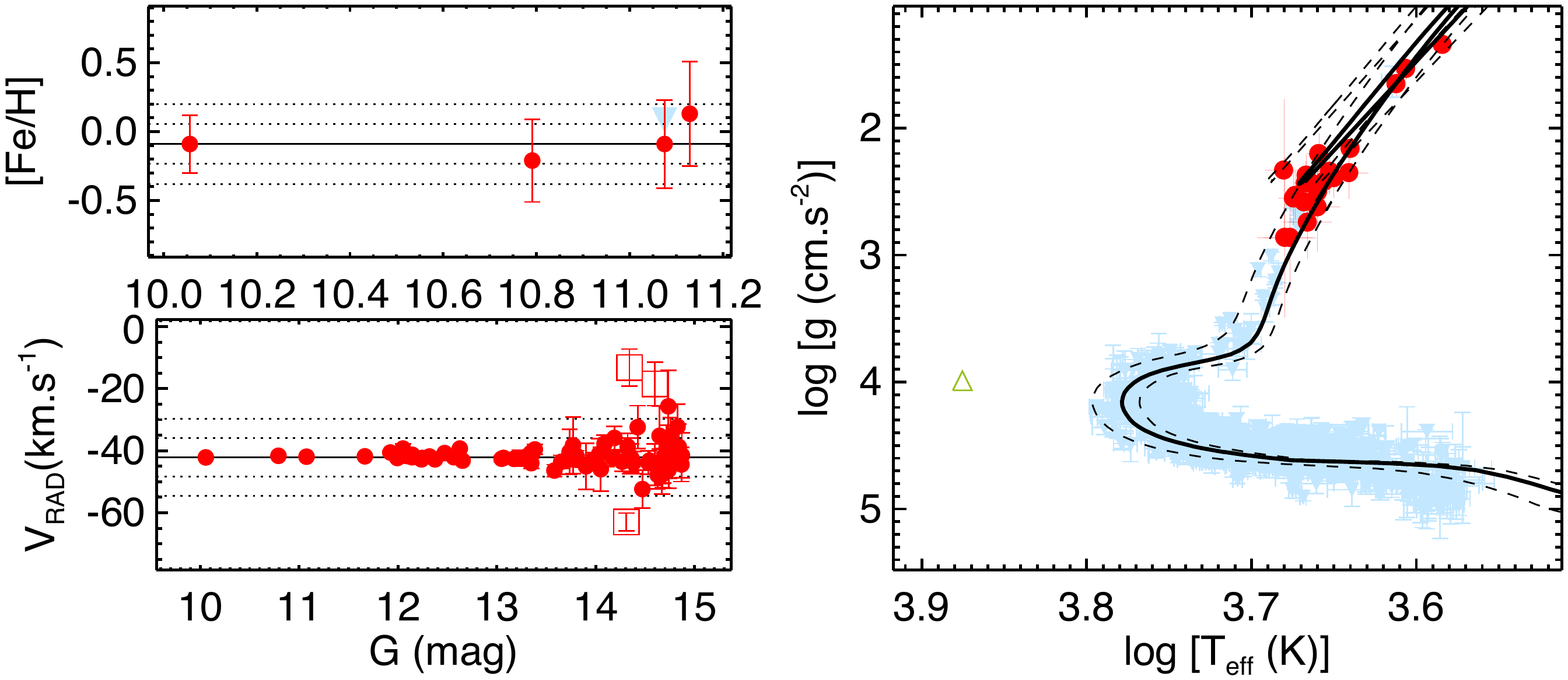}  
    \end{center}    
  }
\caption{ Same as Figure\,5 of the manuscript, but for the OC NGC\,188. The green open triangle represent the member star with \textit{source\_ID} 573968576057423360 ($\alpha_{\textrm{J2016}}=11.08923^{\circ}$; $\delta_{\textrm{J2016}}=85.441673^{\circ}$; $G=13.24\,$mag; $(G_{BP}-G_{RP})=0.45\,$mag), probably a blue straggler. }

\label{fig:HRD_NGC188}
\end{center}
\end{figure*}

\begin{figure*}
\begin{center}

\parbox[c]{0.70\textwidth}
  {
   \begin{center}
    \includegraphics[width=0.70\textwidth]{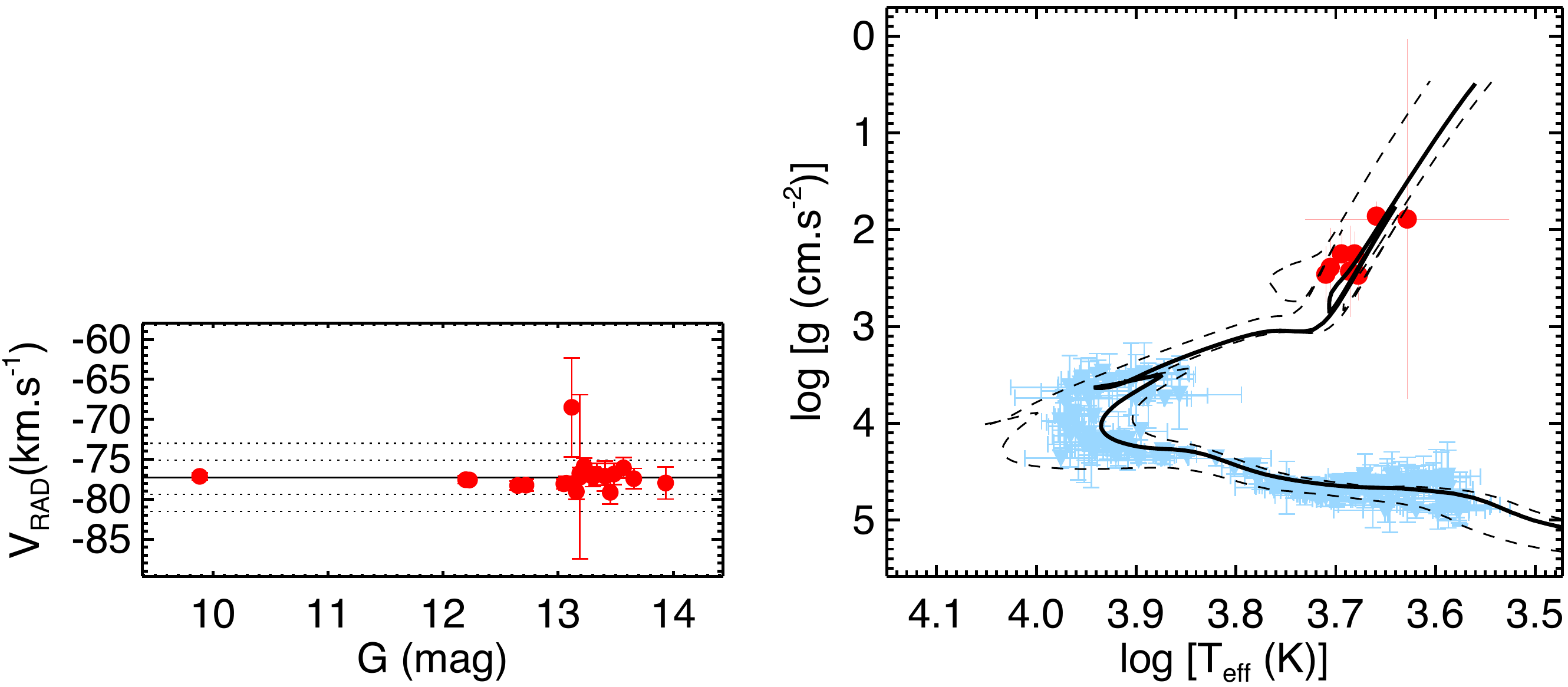}  
    \end{center}    
  }
\caption{ Same as Figure\,5 of the manuscript, but for the OC NGC\,559. No $[Fe/H]$ values available for the set of member stars. }

\label{fig:HRD_NGC559}
\end{center}
\end{figure*}

\begin{figure*}
\begin{center}

\parbox[c]{0.70\textwidth}
  {
   \begin{center}
    \includegraphics[width=0.70\textwidth]{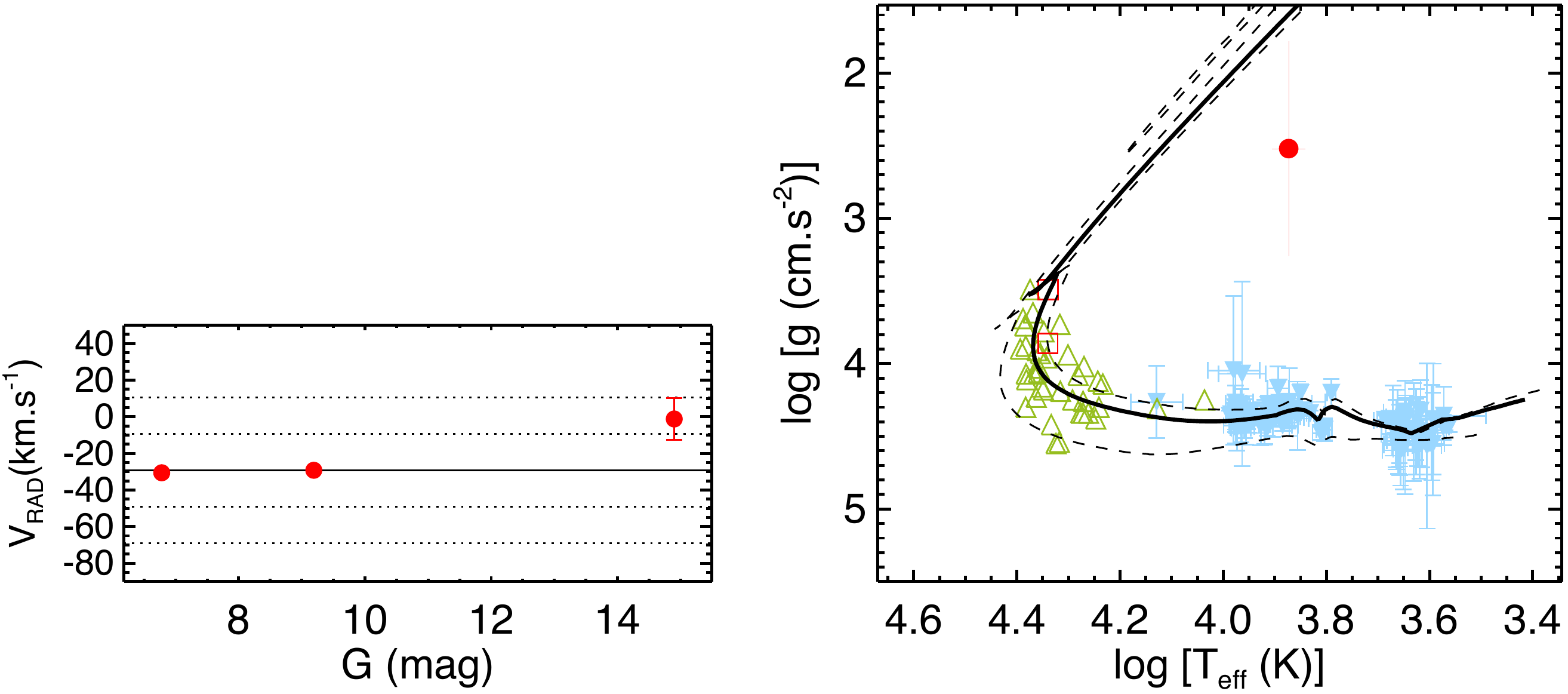}  
    \end{center}    
  }
\caption{ Same as Figure\,5 of the manuscript, but for the OC NGC\,654. No $[Fe/H]$ values available for the set of member stars. }

\label{fig:HRD_NGC654}
\end{center}
\end{figure*}

\begin{figure*}
\begin{center}

\parbox[c]{0.70\textwidth}
  {
   \begin{center}
    \includegraphics[width=0.70\textwidth]{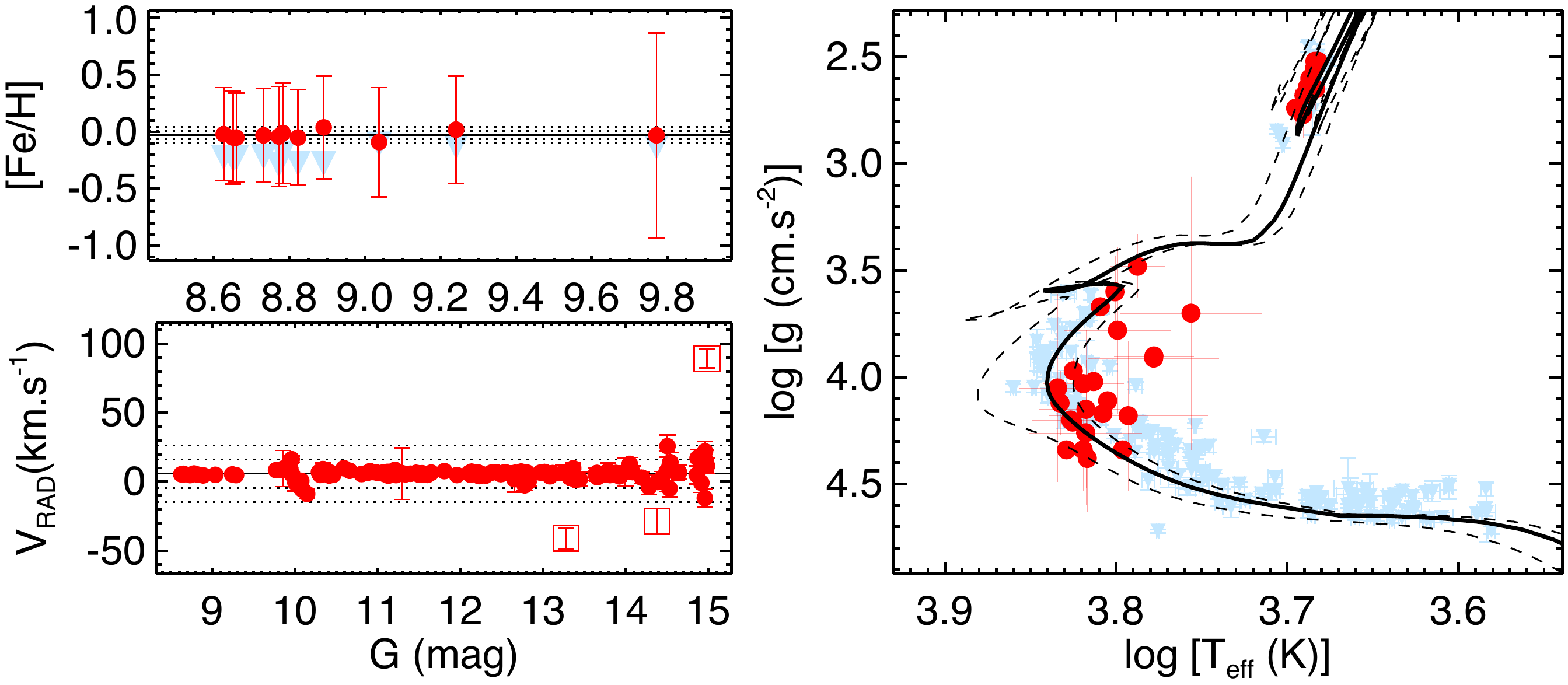}  
    \end{center}    
  }
\caption{ Same as Figure\,5 of the manuscript, but for the OC NGC\,752. }

\label{fig:HRD_NGC752}
\end{center}
\end{figure*}

\begin{figure*}
\begin{center}

\parbox[c]{0.70\textwidth}
  {
   \begin{center}
    \includegraphics[width=0.70\textwidth]{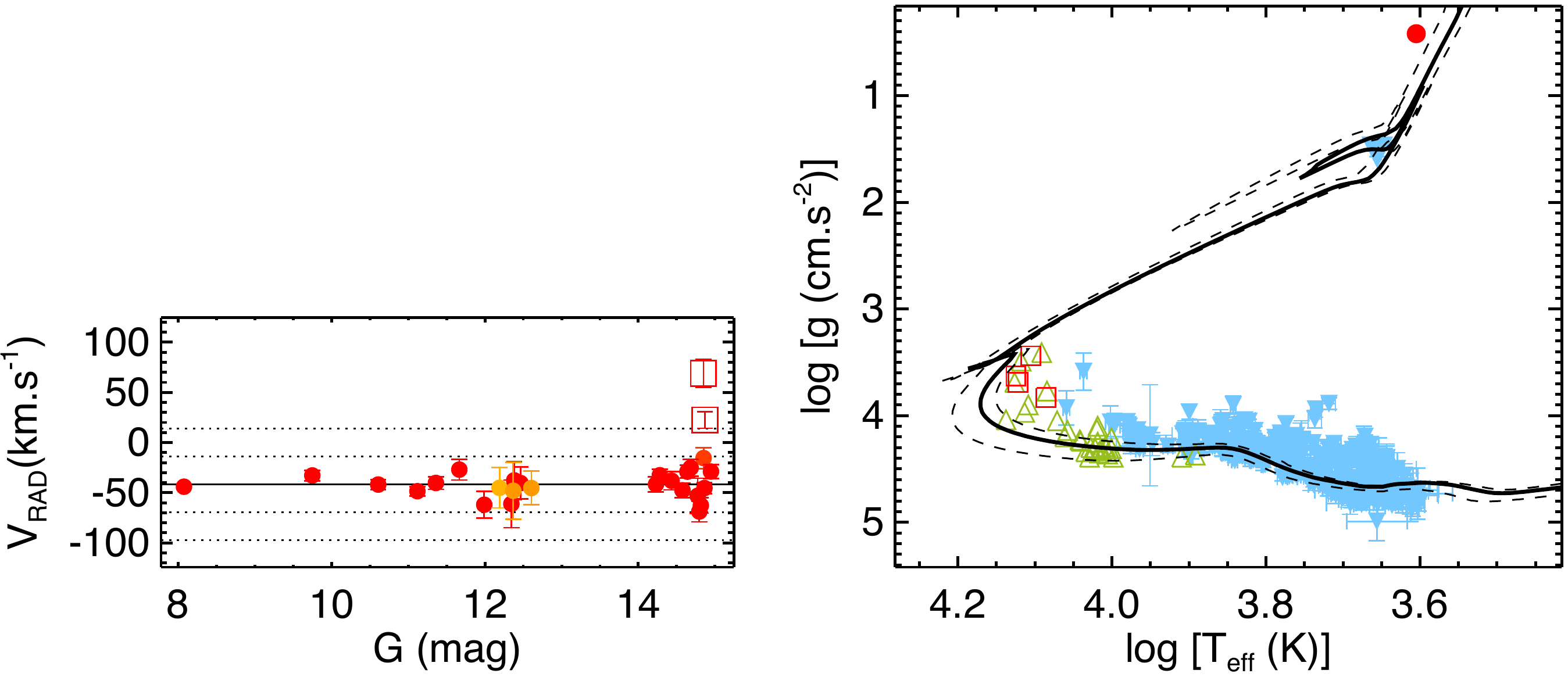}  
    \end{center}    
  }
\caption{ Same as Figure\,5 of the manuscript, but for the OC NGC\,1027. No $[Fe/H]$ values available for the set of member stars. }

\label{fig:HRD_NGC1027}
\end{center}
\end{figure*}

\begin{figure*}
\begin{center}

\parbox[c]{0.70\textwidth}
  {
   \begin{center}
    \includegraphics[width=0.70\textwidth]{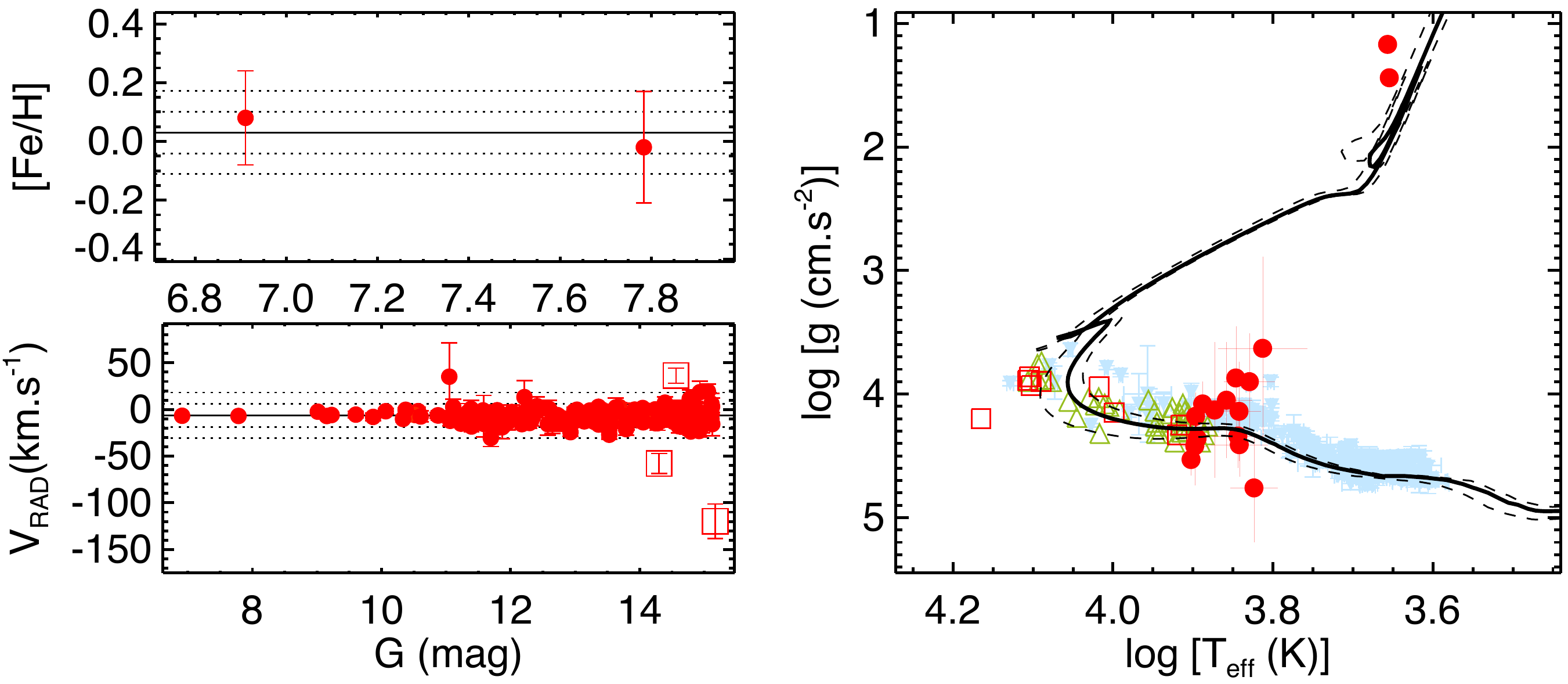}  
    \end{center}    
  }
\caption{ Same as Figure\,5 of the manuscript, but for the OC NGC\,1647. }

\label{fig:HRD_NGC1647}
\end{center}
\end{figure*}

\begin{figure*}
\begin{center}

\parbox[c]{0.70\textwidth}
  {
   \begin{center}
    \includegraphics[width=0.70\textwidth]{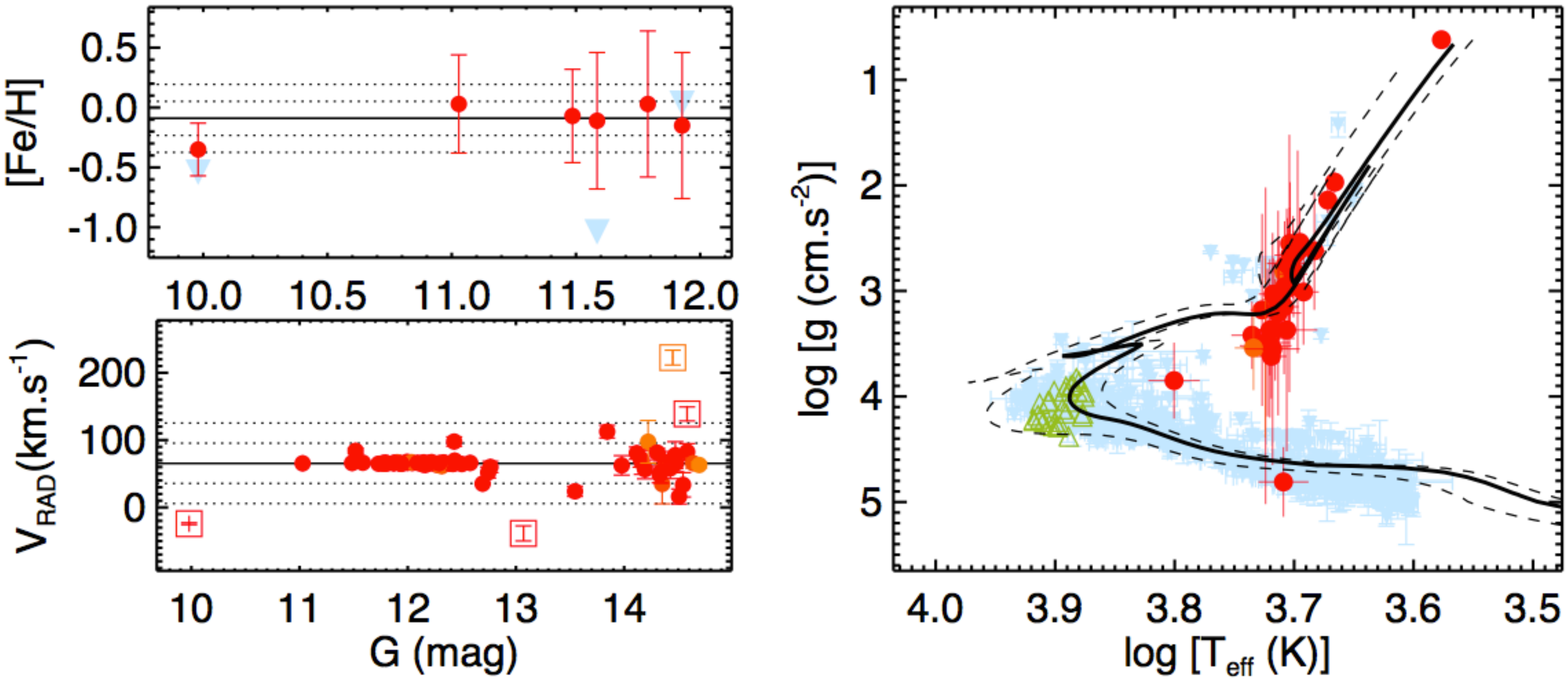}  
    \end{center}    
  }
\caption{ Same as Figure\,5 of the manuscript, but for the OC NGC\,1817. }

\label{fig:HRD_NGC1647}
\end{center}
\end{figure*}

\begin{figure*}
\begin{center}

\parbox[c]{0.70\textwidth}
  {
   \begin{center}
    \includegraphics[width=0.70\textwidth]{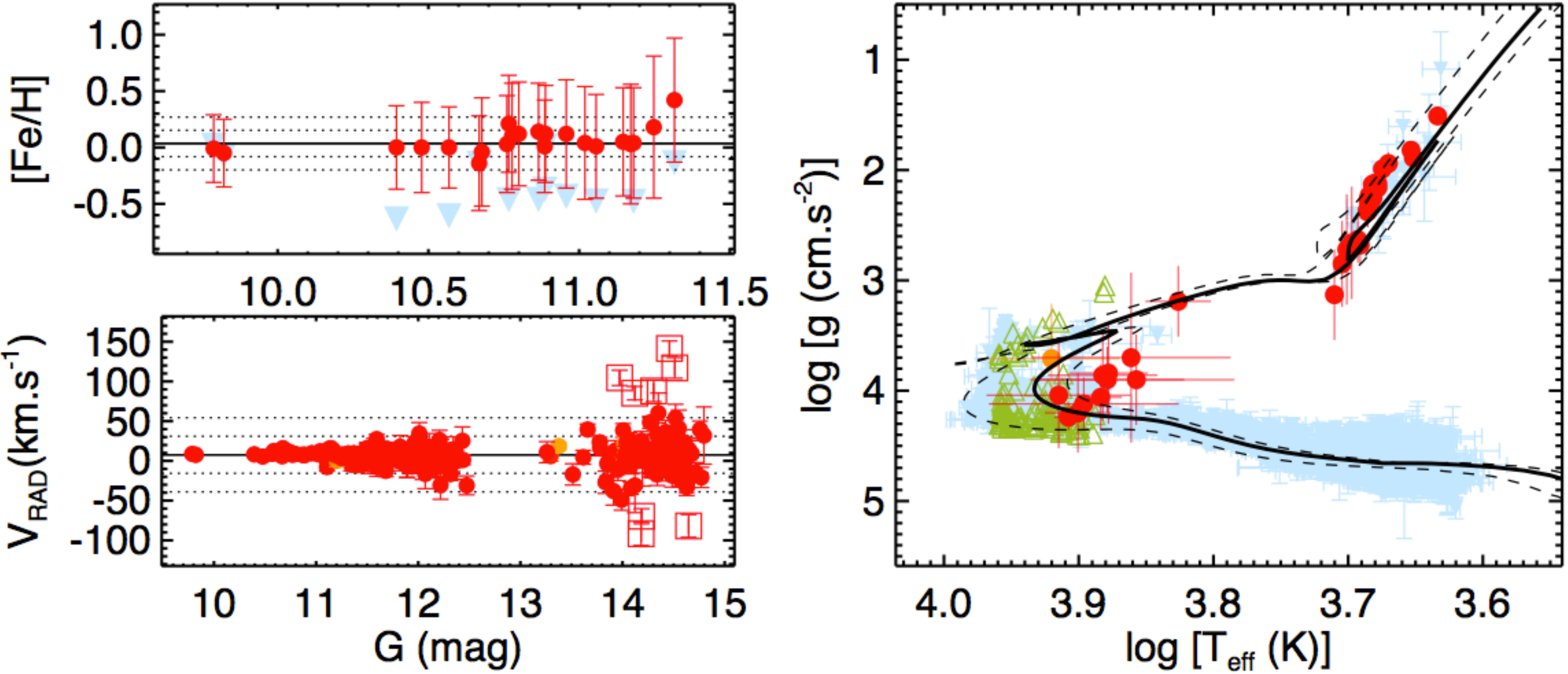}  
    \end{center}    
  }
\caption{ Same as Figure\,5 of the manuscript, but for the OC M\,37. }

\label{fig:HRD_M37}
\end{center}
\end{figure*}

\afterpage{\clearpage}

\begin{figure*}
\begin{center}

\parbox[c]{0.70\textwidth}
  {
   \begin{center}
    \includegraphics[width=0.70\textwidth]{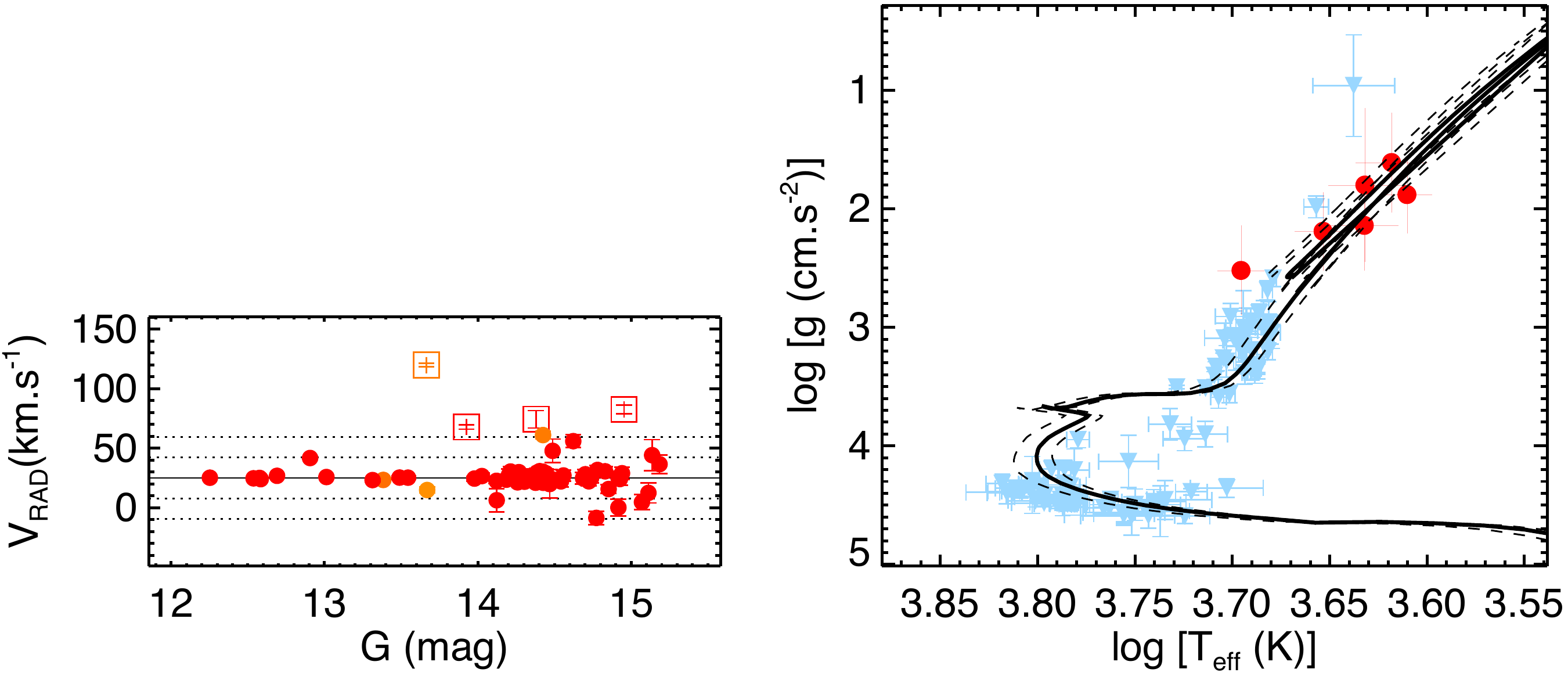}  
    \end{center}    
  }
\caption{ Same as Figure\,5 of the manuscript, but for the OC NGC\,2141. No $[Fe/H]$ values available for the set of member stars.}

\label{fig:HRD_NGC2141}
\end{center}
\end{figure*}

\begin{figure*}
\begin{center}

\parbox[c]{0.70\textwidth}
  {
   \begin{center}
    \includegraphics[width=0.70\textwidth]{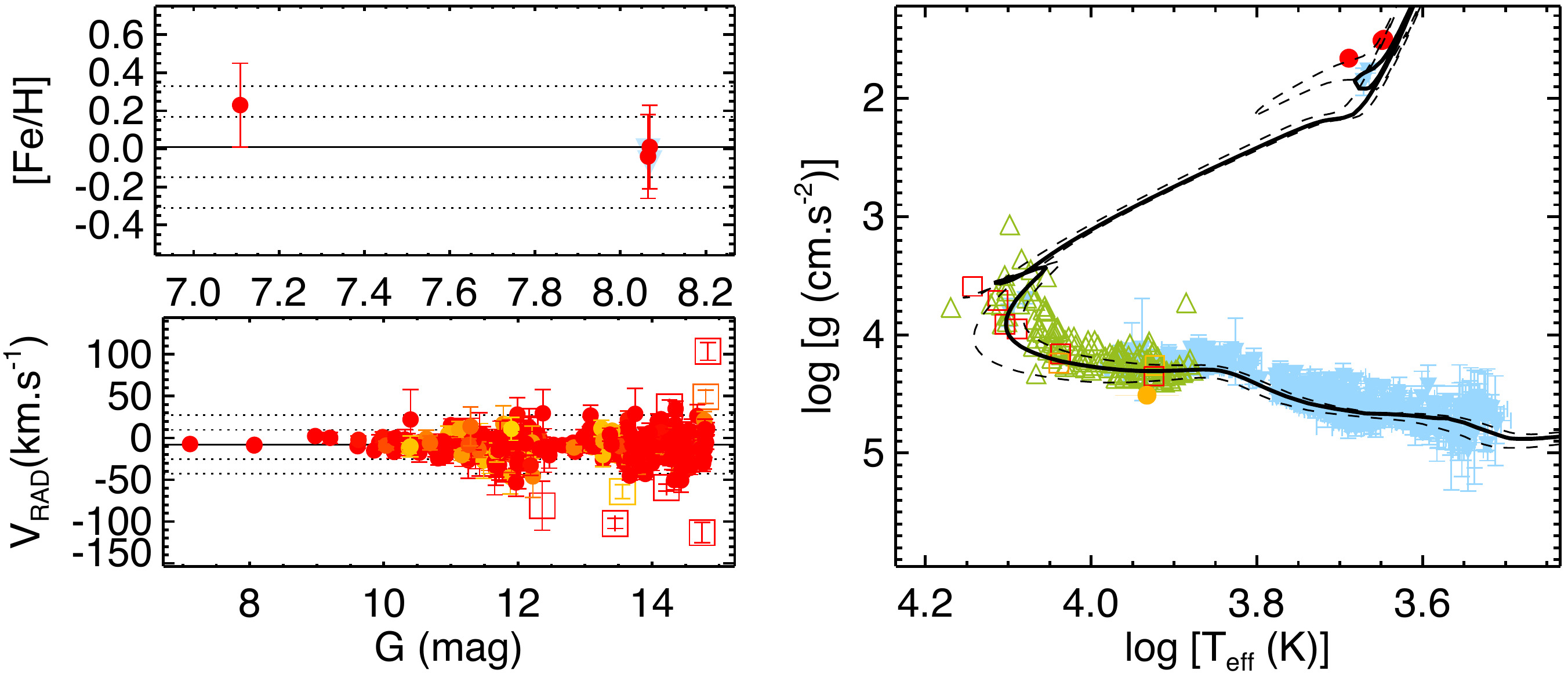}  
    \end{center}    
  }
\caption{ Same as Figure\,5 of the manuscript, but for the OC NGC\,2168. }

\label{fig:HRD_NGC2168}
\end{center}
\end{figure*}

\begin{figure*}
\begin{center}

\parbox[c]{0.70\textwidth}
  {
   \begin{center}
    \includegraphics[width=0.70\textwidth]{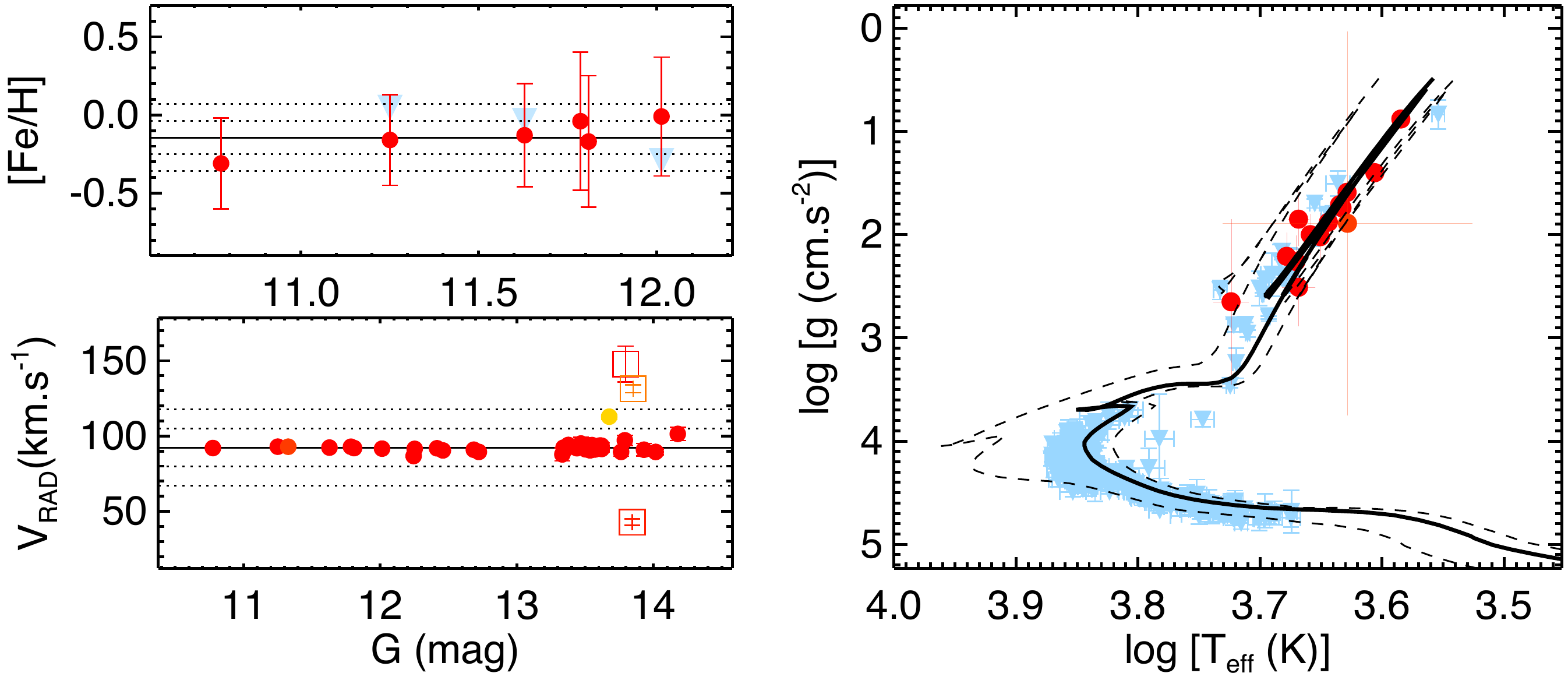}  
    \end{center}    
  }
\caption{ Same as Figure\,5 of the manuscript, but for the OC NGC\,2204. }

\label{fig:HRD_NGC2204}
\end{center}
\end{figure*}

\begin{figure*}
\begin{center}

\parbox[c]{0.70\textwidth}
  {
   \begin{center}
    \includegraphics[width=0.70\textwidth]{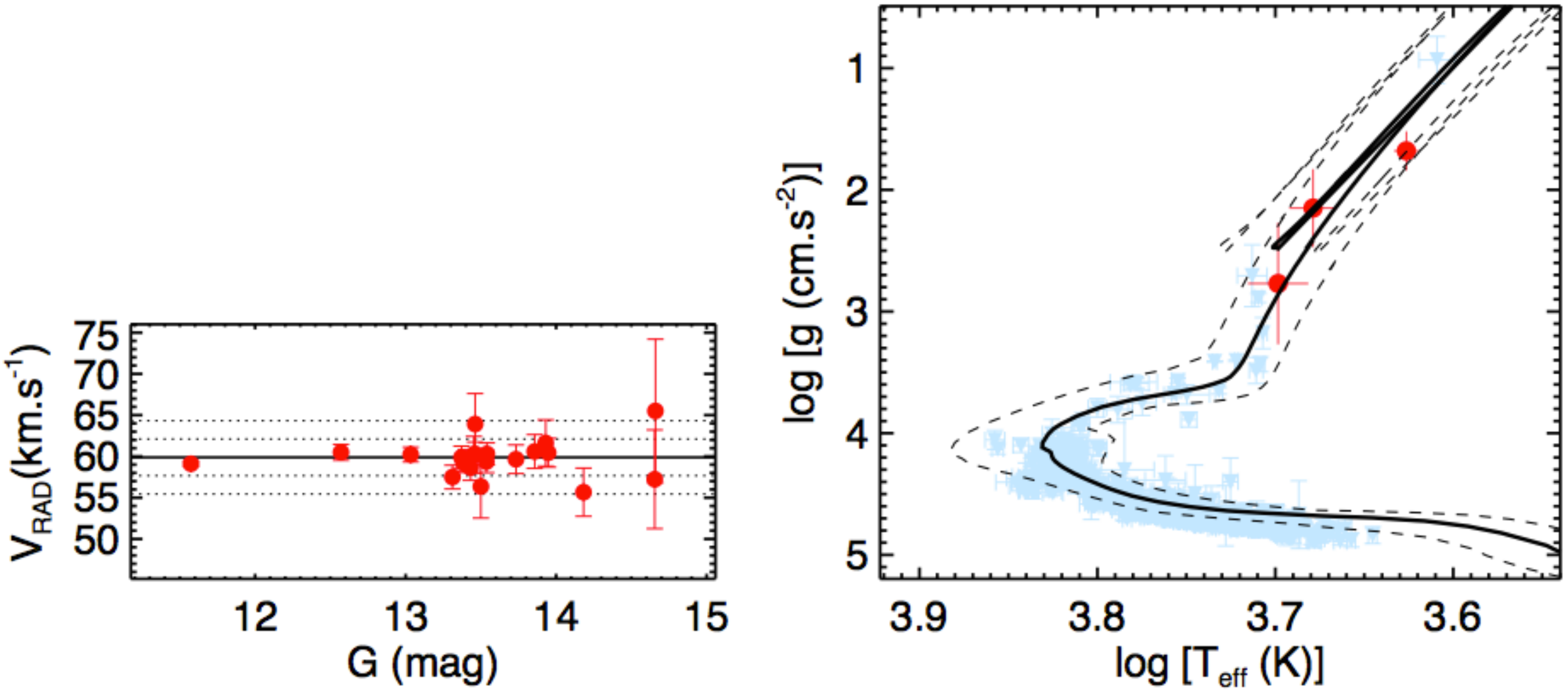}  
    \end{center}    
  }
\caption{ Same as Figure\,5 of the manuscript, but for the OC NGC\,2243. No $[Fe/H]$ values available for the set of member stars. }

\label{fig:HRD_NGC2243}
\end{center}
\end{figure*}

\begin{figure*}
\begin{center}

\parbox[c]{0.70\textwidth}
  {
   \begin{center}
    \includegraphics[width=0.70\textwidth]{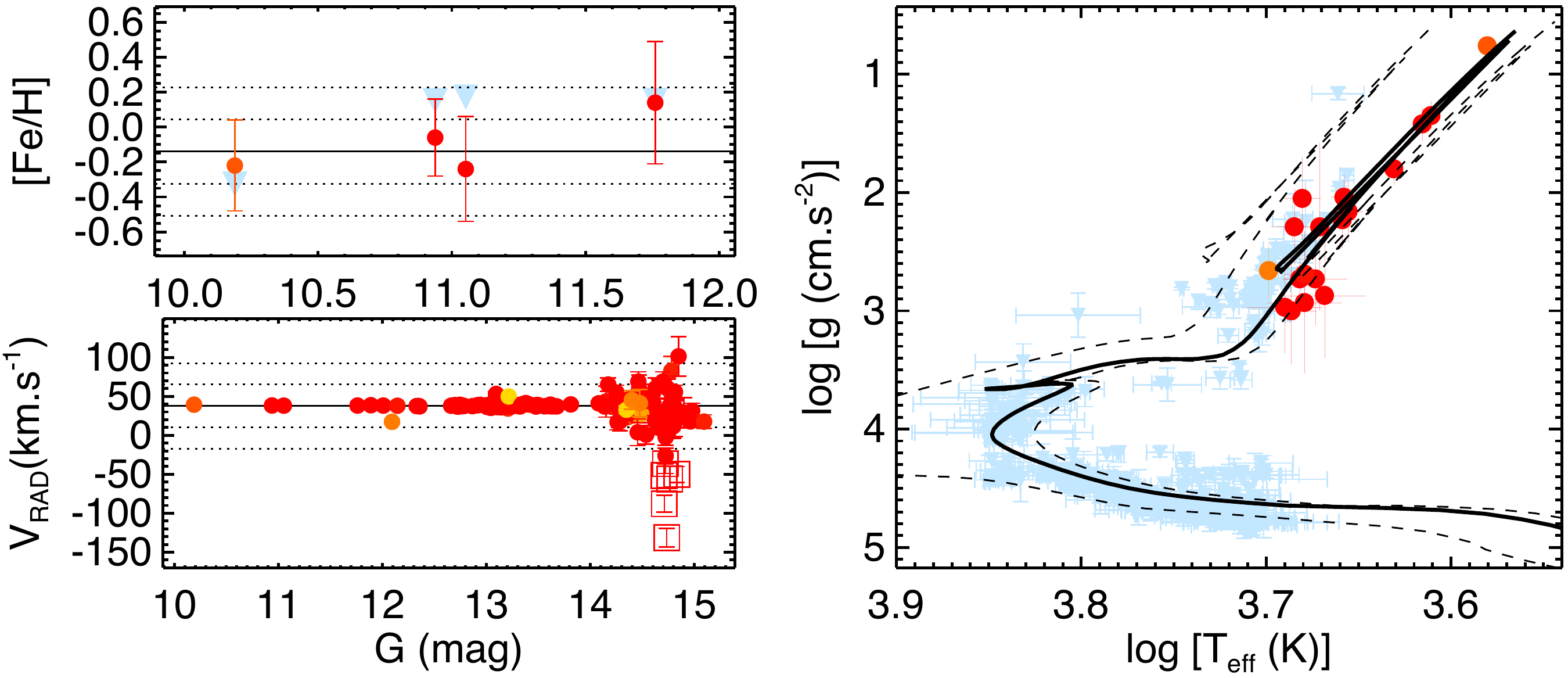}  
    \end{center}    
  }
\caption{ Same as Figure\,5 of the manuscript, but for the OC Collinder\,110. }

\label{fig:HRD_Collinder110}
\end{center}
\end{figure*}

\begin{figure*}
\begin{center}

\parbox[c]{0.70\textwidth}
  {
   \begin{center}
    \includegraphics[width=0.70\textwidth]{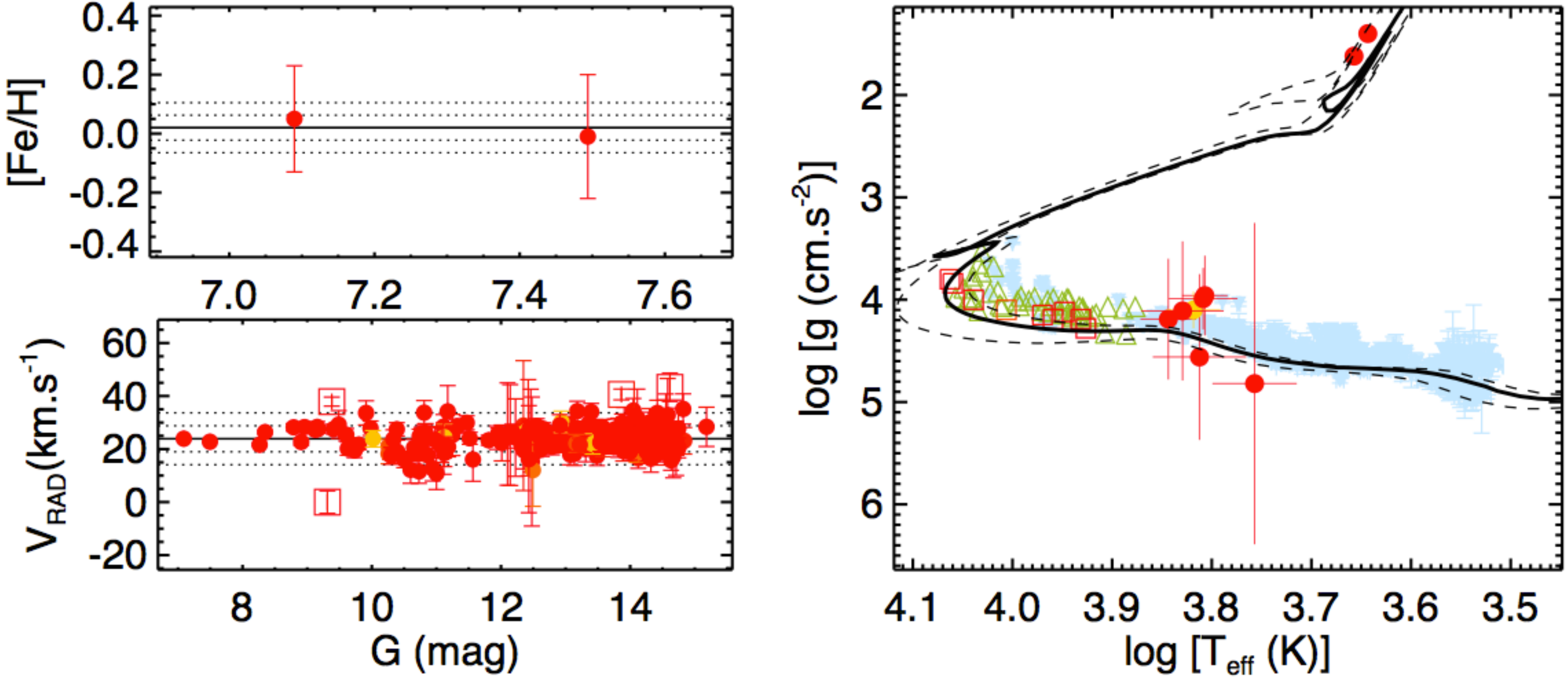}  
    \end{center}    
  }
\caption{ Same as Figure\,5 of the manuscript, but for the OC NGC\,2287. }

\label{fig:HRD_NGC2287}
\end{center}
\end{figure*}

\afterpage{\clearpage}

\begin{figure*}
\begin{center}

\parbox[c]{0.70\textwidth}
  {
   \begin{center}
    \includegraphics[width=0.70\textwidth]{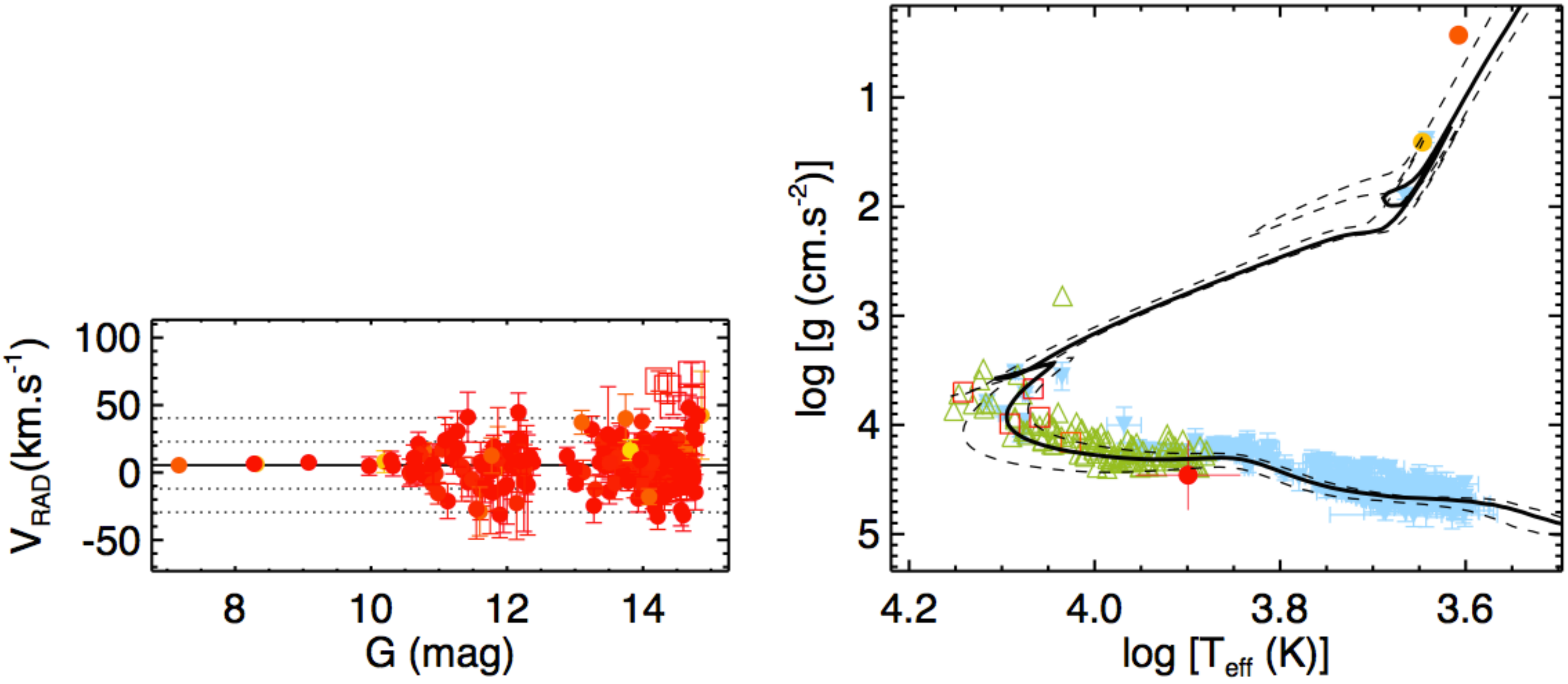}  
    \end{center}    
  }
\caption{ Same as Figure\,5 of the manuscript, but for the OC NGC\,2323. No $[Fe/H]$ values available for the set of member stars. }

\label{fig:HRD_NGC2323}
\end{center}
\end{figure*}

\begin{figure*}
\begin{center}

\parbox[c]{0.70\textwidth}
  {
   \begin{center}
    \includegraphics[width=0.70\textwidth]{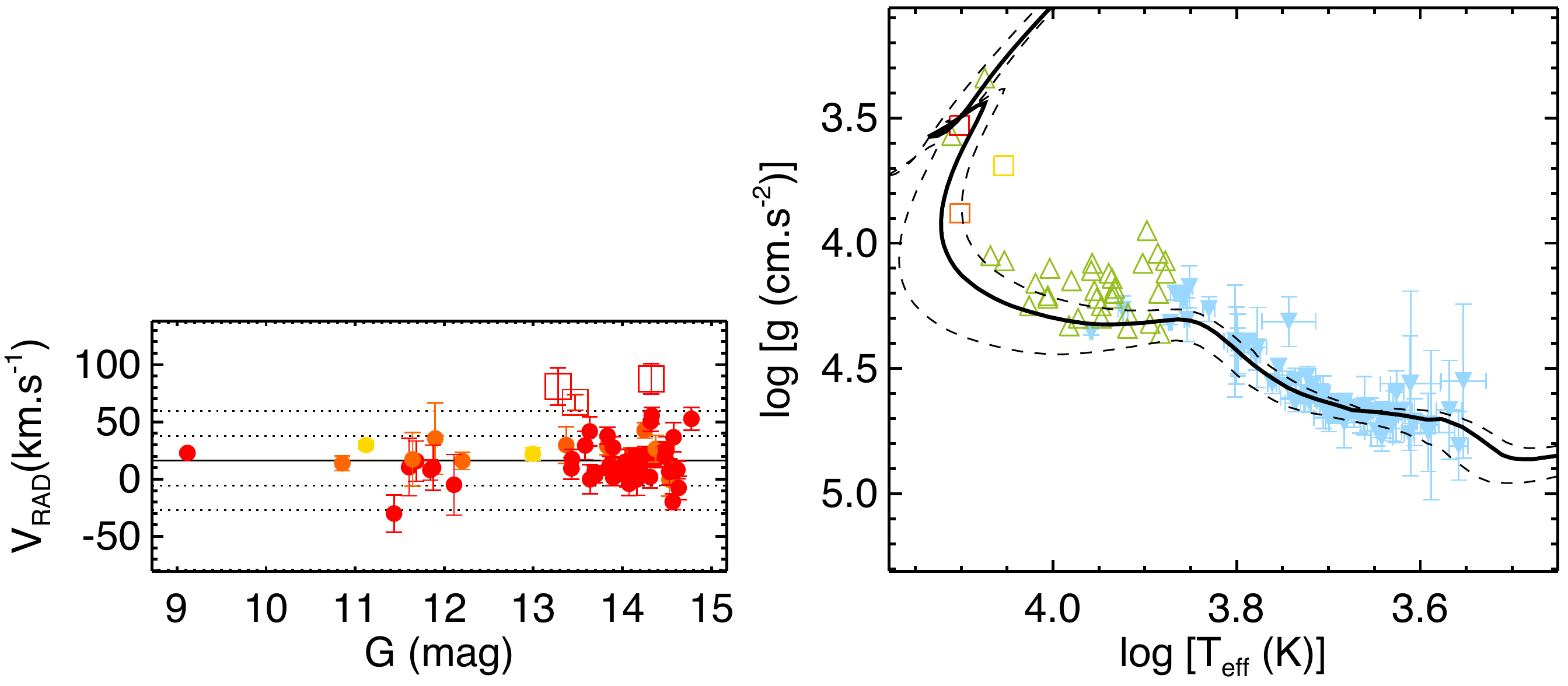}  
    \end{center}    
  }
\caption{ Same as Figure\,5 of the manuscript, but for the OC NGC\,2353. No $[Fe/H]$ values available for the set of member stars. }

\label{fig:HRD_NGC2353}
\end{center}
\end{figure*}

\begin{figure*}
\begin{center}

\parbox[c]{0.70\textwidth}
  {
   \begin{center}
    \includegraphics[width=0.70\textwidth]{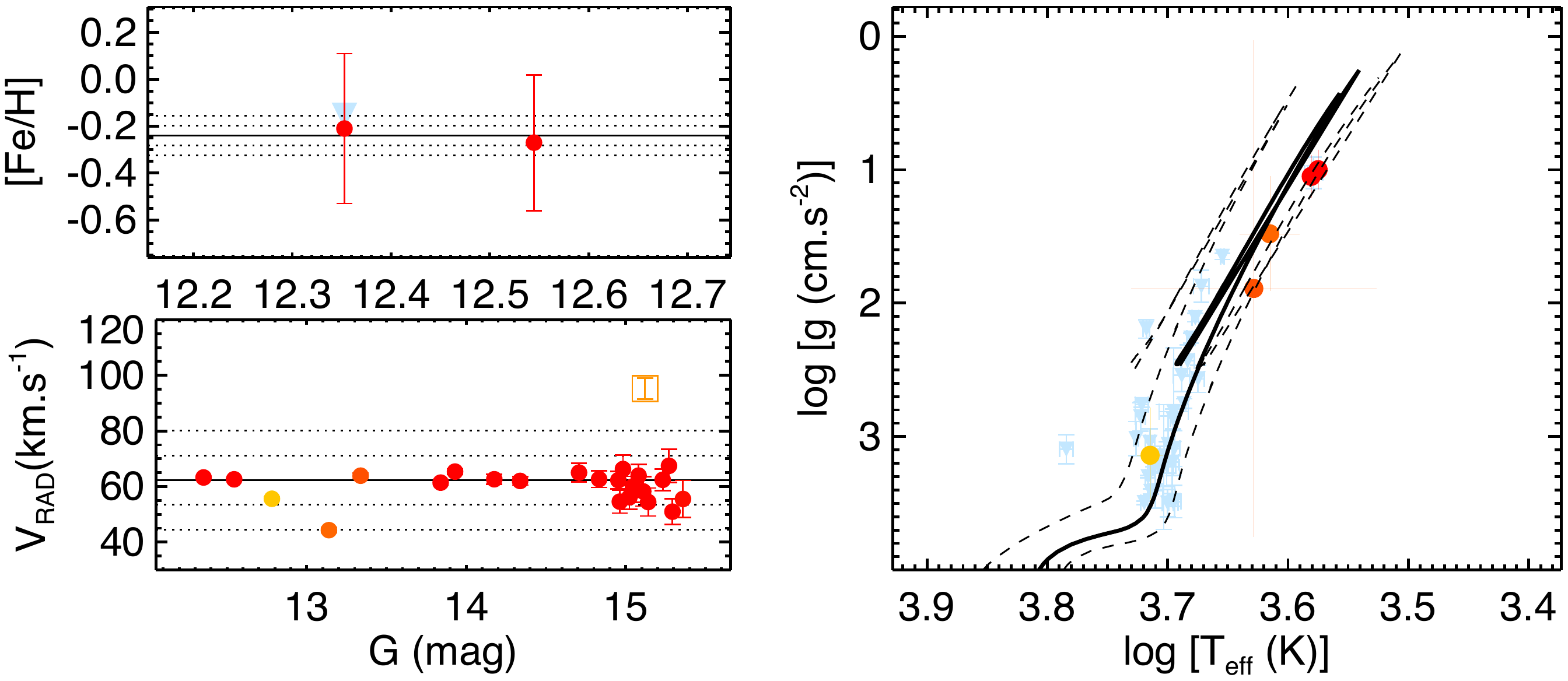}  
    \end{center}    
  }
\caption{ Same as Figure\,5 of the manuscript, but for the OC Berkeley\,36. }

\label{fig:HRD_Berkeley36}
\end{center}
\end{figure*}

\begin{figure*}
\begin{center}

\parbox[c]{0.70\textwidth}
  {
   \begin{center}
    \includegraphics[width=0.70\textwidth]{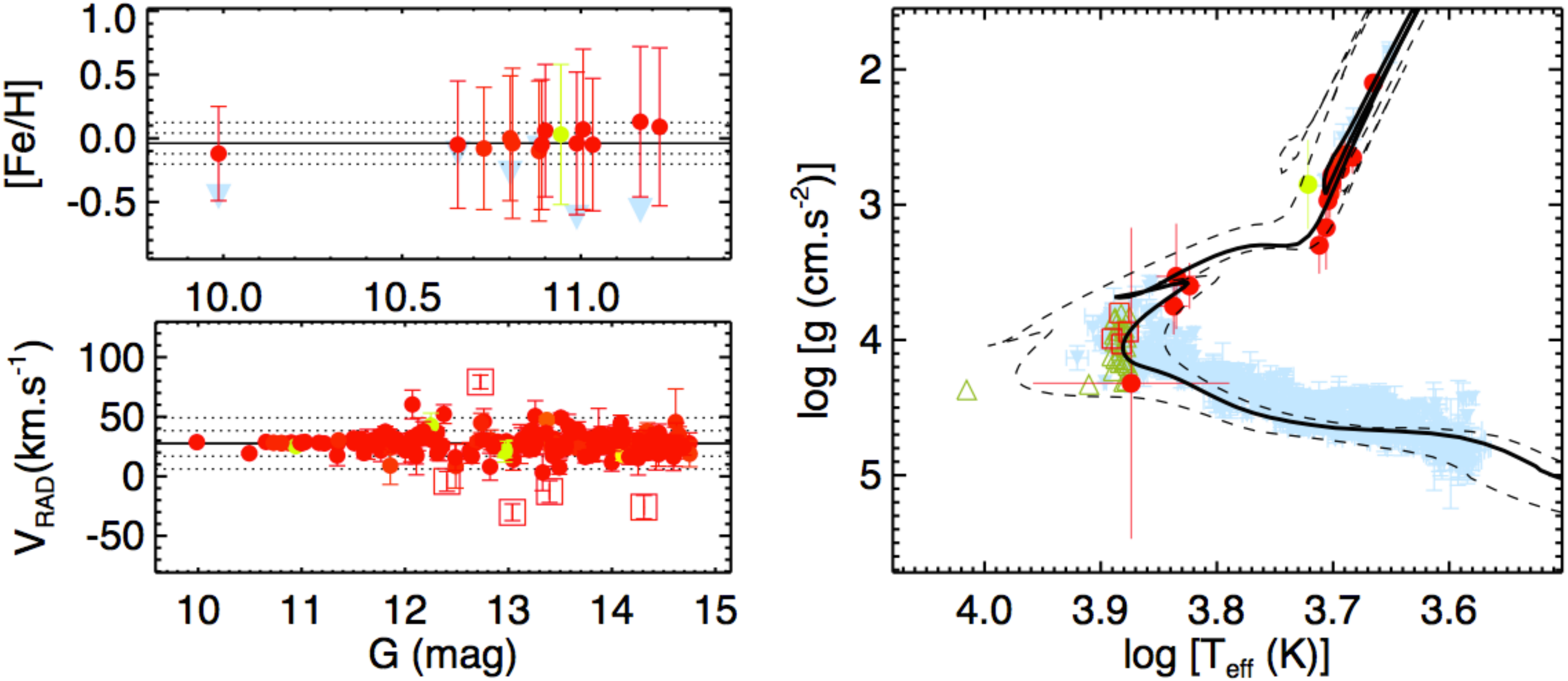}  
    \end{center}    
  }
\caption{ Same as Figure\,5 of the manuscript, but for the OC NGC\,2360. }

\label{fig:HRD_NGC2360}
\end{center}
\end{figure*}

\begin{figure*}
\begin{center}

\parbox[c]{0.70\textwidth}
  {
   \begin{center}
    \includegraphics[width=0.70\textwidth]{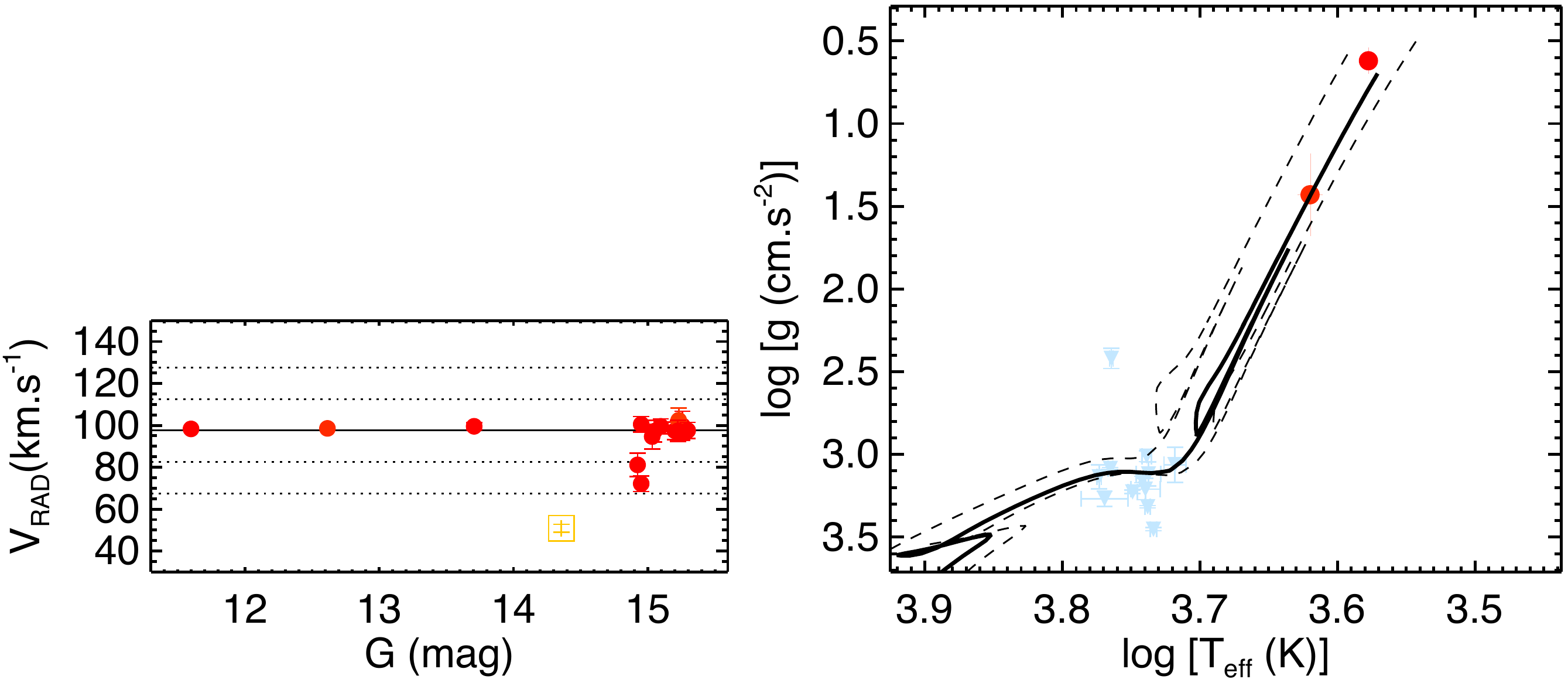}  
    \end{center}    
  }
\caption{ Same as Figure\,5 of the manuscript, but for the OC Haffner\,11. No $[Fe/H]$ values available for the set of member stars. }

\label{fig:HRD_Haffner11}
\end{center}
\end{figure*}

\afterpage{\clearpage}

\begin{figure*}
\begin{center}

\parbox[c]{0.70\textwidth}
  {
   \begin{center}
    \includegraphics[width=0.70\textwidth]{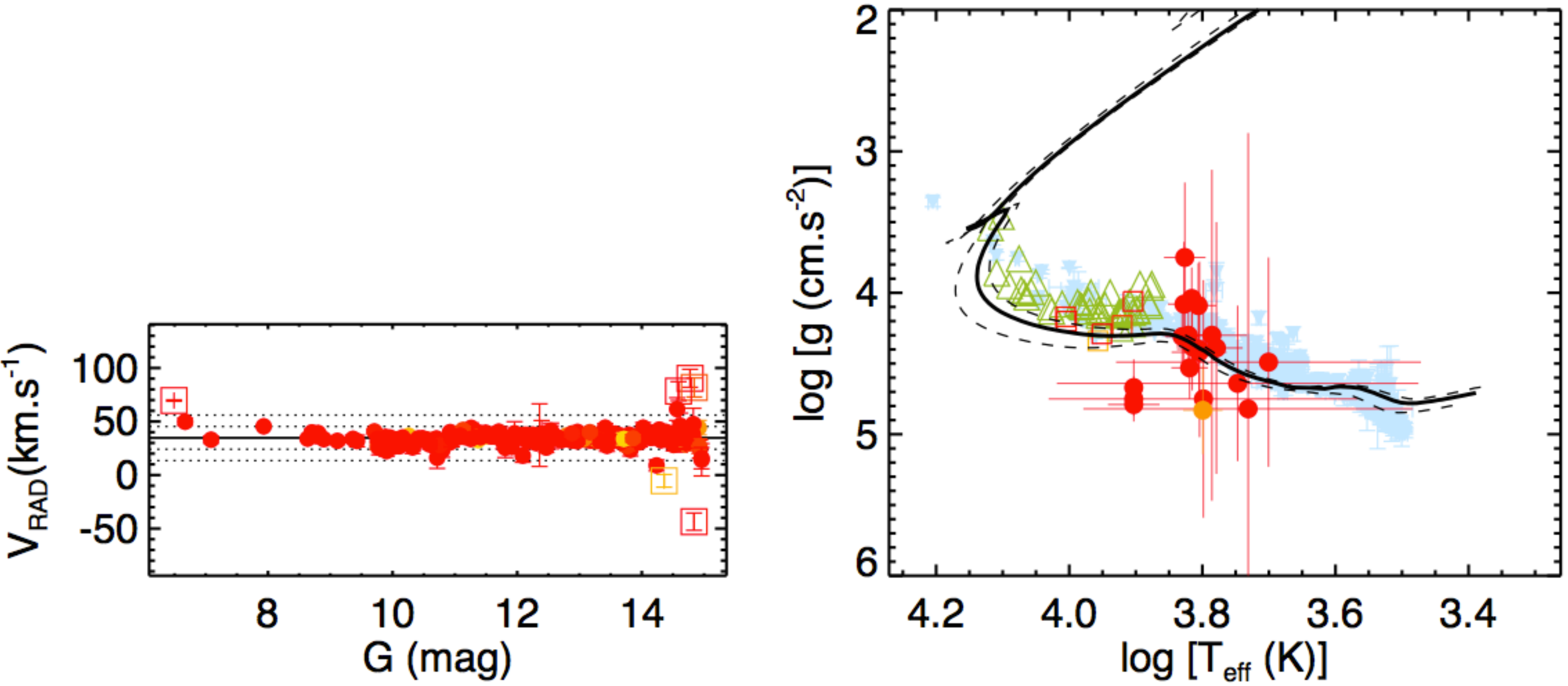}  
    \end{center}    
  }
\caption{ Same as Figure\,5 of the manuscript, but for the OC NGC\,2422. No $[Fe/H]$ values available for the set of member stars. }

\label{fig:HRD_NGC2422}
\end{center}
\end{figure*}

\begin{figure*}
\begin{center}

\parbox[c]{0.70\textwidth}
  {
   \begin{center}
    \includegraphics[width=0.70\textwidth]{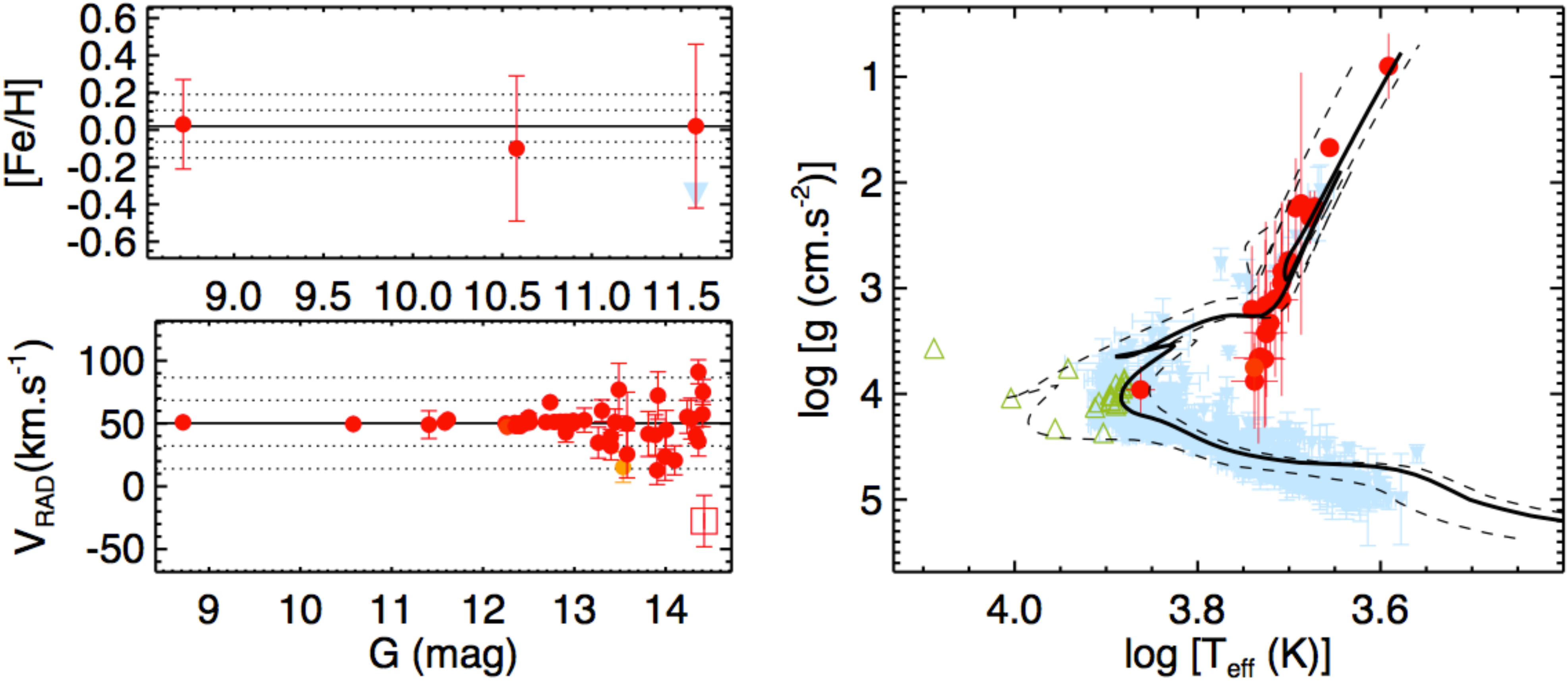}  
    \end{center}    
  }
\caption{ Same as Figure\,5 of the manuscript, but for the OC Melotte\,71. The hottest star (open green triangle with log\,($T_{\textrm{eff}}$)=4.09, log\,$g$=3.57) represent the member star with \textit{source\_ID} 3033959198481332736 ($\alpha_{\textrm{J2016}}=114.40683^{\circ}$; $\delta_{\textrm{J2016}}=-12.064518^{\circ}$; $G=11.41\,$mag; $(G_{BP}-G_{RP})=0.15\,$mag), probably a blue straggler. }

\label{fig:HRD_Melotte71}
\end{center}
\end{figure*}

\begin{figure*}
\begin{center}

\parbox[c]{0.70\textwidth}
  {
   \begin{center}
    \includegraphics[width=0.70\textwidth]{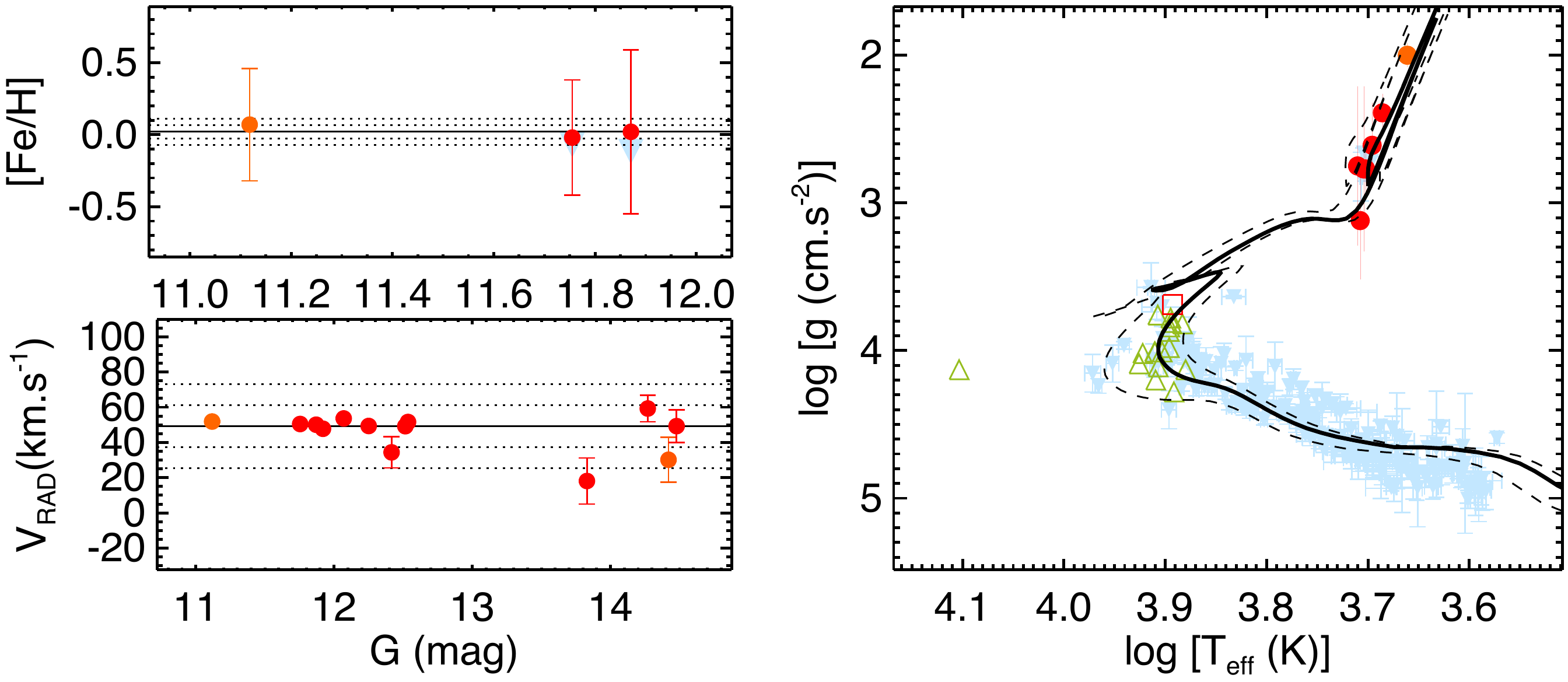}  
    \end{center}    
  }
\caption{ Same as Figure\,5 of the manuscript, but for the OC NGC\,2432. The hottest star (open green triangle with log\,($T_{\textrm{eff}}$)=4.10, log\,$g$=4.13) represent the member star with \textit{source\_ID} 5716778737170678912 ($\alpha_{\textrm{J2016}}=115.22637^{\circ}$; $\delta_{\textrm{J2016}}=-19.09102^{\circ}$; $G=12.03\,$mag; $(G_{BP}-G_{RP})=0.35\,$mag), probably a blue straggler. }

\label{fig:HRD_NGC2432}
\end{center}
\end{figure*}

\begin{figure*}
\begin{center}

\parbox[c]{0.70\textwidth}
  {
   \begin{center}
    \includegraphics[width=0.70\textwidth]{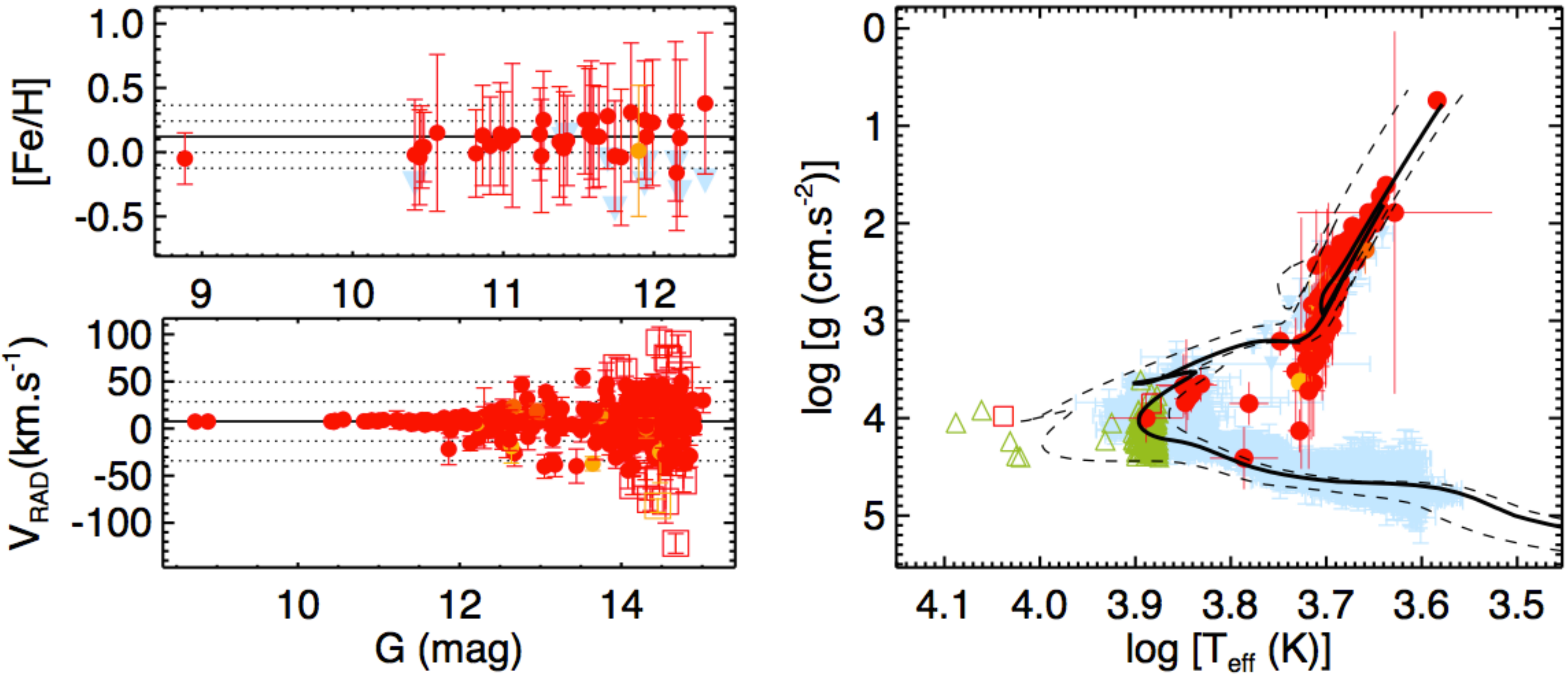}  
    \end{center}    
  }
\caption{ Same as Figure\,5 of the manuscript, but for the OC NGC\,2477. }

\label{fig:HRD_NGC2477}
\end{center}
\end{figure*}

\begin{figure*}
\begin{center}

\parbox[c]{0.70\textwidth}
  {
   \begin{center}
    \includegraphics[width=0.70\textwidth]{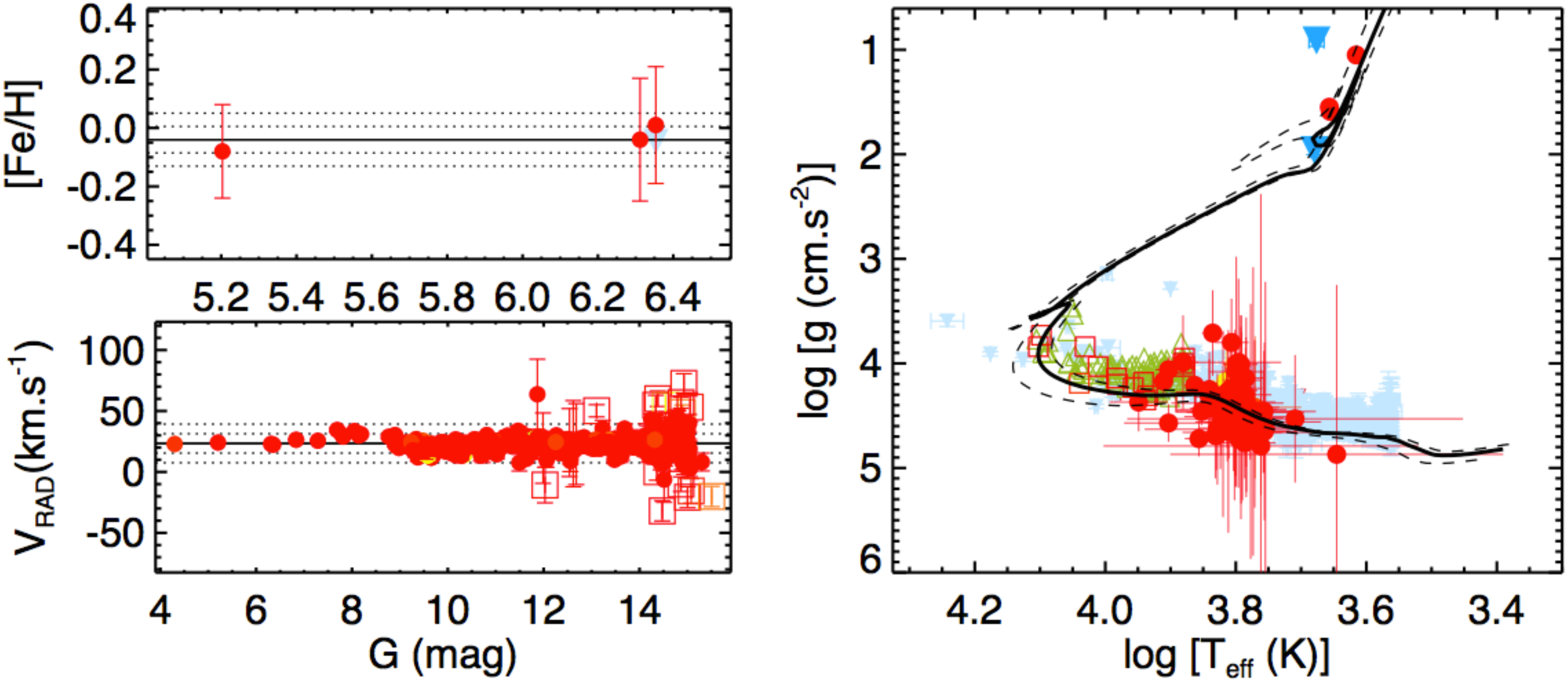}  
    \end{center}    
  }
\caption{ Same as Figure\,5 of the manuscript, but for the OC NGC\,2516. }

\label{fig:HRD_NGC2516}
\end{center}
\end{figure*}

\begin{figure*}
\begin{center}

\parbox[c]{0.70\textwidth}
  {
   \begin{center}
    \includegraphics[width=0.70\textwidth]{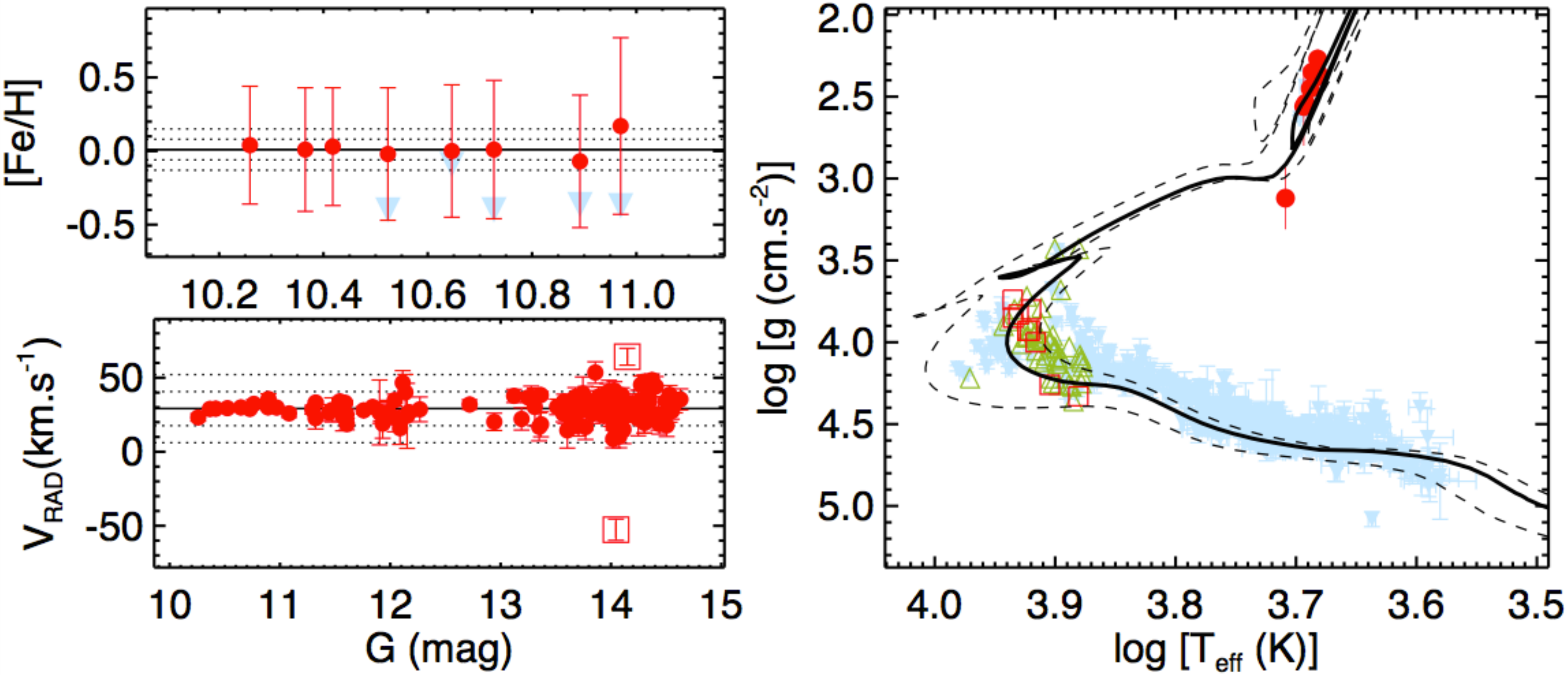}  
    \end{center}    
  }
\caption{ Same as Figure\,5 of the manuscript, but for the OC NGC\,2539. }

\label{fig:HRD_NGC2539}
\end{center}
\end{figure*}

\begin{figure*}
\begin{center}

\parbox[c]{0.70\textwidth}
  {
   \begin{center}
    \includegraphics[width=0.70\textwidth]{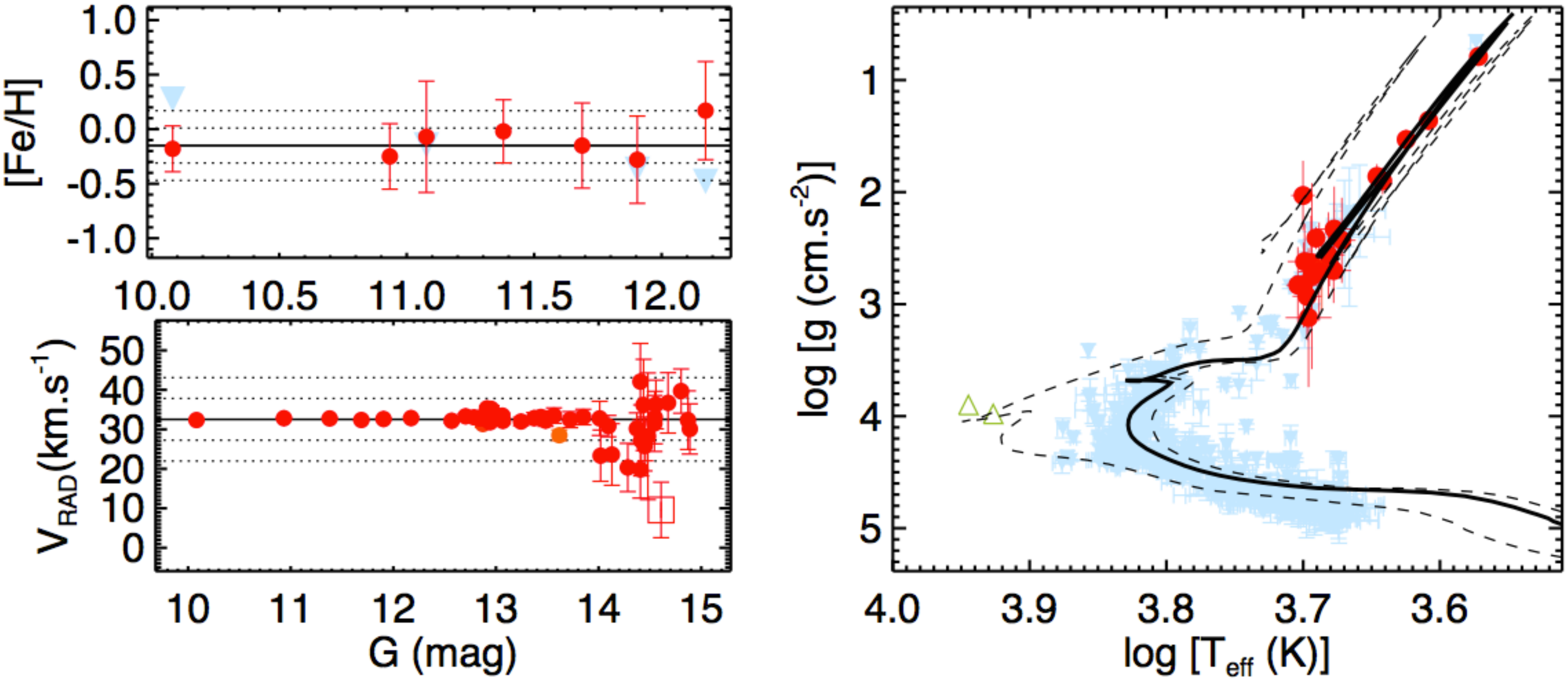}  
    \end{center}    
  }
\caption{ Same as Figure\,5 of the manuscript, but for the OC Haffner\,22. The 2 hottest stars (open green triangles with log\,$T_{\textrm{eff}}$=3.93, log\,$g$=3.98  and log\,$T_{\textrm{eff}}$=3.95, log\,$g$=3.90) represent member stars with \textit{source\_ID} 5693063199077161984 ($\alpha_{\textrm{J2016}}=123.32606^{\circ}$; $\delta_{\textrm{J2016}}=-27.95939^{\circ}$; $G=13.29\,$mag; $(G_{BP}-G_{RP})=0.42\,$mag) and 5693065398100156800 ($\alpha_{\textrm{J2016}}=123.14115^{\circ}$; $\delta_{\textrm{J2016}}=-27.95287^{\circ}$; $G=13.24\,$mag; $(G_{BP}-G_{RP})=0.38\,$mag), respectively. Both are blue straggler candidates. }

\label{fig:HRD_Haffner22}
\end{center}
\end{figure*}

\begin{figure*}
\begin{center}

\parbox[c]{0.70\textwidth}
  {
   \begin{center}
    \includegraphics[width=0.70\textwidth]{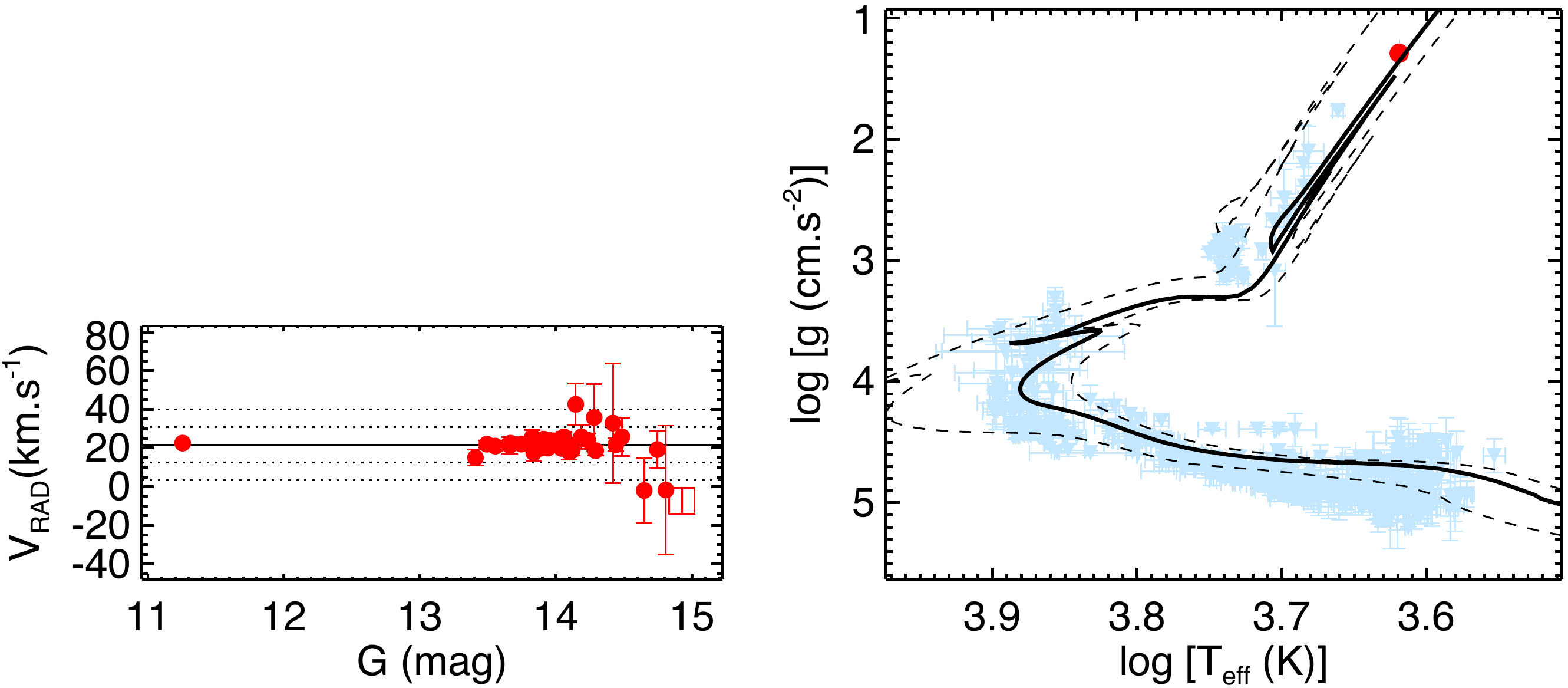}  
    \end{center}    
  }
\caption{ Same as Figure\,5 of the manuscript, but for the OC NGC\,2660. No $[Fe/H]$ values available for the set of member stars. There is only one member star with parameters available from the GSP-Spec modulus (\textit{source\_ID} 5329369865272032640, $\alpha_{\textrm{J2016}}=130.70586^{\circ}$; $\delta_{\textrm{J2016}}=-47.21597^{\circ}$; $G=11.26\,$mag; $(G_{BP}-G_{RP})=2.06\,$mag). }

\label{fig:HRD_NGC2660}
\end{center}
\end{figure*}

\clearpage

\begin{figure*}
\begin{center}

\parbox[c]{0.70\textwidth}
  {
   \begin{center}
    \includegraphics[width=0.70\textwidth]{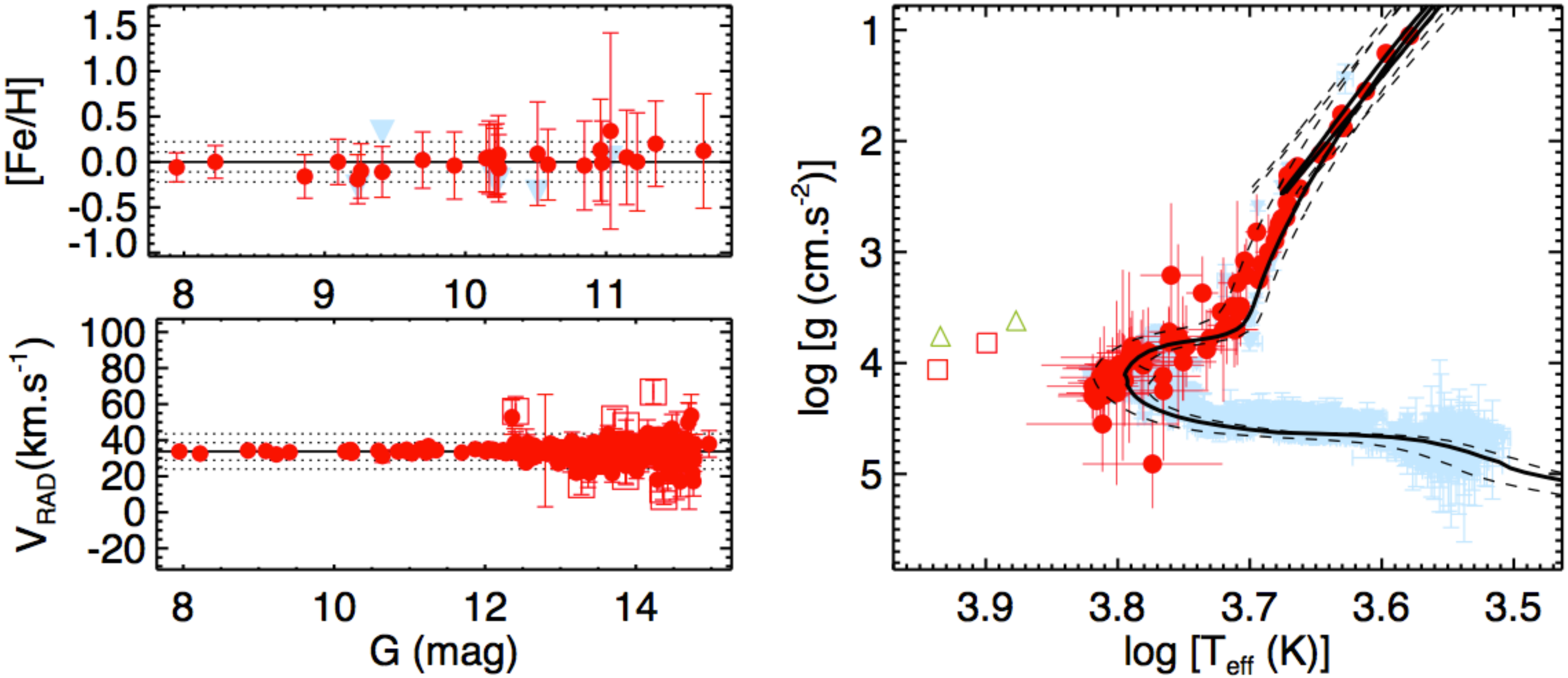}  
    \end{center}    
  }
\caption{ Same as Figure\,5 of the manuscript, but for the OC M\,67. }

\label{fig:HRD_M67}
\end{center}
\end{figure*}

\begin{figure*}
\begin{center}

\parbox[c]{0.70\textwidth}
  {
   \begin{center}
    \includegraphics[width=0.70\textwidth]{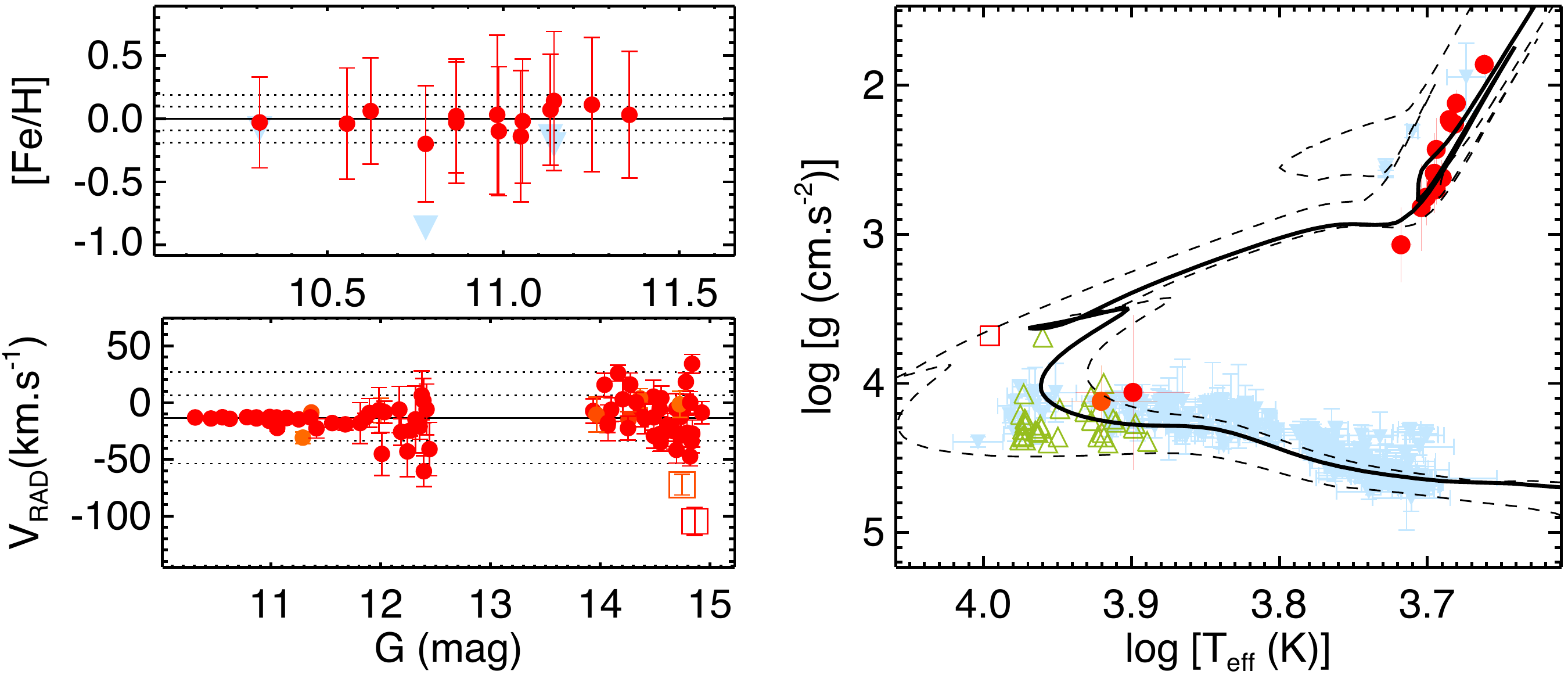}  
    \end{center}    
  }
\caption{ Same as Figure\,5 of the manuscript, but for the OC IC\,2714. }

\label{fig:HRD_IC2714}
\end{center}
\end{figure*}

\begin{figure*}
\begin{center}

\parbox[c]{0.70\textwidth}
  {
   \begin{center}
    \includegraphics[width=0.70\textwidth]{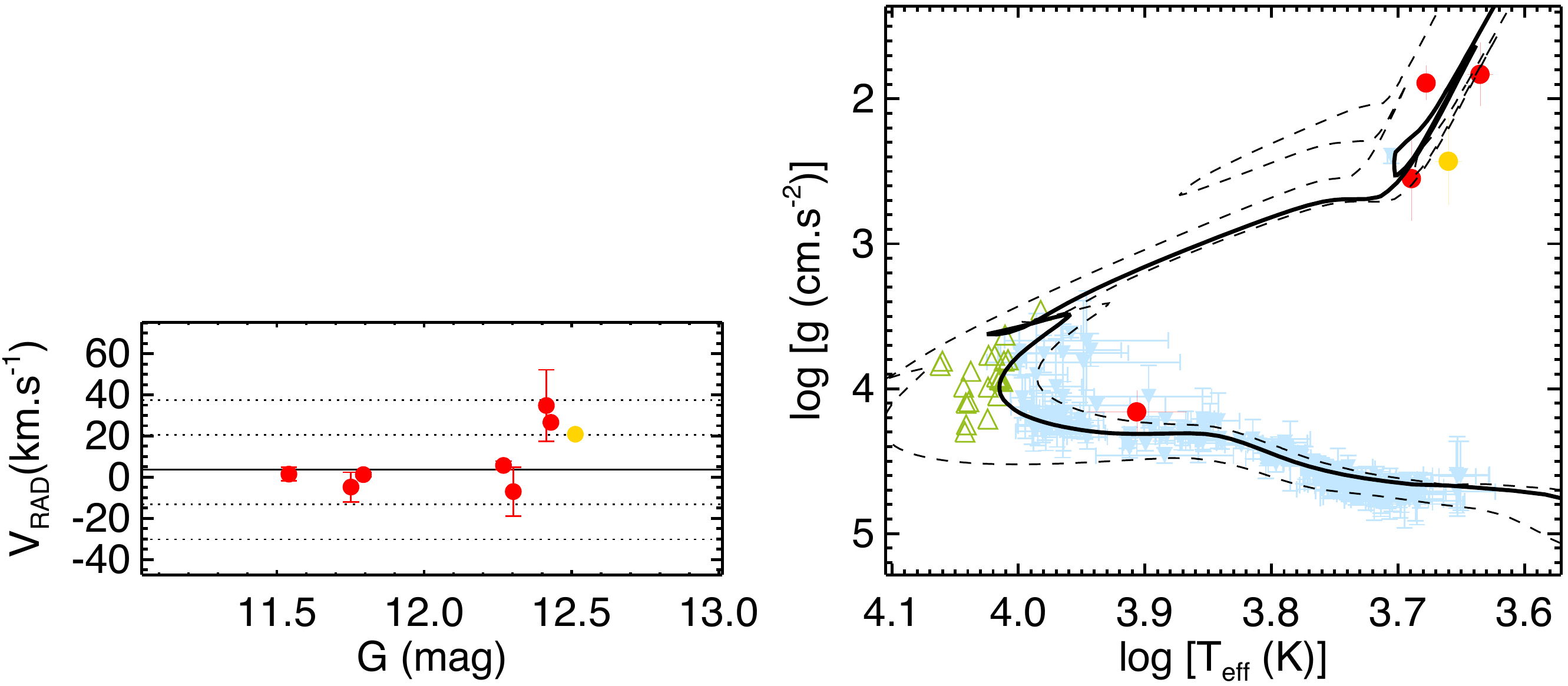}  
    \end{center}    
  }
\caption{ Same as Figure\,5 of the manuscript, but for the OC Melotte\,105. No $[Fe/H]$ values available for the set of member stars. }

\label{fig:HRD_Melotte105}
\end{center}
\end{figure*}

\begin{figure*}
\begin{center}

\parbox[c]{0.70\textwidth}
  {
   \begin{center}
    \includegraphics[width=0.70\textwidth]{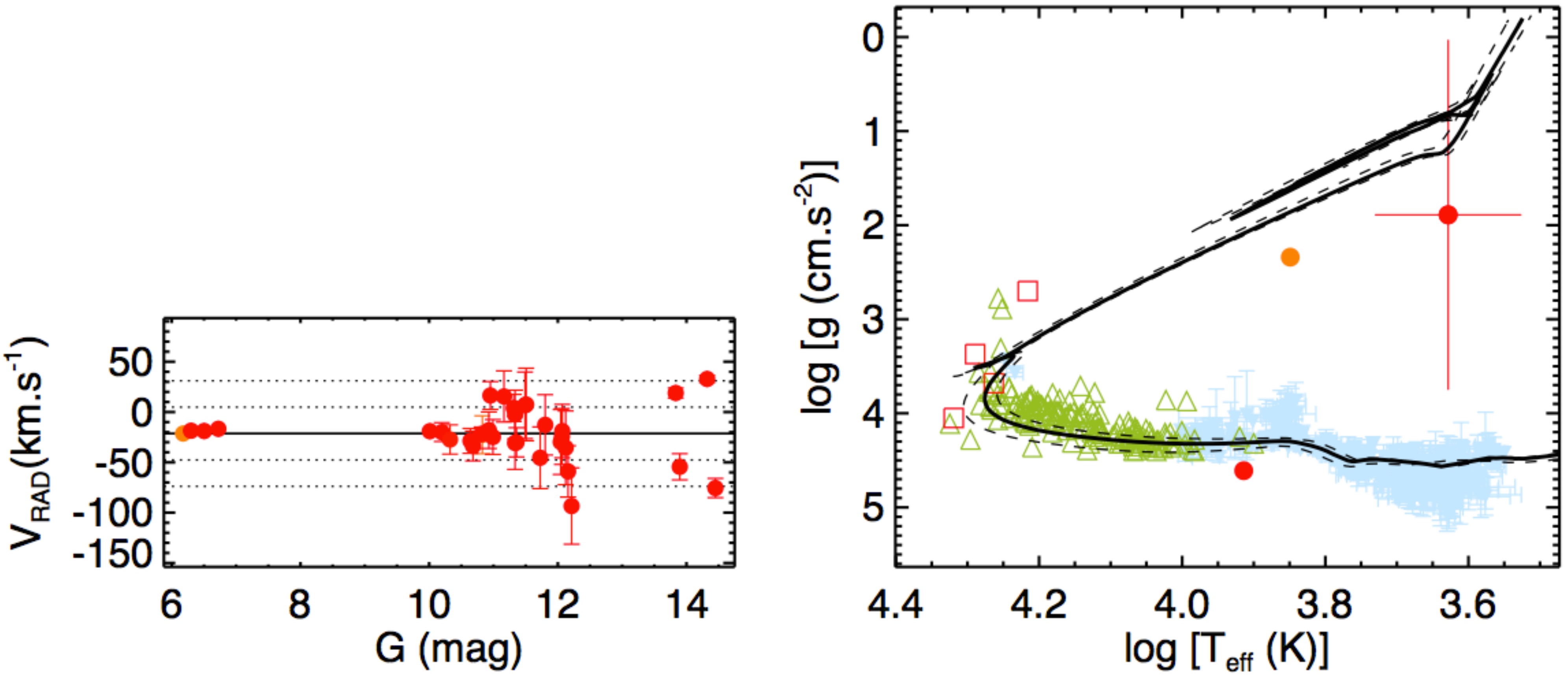}  
    \end{center}    
  }
\caption{ Same as Figure\,5 of the manuscript, but for the OC NGC\,3766. No $[Fe/H]$ values available for the set of member stars. }

\label{fig:HRD_NGC3766}
\end{center}
\end{figure*}

\begin{figure*}
\begin{center}

\parbox[c]{0.70\textwidth}
  {
   \begin{center}
    \includegraphics[width=0.70\textwidth]{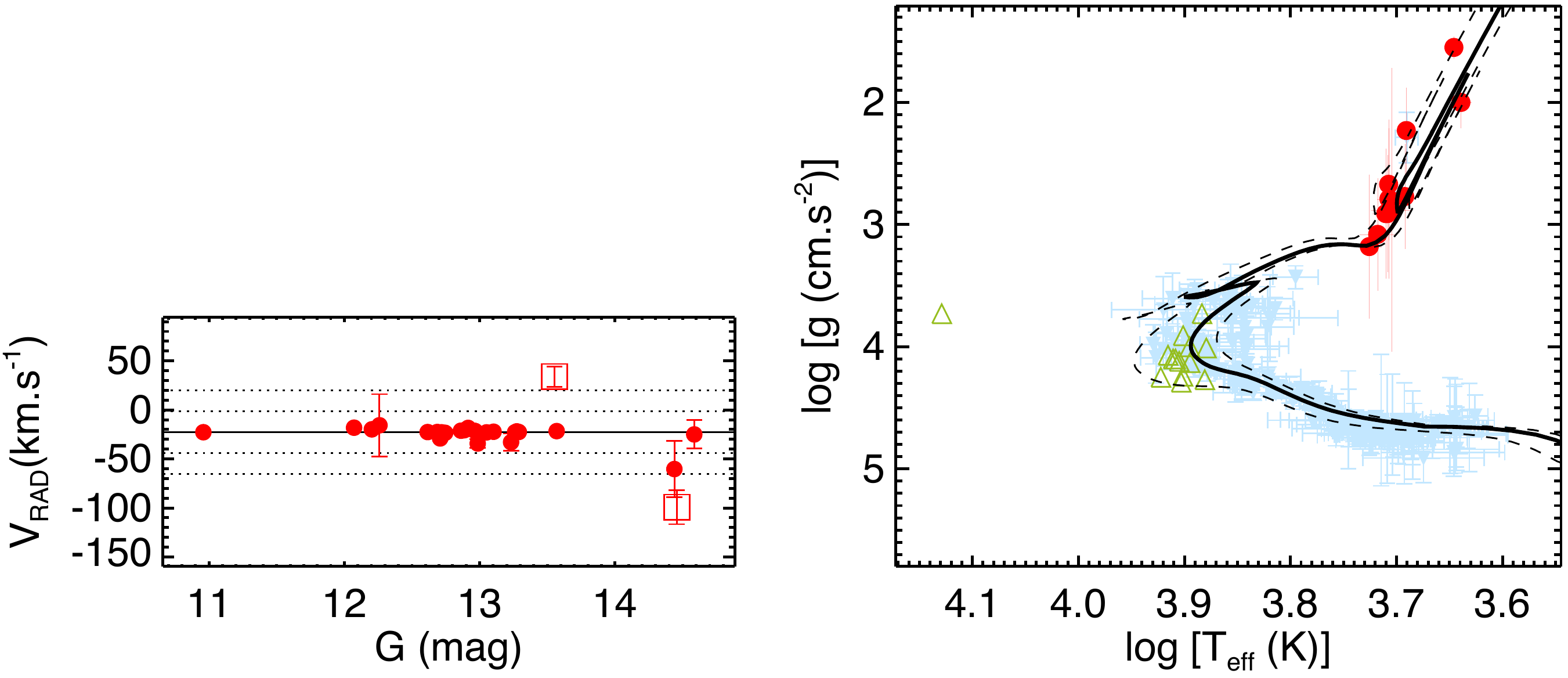}  
    \end{center}    
  }
\caption{ Same as Figure\,5 of the manuscript, but for the OC NGC\,3960. No $[Fe/H]$ values available for the set of member stars. The hottest star (open triangle with log\,$T_{\textrm{eff}}=4.13$, log\,$g$=3.73) has \textit{source\_ID} 5343855106334040704 ($\alpha_{\textrm{J2016}}=177.61861^{\circ}$; $\delta_{\textrm{J2016}}=-55.71043^{\circ}$; $G=12.26\,$mag; $(G_{BP}-G_{RP})=0.40\,$mag), probably a blue straggler. }

\label{fig:HRD_NGC3960}
\end{center}
\end{figure*}

\begin{figure*}
\begin{center}

\parbox[c]{0.70\textwidth}
  {
   \begin{center}
    \includegraphics[width=0.70\textwidth]{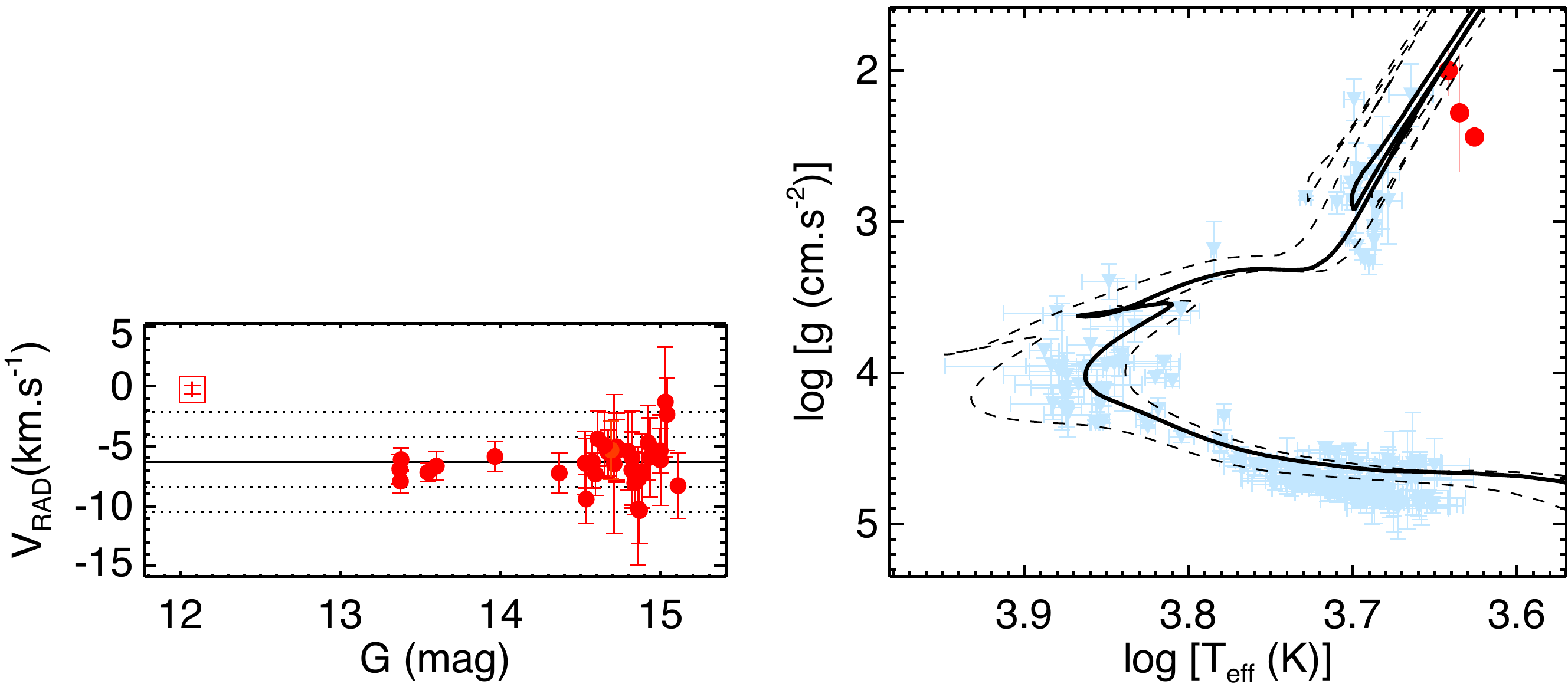}  
    \end{center}    
  }
\caption{ Same as Figure\,5 of the manuscript, but for the OC Juchert\,13. No $[Fe/H]$ values available for the set of member stars. }

\label{fig:HRD_Juchert13}
\end{center}
\end{figure*}

\begin{figure*}
\begin{center}

\parbox[c]{0.70\textwidth}
  {
   \begin{center}
    \includegraphics[width=0.70\textwidth]{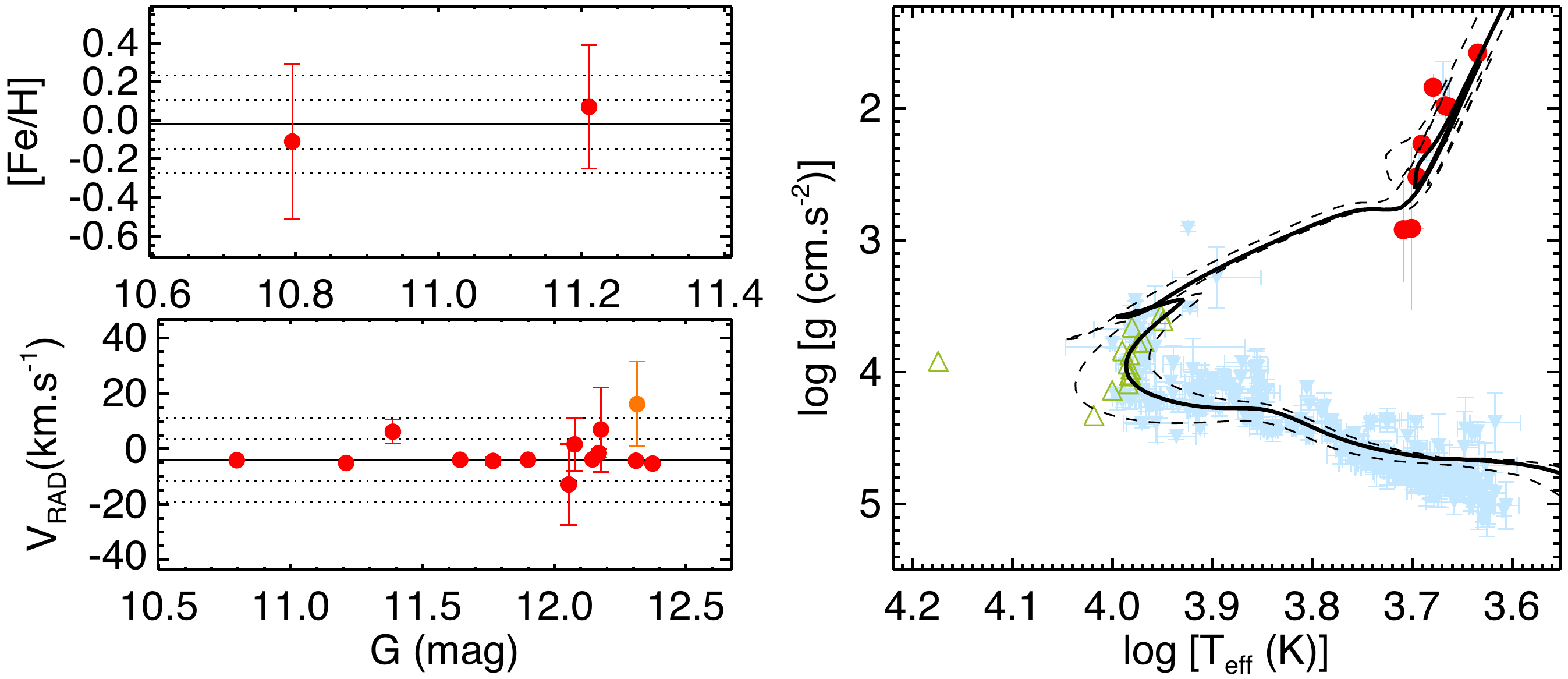}  
    \end{center}    
  }
\caption{ Same as Figure\,5 of the manuscript, but for the OC NGC\,4052. The hottest star (open triangle with log\,$T_{\textrm{eff}}=4.17$, log\,$g$=3.92) has \textit{source\_ID} 6057345793780687232 ($\alpha_{\textrm{J2016}}=180.59343^{\circ}$; $\delta_{\textrm{J2016}}=-63.18917^{\circ}$; $G=12.23\,$mag; $(G_{BP}-G_{RP})=0.22\,$mag), probably a blue straggler. }

\label{fig:HRD_NGC4052}
\end{center}
\end{figure*}

\begin{figure*}
\begin{center}

\parbox[c]{0.70\textwidth}
  {
   \begin{center}
    \includegraphics[width=0.70\textwidth]{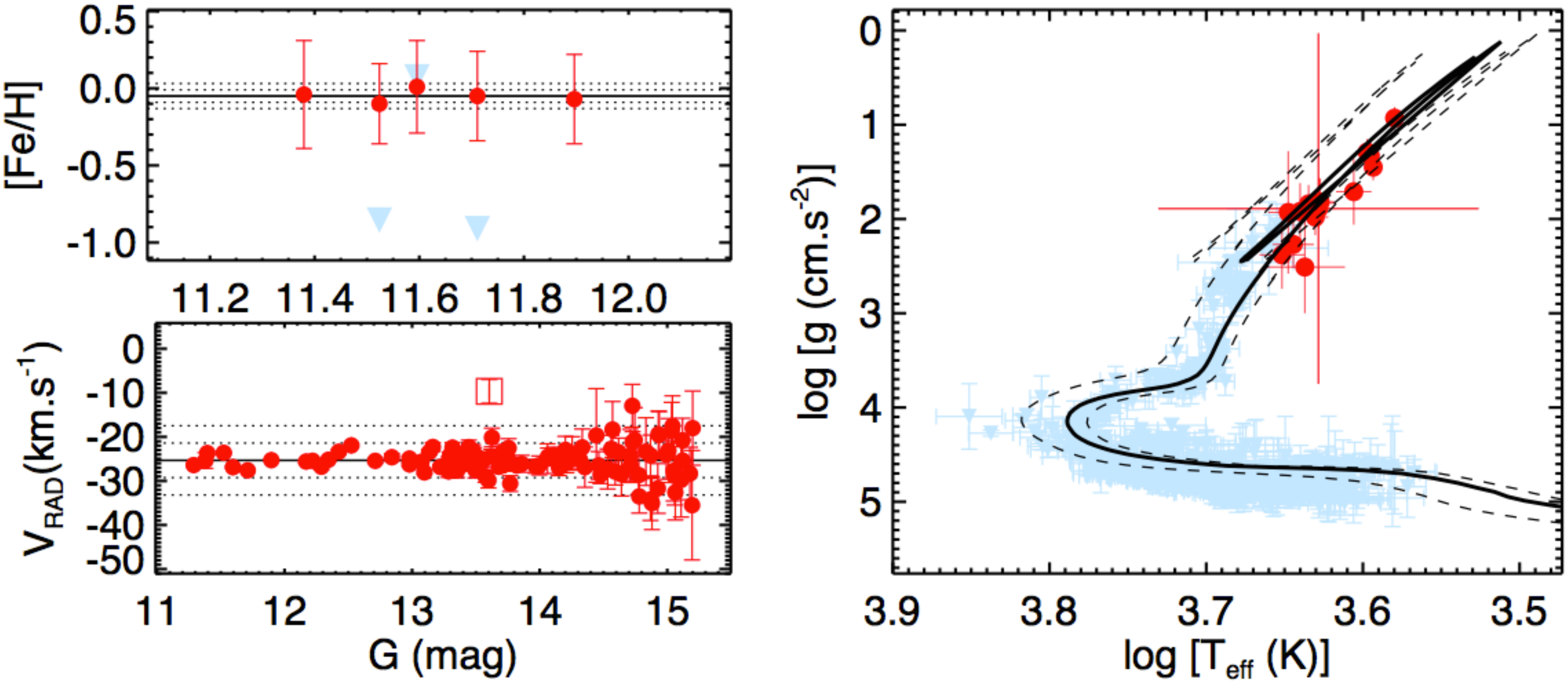}  
    \end{center}    
  }
\caption{ Same as Figure\,5 of the manuscript, but for the OC Collinder\,261. }

\label{fig:HRD_Collinder261}
\end{center}
\end{figure*}

\begin{figure*}
\begin{center}

\parbox[c]{0.70\textwidth}
  {
   \begin{center}
    \includegraphics[width=0.70\textwidth]{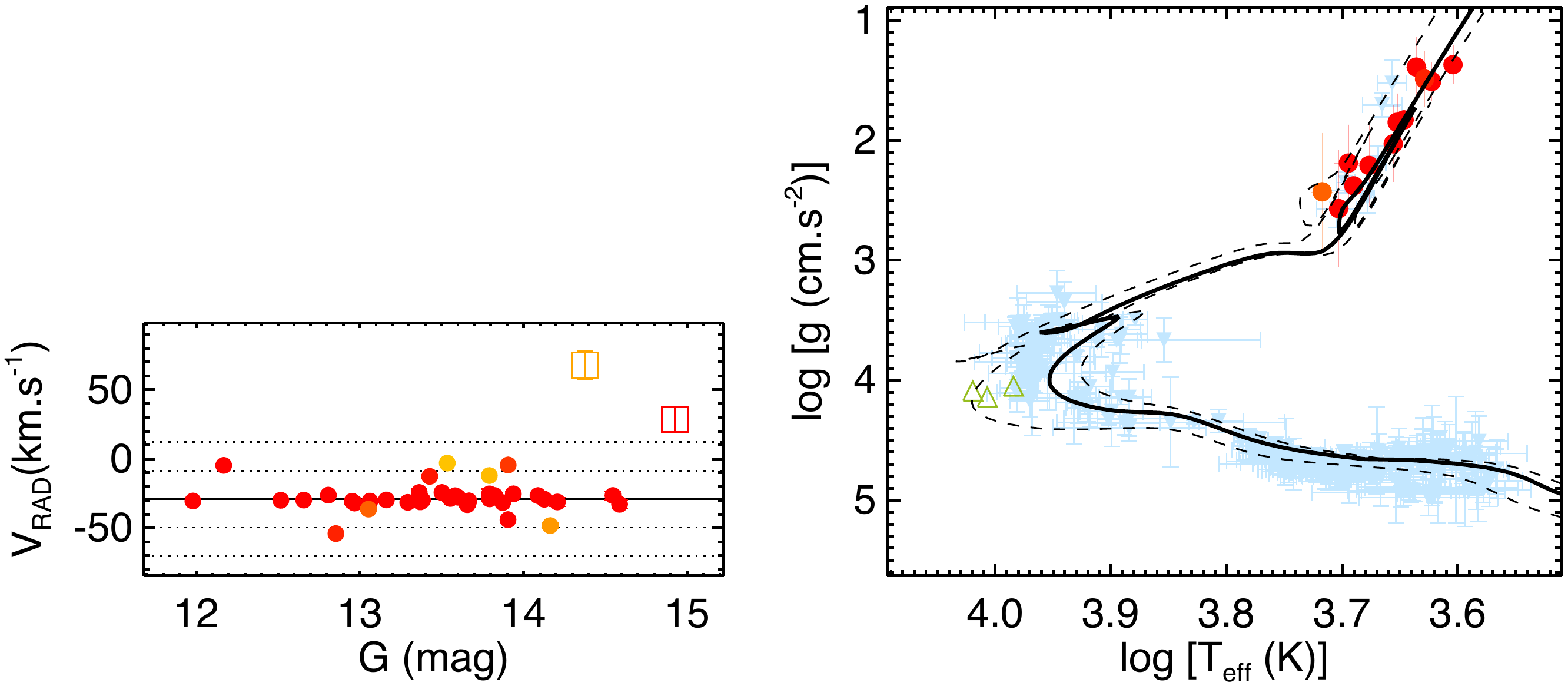}  
    \end{center}    
  }
\caption{ Same as Figure\,5 of the manuscript, but for the OC NGC\,4815. No $[Fe/H]$ values available for the set of member stars. }

\label{fig:HRD_NGC4815}
\end{center}
\end{figure*}

\begin{figure*}
\begin{center}

\parbox[c]{0.70\textwidth}
  {
   \begin{center}
    \includegraphics[width=0.70\textwidth]{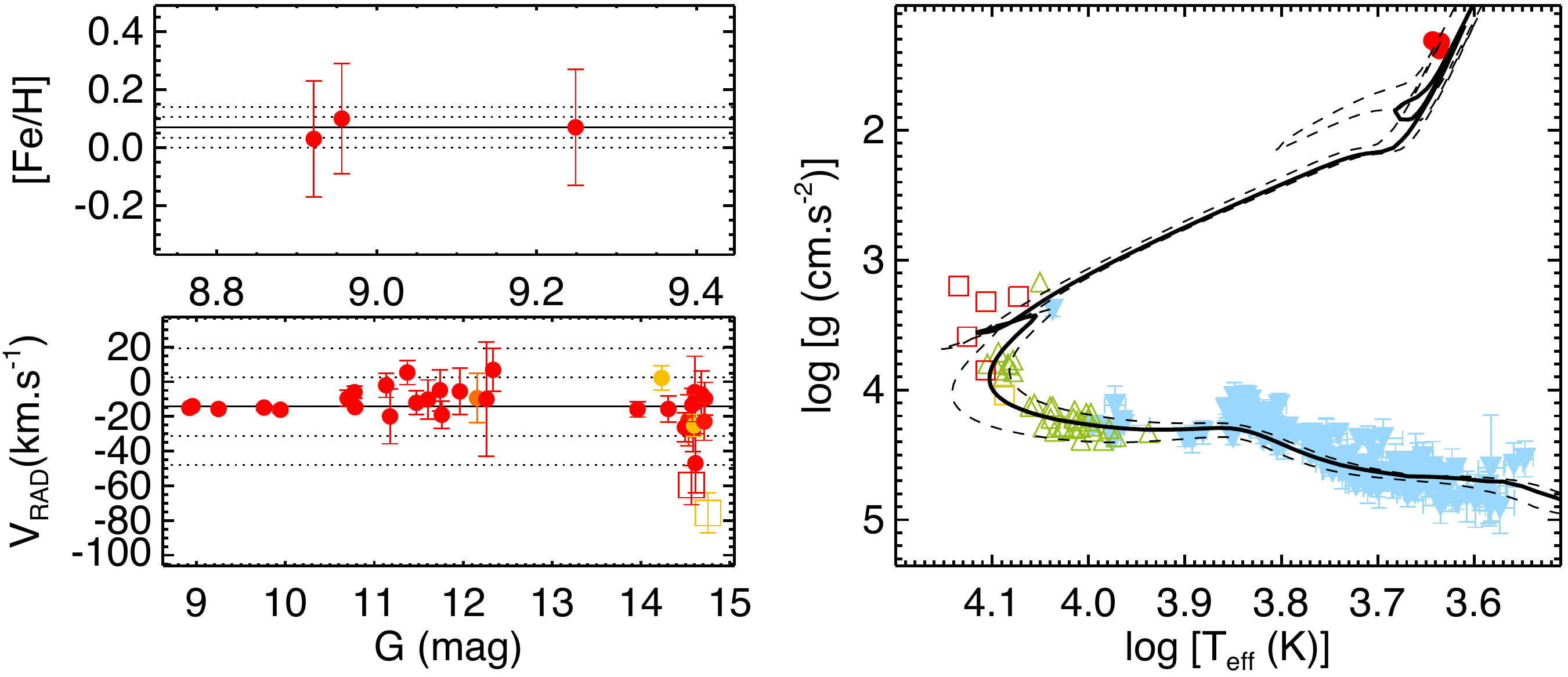}  
    \end{center}    
  }
\caption{ Same as Figure\,5 of the manuscript, but for the OC NGC\,5316. }

\label{fig:HRD_NGC5316}
\end{center}
\end{figure*}

\begin{figure*}
\begin{center}

\parbox[c]{0.70\textwidth}
  {
   \begin{center}
    \includegraphics[width=0.70\textwidth]{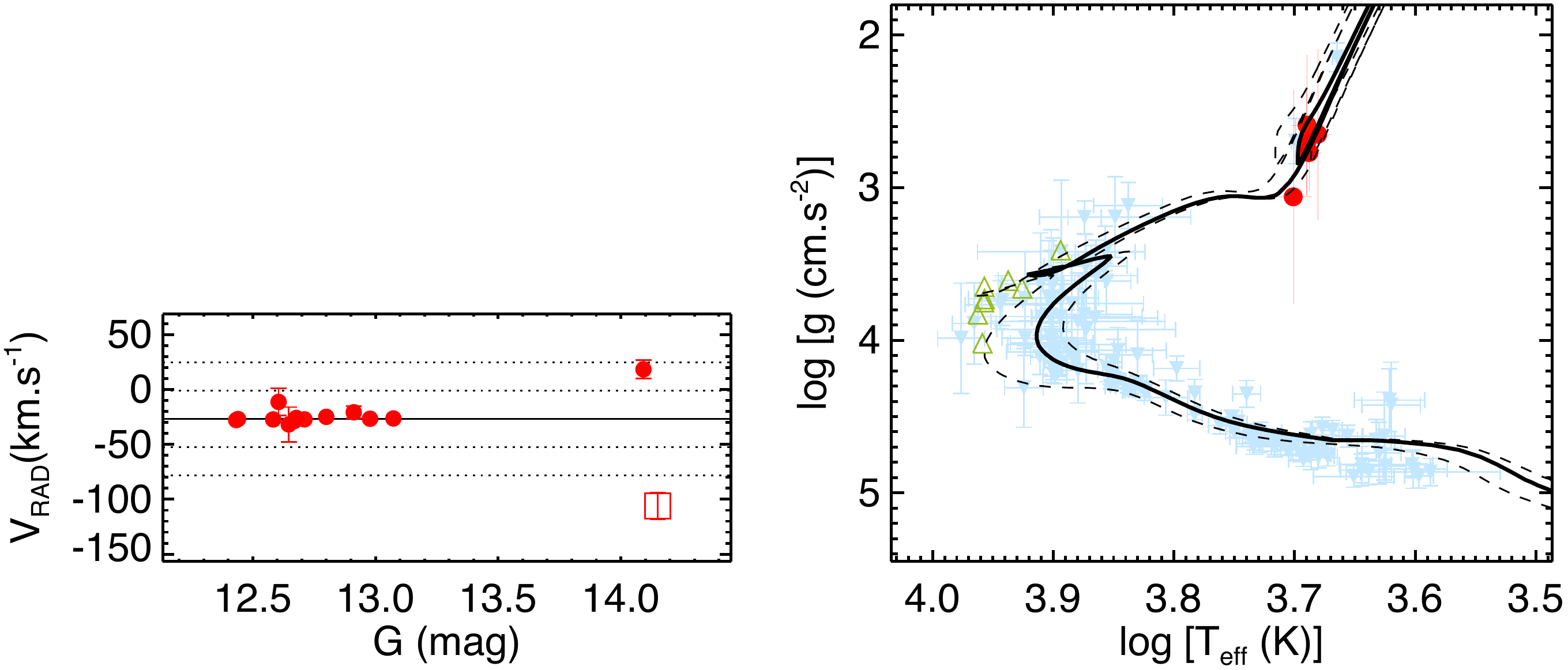}  
    \end{center}    
  }
\caption{ Same as Figure\,5 of the manuscript, but for the OC NGC\,5715. No $[Fe/H]$ values available for the set of member stars. }

\label{fig:HRD_NGC5715}
\end{center}
\end{figure*}

\begin{figure*}
\begin{center}

\parbox[c]{0.70\textwidth}
  {
   \begin{center}
    \includegraphics[width=0.70\textwidth]{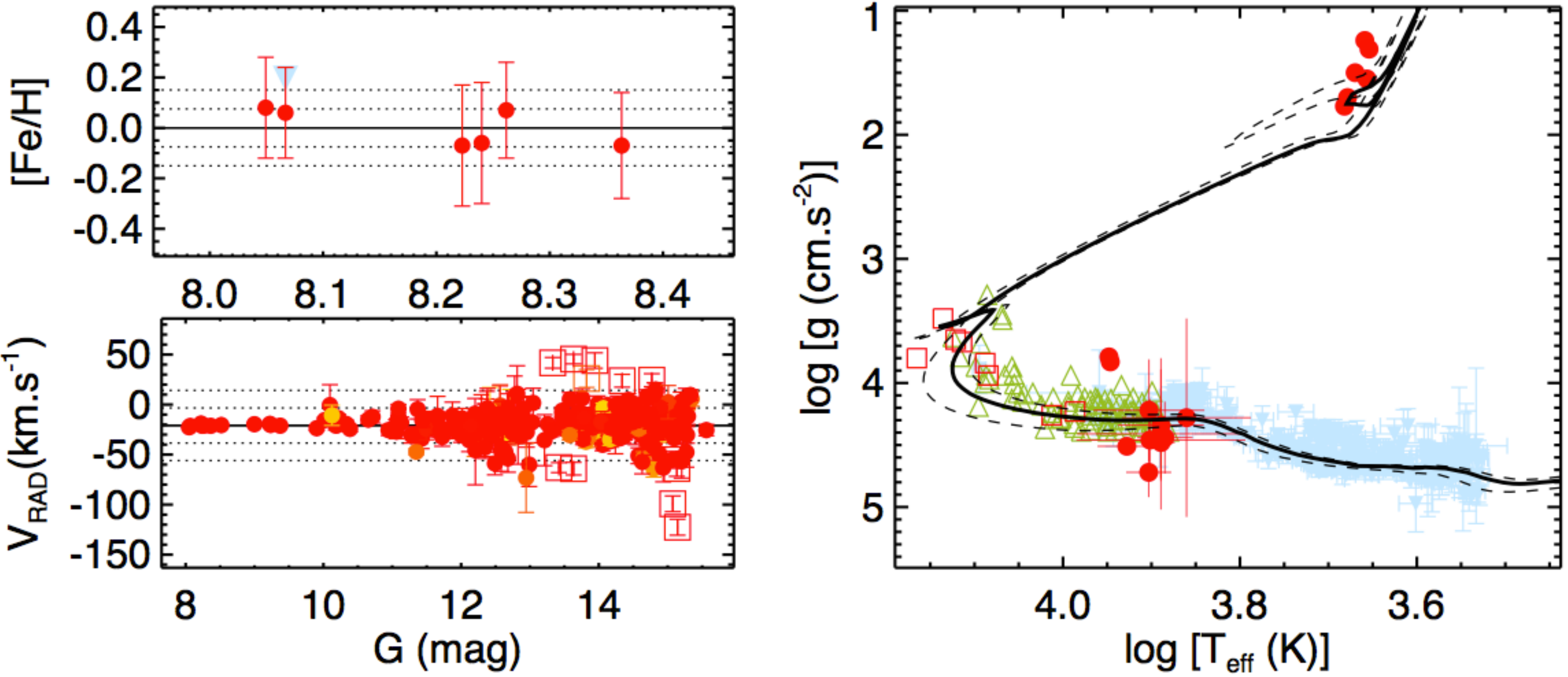}  
    \end{center}    
  }
\caption{ Same as Figure\,5 of the manuscript, but for the OC NGC\,6124. }

\label{fig:HRD_NGC6124}
\end{center}
\end{figure*}

\begin{figure*}
\begin{center}

\parbox[c]{0.70\textwidth}
  {
   \begin{center}
    \includegraphics[width=0.70\textwidth]{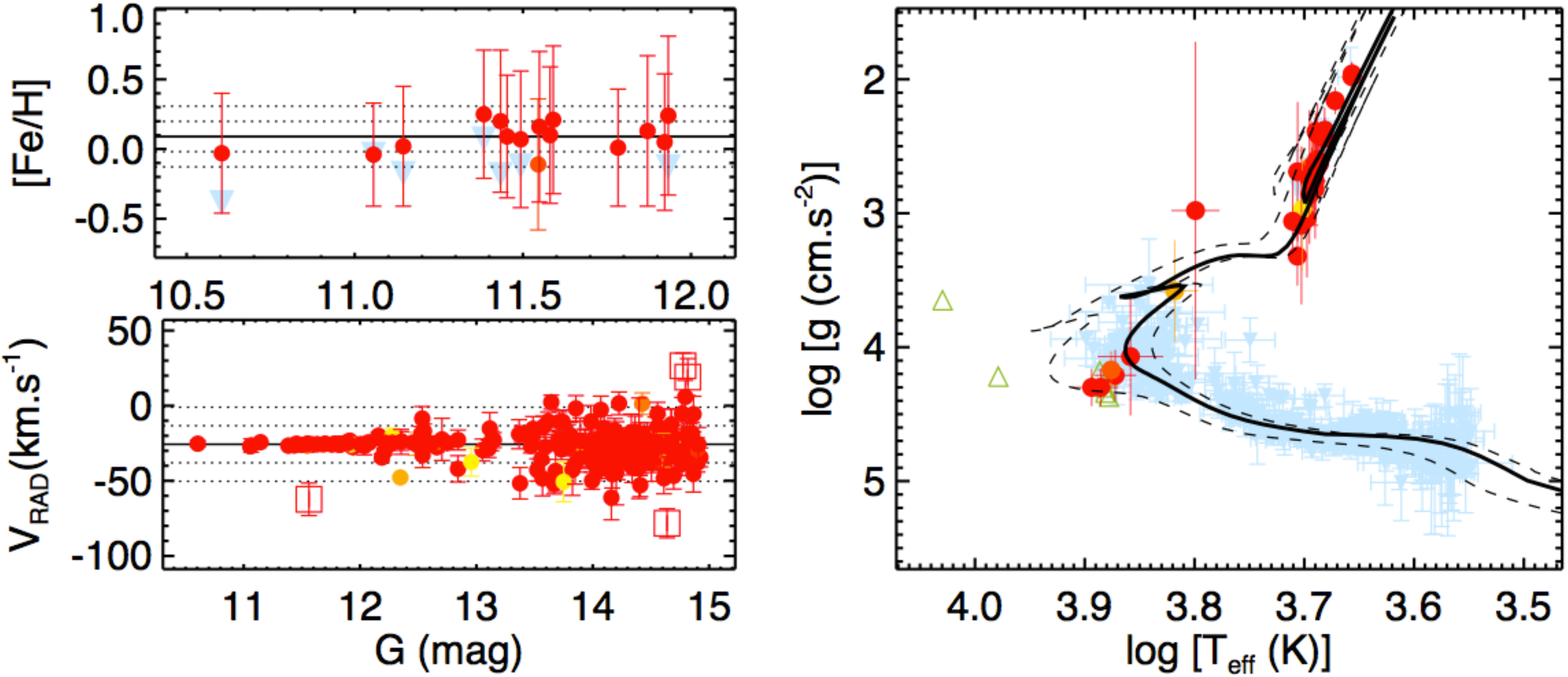}  
    \end{center}    
  }
\caption{ Same as Figure\,5 of the manuscript, but for the OC NGC\,6134. The 2 hottest stars (open green triangles with log\,$T_{\textrm{eff}}$=3.98, log\,$g$=4.22  and log\,$T_{\textrm{eff}}$=4.03, log\,$g$=3.65) represent member stars with \textit{source\_ID} 5941411299209346688 ($\alpha_{\textrm{J2016}}=246.90752^{\circ}$; $\delta_{\textrm{J2016}}=-49.17845^{\circ}$; $G=11.56\,$mag; $(G_{BP}-G_{RP})=0.60\,$mag) and 5941411402288564224 ($\alpha_{\textrm{J2016}}=246.92978^{\circ}$; $\delta_{\textrm{J2016}}=-49.16535^{\circ}$; $G=11.06\,$mag; $(G_{BP}-G_{RP})=0.52\,$mag), respectively. Both are blue straggler candidates. }

\label{fig:HRD_NGC6134}
\end{center}
\end{figure*}

\begin{figure*}
\begin{center}

\parbox[c]{0.70\textwidth}
  {
   \begin{center}
    \includegraphics[width=0.70\textwidth]{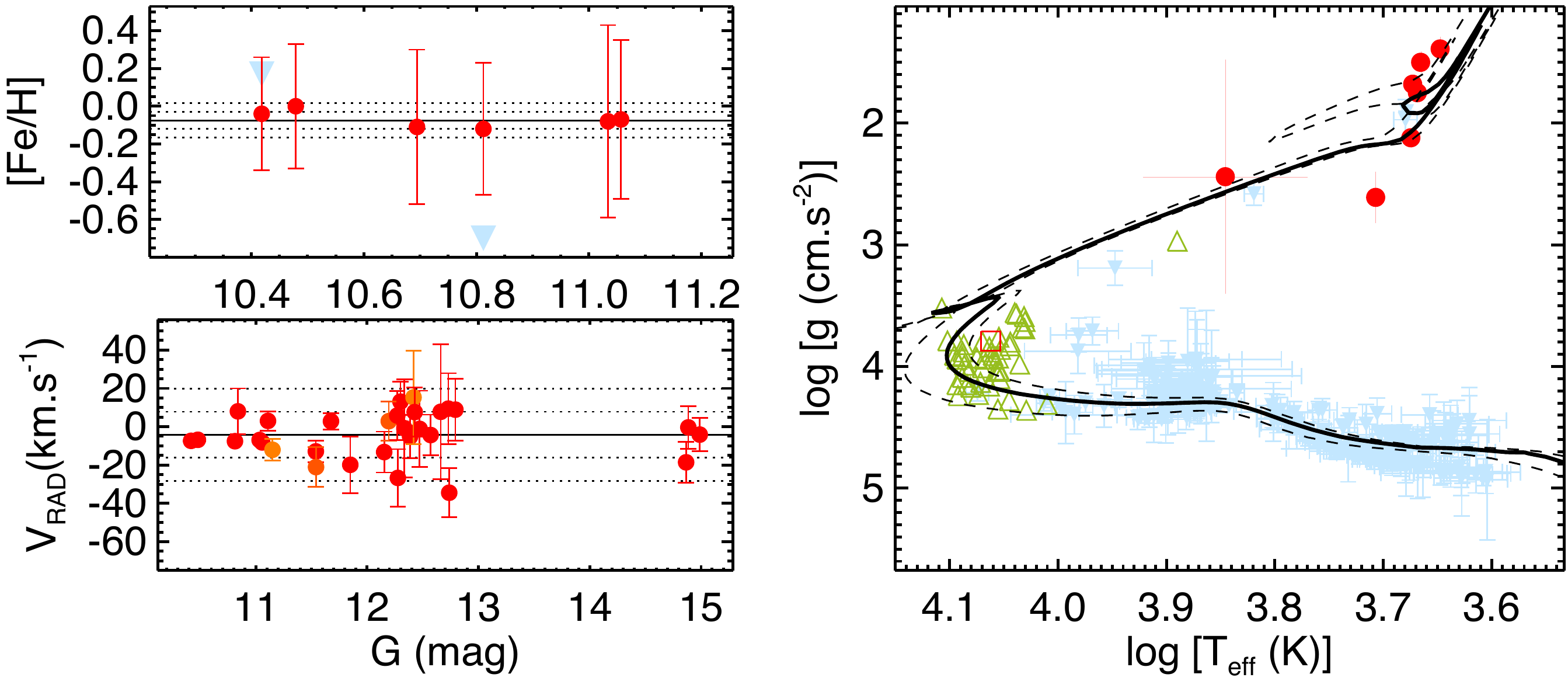}  
    \end{center}    
  }
\caption{ Same as Figure\,5 of the manuscript, but for the OC NGC\,6192. }

\label{fig:HRD_NGC6192}
\end{center}
\end{figure*}

\begin{figure*}
\begin{center}

\parbox[c]{0.70\textwidth}
  {
   \begin{center}
    \includegraphics[width=0.70\textwidth]{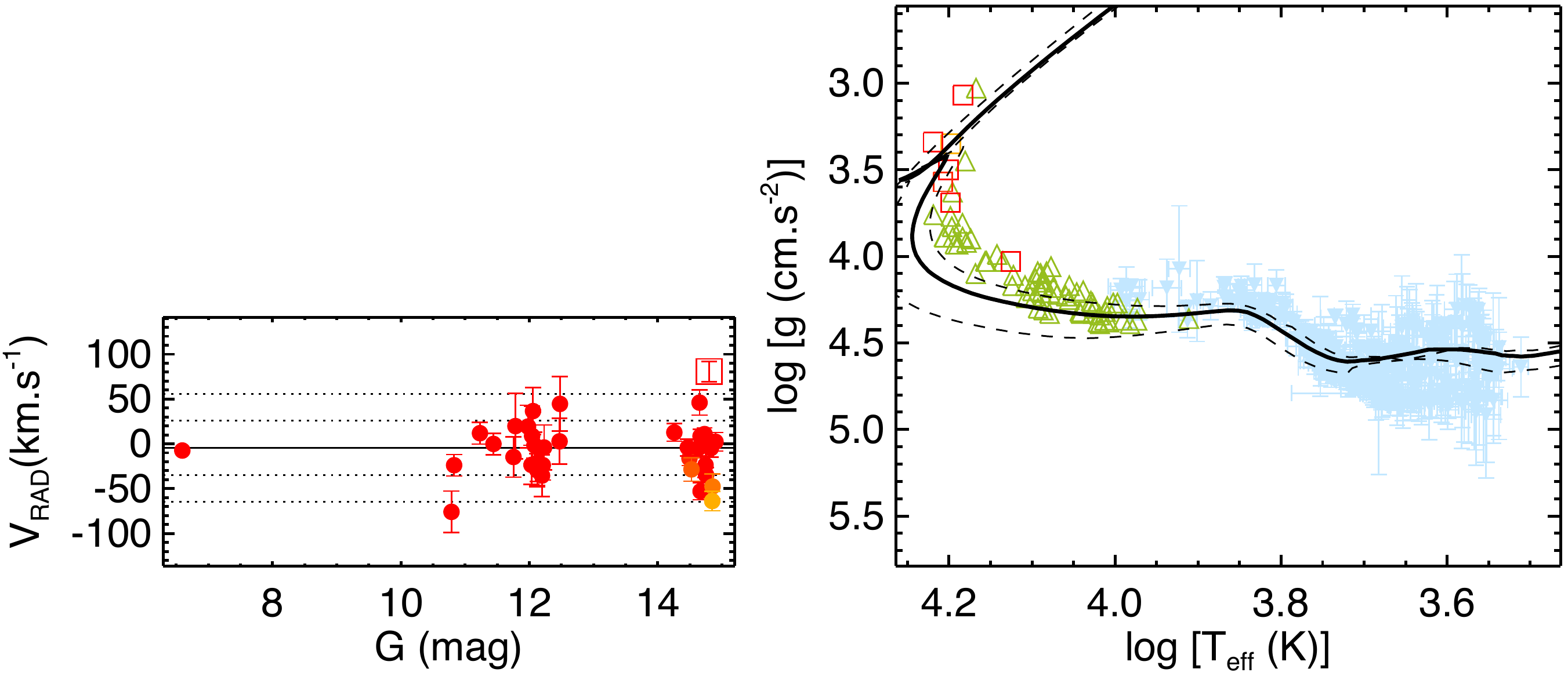}  
    \end{center}    
  }
\caption{ Same as Figure\,5 of the manuscript, but for the OC NGC\,6242. No $[Fe/H]$ values available for the set of member stars. The brightest member star ($G=6.6\,$mag; \textit{source\_ID}=5970176928558756864; see Figure~\ref{fig:CMD_SupplMater5}) does not have valid atmospheric parameters in \textit{Gaia} DR3. }

\label{fig:HRD_NGC6242}
\end{center}
\end{figure*}

\begin{figure*}
\begin{center}

\parbox[c]{0.70\textwidth}
  {
   \begin{center}
    \includegraphics[width=0.70\textwidth]{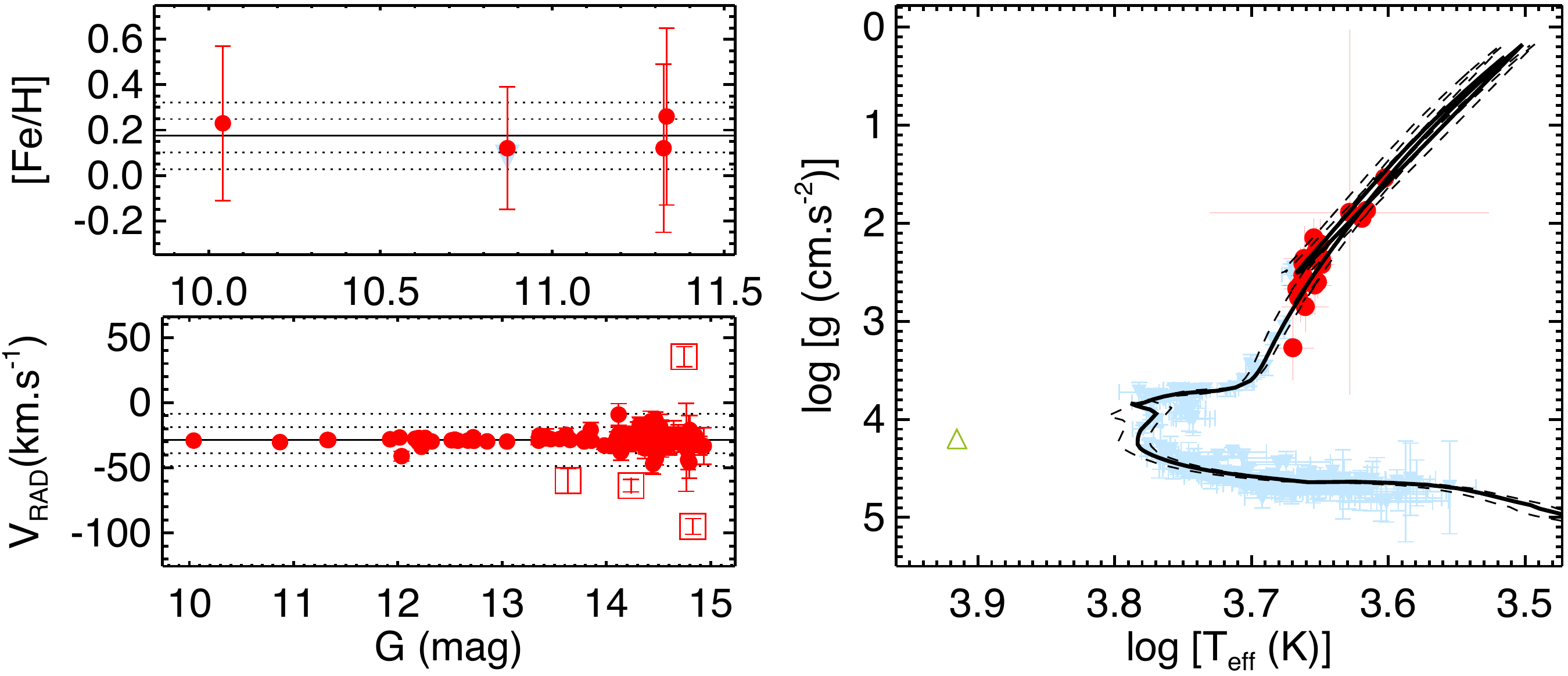}  
    \end{center}    
  }
\caption{ Same as Figure\,5 of the manuscript, but for the OC NGC\,6253. The hottest star (open green triangle with log\,$T_{\textrm{eff}}$=3.92, log\,$g$=4.20) represents the member star with \textit{source\_ID} 5935943324834287872 ($\alpha_{\textrm{J2016}}=254.78925^{\circ}$; $\delta_{\textrm{J2016}}=-52.83548^{\circ}$; $G=12.87\,$mag; $(G_{BP}-G_{RP})=0.52\,$mag), a blue straggler candidate. }

\label{fig:HRD_NGC6253}
\end{center}
\end{figure*}

\begin{figure*}
\begin{center}

\parbox[c]{0.70\textwidth}
  {
   \begin{center}
    \includegraphics[width=0.70\textwidth]{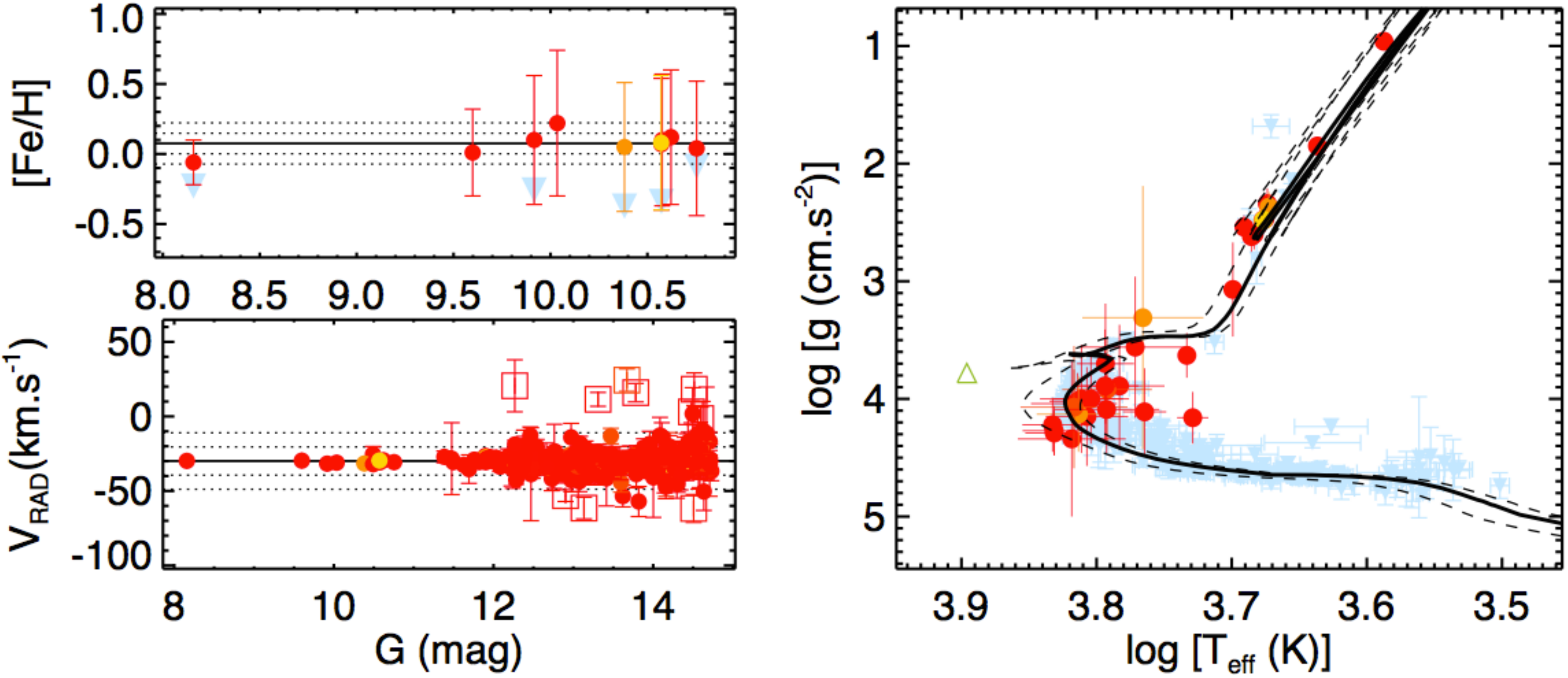}  
    \end{center}    
  }
\caption{ Same as Figure\,5 of the manuscript, but for the OC IC\,4651. The hottest star (open green triangle with log\,$T_{\textrm{eff}}$=3.90, log\,$g$=3.78) represents the member star with \textit{source\_ID} 5949553835654195968 ($\alpha_{\textrm{J2016}}=261.27313^{\circ}$; $\delta_{\textrm{J2016}}=-49.945927^{\circ}$; $G=10.49\,$mag; $(G_{BP}-G_{RP})=0.41\,$mag), a blue straggler candidate. }

\label{fig:HRD_IC4651}
\end{center}
\end{figure*}

\begin{figure*}
\begin{center}

\parbox[c]{0.70\textwidth}
  {
   \begin{center}
    \includegraphics[width=0.70\textwidth]{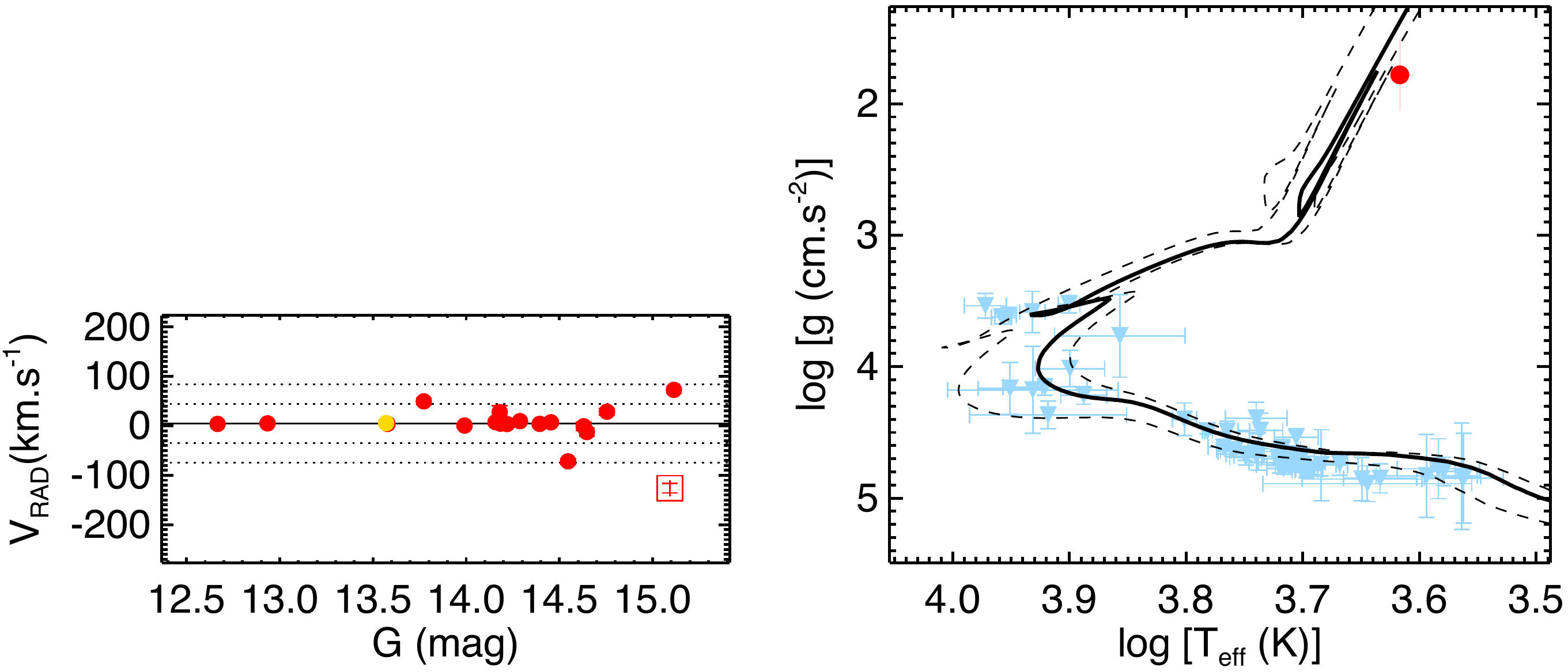}  
    \end{center}    
  }
\caption{ Same as Figure\,5 of the manuscript, but for the OC Dias\,6. There is only one member star with parameters available from the GSP-Spec modulus (\textit{source\_ID}: 4153043881274938240, $\alpha_{\textrm{J2016}}=277.60817^{\circ}$; $\delta_{\textrm{J2016}}=-12.306640^{\circ}$; $G=12.67\,$mag; $(G_{BP}-G_{RP})=2.47\,$mag). }

\label{fig:HRD_Dias6}
\end{center}
\end{figure*}

\afterpage{\clearpage}

\begin{figure*}
\begin{center}

\parbox[c]{0.70\textwidth}
  {
   \begin{center}
    \includegraphics[width=0.70\textwidth]{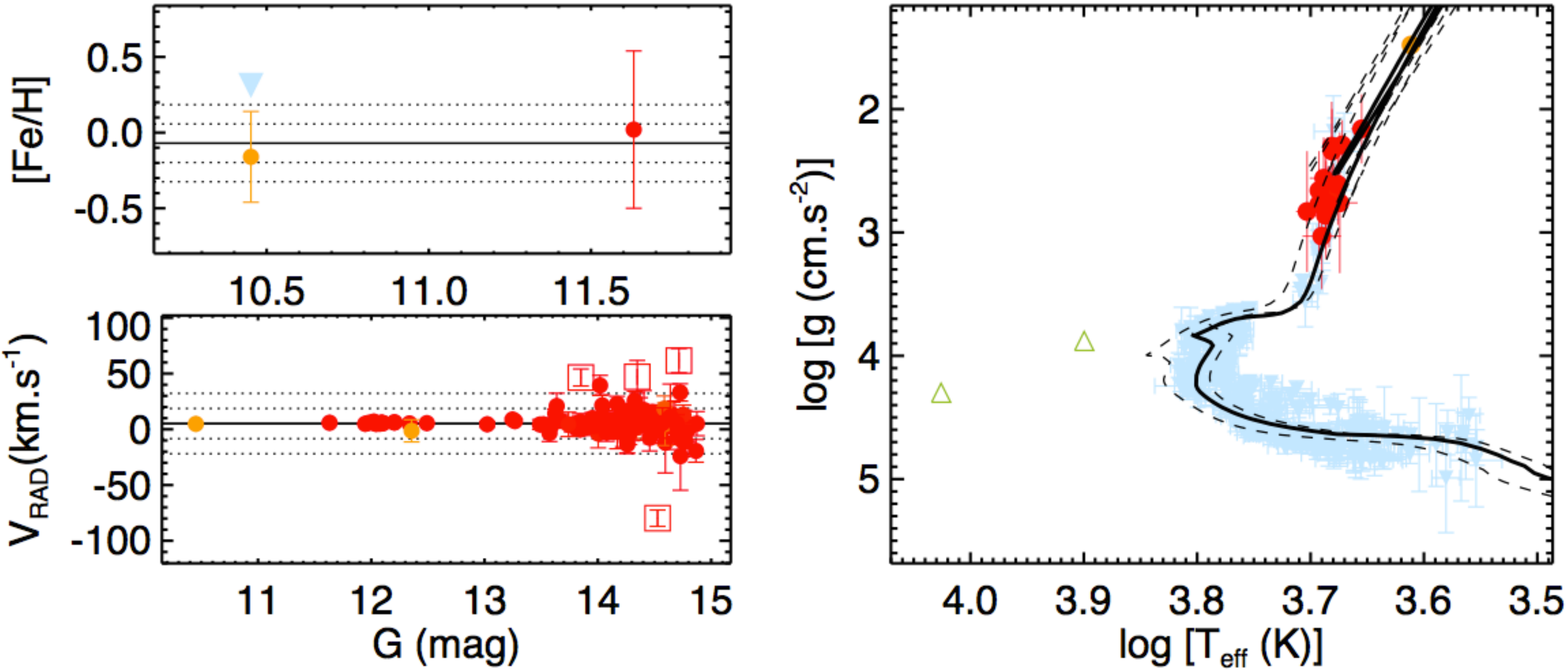}  
    \end{center}    
  }
\caption{ Same as Figure\,5 of the manuscript, but for the OC Ruprecht\,171. The 2 hottest stars (open green triangles with log\,$T_{\textrm{eff}}$=4.03, log\,$g$=4.30  and log\,$T_{\textrm{eff}}$=3.90, log\,$g$=3.88) represent member stars with \textit{source\_ID} 4103071730634235136 ($\alpha_{\textrm{J2016}}=277.99646^{\circ}$; $\delta_{\textrm{J2016}}=-16.09908^{\circ}$; $G=12.73\,$mag; $(G_{BP}-G_{RP})=0.39\,$mag) and 4103072520908233600 ($\alpha_{\textrm{J2016}}=278.07070^{\circ}$; $\delta_{\textrm{J2016}}=-16.01338^{\circ}$; $G=12.36\,$mag; $(G_{BP}-G_{RP})=0.64\,$mag), respectively. Both are blue straggler candidates. }

\label{fig:HRD_Ruprecht171}
\end{center}
\end{figure*}

\begin{figure*}
\begin{center}

\parbox[c]{0.70\textwidth}
  {
   \begin{center}
    \includegraphics[width=0.70\textwidth]{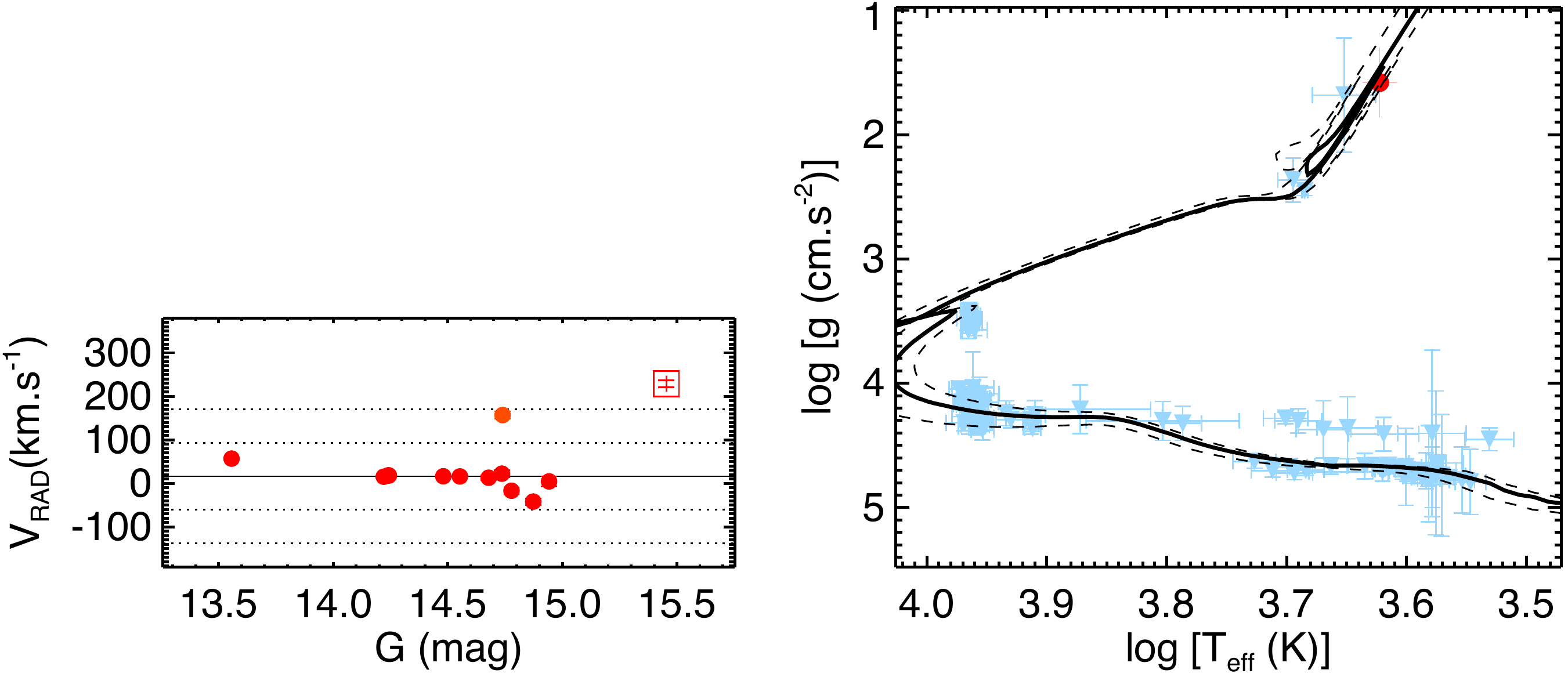}  
    \end{center}    
  }
\caption{ Same as Figure\,5 of the manuscript, but for the OC Czernik\,38. No $[Fe/H]$ values available for the set of member stars. There is only one member star with parameters available from the GSP-Spec modulus (\textit{source\_ID} 4282237910591803648, $\alpha_{\textrm{J2016}}=282.54130^{\circ}$; $\delta_{\textrm{J2016}}=5.00413^{\circ}$; $G=13.56\,$mag; $(G_{BP}-G_{RP})=3.63\,$mag). }

\label{fig:HRD_Czernik38}
\end{center}
\end{figure*}

\begin{figure*}
\begin{center}

\parbox[c]{0.70\textwidth}
  {
   \begin{center}
    \includegraphics[width=0.70\textwidth]{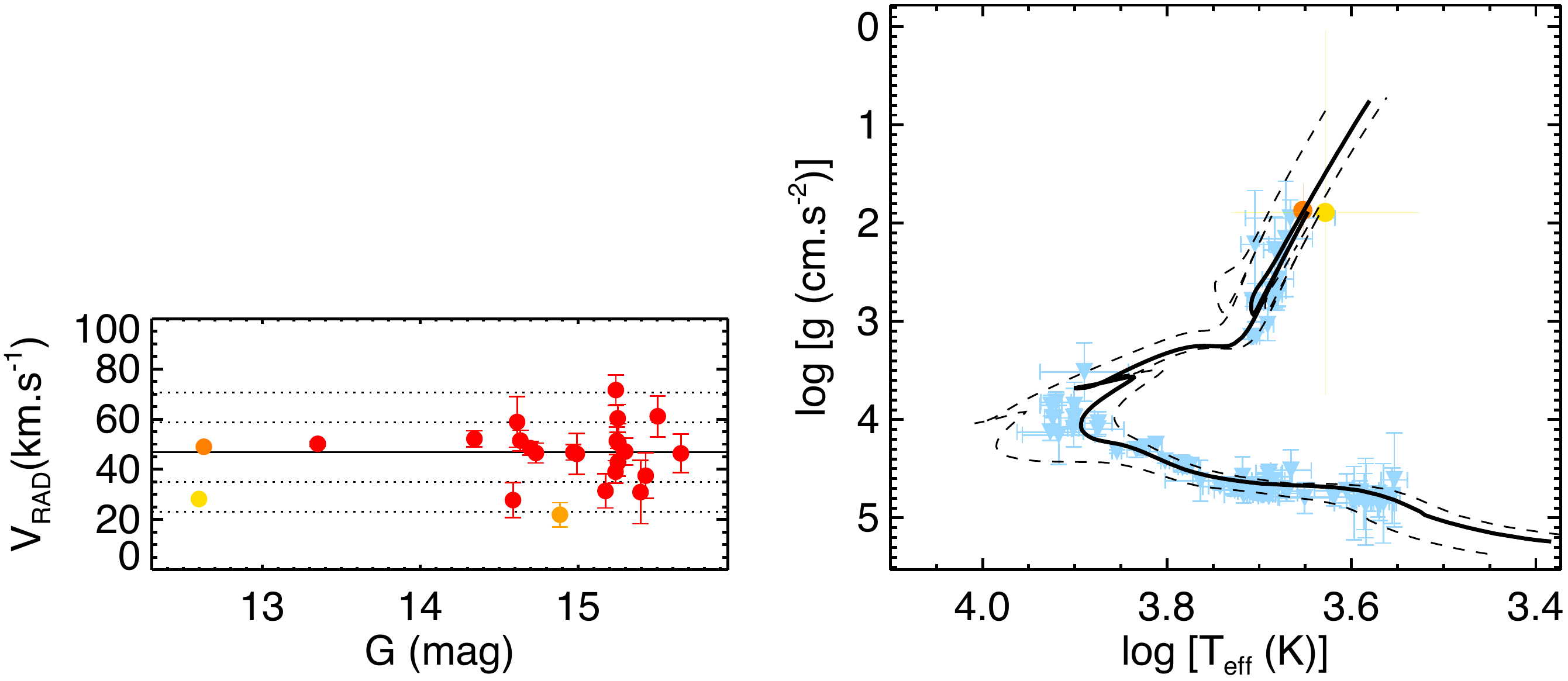}  
    \end{center}    
  }
\caption{ Same as Figure\,5 of the manuscript, but for the OC Berkeley\,81. No $[Fe/H]$ values available for the set of member stars. }

\label{fig:HRD_Berkeley81}
\end{center}
\end{figure*}

\begin{figure*}
\begin{center}

\parbox[c]{0.70\textwidth}
  {
   \begin{center}
    \includegraphics[width=0.70\textwidth]{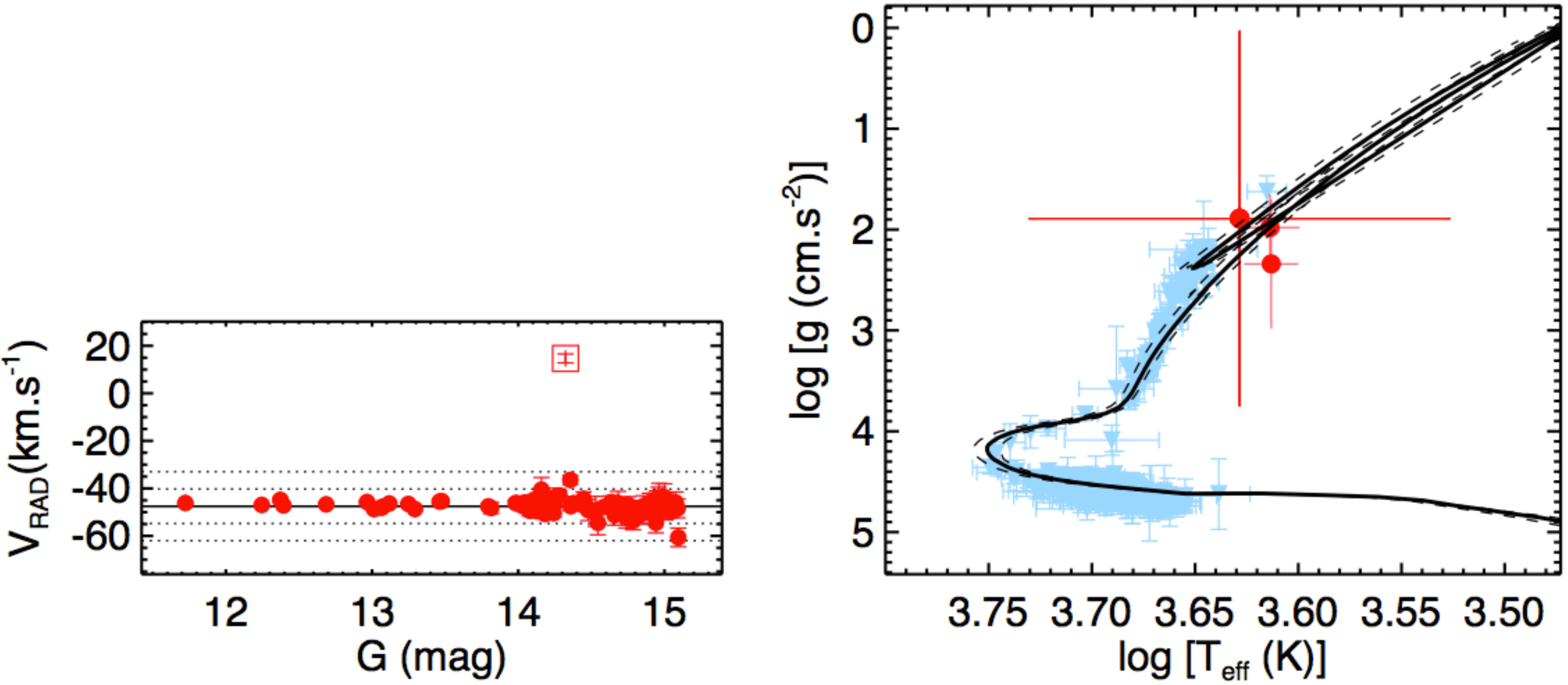}  
    \end{center}    
  }
\caption{ Same as Figure\,5 of the manuscript, but for the OC NGC\,6791. No $[Fe/H]$ values available for the set of member stars. }

\label{fig:HRD_NGC6791}
\end{center}
\end{figure*}

\begin{figure*}
\begin{center}

\parbox[c]{0.70\textwidth}
  {
   \begin{center}
    \includegraphics[width=0.70\textwidth]{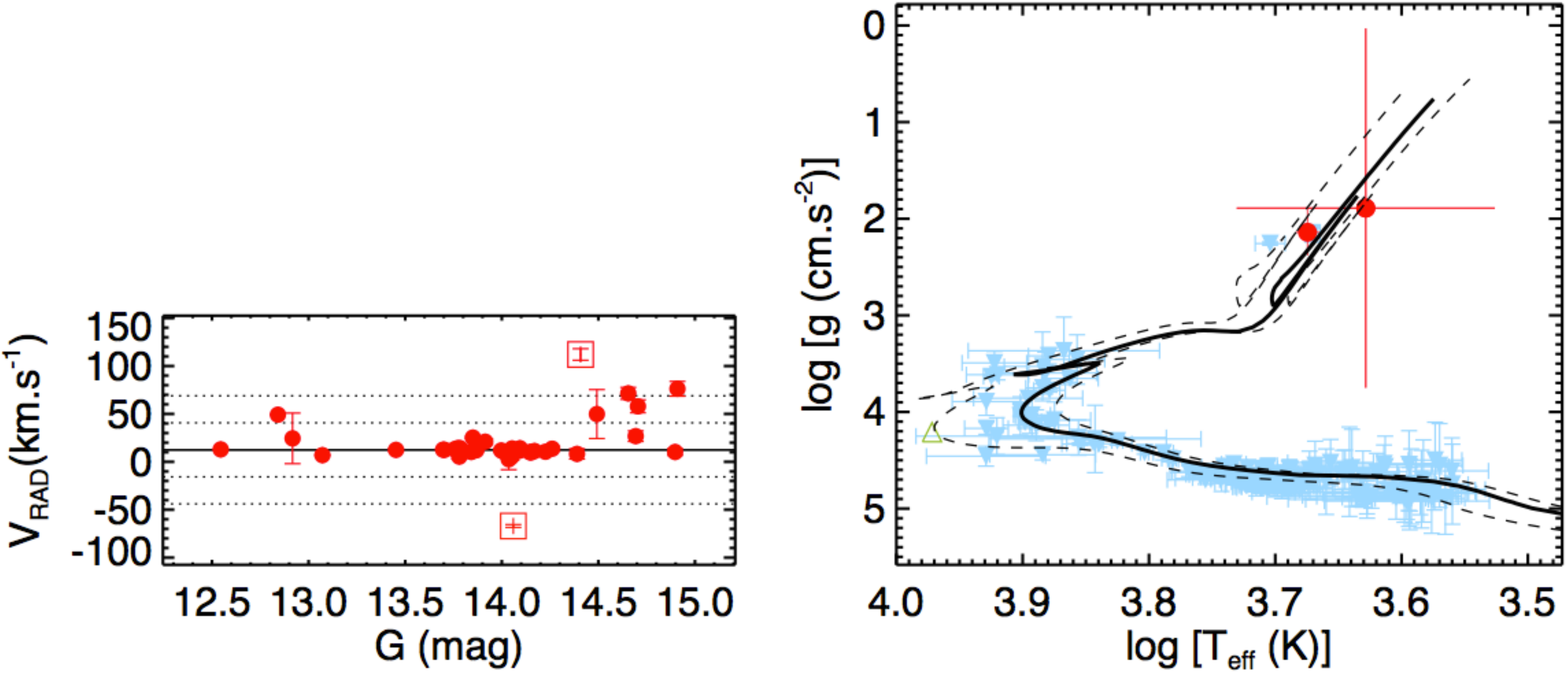}  
    \end{center}    
  }
\caption{ Same as Figure\,5 of the manuscript, but for the OC NGC\,6802. No $[Fe/H]$ values available for the set of member stars. }

\label{fig:HRD_NGC6802}
\end{center}
\end{figure*}

\begin{figure*}
\begin{center}

\parbox[c]{0.70\textwidth}
  {
   \begin{center}
    \includegraphics[width=0.70\textwidth]{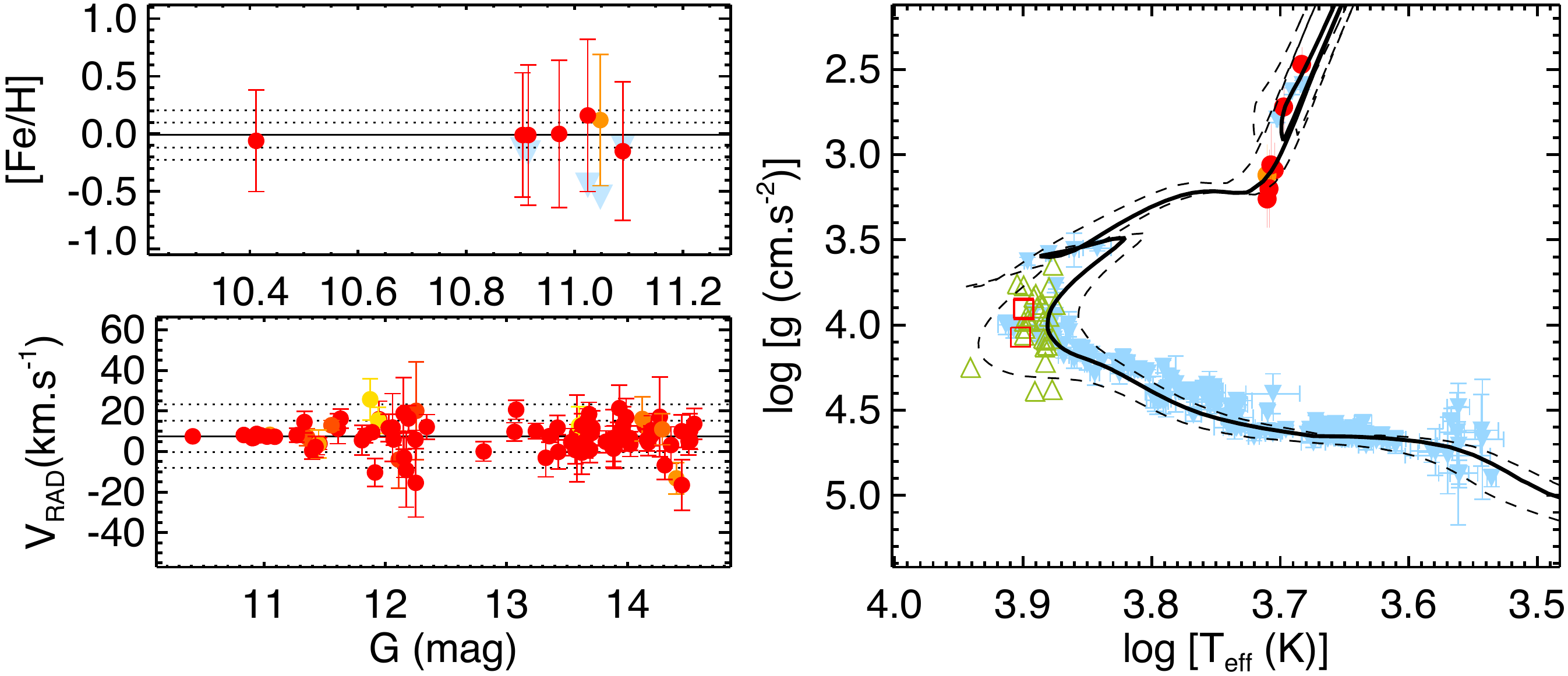}  
    \end{center}    
  }
\caption{ Same as Figure\,5 of the manuscript, but for the OC NGC\,6811. }

\label{fig:HRD_NGC6811}
\end{center}
\end{figure*}

\begin{figure*}
\begin{center}

\parbox[c]{0.70\textwidth}
  {
   \begin{center}
    \includegraphics[width=0.70\textwidth]{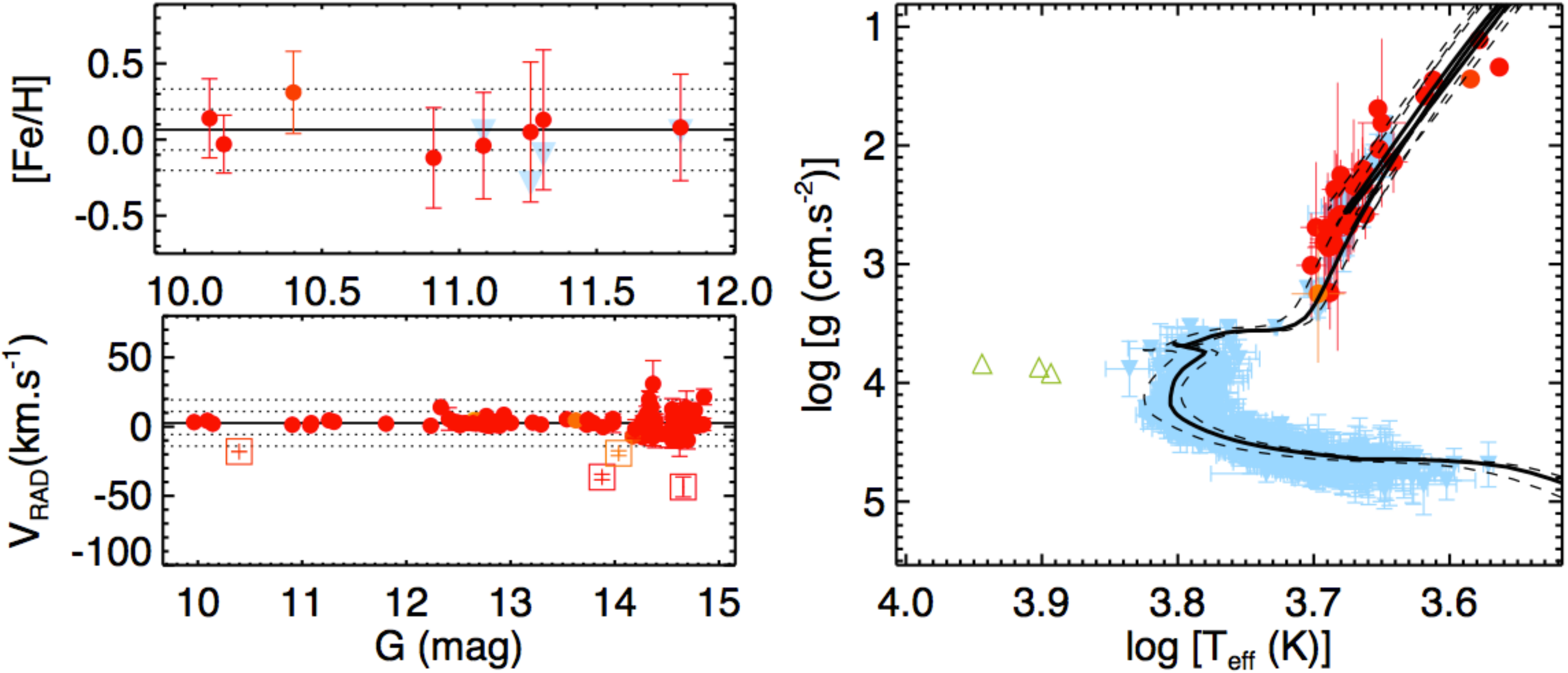}  
    \end{center}    
  }
\caption{ Same as Figure\,5 of the manuscript, but for the OC NGC\,6819. The open green squares identify 3 blue straggler star candidates. Their \textit{source\_{ID}} are 2076299688981549312, 2076299826420542080 and  2076392769511459712. Their equatorial coordinates ($\alpha_{\textrm{J2016}}$, $\delta_{\textrm{J2016}}$) are, respectively: (295.29882$^{\circ}$, 40.16309$^{\circ}$), (295.32275$^{\circ}$, 40.18429$^{\circ}$), (295.42766$^{\circ}$, 40.19446$^{\circ}$). The photometric data ($G$, ($G_{BP}$-$G_{RP}$)) is, respectively: (12.96, 0.48)\,mag, (12.90, 0.56)\,mag, (12.78, 0.32)\,mag. The atmospheric parametes (log\,$T_{\textrm{eff}}$, log\,$g$) are, respectively: (3.90, 3.87), (3.89, 3.92) and (3.94, 3.84). }

\label{fig:HRD_NGC6819}
\end{center}
\end{figure*}

\begin{figure*}
\begin{center}

\parbox[c]{0.70\textwidth}
  {
   \begin{center}
    \includegraphics[width=0.70\textwidth]{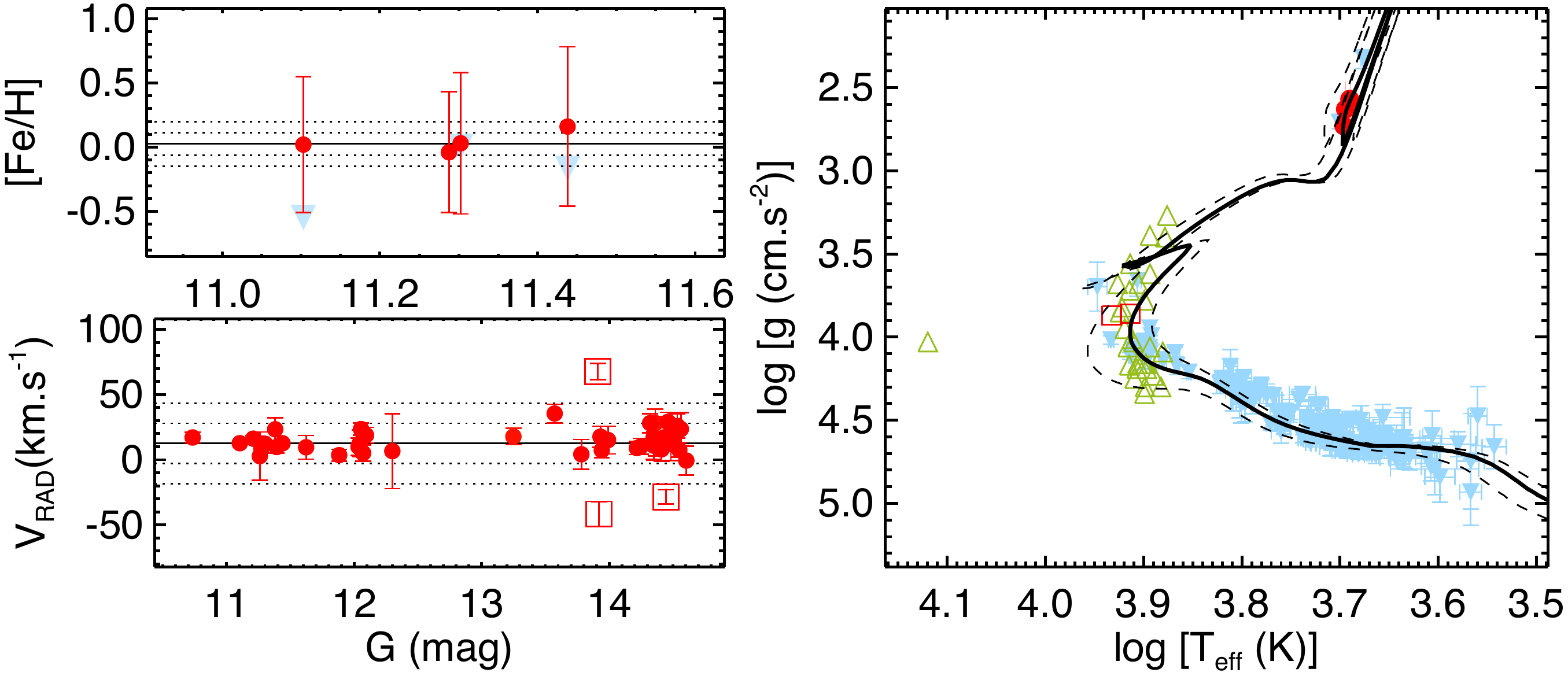}  
    \end{center}    
  }
\caption{ Same as Figure\,5 of the manuscript, but for the OC NGC\,6866. The hottest star (open green triangle with log\,$T_{\textrm{eff}}$=4.12, log\,$g$=4.03) represents the member star with \textit{source\_ID} 2076065699159591296 ($\alpha_{\textrm{J2016}}=300.91939^{\circ}$; $\delta_{\textrm{J2016}}=44.195388^{\circ}$; $G=11.26\,$mag; $(G_{BP}-G_{RP})=0.047\,$mag), a blue straggler candidate. }

\label{fig:HRD_NGC6866}
\end{center}
\end{figure*}

\begin{figure*}
\begin{center}

\parbox[c]{0.70\textwidth}
  {
   \begin{center}
    \includegraphics[width=0.70\textwidth]{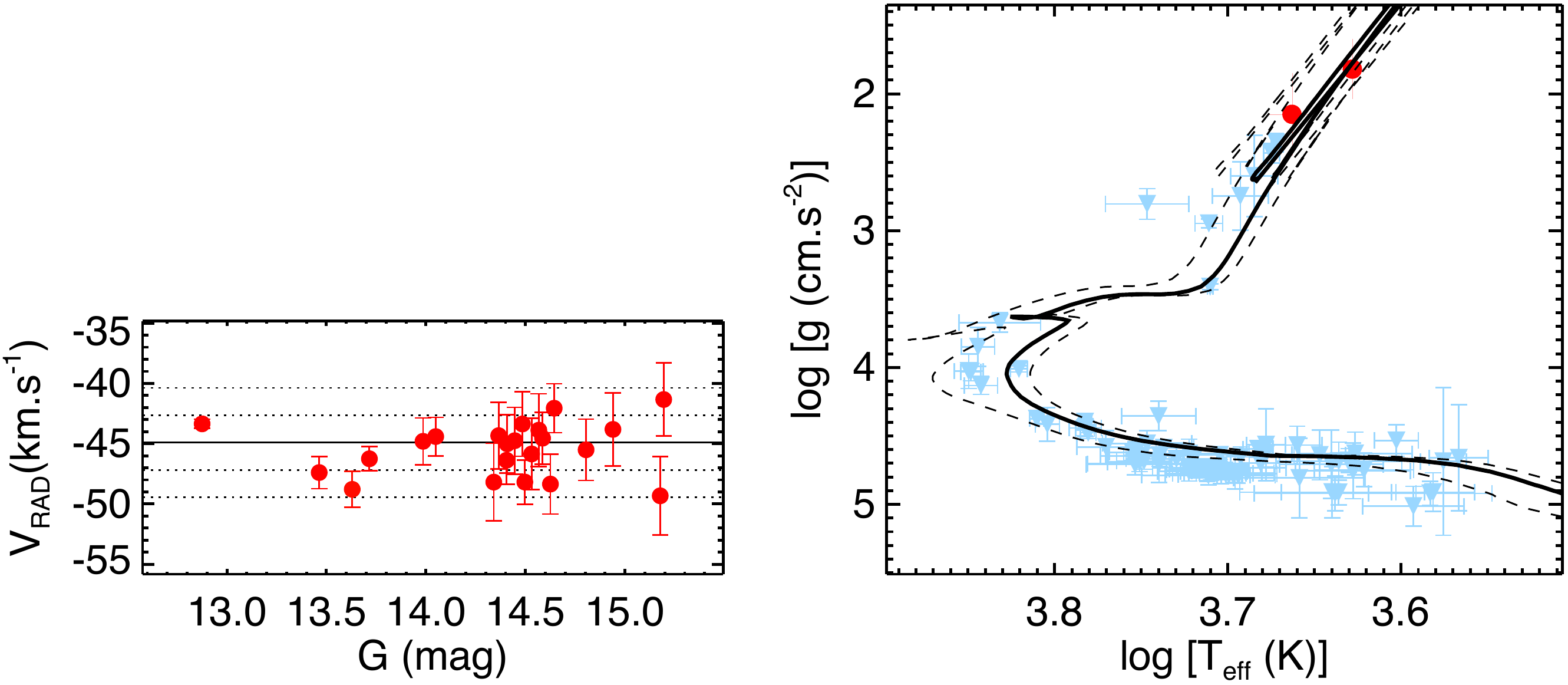}  
    \end{center}    
  }
\caption{ Same as Figure\,5 of the manuscript, but for the OC Berkeley\,89. No $[Fe/H]$ values available for the set of member stars. }

\label{fig:HRD_Berkeley89}
\end{center}
\end{figure*}

\afterpage{\clearpage}

\begin{figure*}
\begin{center}

\parbox[c]{0.70\textwidth}
  {
   \begin{center}
    \includegraphics[width=0.70\textwidth]{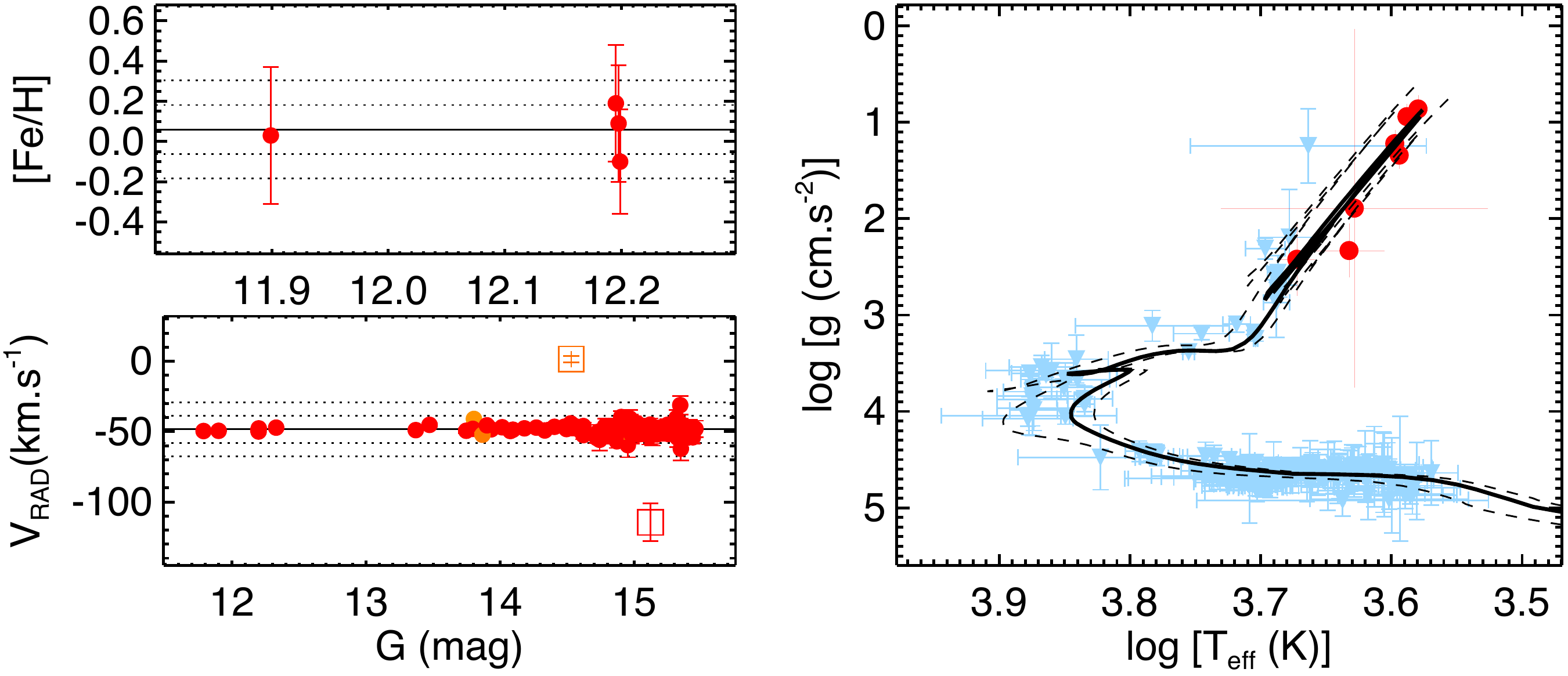}  
    \end{center}    
  }
\caption{ Same as Figure\,5 of the manuscript, but for the OC NGC\,7044. }

\label{fig:HRD_NGC7044}
\end{center}
\end{figure*}

\begin{figure*}
\begin{center}

\parbox[c]{0.70\textwidth}
  {
   \begin{center}
    \includegraphics[width=0.70\textwidth]{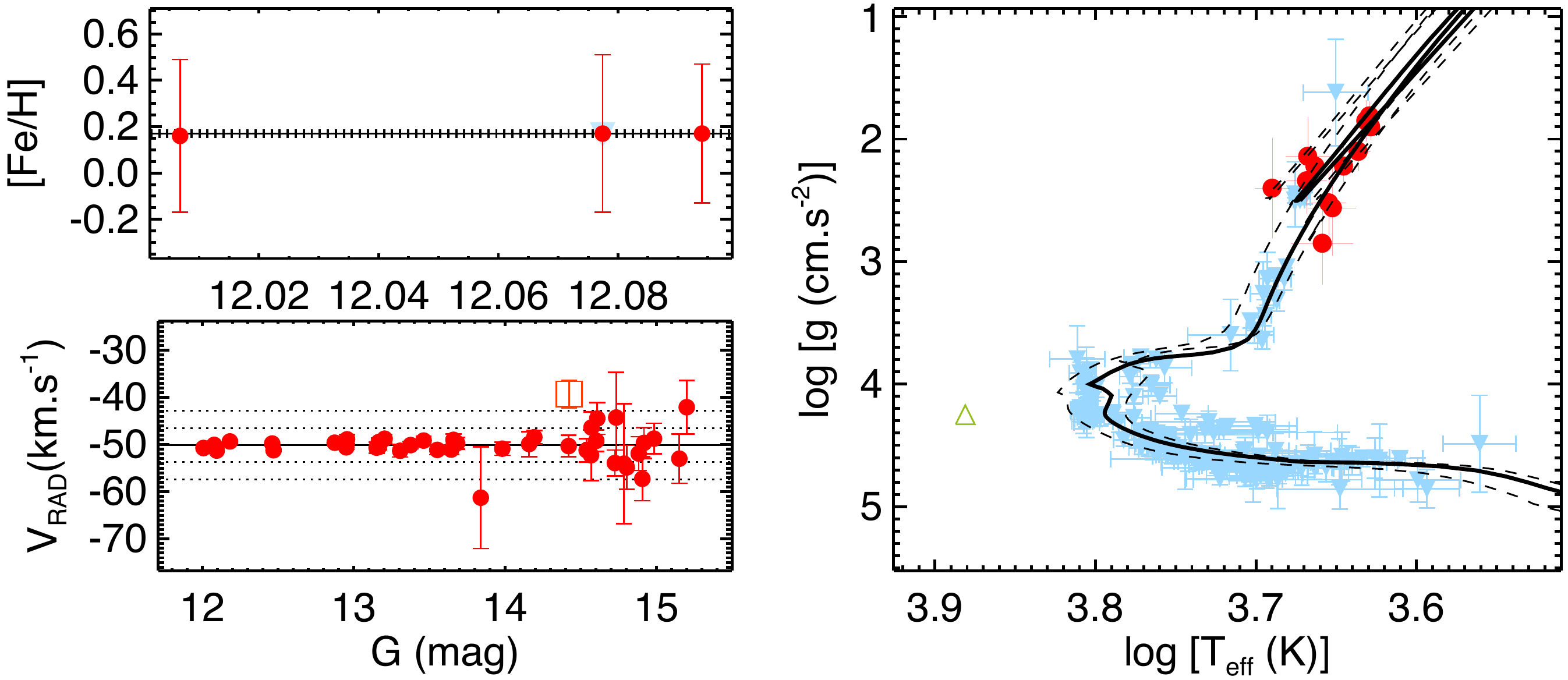}  
    \end{center}    
  }
\caption{ Same as Figure\,5 of the manuscript, but for the OC NGC\,7142. The hottest star (open green triangle with log\,$T_{\textrm{eff}}$=3.88, log\,$g$=4.25) represents the member star with \textit{source\_ID} 2217989766712690432 ($\alpha_{\textrm{J2016}}=325.95112^{\circ}$; $\delta_{\textrm{J2016}}=65.85986^{\circ}$; $G=13.84\,$mag; $(G_{BP}-G_{RP})=1.00\,$mag), a blue straggler candidate. }

\label{fig:HRD_NGC7142}
\end{center}
\end{figure*}

\begin{figure*}
\begin{center}

\parbox[c]{0.70\textwidth}
  {
   \begin{center}
    \includegraphics[width=0.70\textwidth]{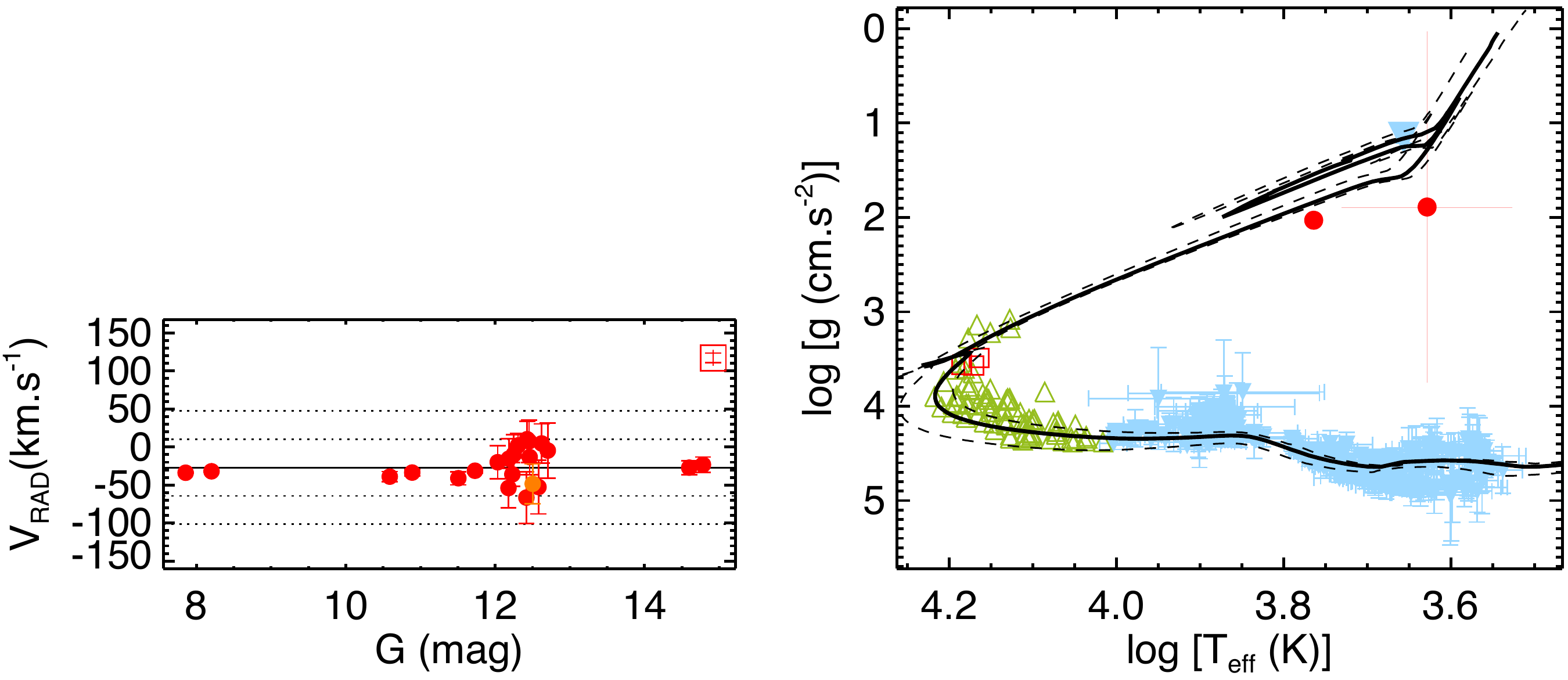}  
    \end{center}    
  }
\caption{ Same as Figure\,5 of the manuscript, but for the OC NGC\,7654. No $[Fe/H]$ values available for the set of member stars. }

\label{fig:HRD_NGC7654}
\end{center}
\end{figure*}

\begin{figure*}
\begin{center}

\parbox[c]{0.70\textwidth}
  {
   \begin{center}
    \includegraphics[width=0.70\textwidth]{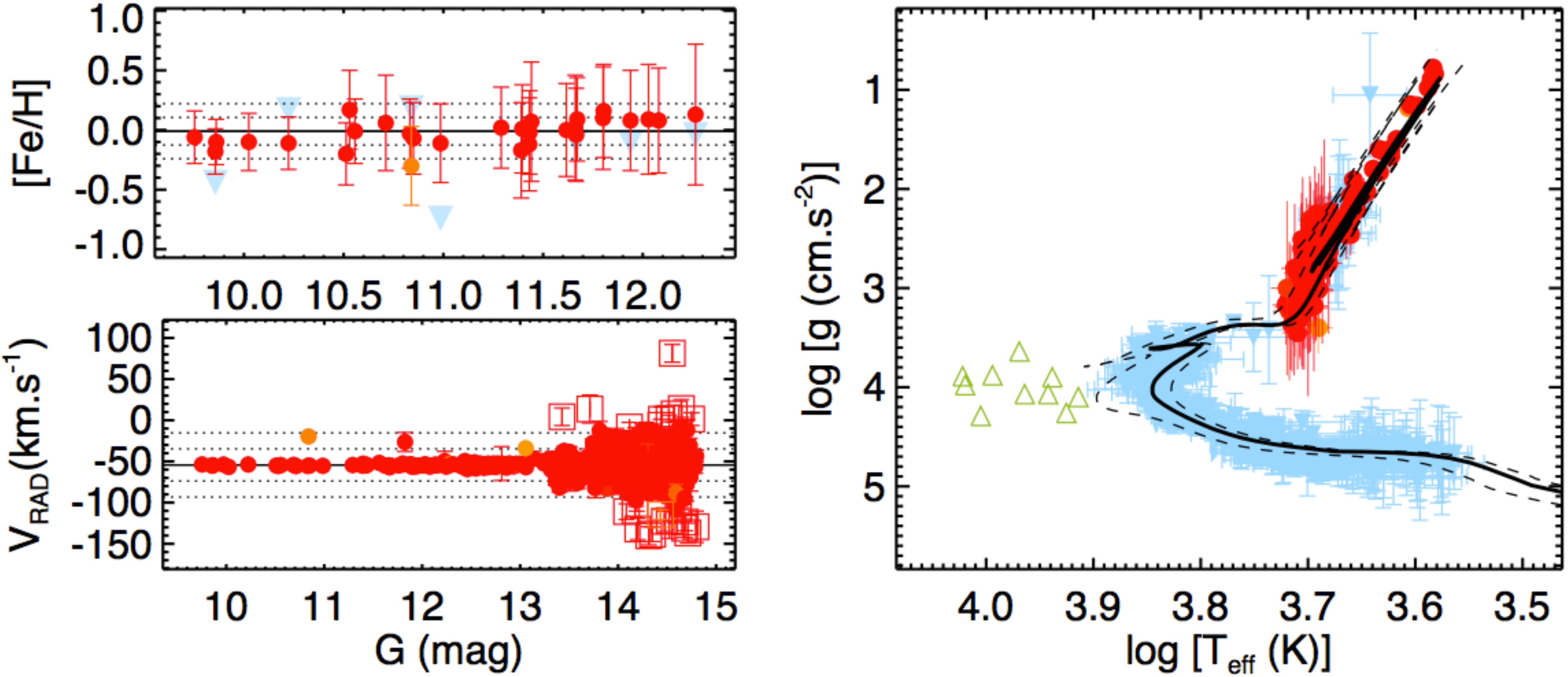}  
    \end{center}    
  }
\caption{ Same as Figure\,5 of the manuscript, but for the OC NGC\,7789. The set of stars plotted as green triangles are blue straggler candidates. They are located in the ranges $G<14\,$mag, $(G_{BP}-G_{RP})<0.7\,$mag in the cluster CMD (Figure~\ref{fig:CMD_SupplMater7}).  }

\label{fig:HRD_NGC7789}
\end{center}
\end{figure*}

\clearpage

\section{Supplementary figures - Skymaps}
This Appendix shows the skymaps for the 60 investigated OCs (Figures\,H1 to H5).

\begin{figure*}
\begin{center}

\parbox[c]{1.00\textwidth}
  {
    \includegraphics[width=0.333\textwidth]{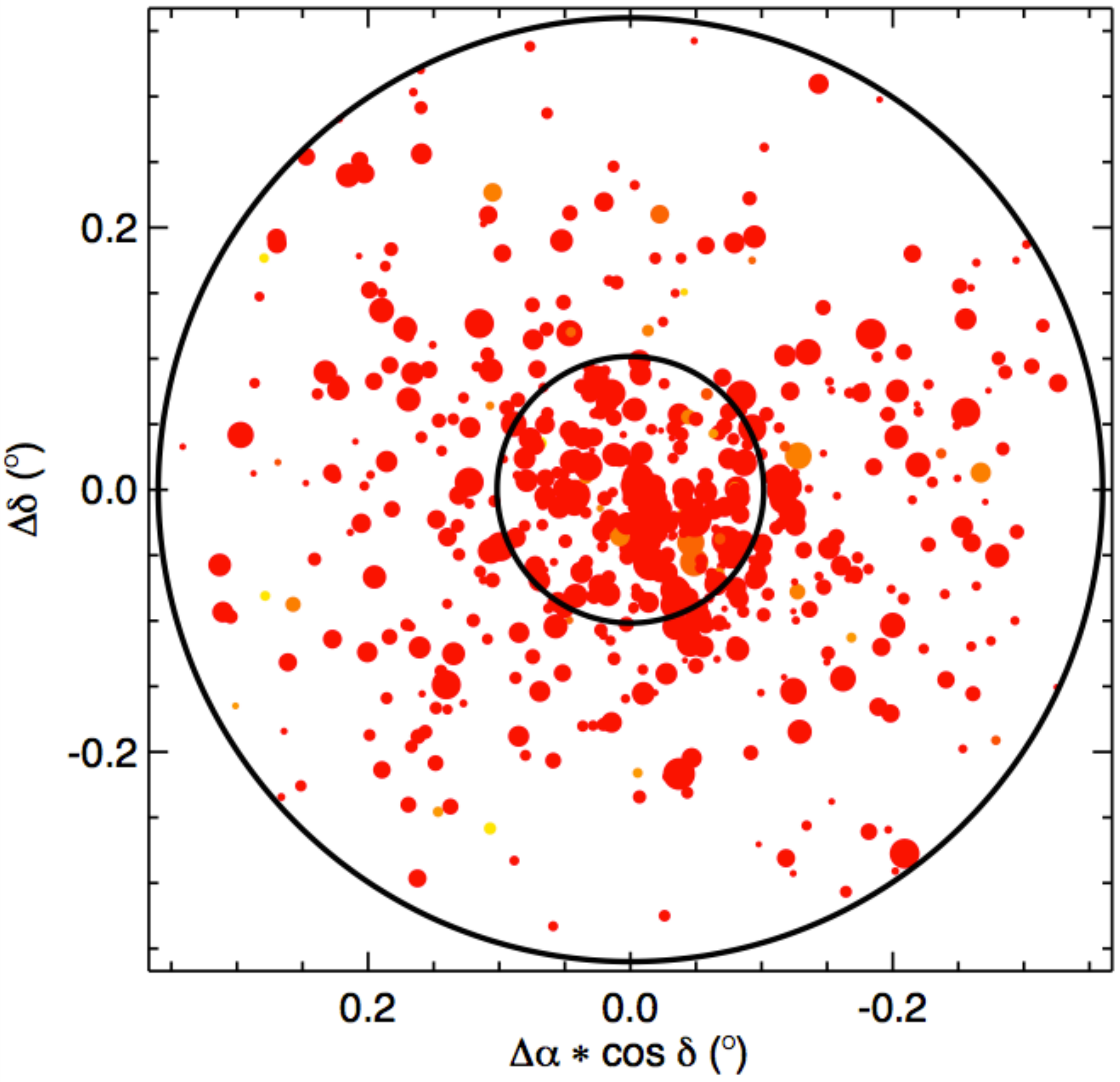}  
    \includegraphics[width=0.333\textwidth]{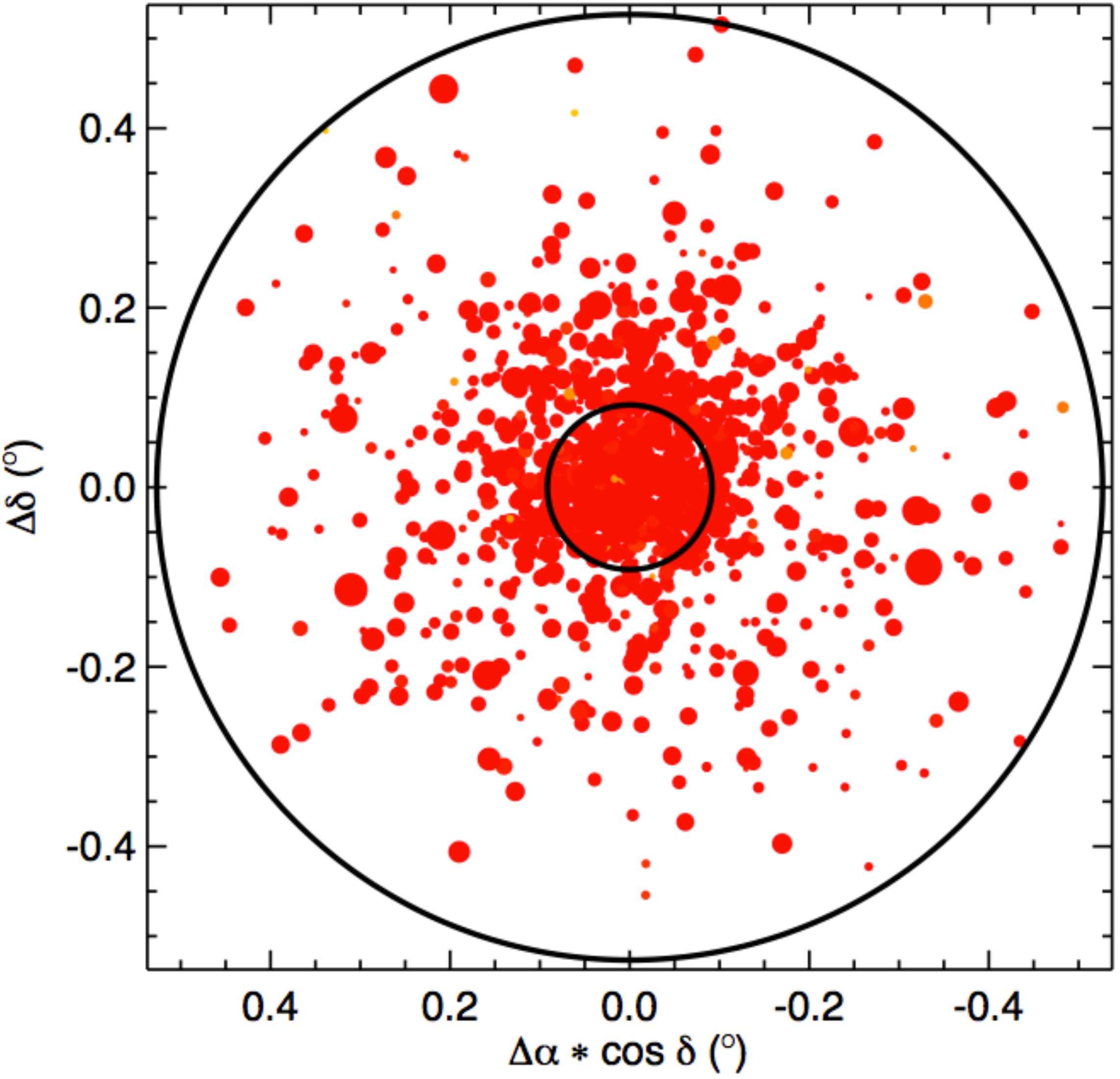}
    \includegraphics[width=0.333\textwidth]{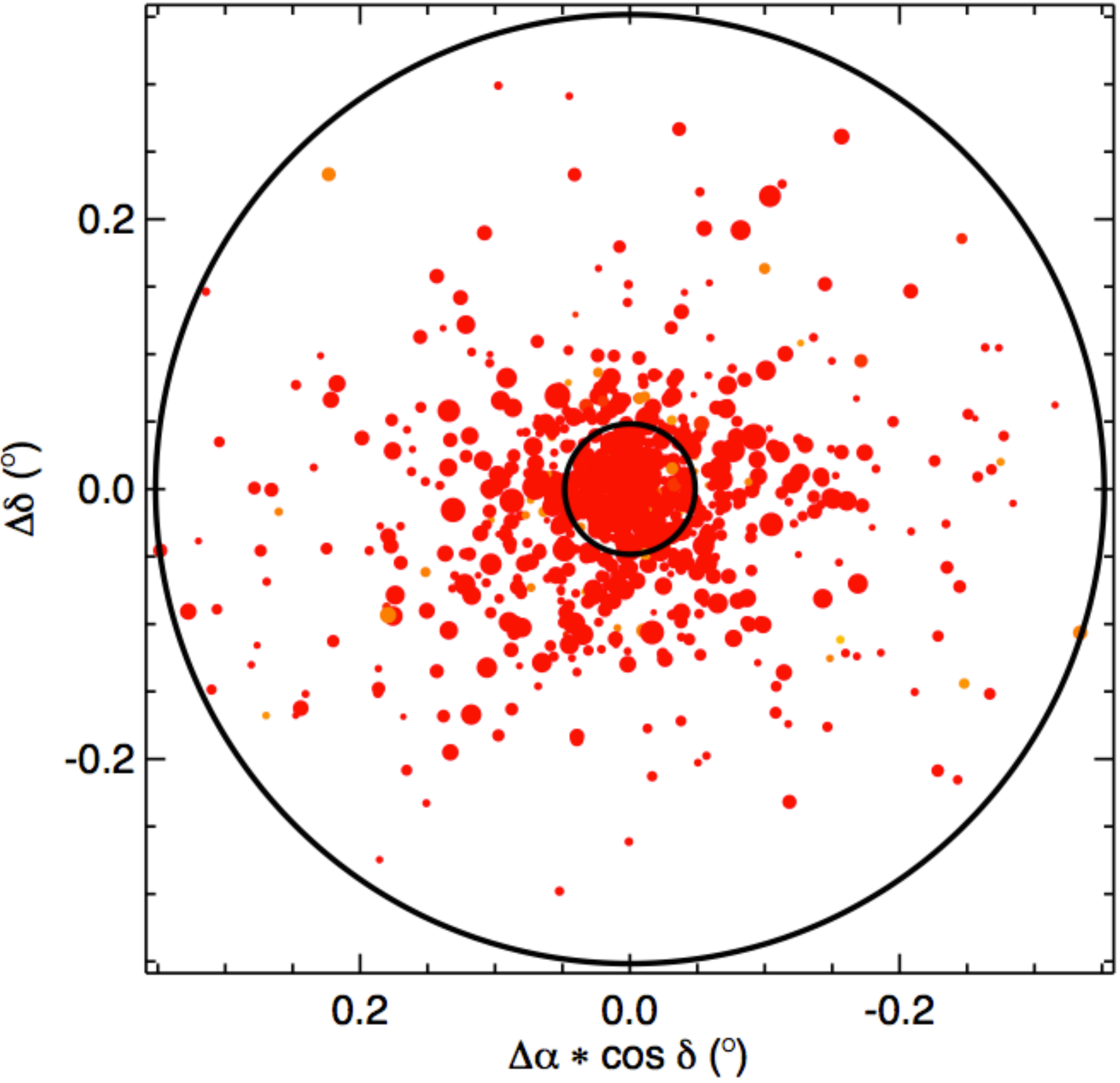}   
 
    \includegraphics[width=0.333\textwidth]{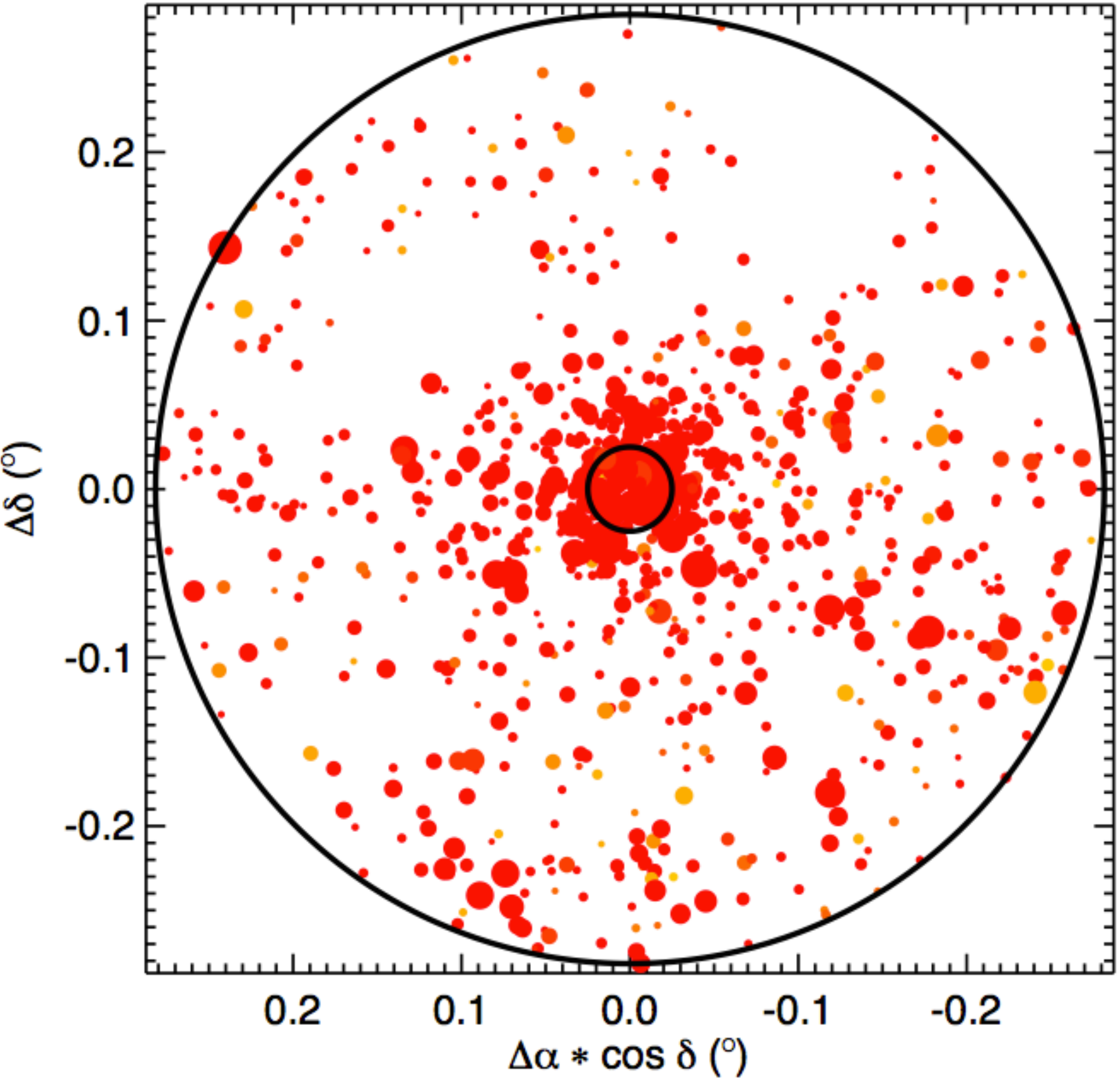}     
    \includegraphics[width=0.333\textwidth]{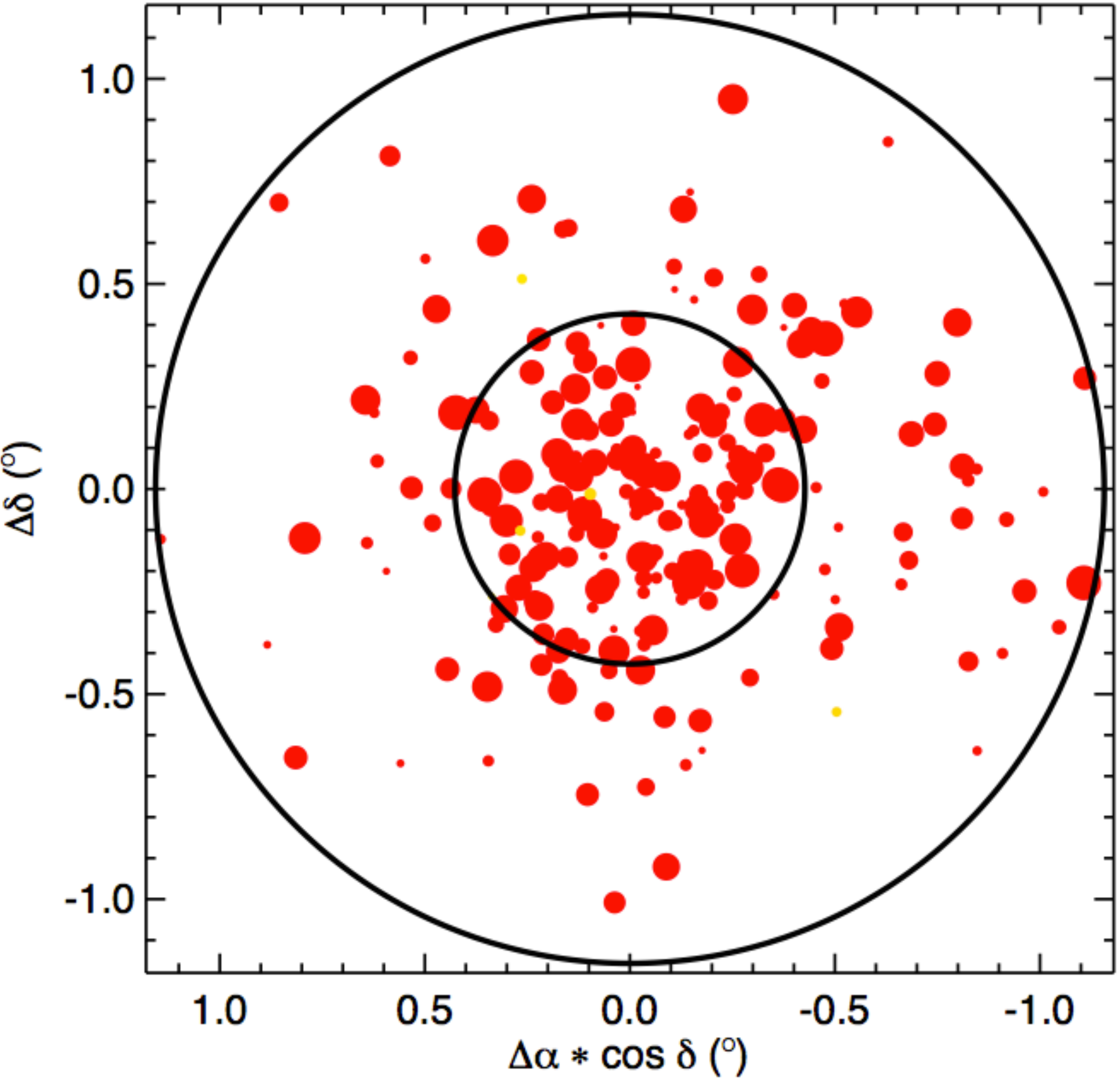}     
    \includegraphics[width=0.333\textwidth]{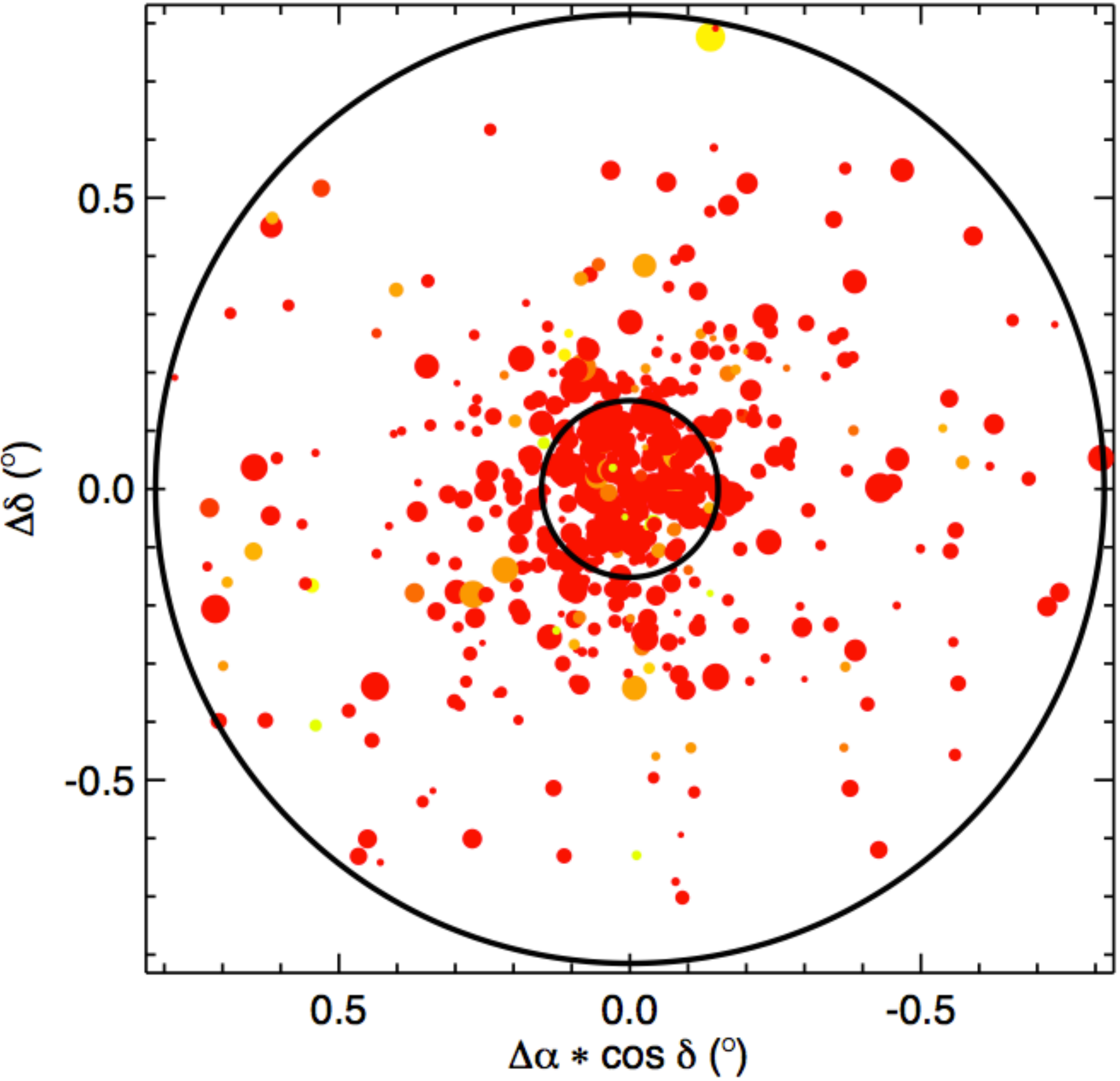}     

    \includegraphics[width=0.333\textwidth]{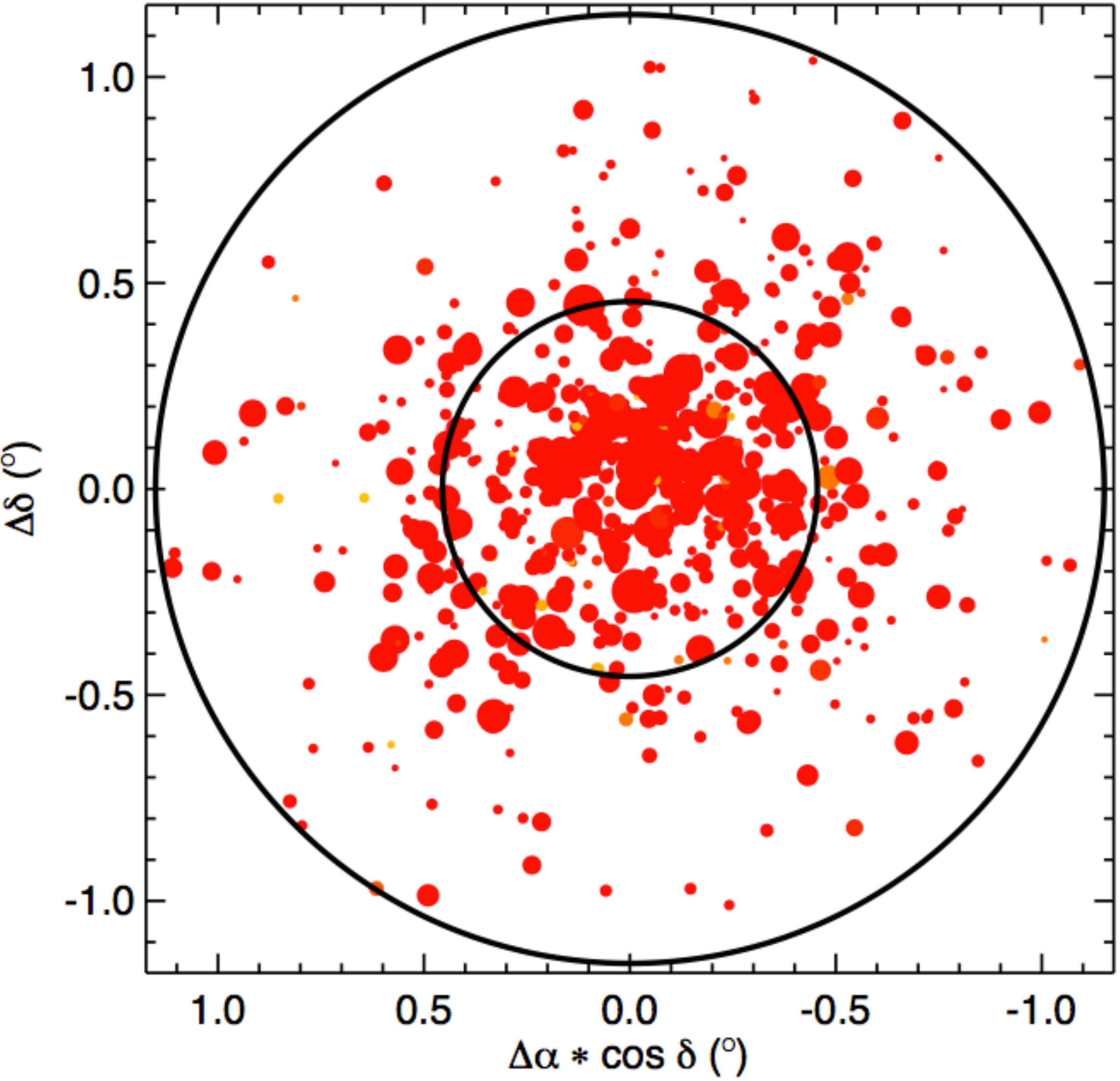}     
    \includegraphics[width=0.333\textwidth]{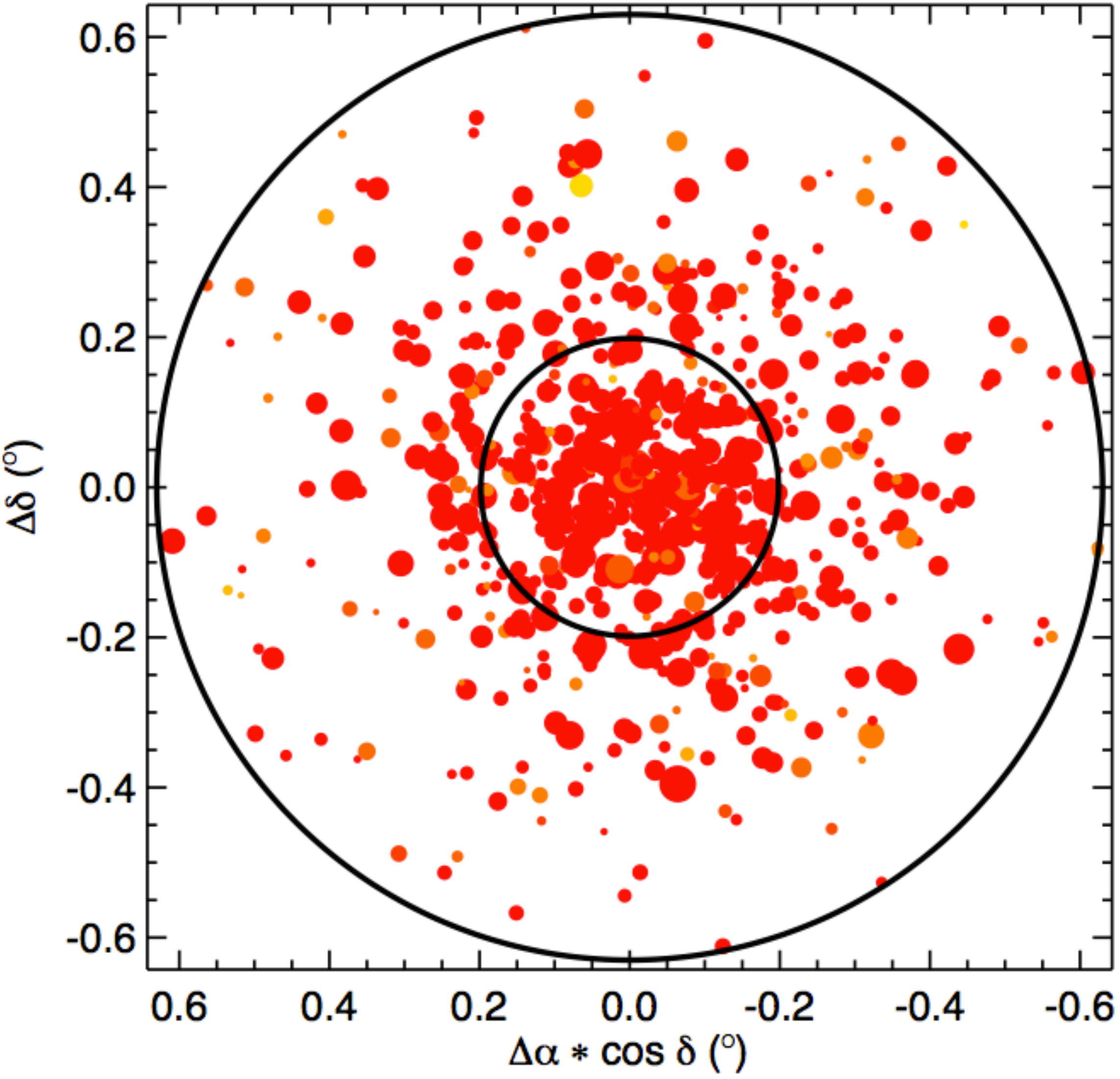}     
    \includegraphics[width=0.333\textwidth]{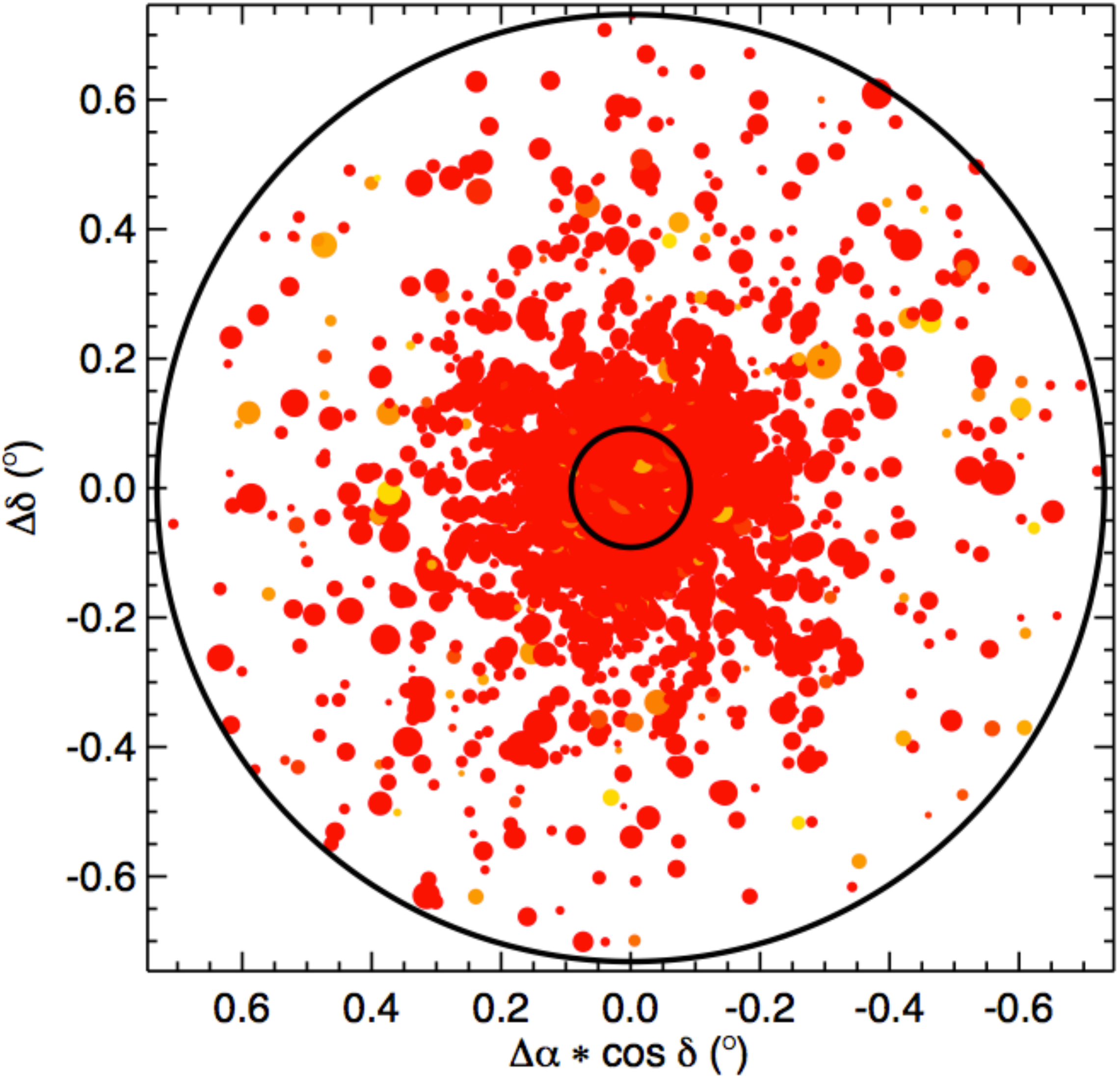}     

    \includegraphics[width=0.333\textwidth]{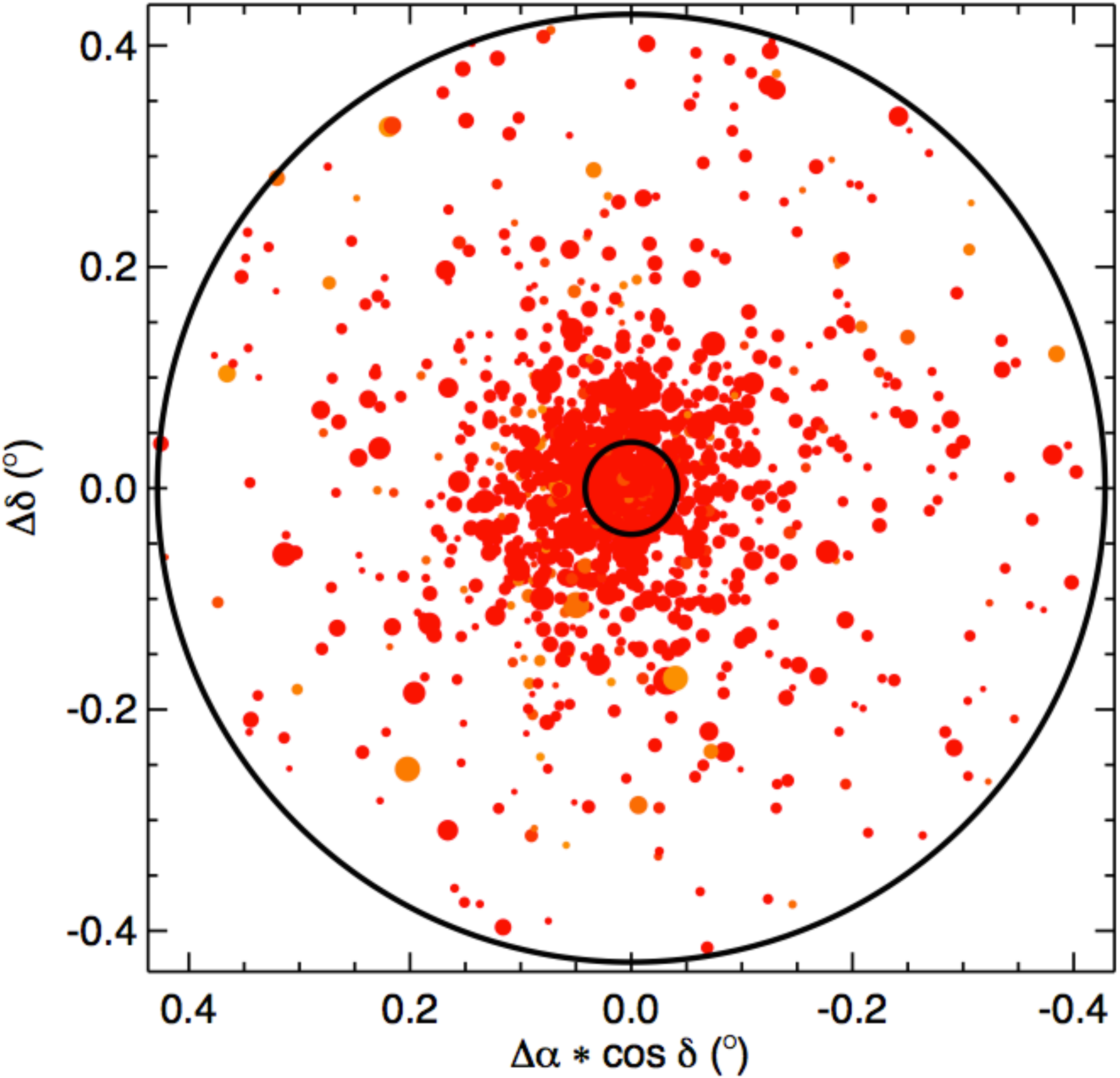}     
    \includegraphics[width=0.333\textwidth]{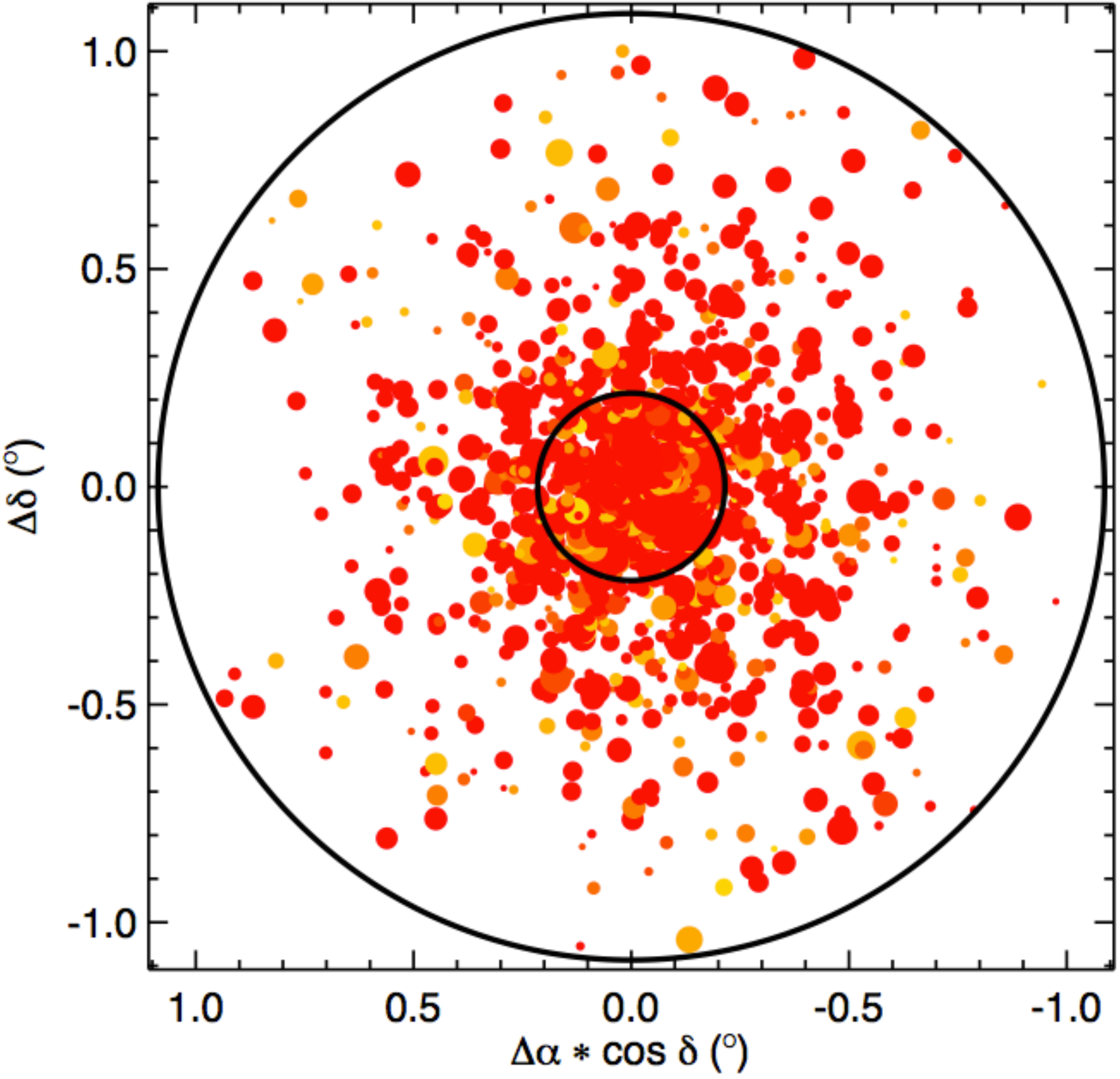}     
    \includegraphics[width=0.333\textwidth]{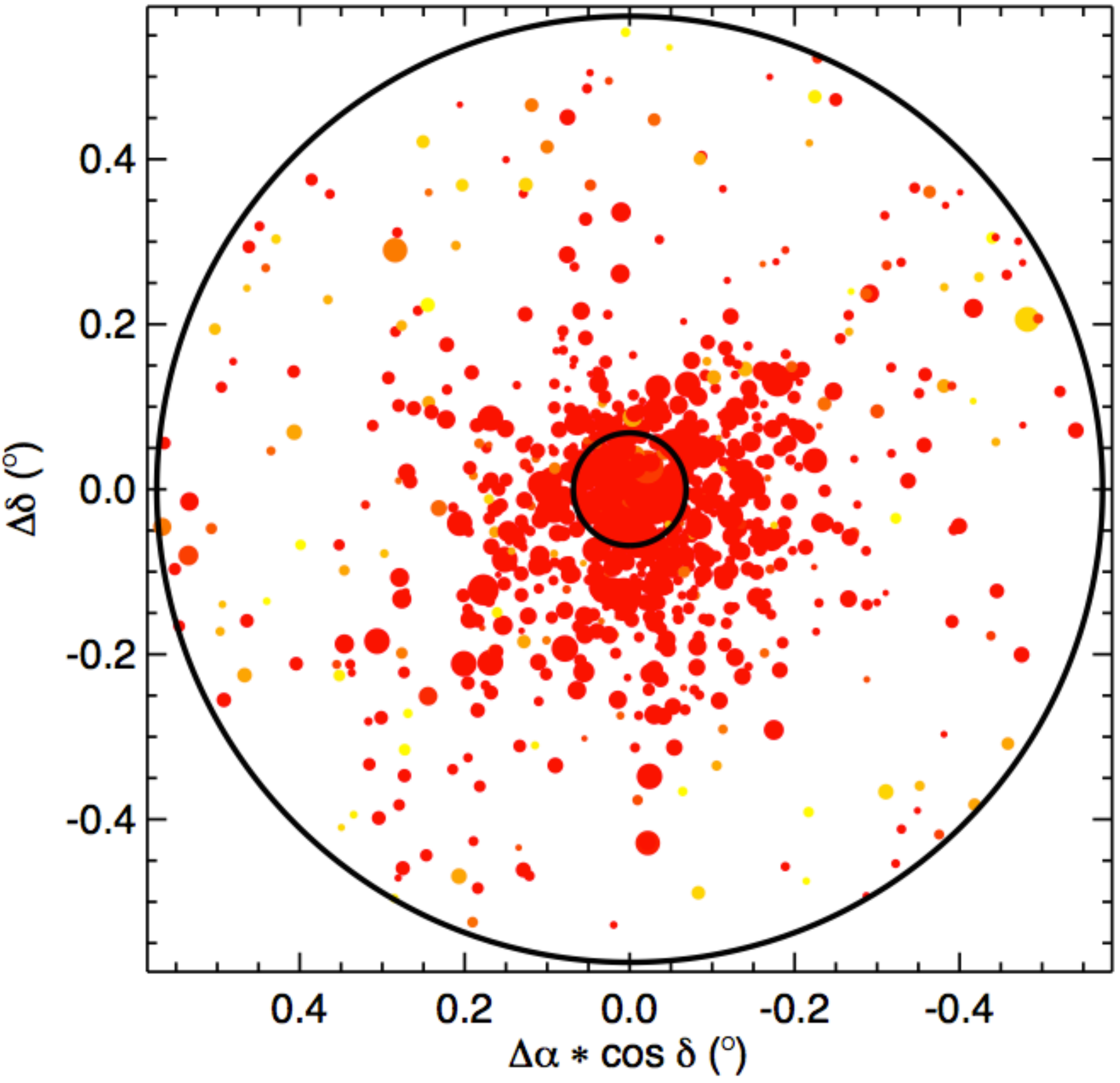}     
     
  }
\caption{ Skymap for the OCs (from top left to bottom right): NGC\,129, NGC\,188, NGC\,559 (top line), NGC\,654, NGC\,752, NGC\,1027 (second line), NGC\,1647, NGC\,1817, M\,37 (third line), NGC\,2141, NGC\,2168, NGC\,2204 (bottom line). }

\label{fig:skymaps_1_12}
\end{center}
\end{figure*}

\begin{figure*}
\begin{center}

\parbox[c]{1.00\textwidth}
  {
    \includegraphics[width=0.333\textwidth]{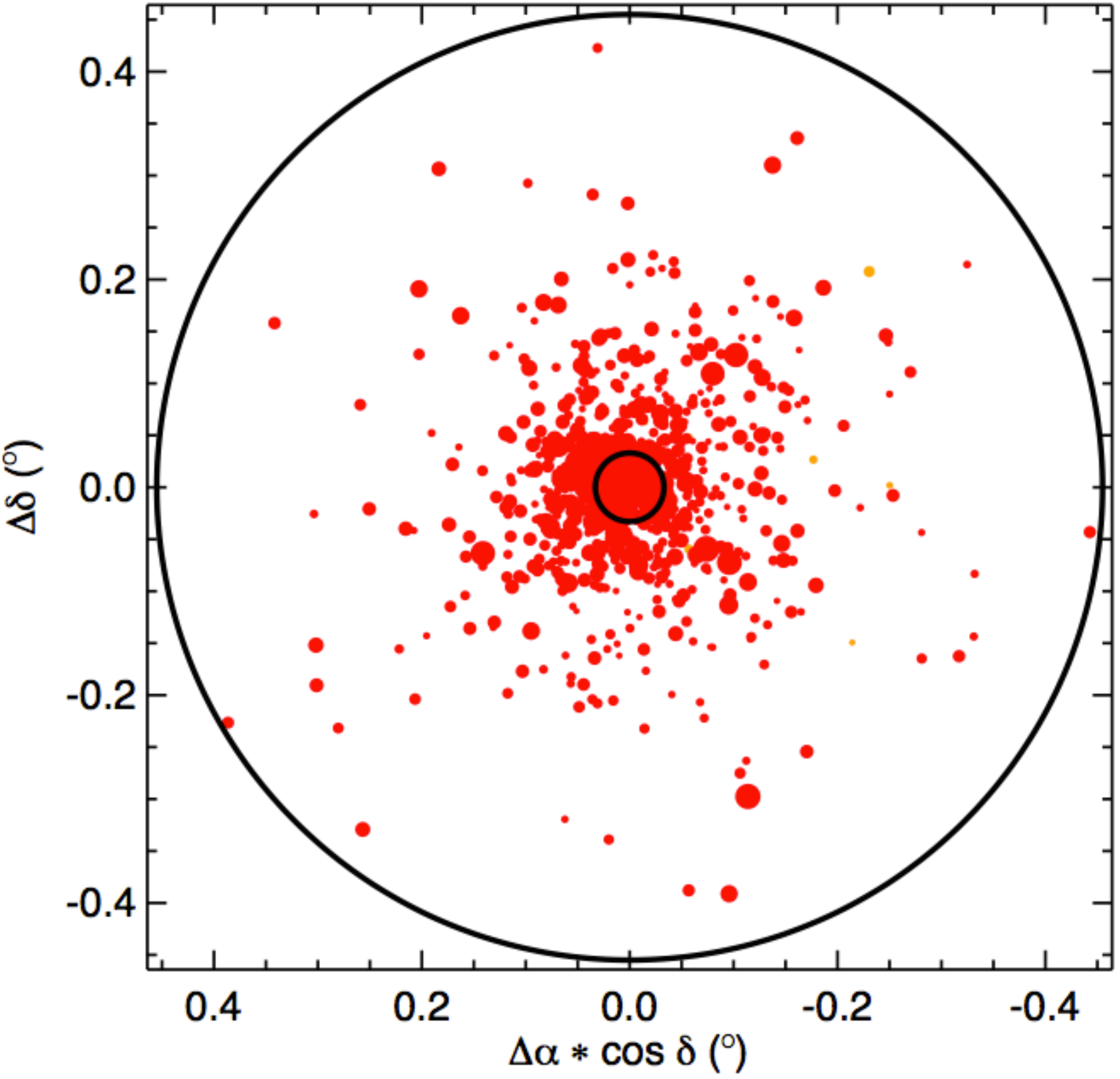}     
    \includegraphics[width=0.333\textwidth]{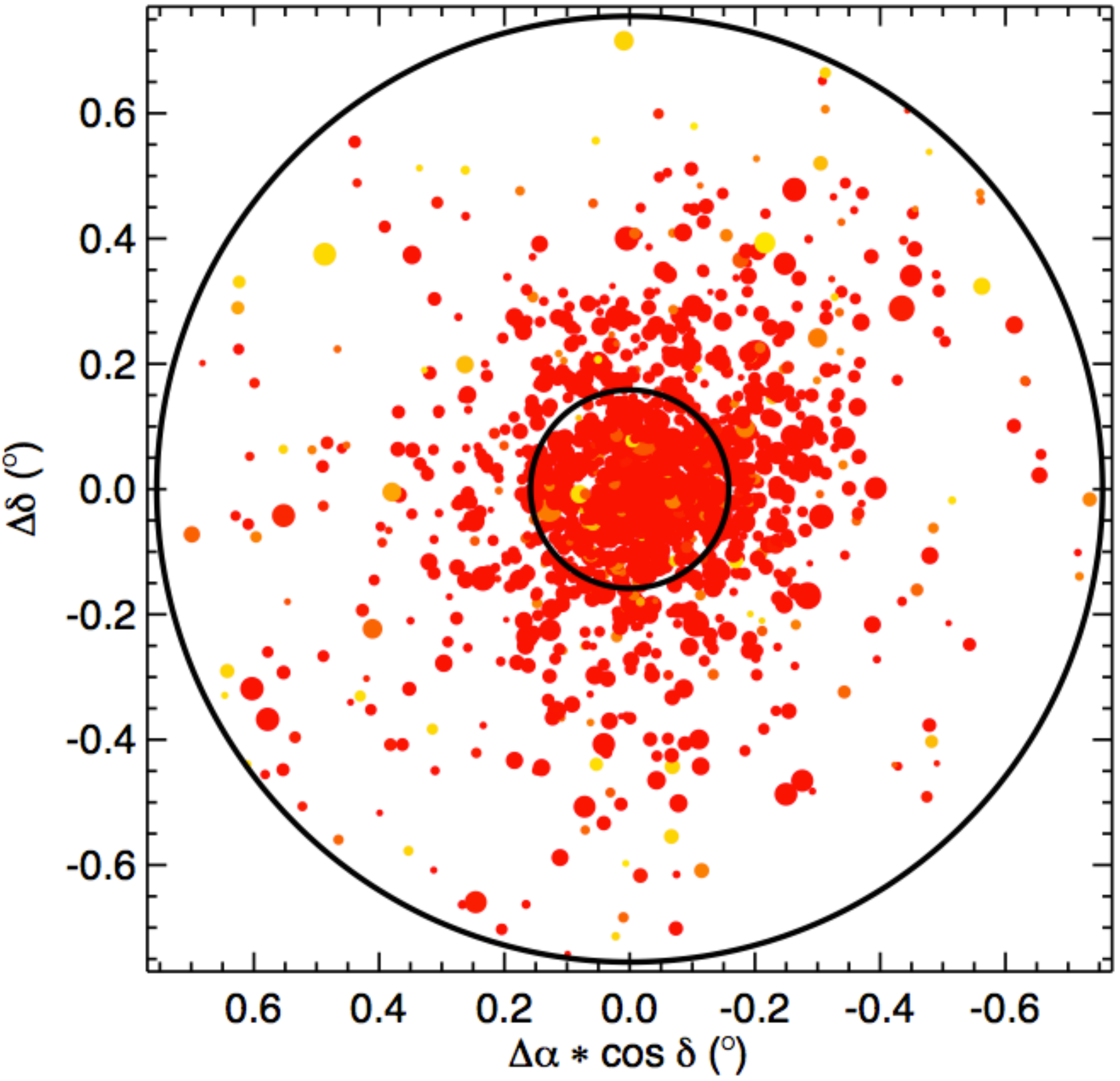}     
    \includegraphics[width=0.333\textwidth]{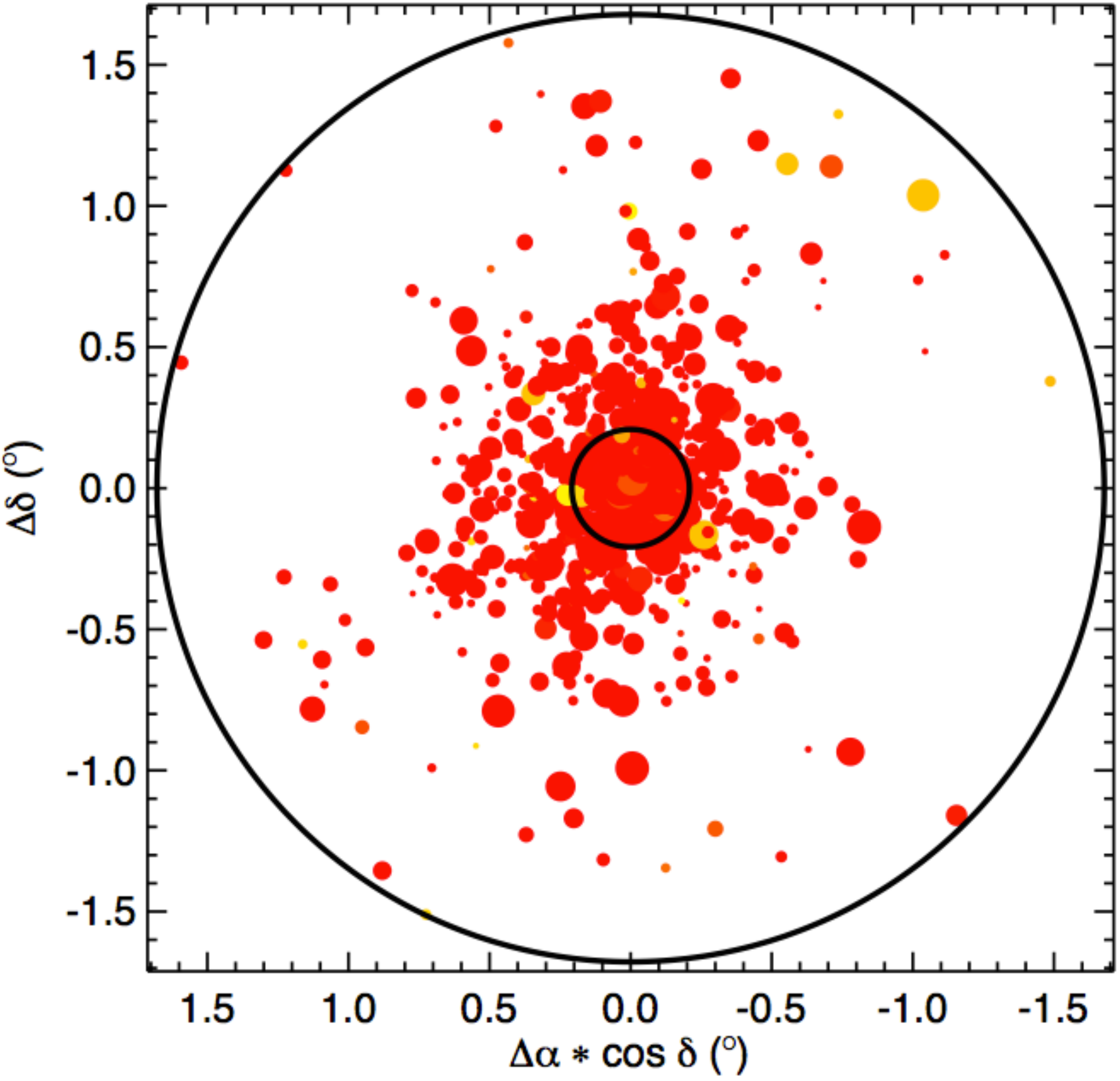}     
 
    \includegraphics[width=0.333\textwidth]{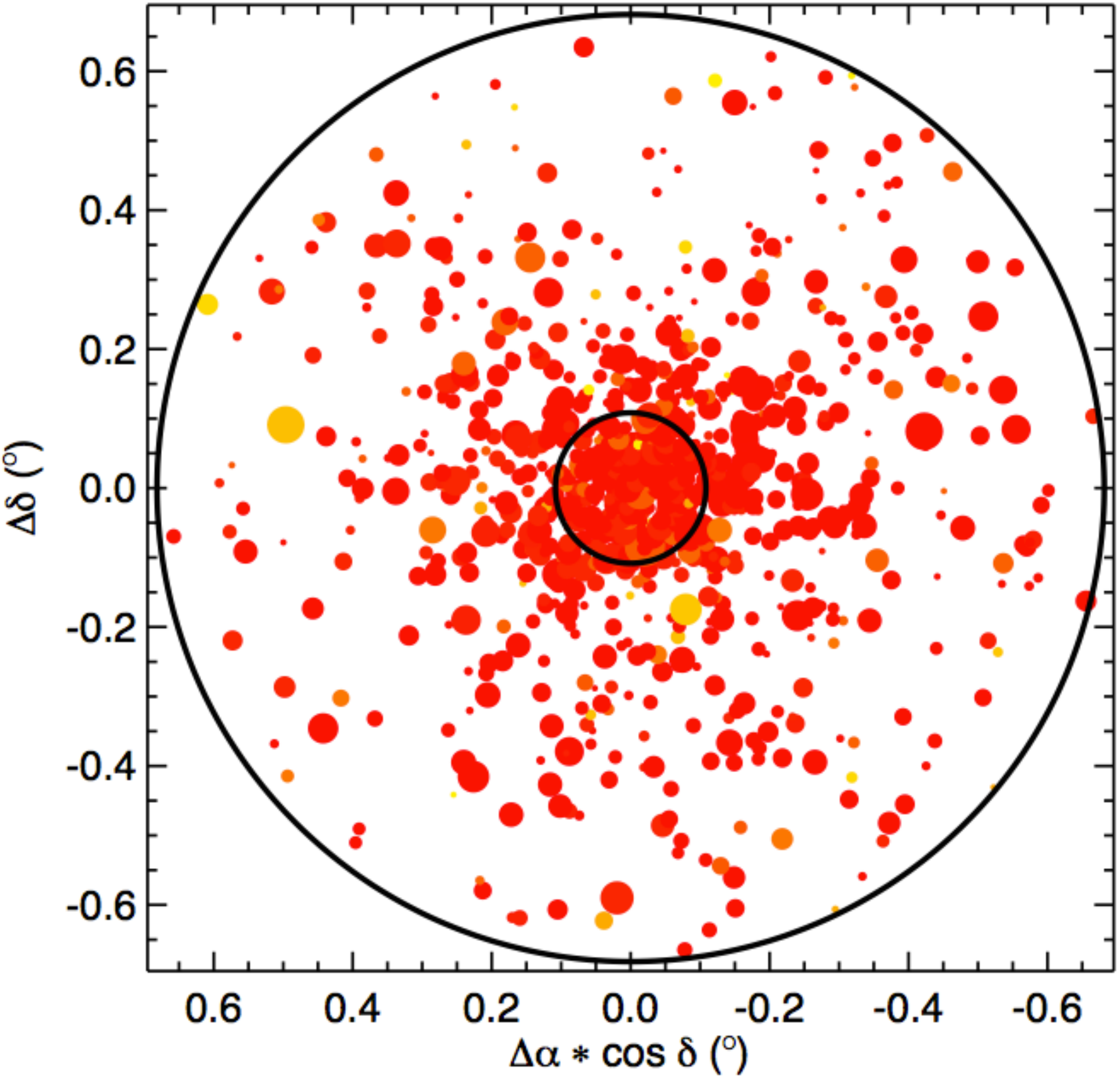}     
    \includegraphics[width=0.333\textwidth]{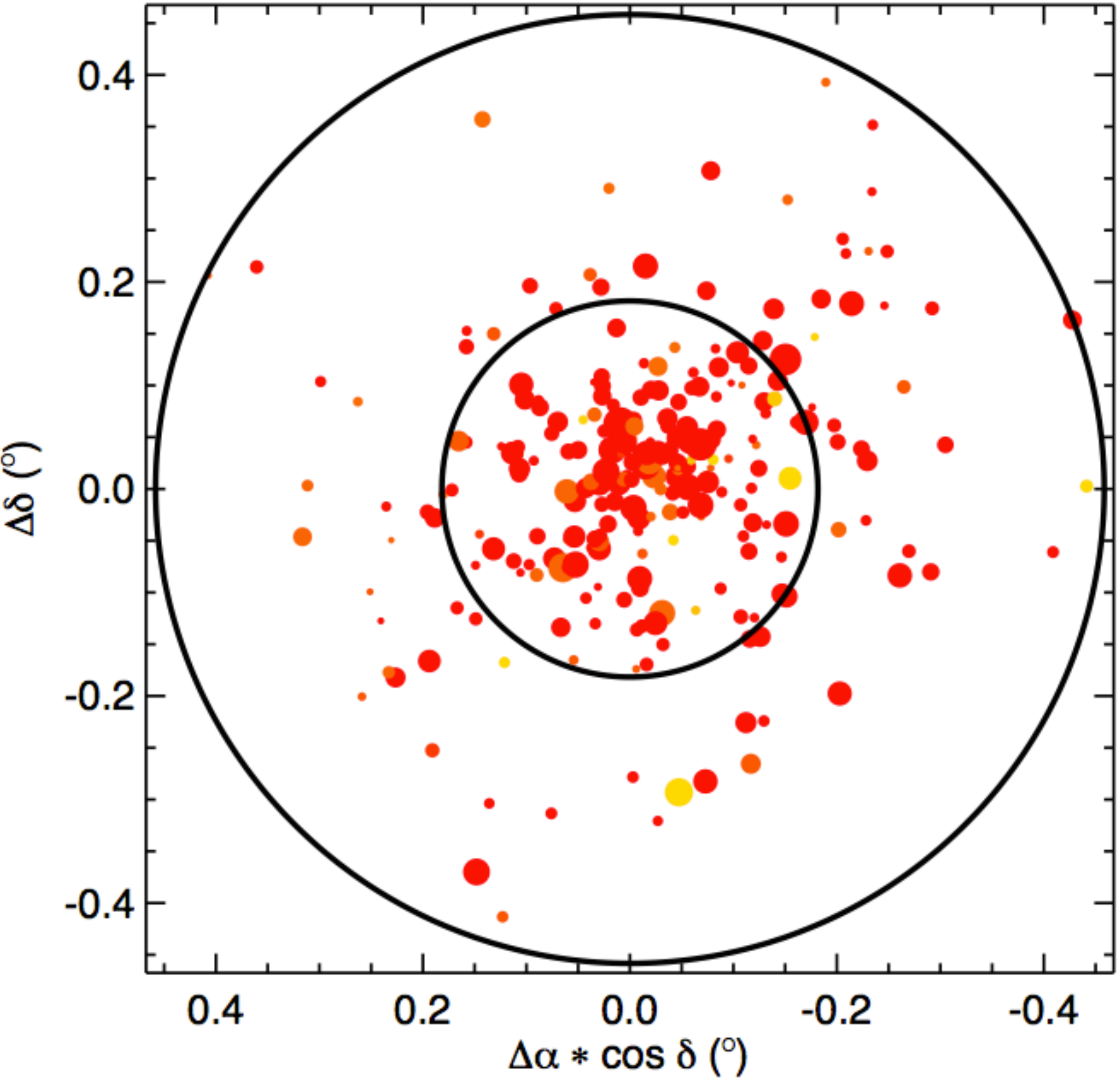}     
    \includegraphics[width=0.333\textwidth]{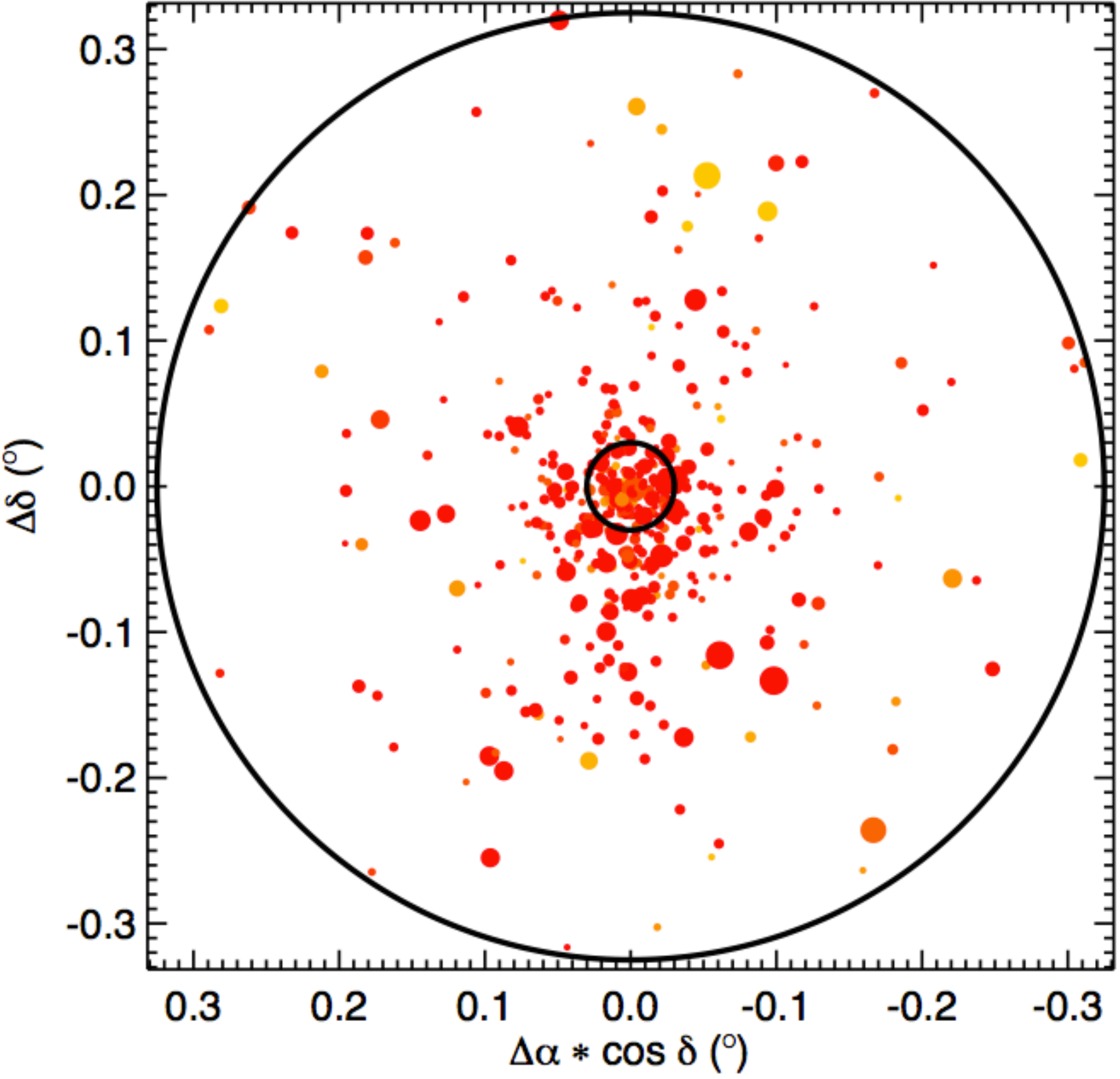}     

    \includegraphics[width=0.333\textwidth]{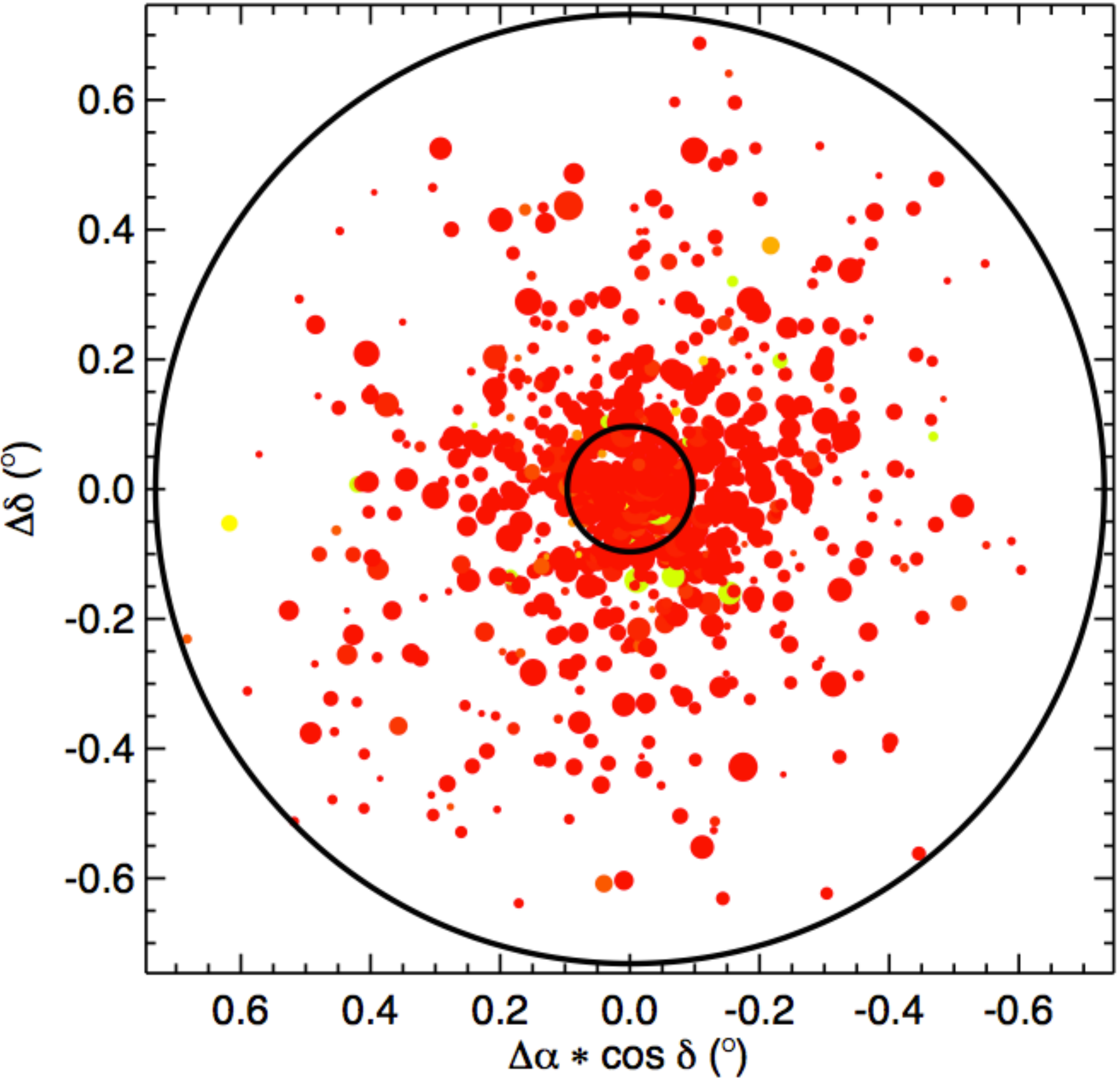}     
    \includegraphics[width=0.333\textwidth]{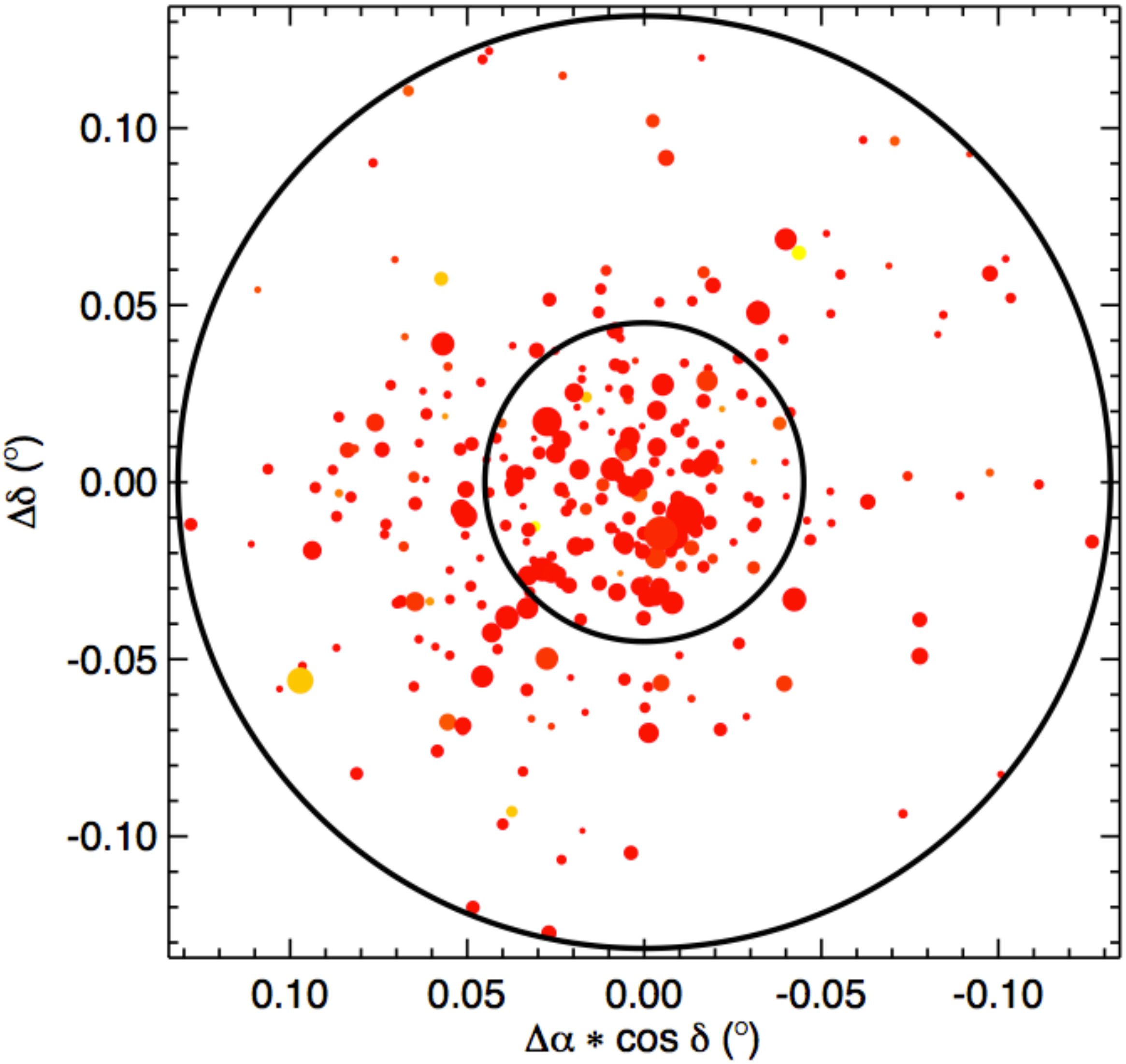}     
    \includegraphics[width=0.333\textwidth]{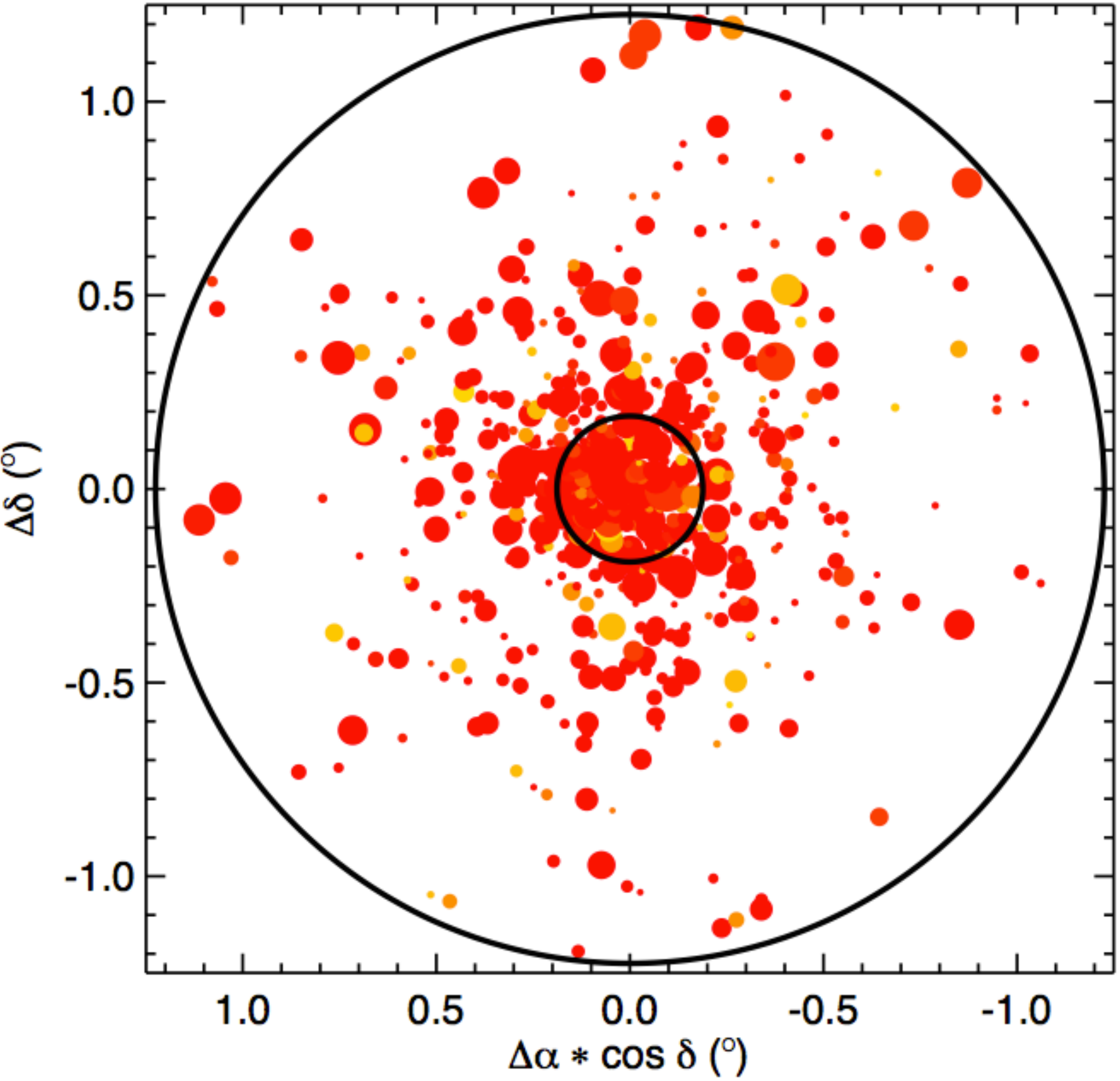}     

    \includegraphics[width=0.333\textwidth]{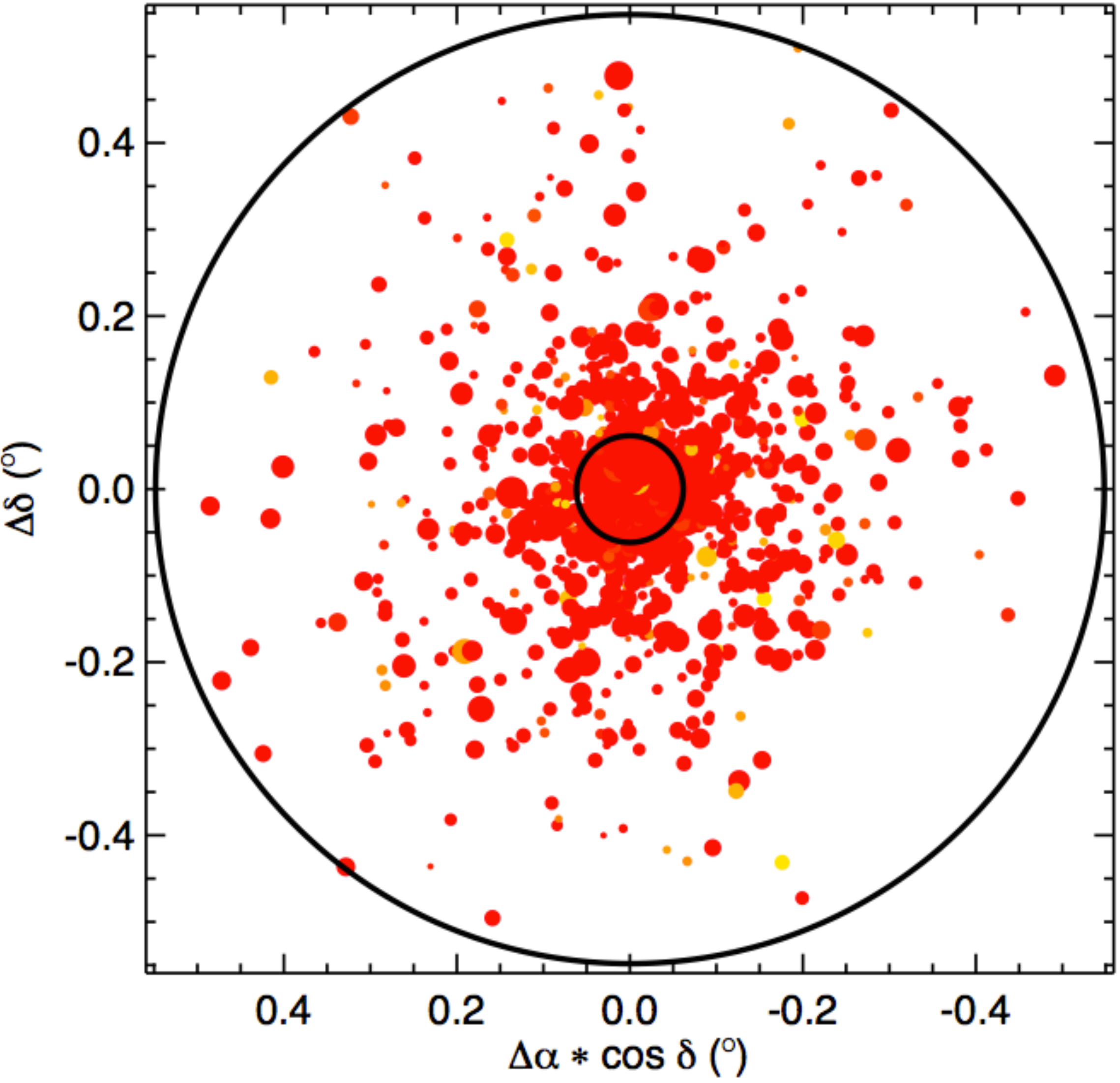}     
    \includegraphics[width=0.333\textwidth]{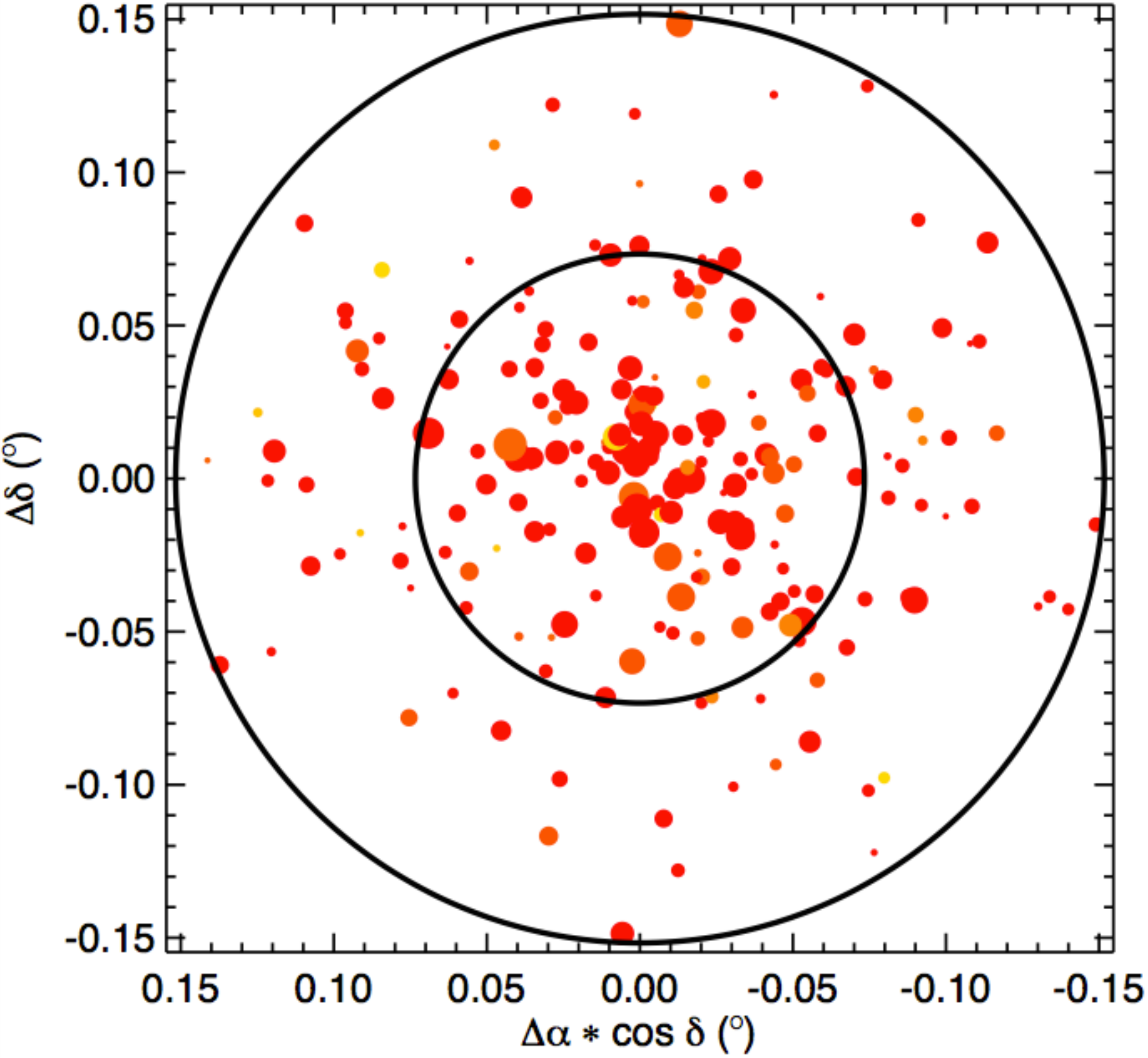}     
    \includegraphics[width=0.333\textwidth]{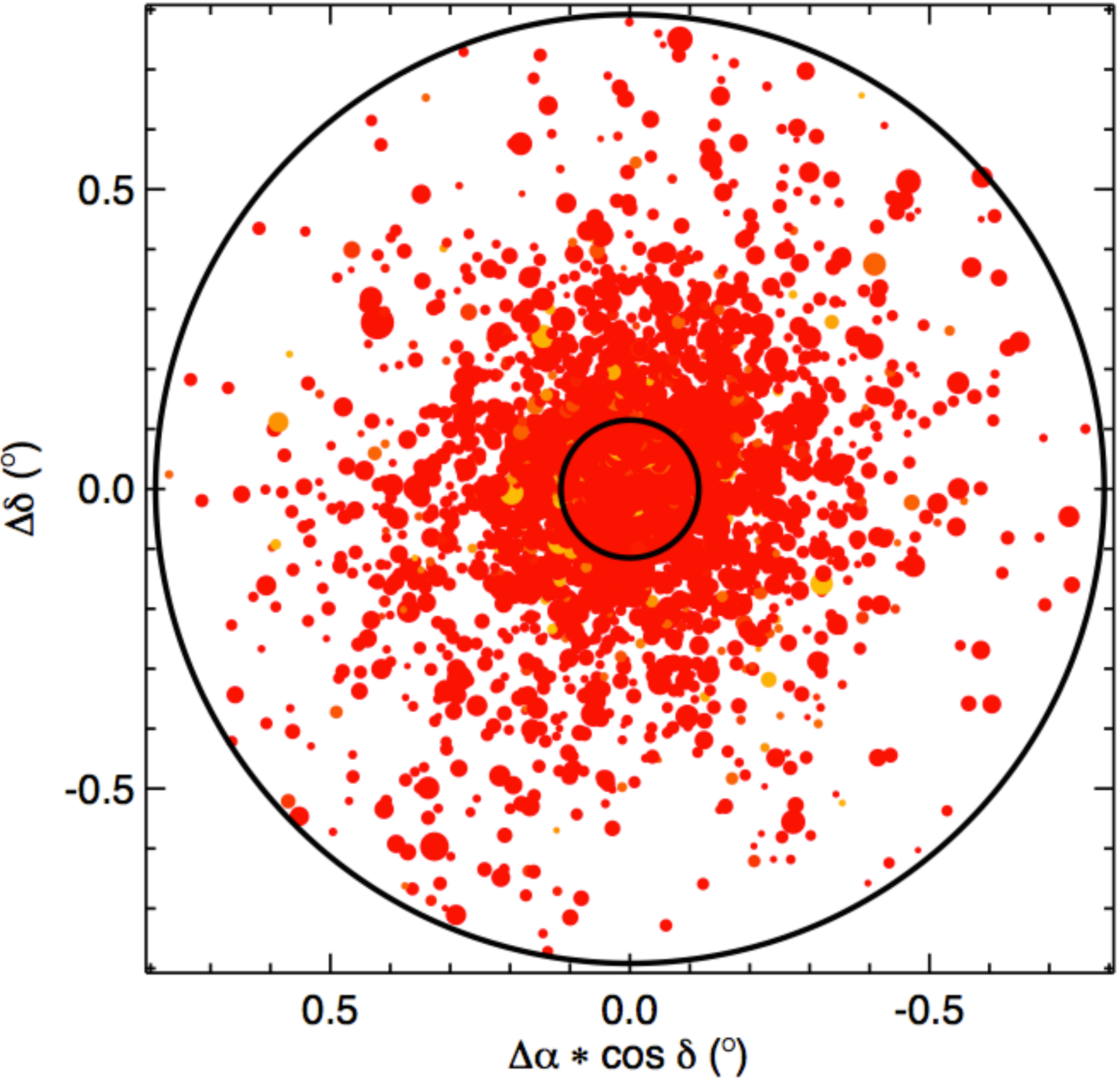}

  }
\caption{ Skymap for the OCs (from top left to bottom right): NGC\,2243, Collinder\,110, NGC\,2287 (top line), NGC\,2323, NGC\,2353, Berkeley\,36 (second line), NGC\,2360, Haffner\,11, NGC\,2422 (third line), Melotte\,71, NGC\,2432, NGC\,2477 (bottom line).  }

\label{fig:skymaps_13_24}
\end{center}
\end{figure*}

\begin{figure*}
\begin{center}

\parbox[c]{1.00\textwidth}
  {
    \includegraphics[width=0.328\textwidth]{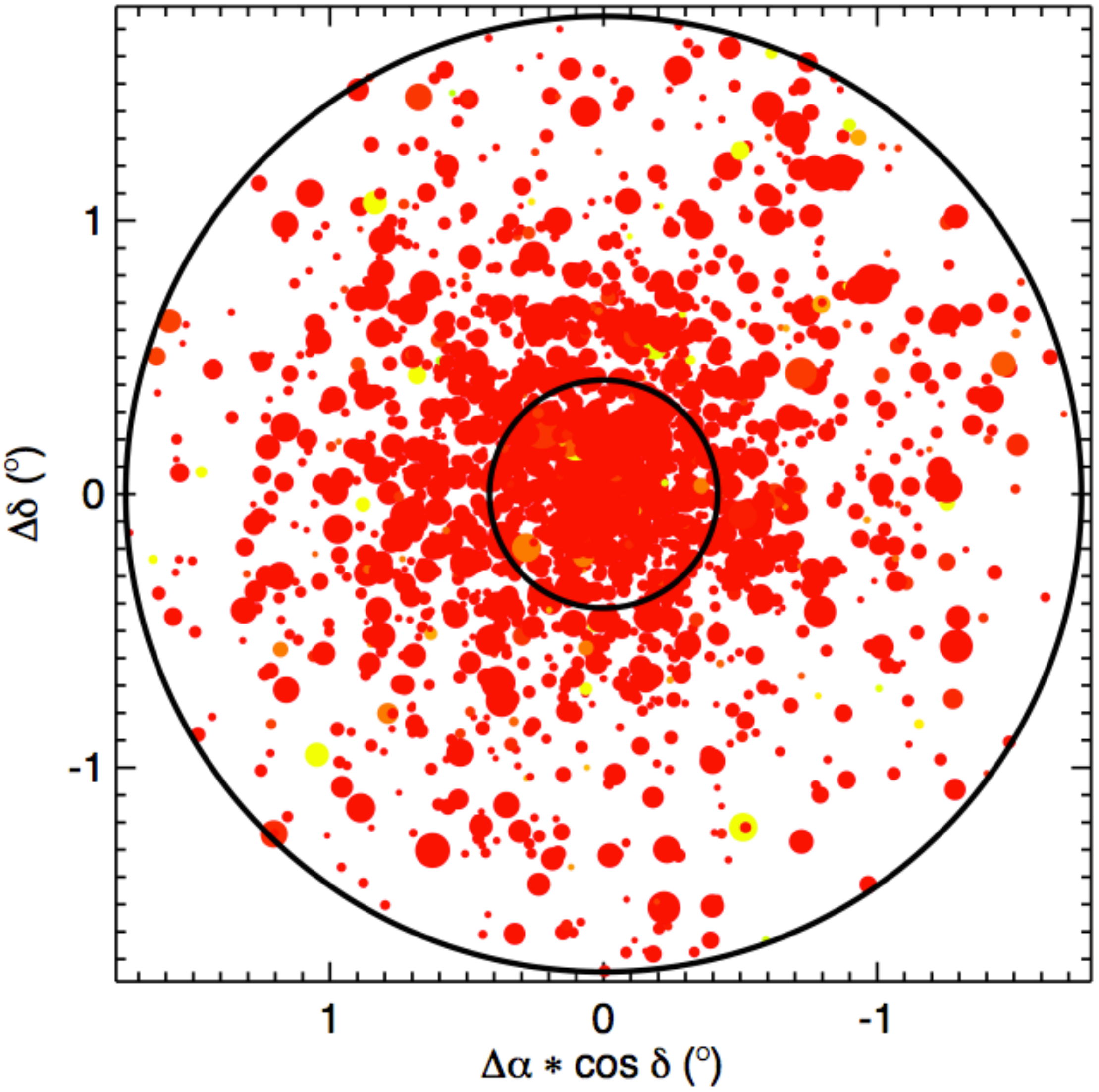}     
    \includegraphics[width=0.333\textwidth]{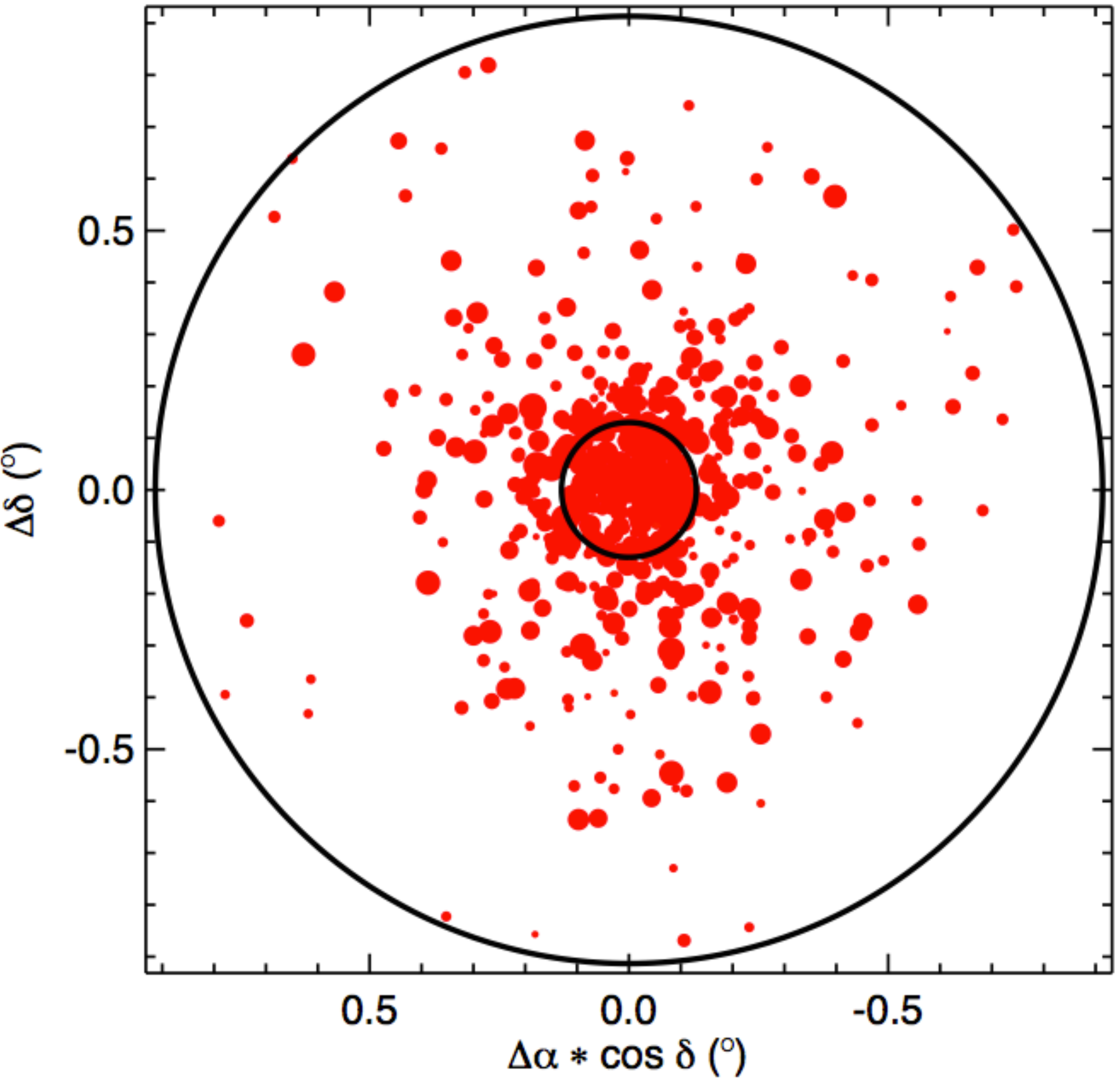}     
    \includegraphics[width=0.333\textwidth]{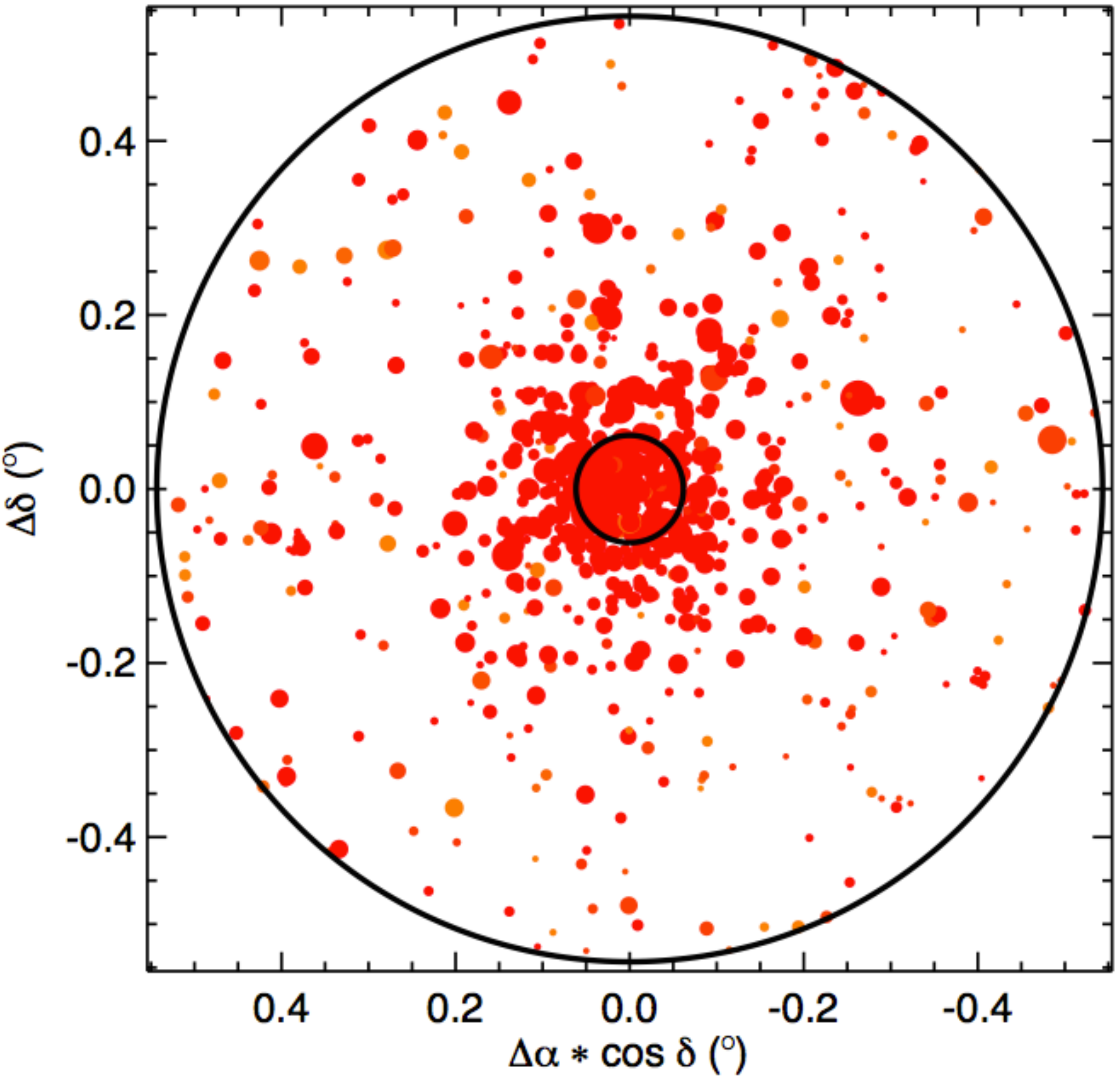}     
 
    \includegraphics[width=0.333\textwidth]{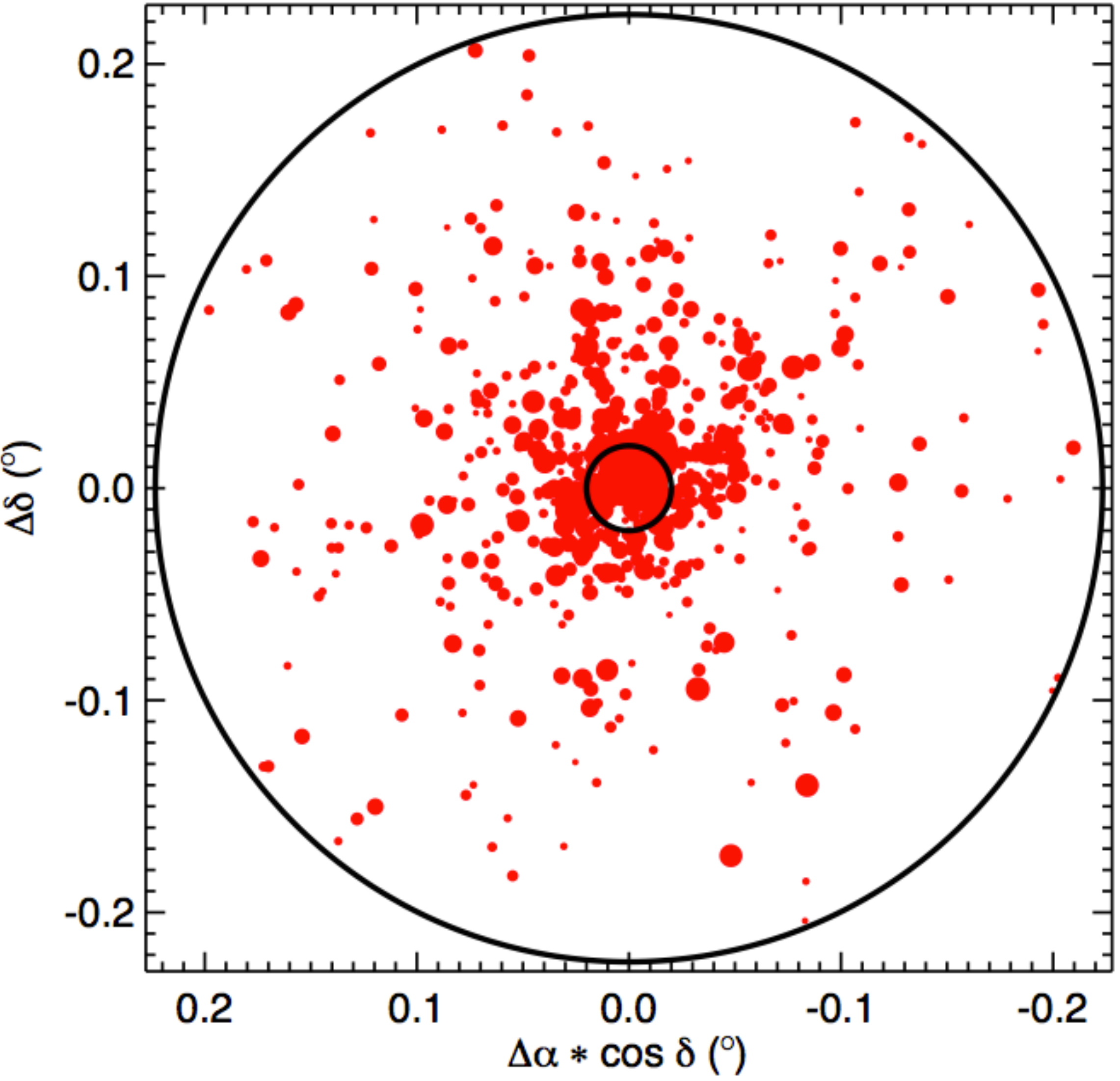}     
    \includegraphics[width=0.333\textwidth]{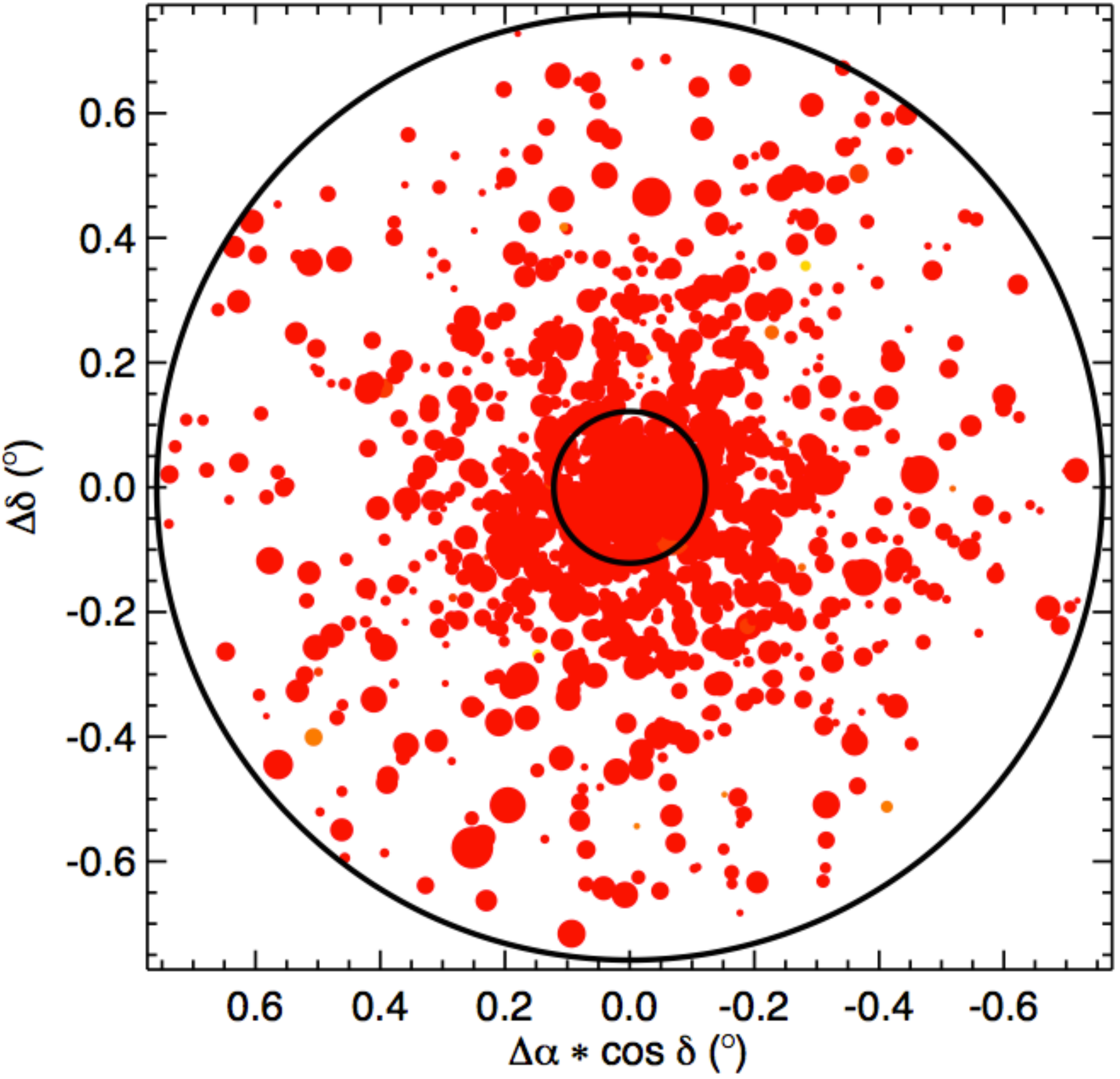}     
    \includegraphics[width=0.333\textwidth]{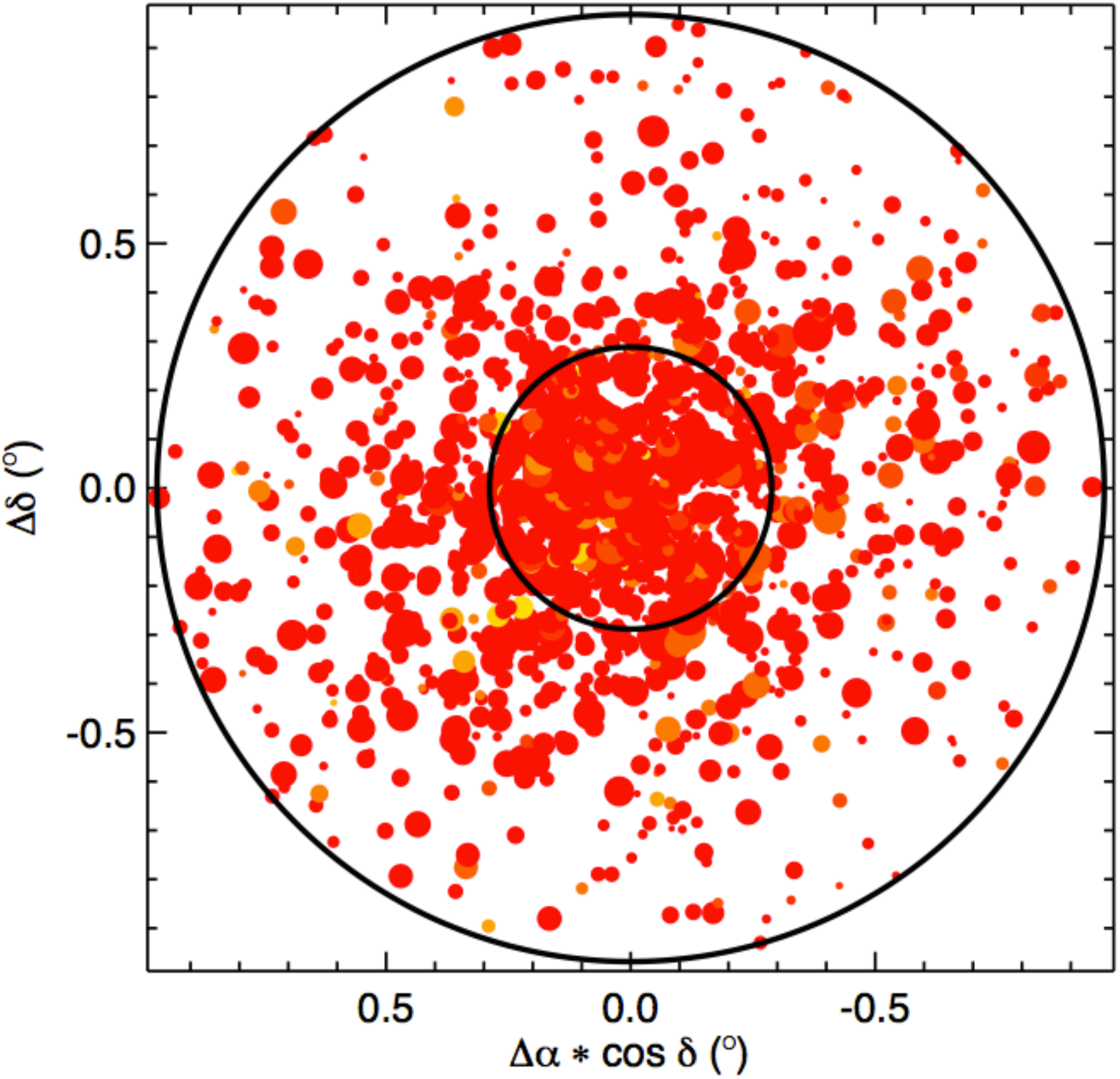}     

    \includegraphics[width=0.333\textwidth]{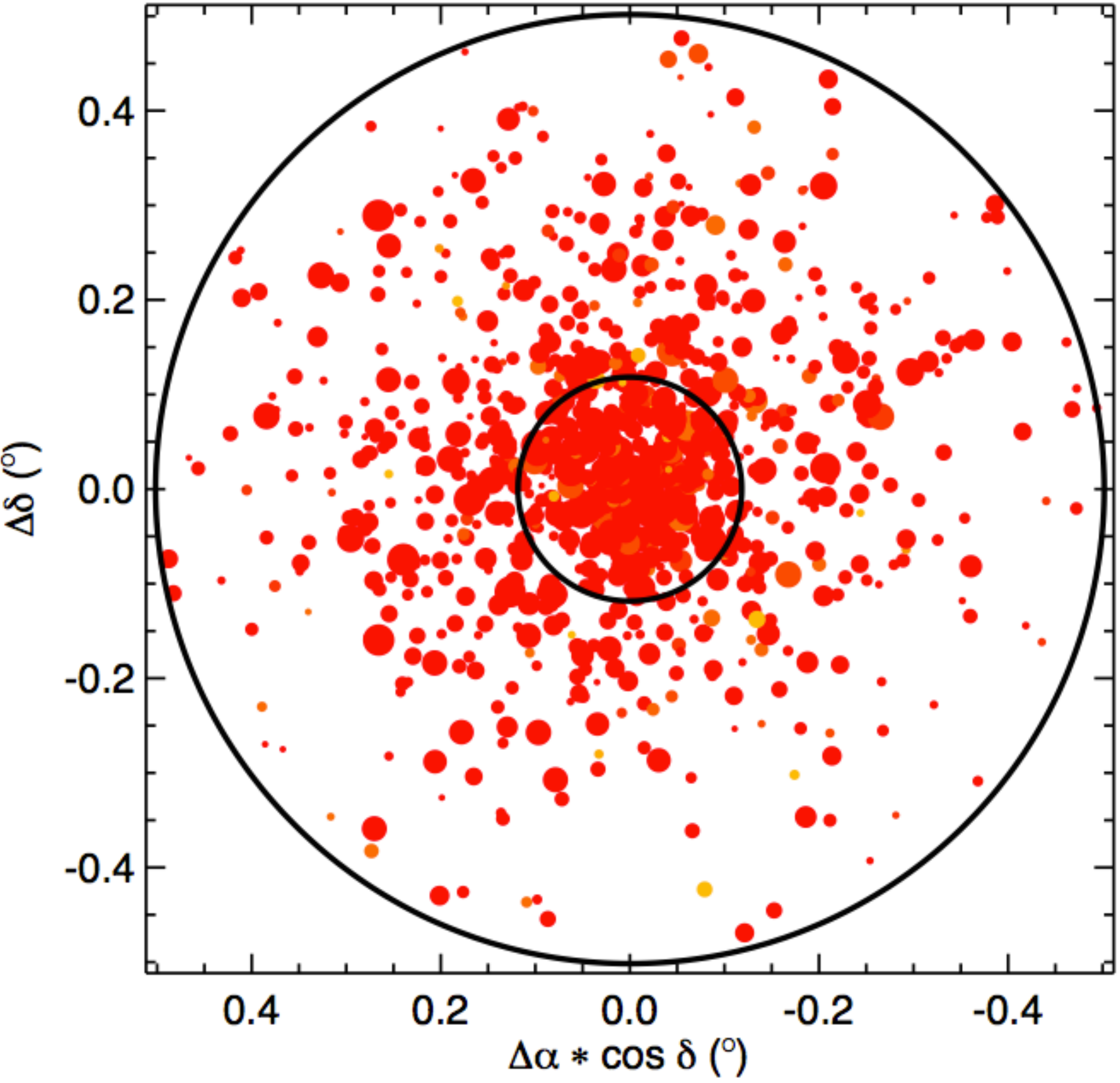}     
    \includegraphics[width=0.333\textwidth]{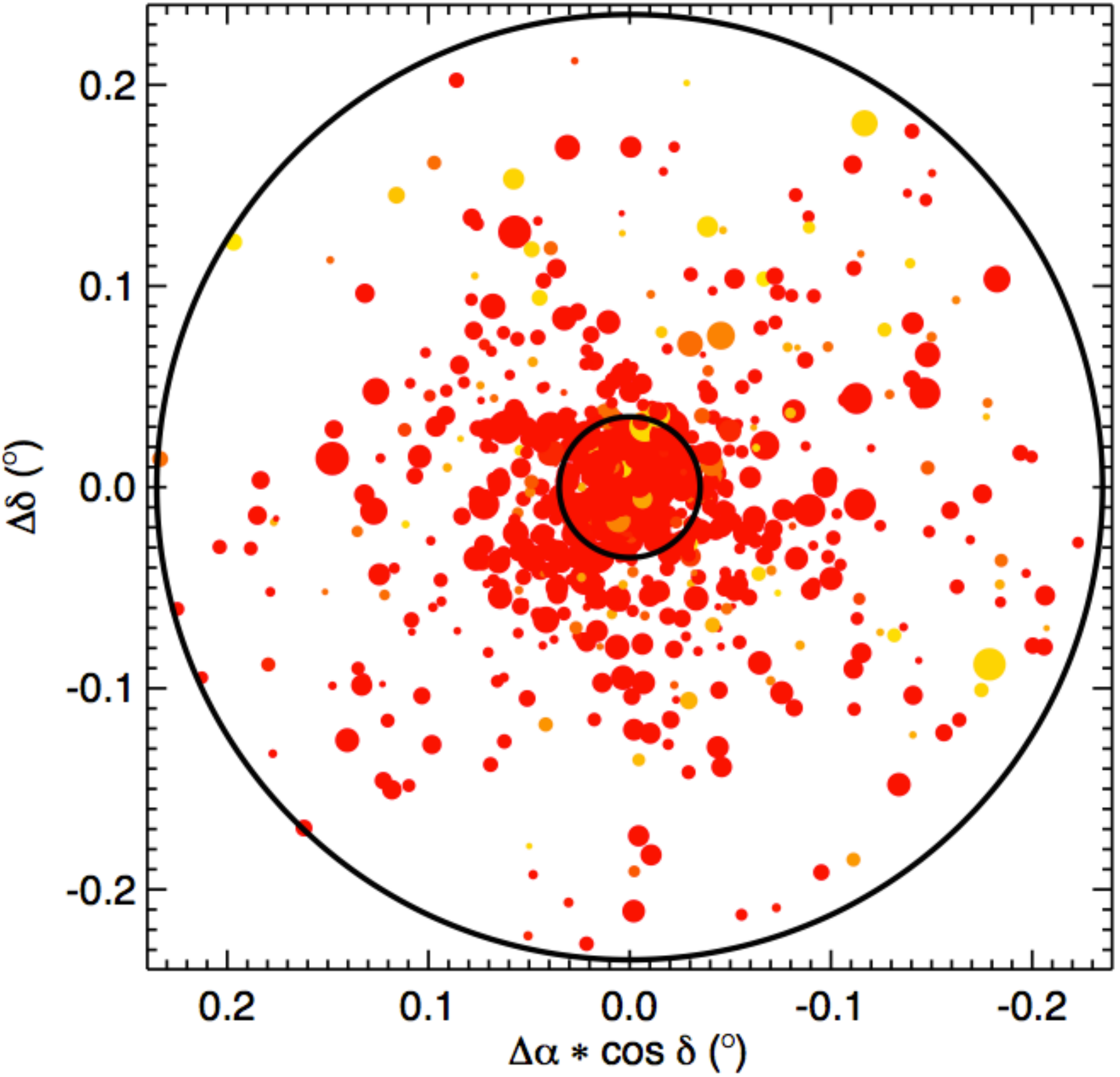}     
    \includegraphics[width=0.333\textwidth]{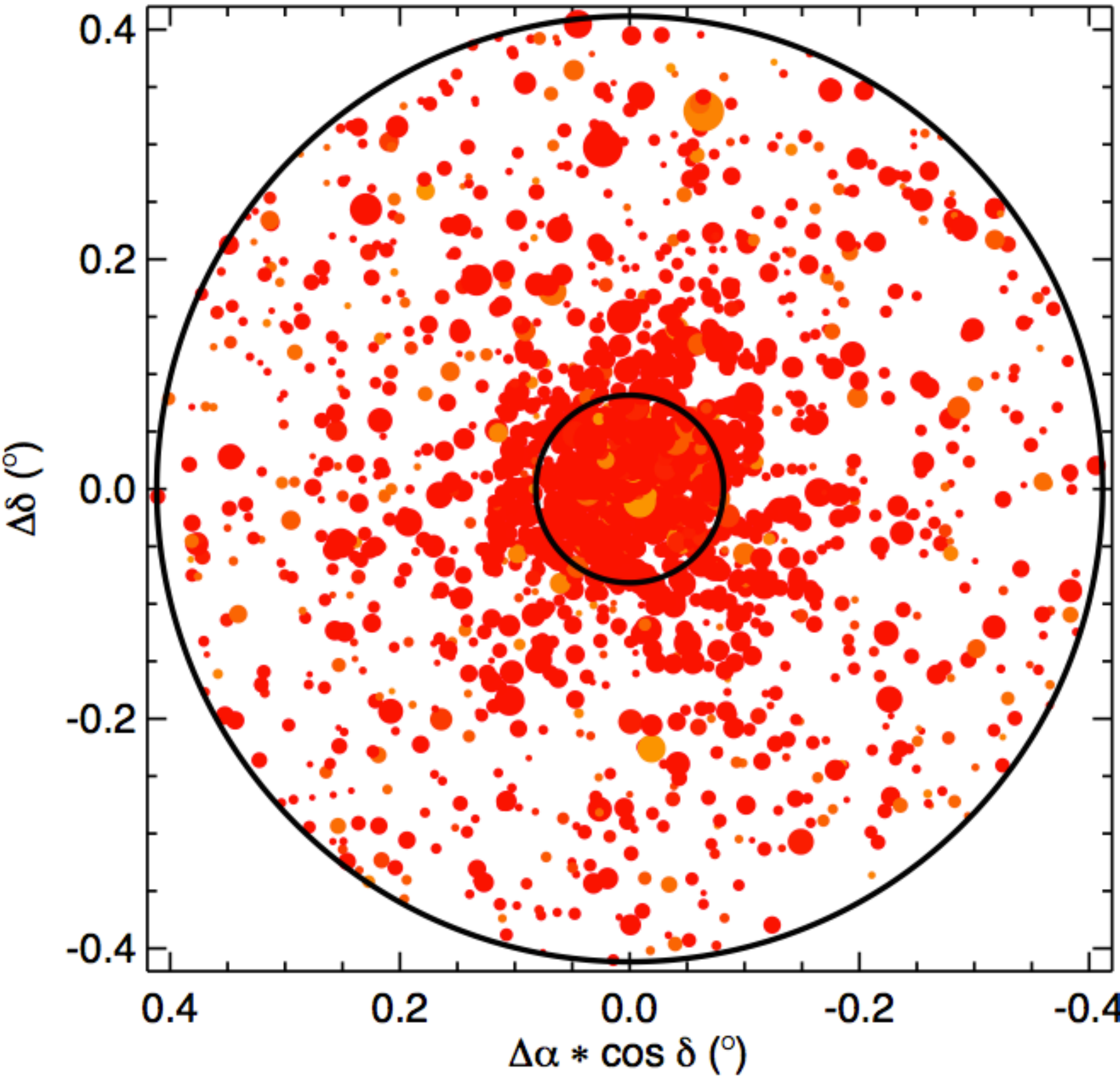}     

    \includegraphics[width=0.333\textwidth]{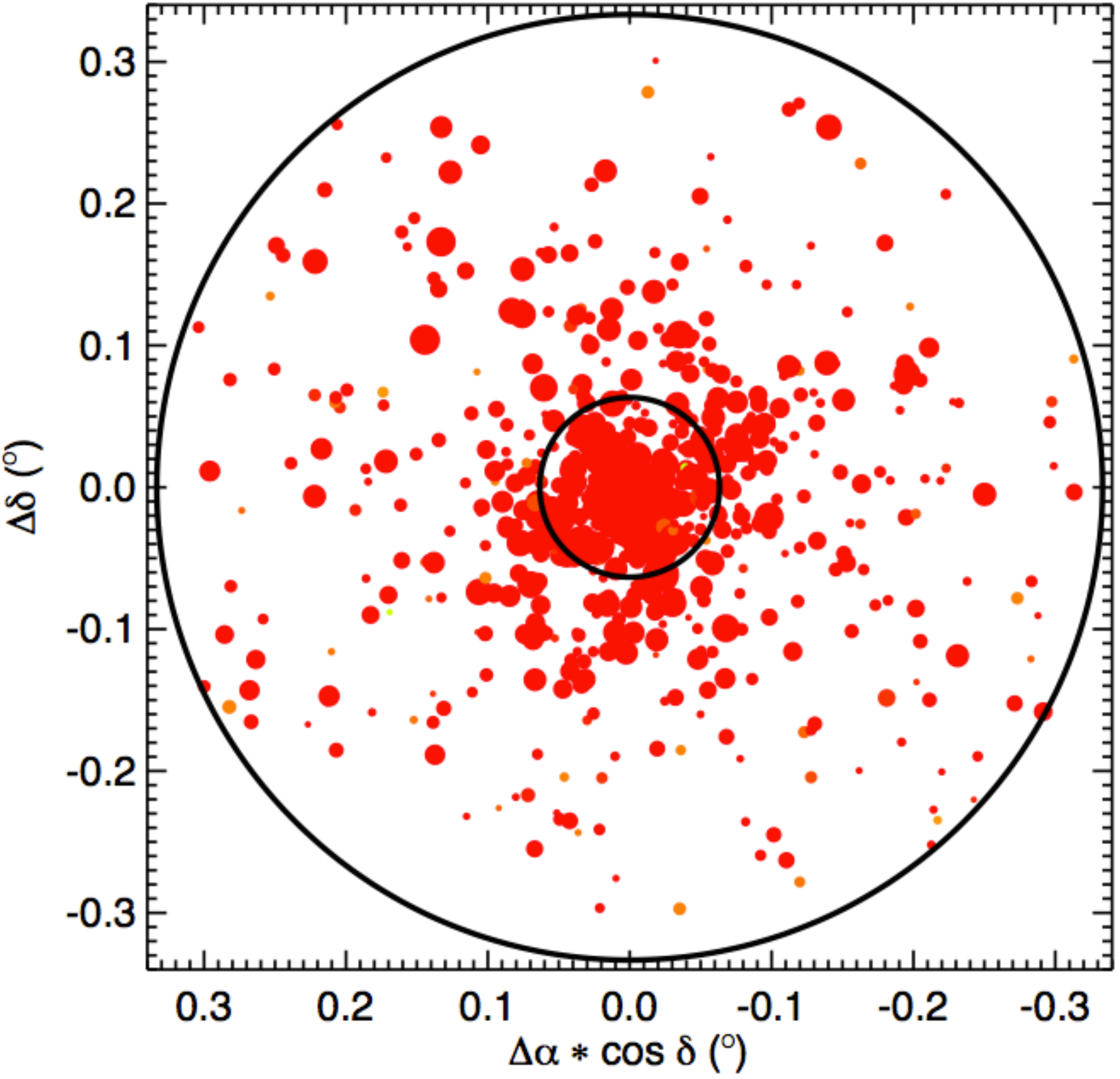}     
    \includegraphics[width=0.333\textwidth]{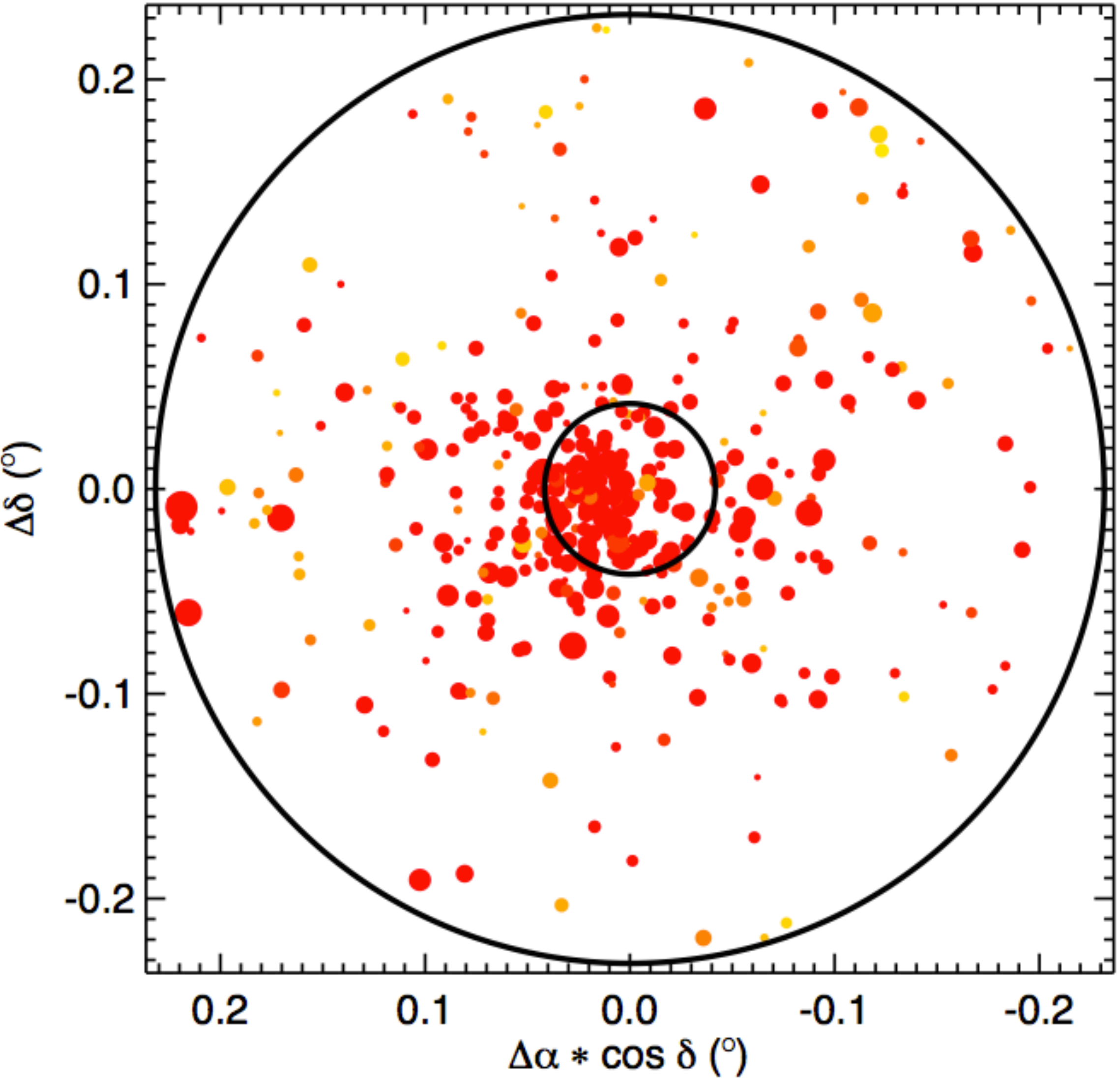}     
    \includegraphics[width=0.333\textwidth]{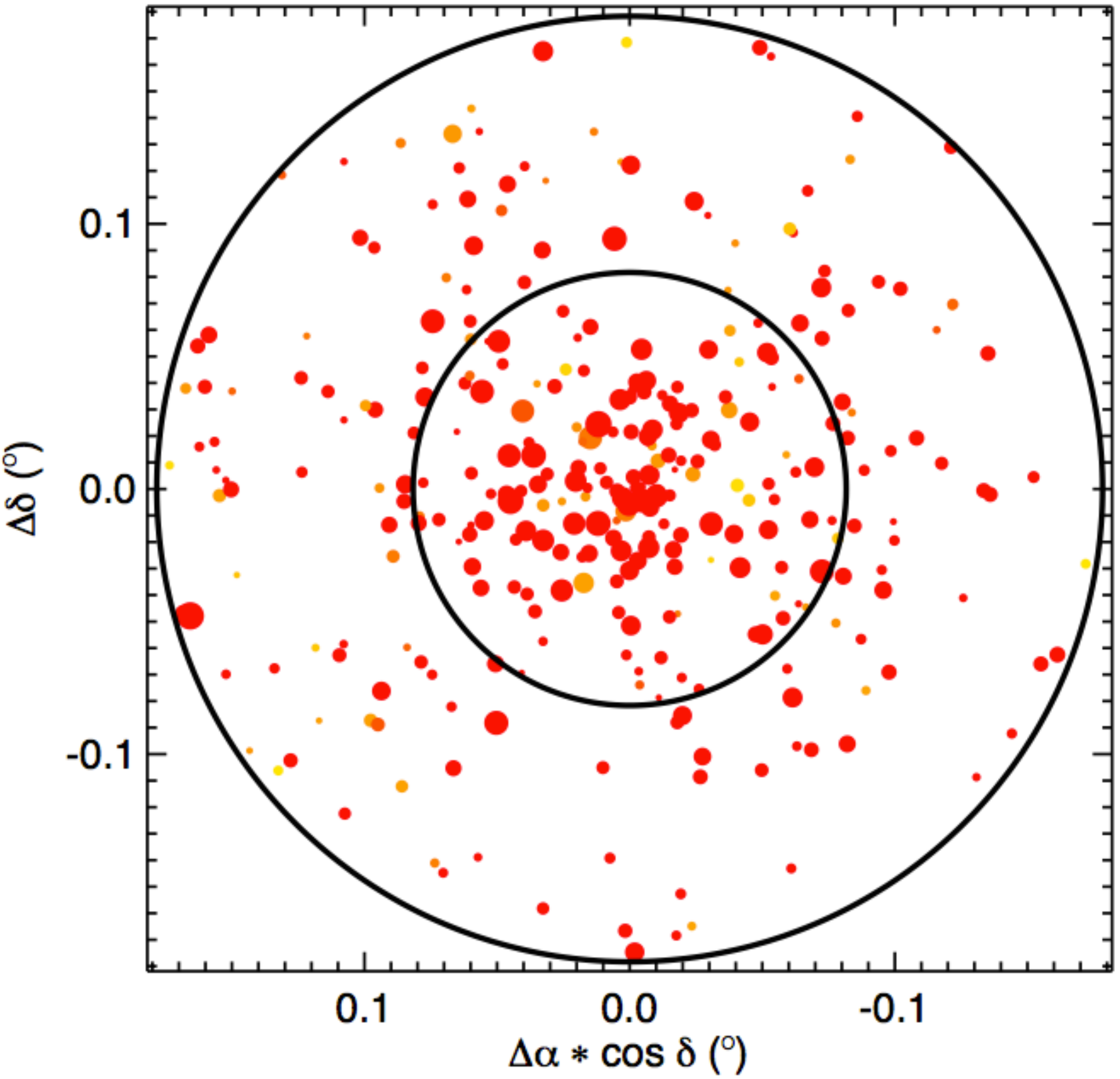}

  }
\caption{ Skymap for the OCs (from top left to bottom right): NGC\,2516, NGC\,2539, Haffner\,22 (top line), NGC\,2660, M\,67, NGC\,3114 (second line), IC\,2714, Melotte\,105, NGC\,3766 (third line), NGC\,3960, Juchert\,13, NGC\,4052 (bottom line).  }

\label{fig:skymaps_25_36}
\end{center}
\end{figure*}

\begin{figure*}
\begin{center}

\parbox[c]{1.00\textwidth}
  {
    \includegraphics[width=0.333\textwidth]{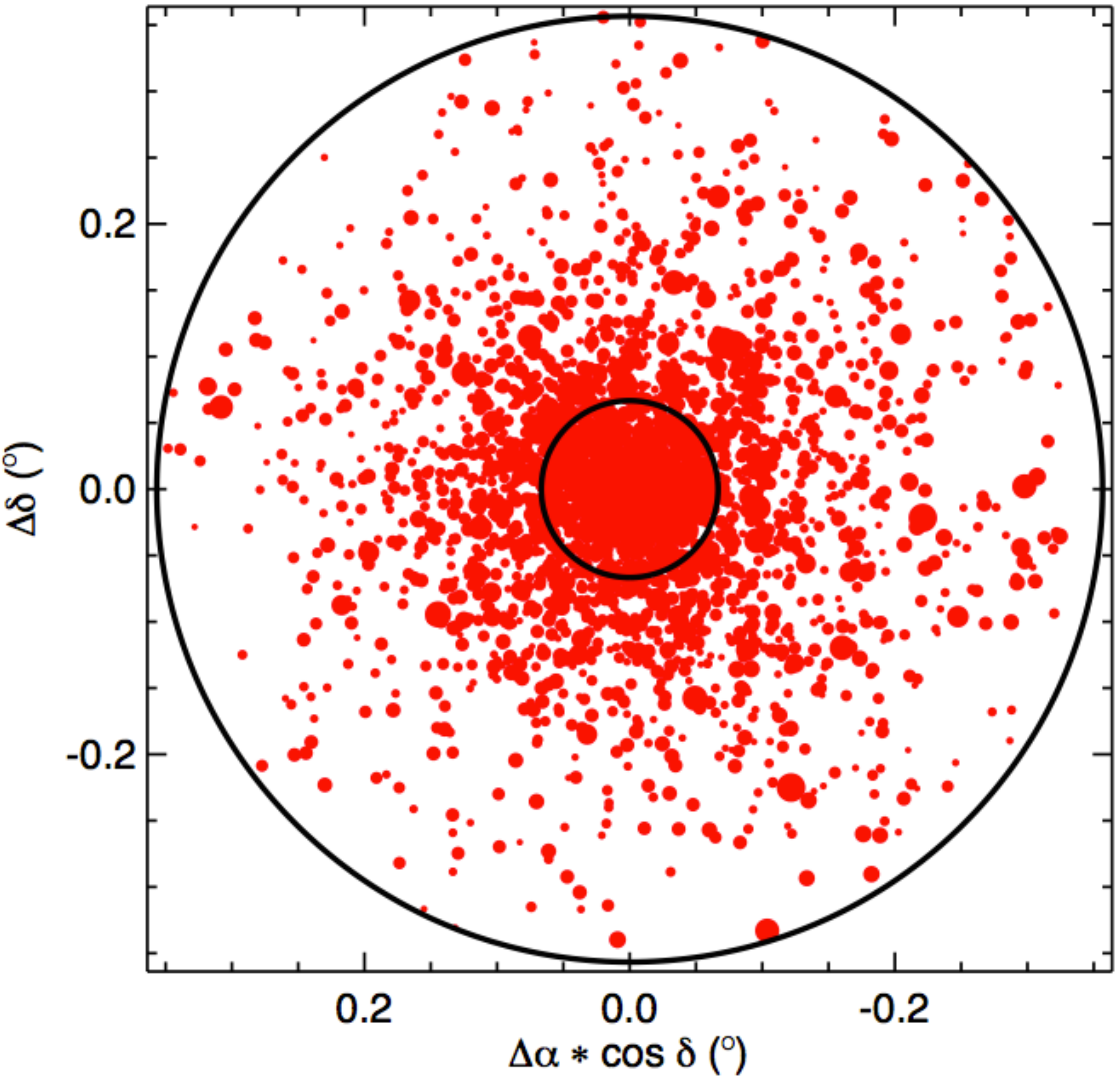}     
    \includegraphics[width=0.333\textwidth]{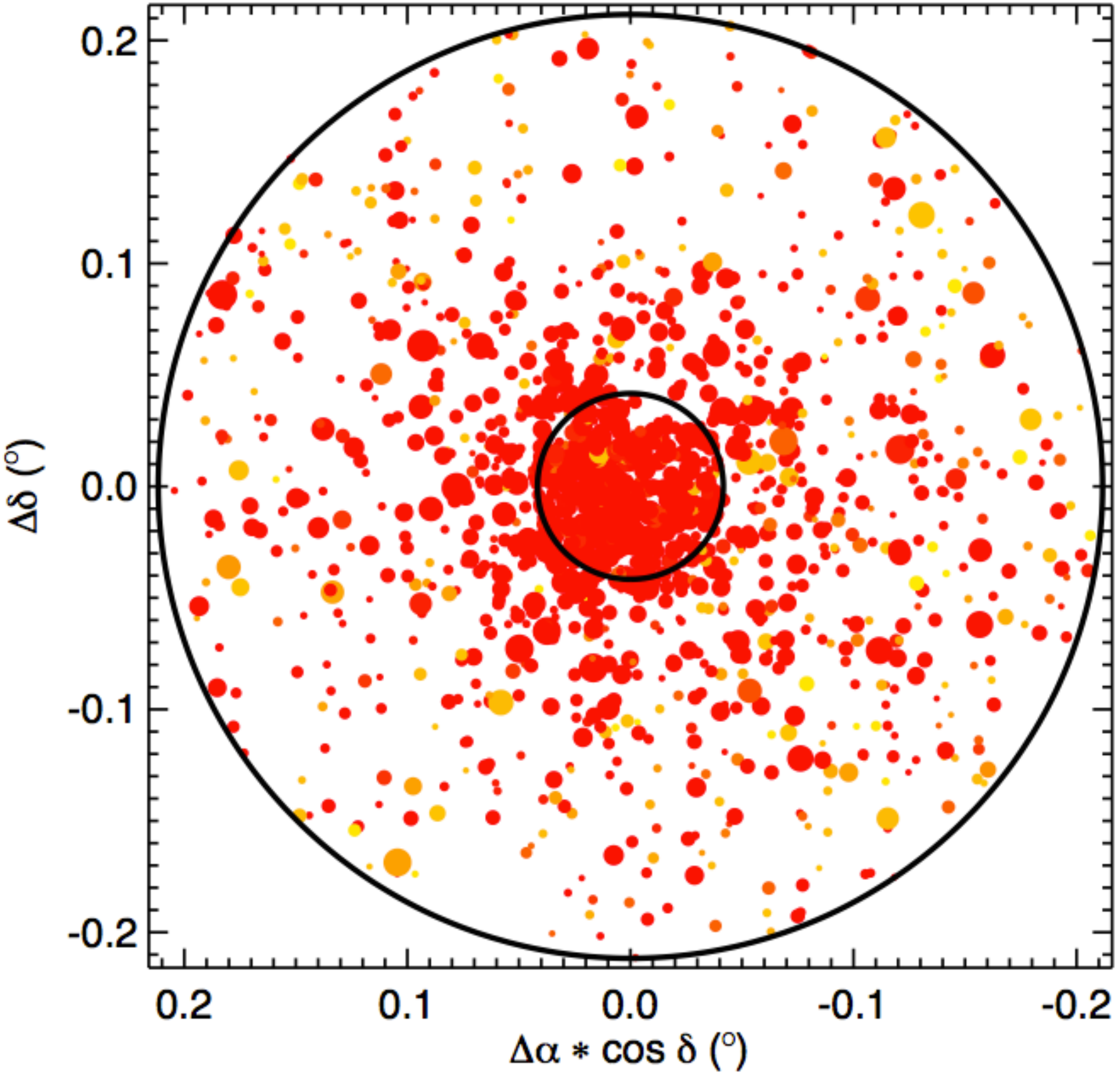}     
    \includegraphics[width=0.333\textwidth]{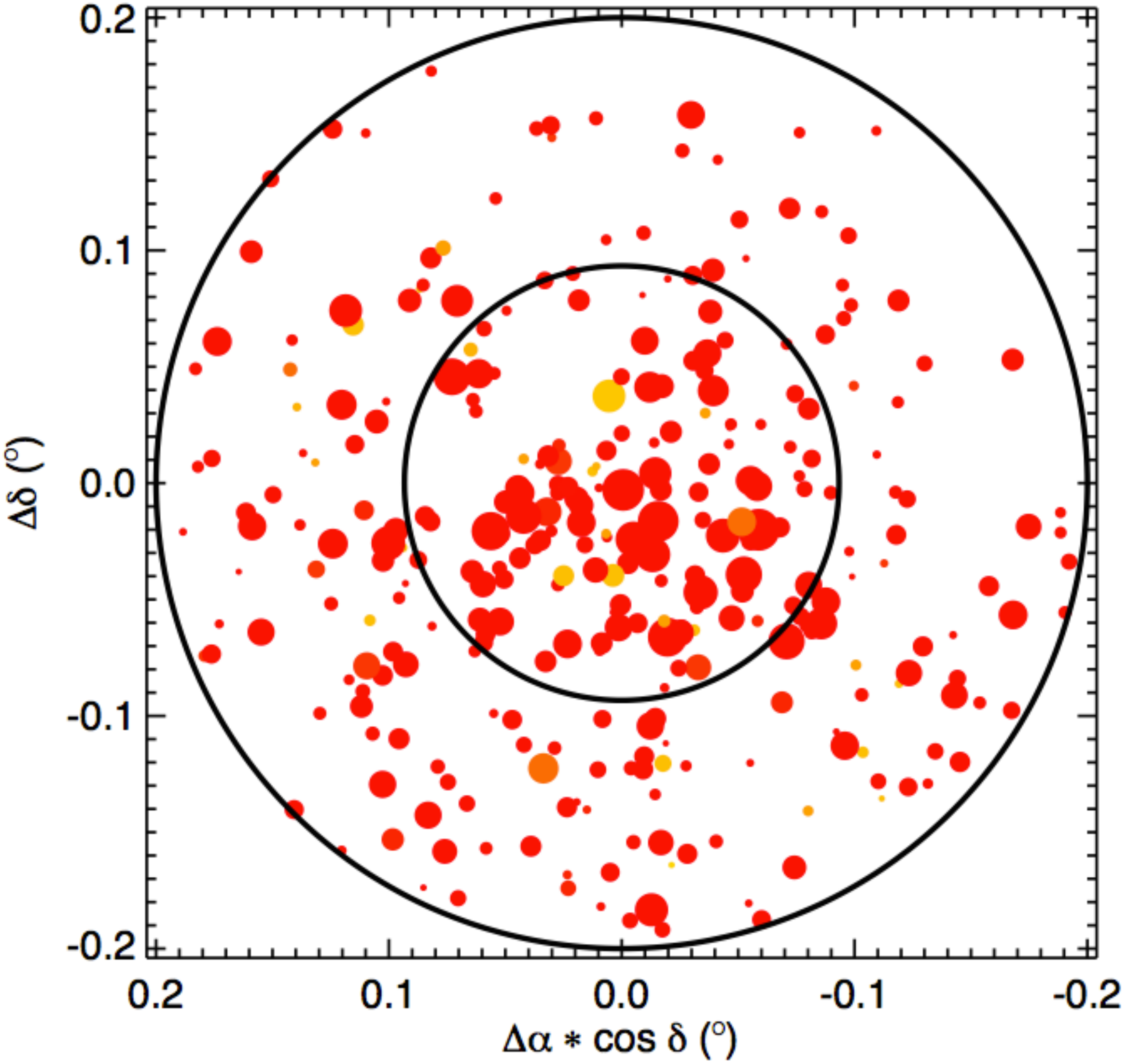}     
 
    \includegraphics[width=0.333\textwidth]{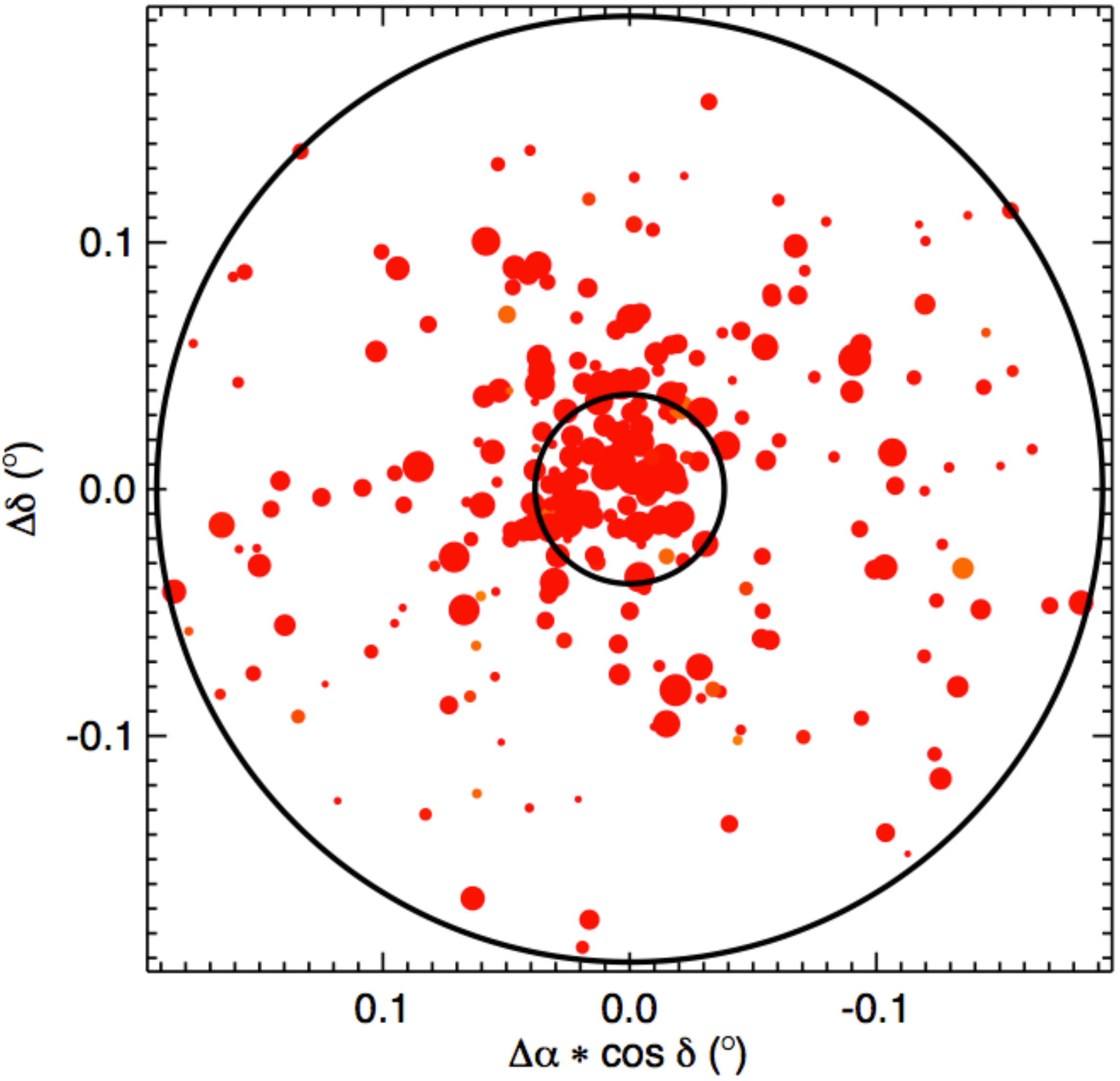}     
    \includegraphics[width=0.333\textwidth]{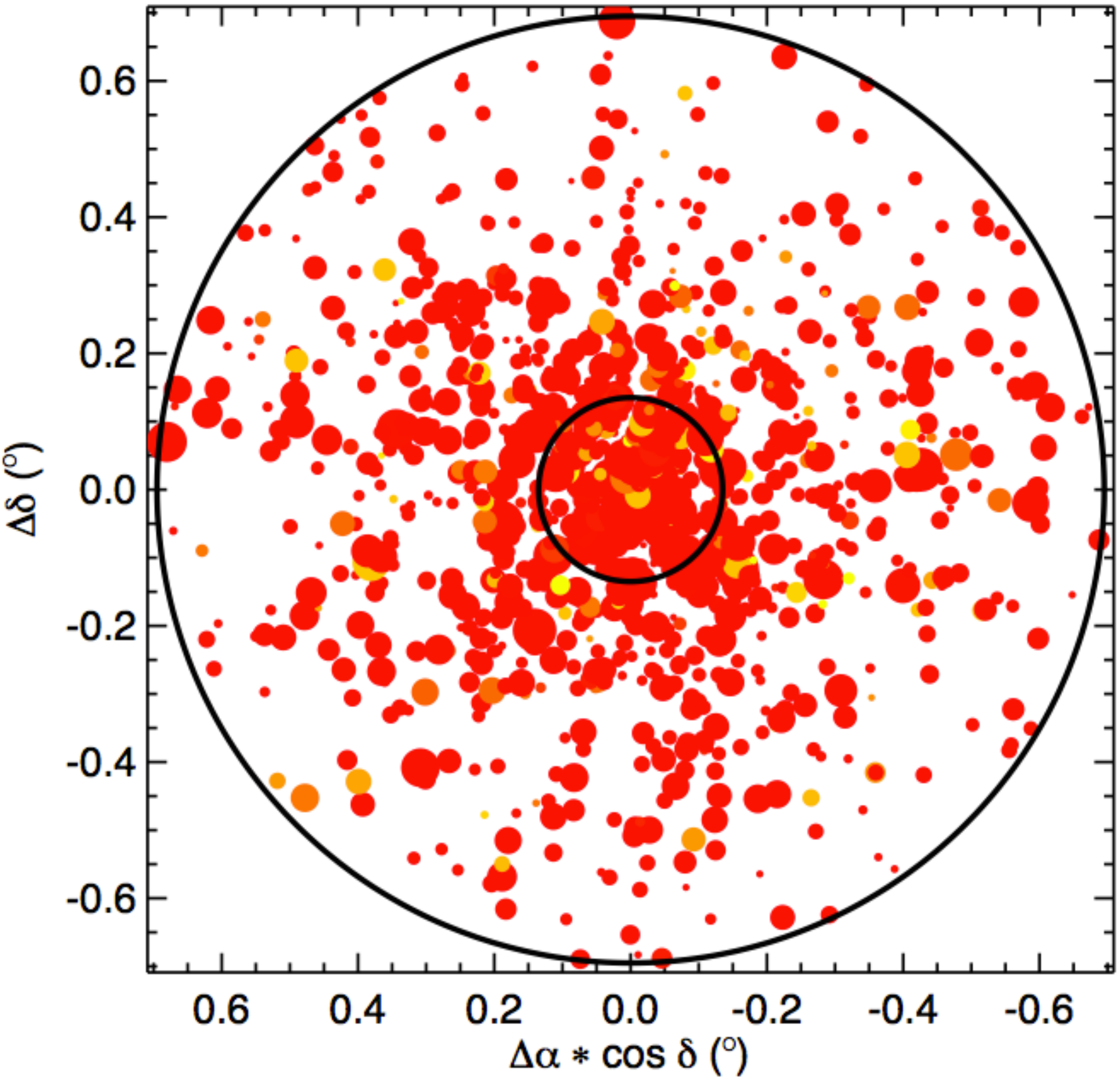}     
    \includegraphics[width=0.333\textwidth]{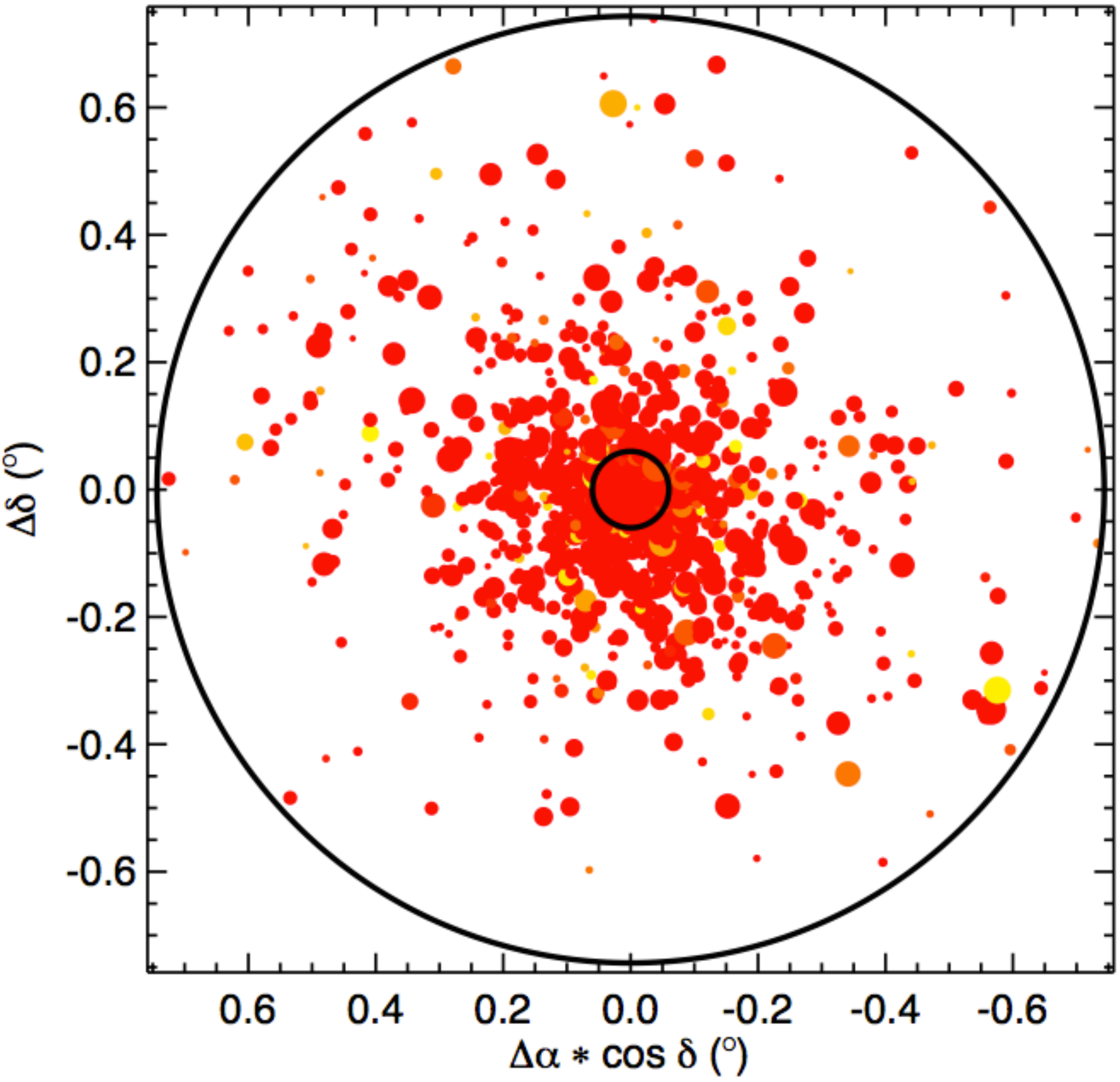}     

    \includegraphics[width=0.333\textwidth]{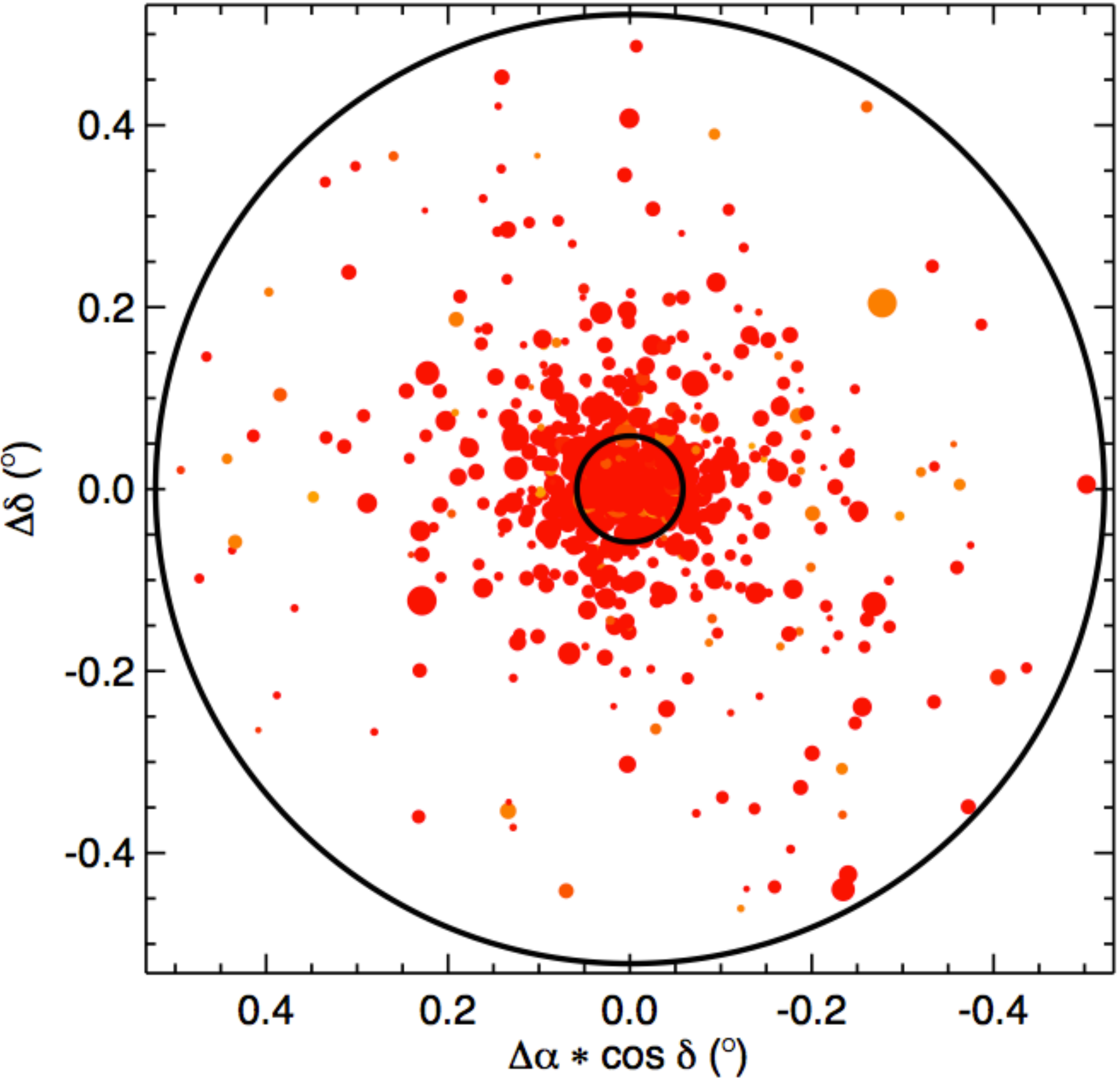}     
    \includegraphics[width=0.333\textwidth]{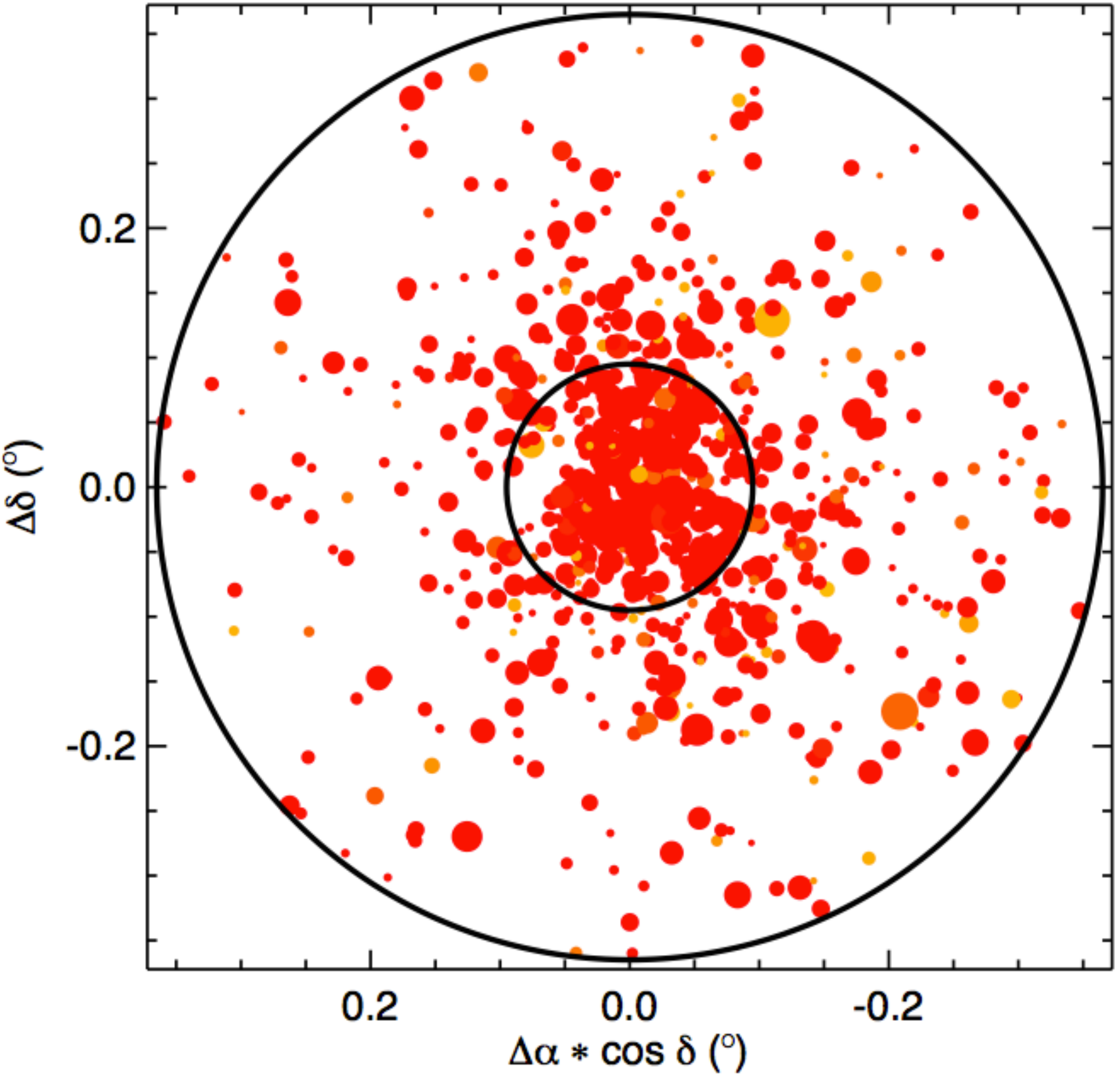}     
    \includegraphics[width=0.333\textwidth]{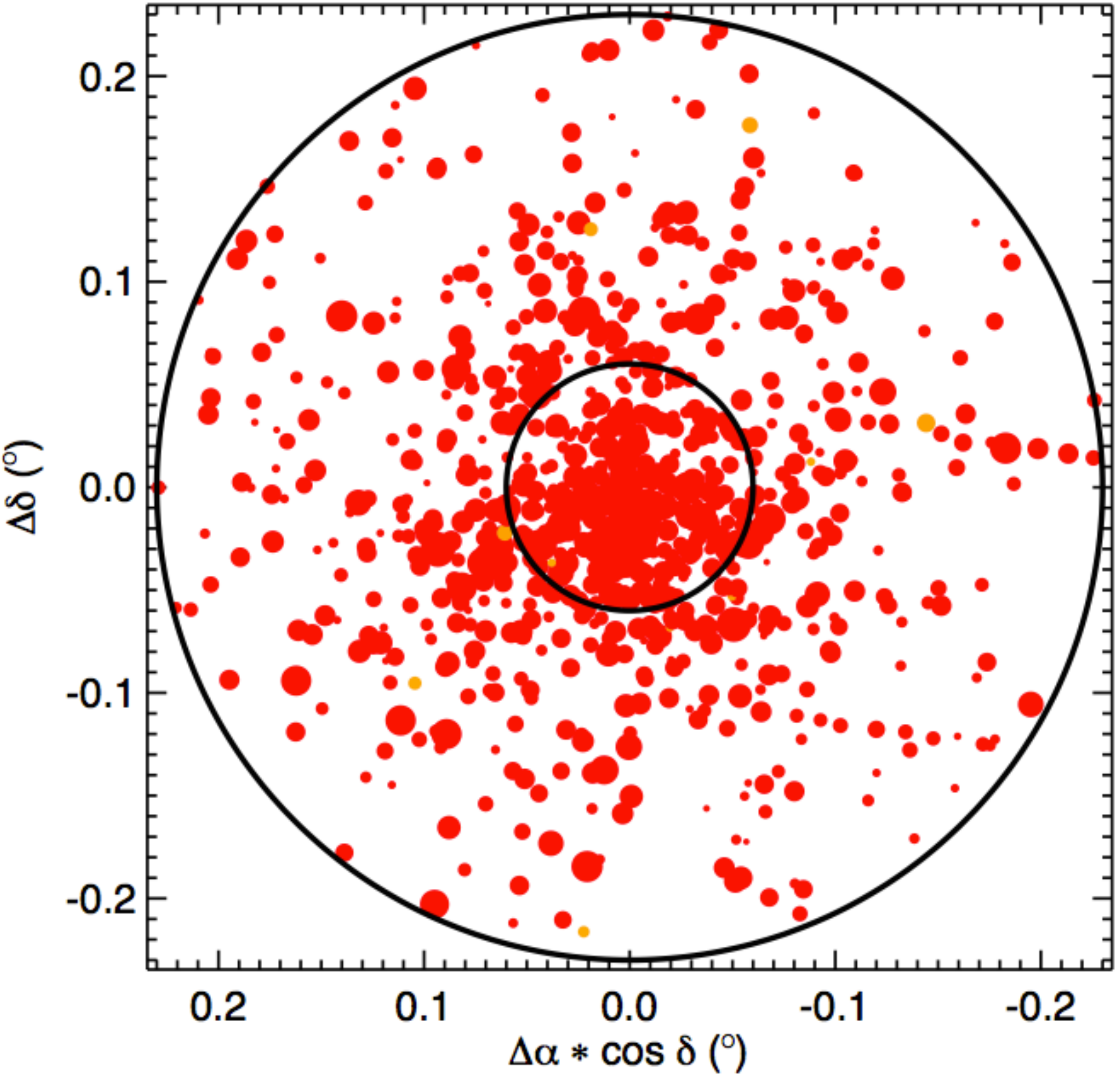}     

    \includegraphics[width=0.333\textwidth]{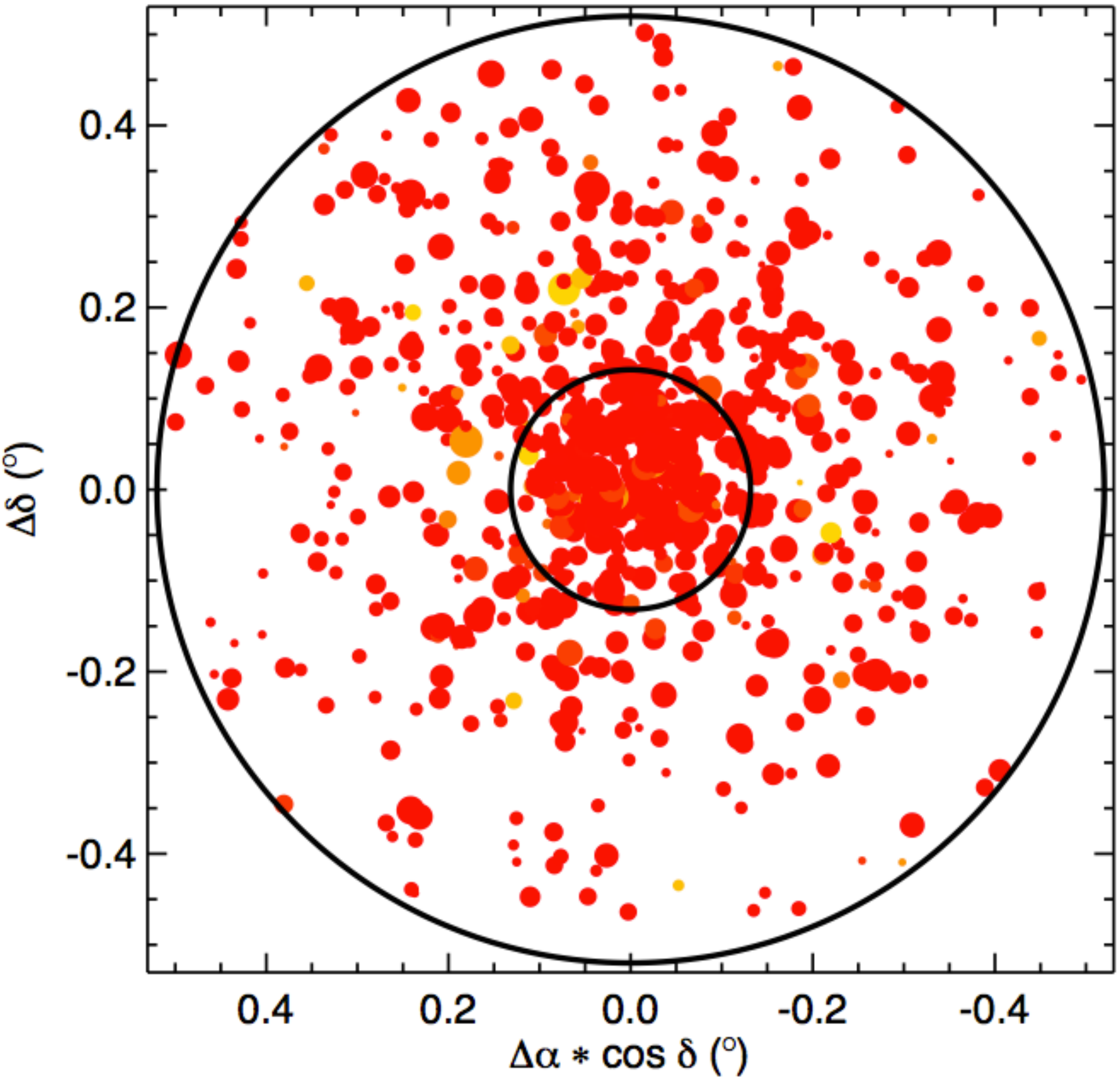}     
    \includegraphics[width=0.333\textwidth]{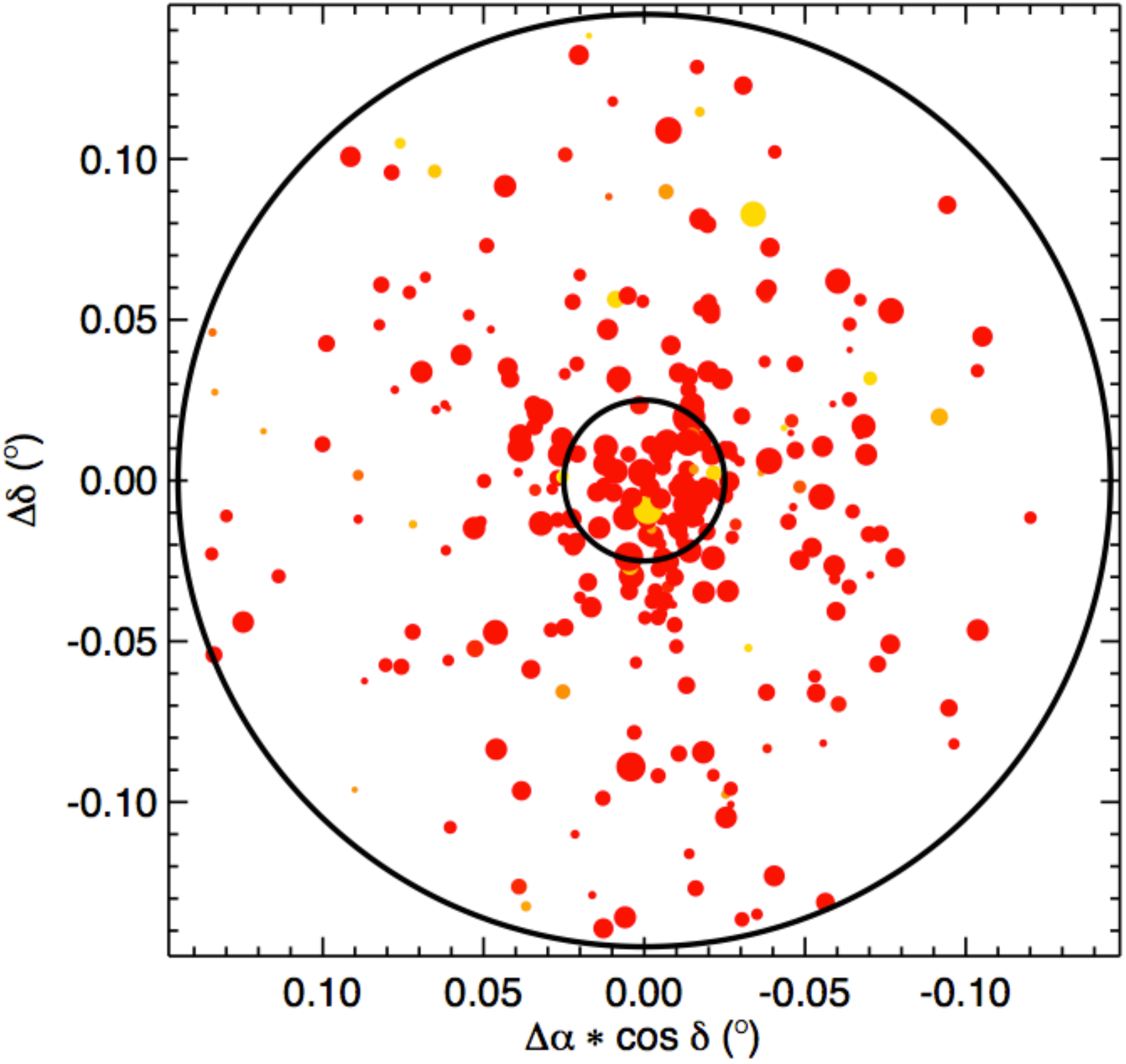}     
    \includegraphics[width=0.333\textwidth]{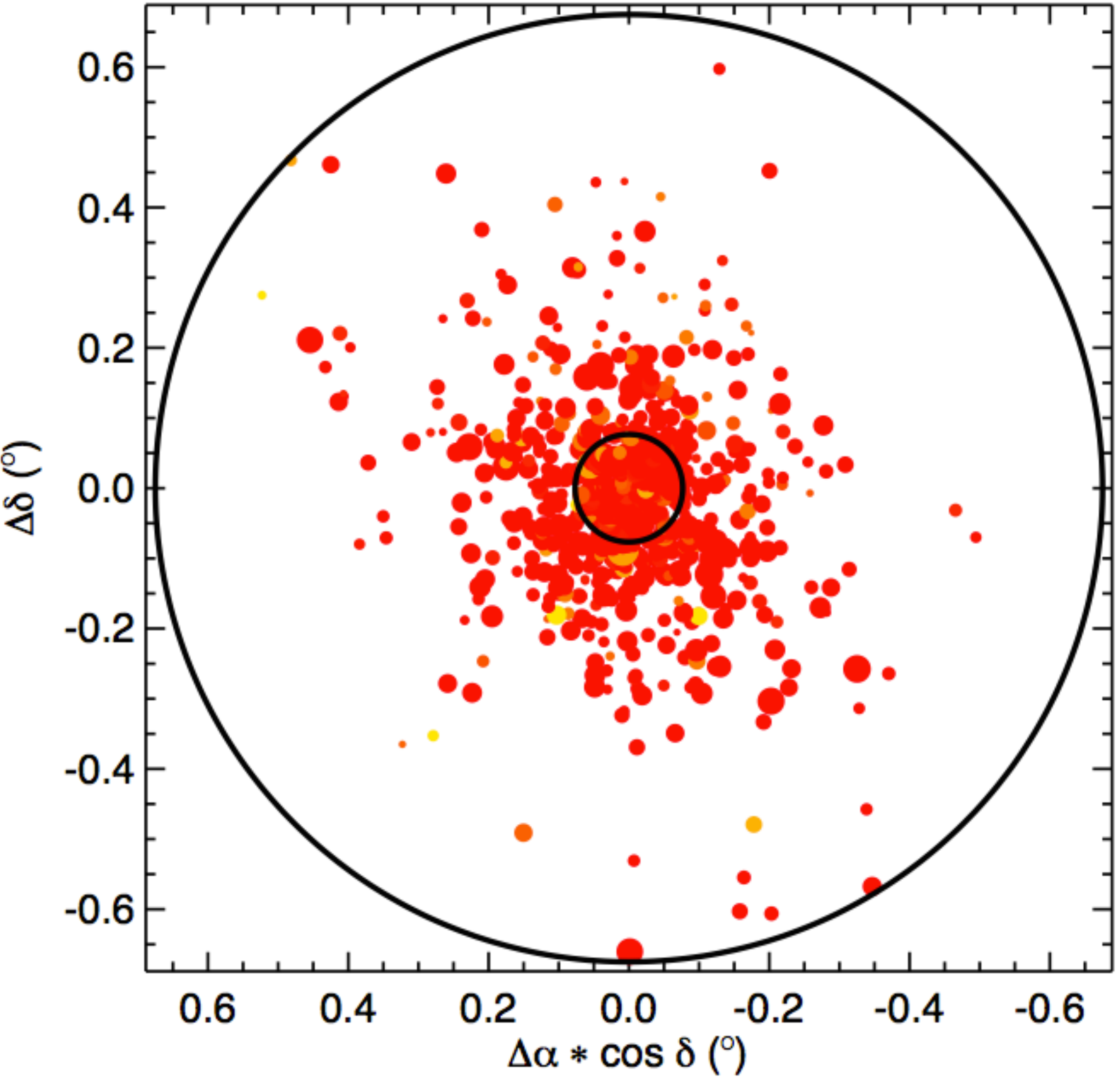}     
     
  }
\caption{ Skymap for the OCs (from top left to bottom right): Collinder\,261, NGC\,4815, NGC\,5316 (top line), NGC\,5715, NGC\,6124, NGC\,6134 (second line), NGC\,6192, NGC\,6242, NGC\,6253 (third line), IC\,4651, Dias\,6, Ruprecht\,171 (bottom line).  }

\label{fig:skymaps_37_48}
\end{center}
\end{figure*}

\begin{figure*}
\begin{center}

\parbox[c]{1.00\textwidth}
  {
    \includegraphics[width=0.325\textwidth]{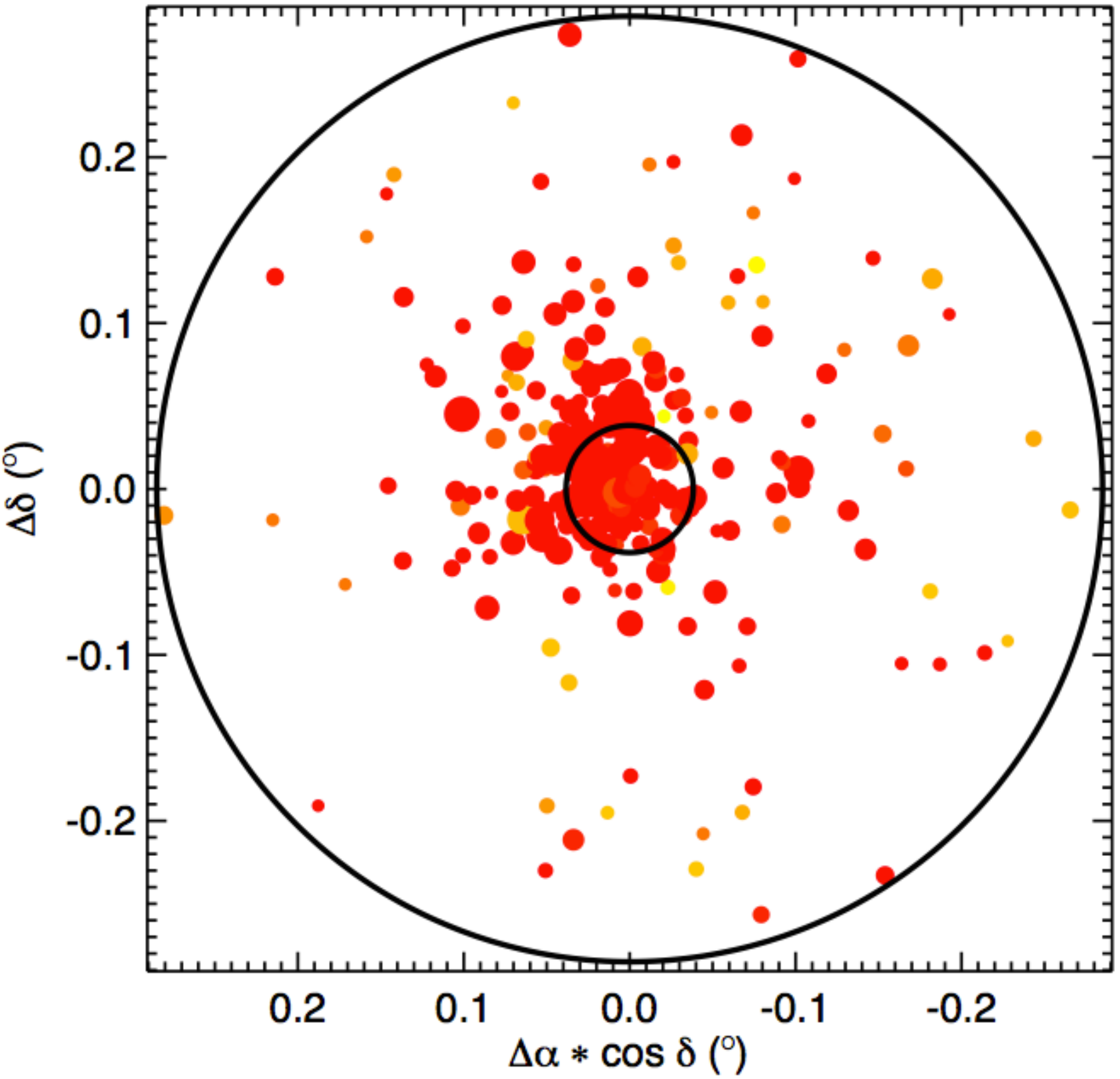}     
    \includegraphics[width=0.333\textwidth]{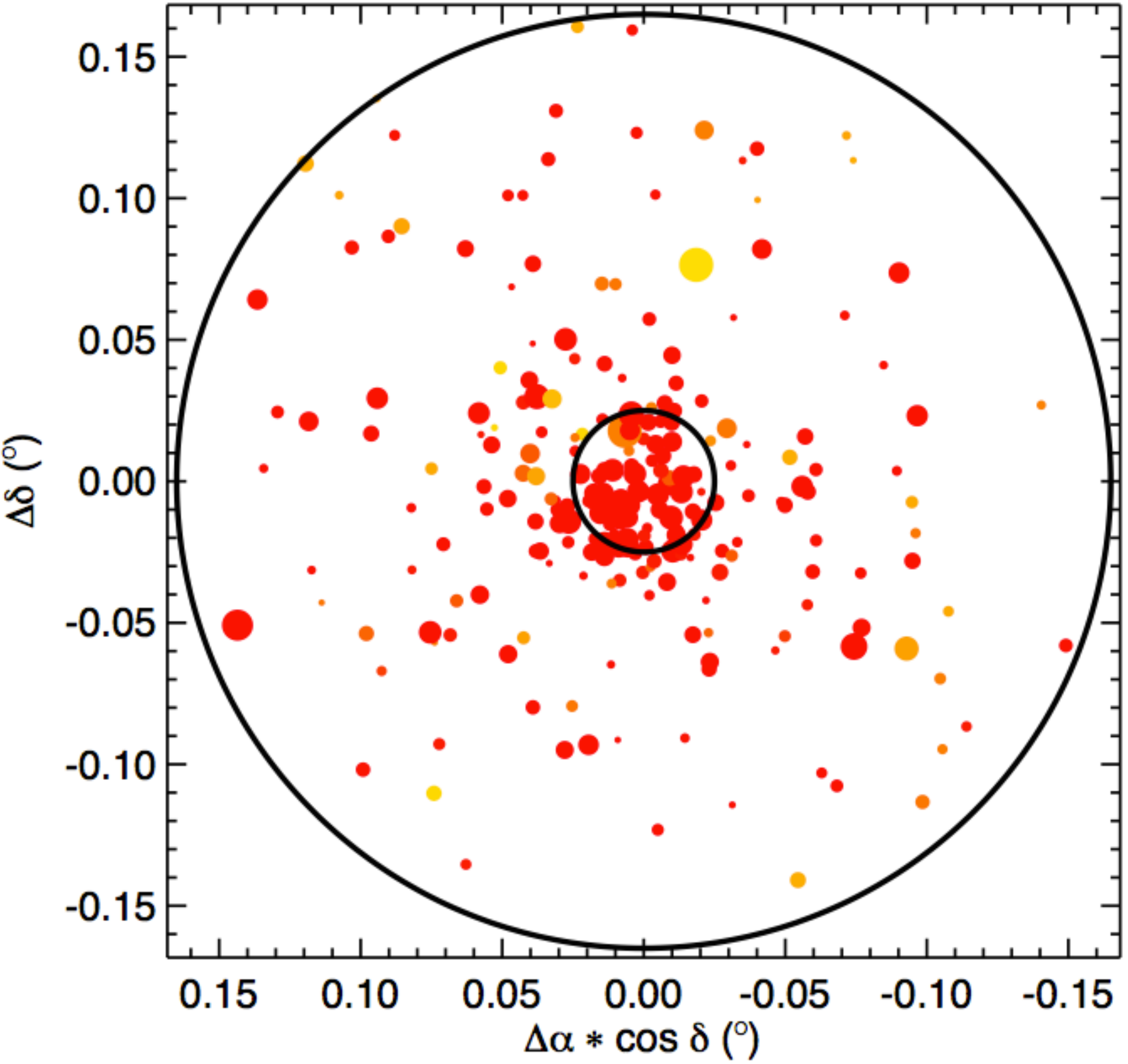}     
    \includegraphics[width=0.333\textwidth]{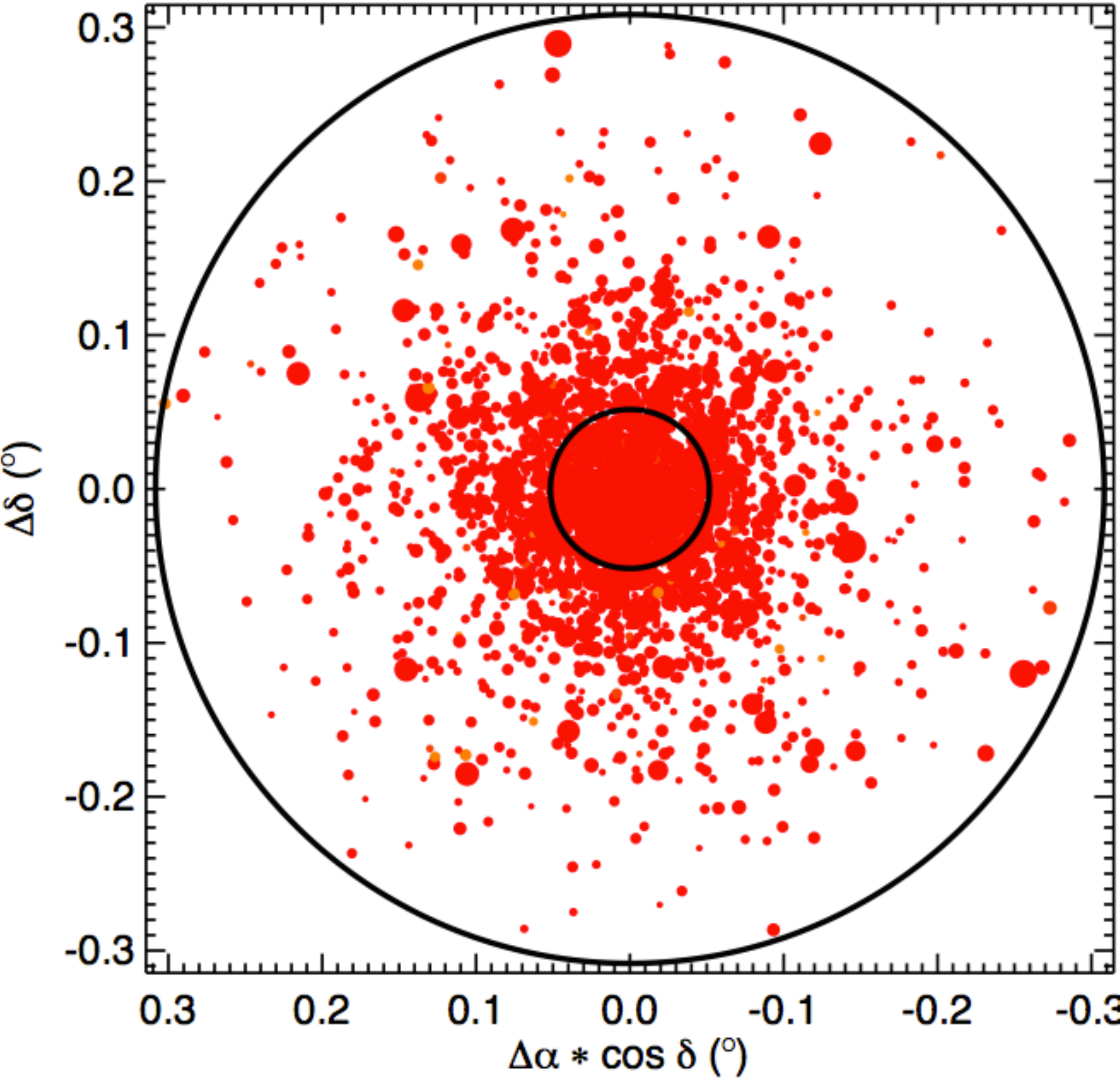}     
 
    \includegraphics[width=0.333\textwidth]{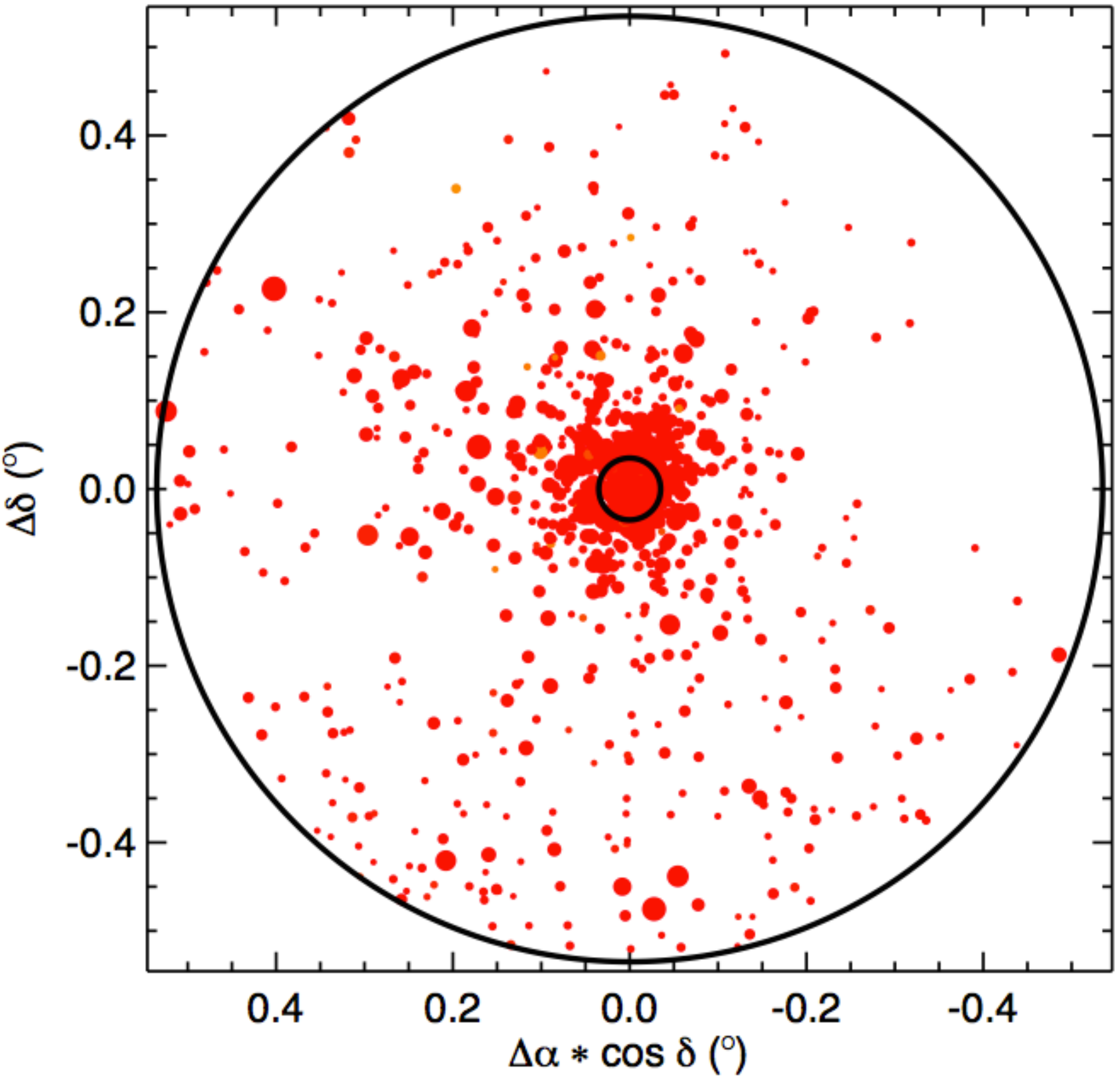}     
    \includegraphics[width=0.333\textwidth]{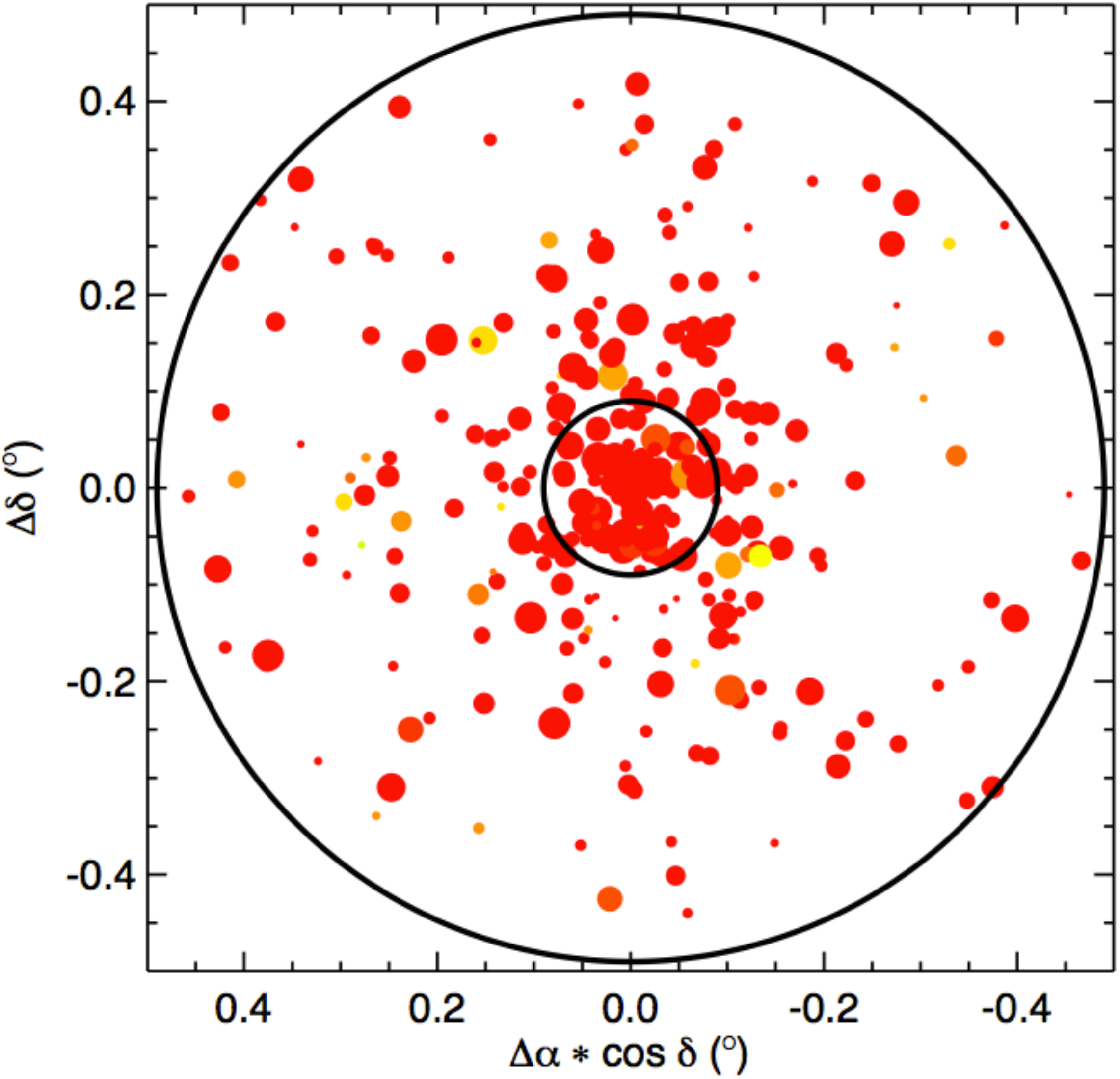}     
    \includegraphics[width=0.333\textwidth]{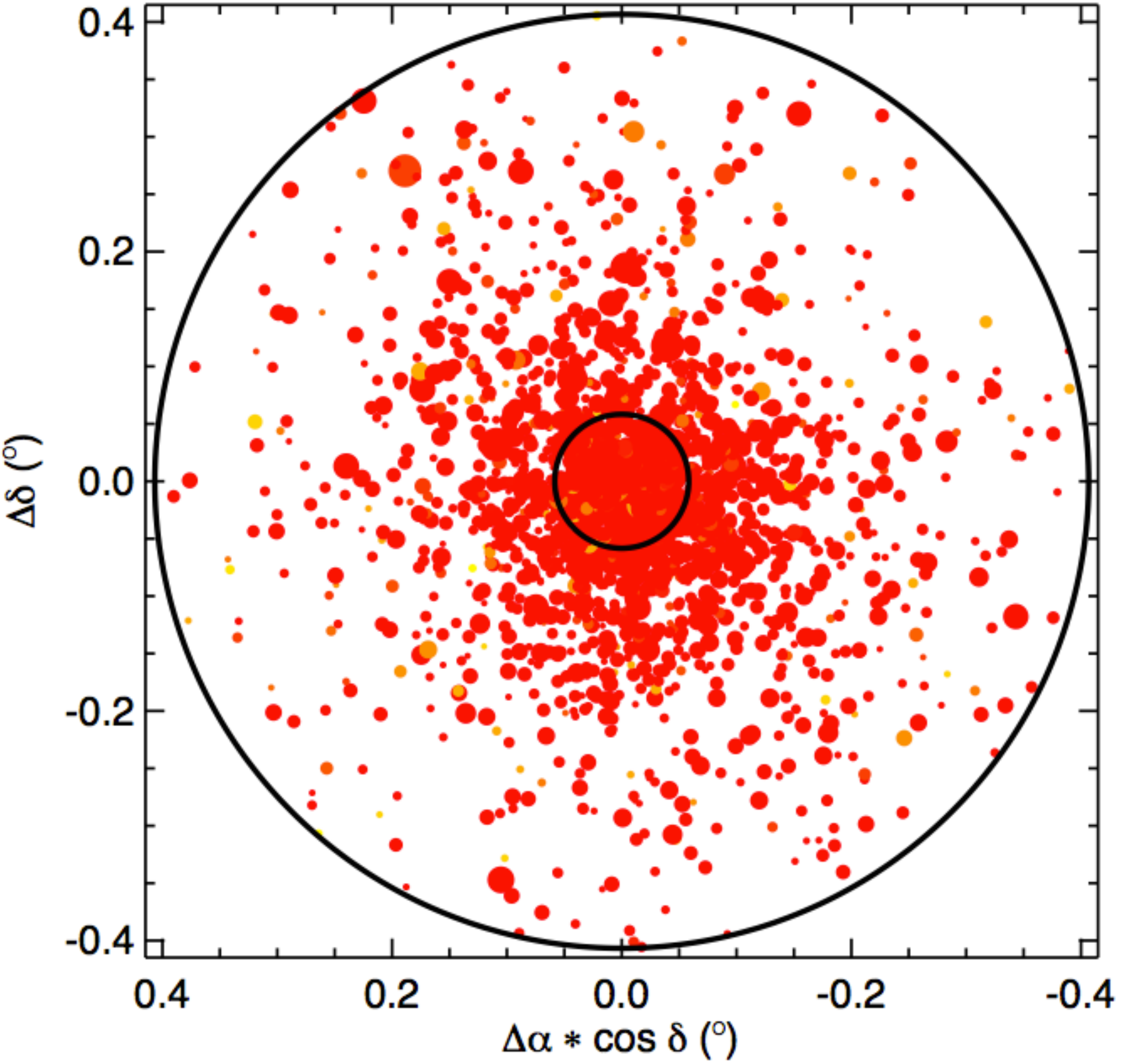}     

    \includegraphics[width=0.333\textwidth]{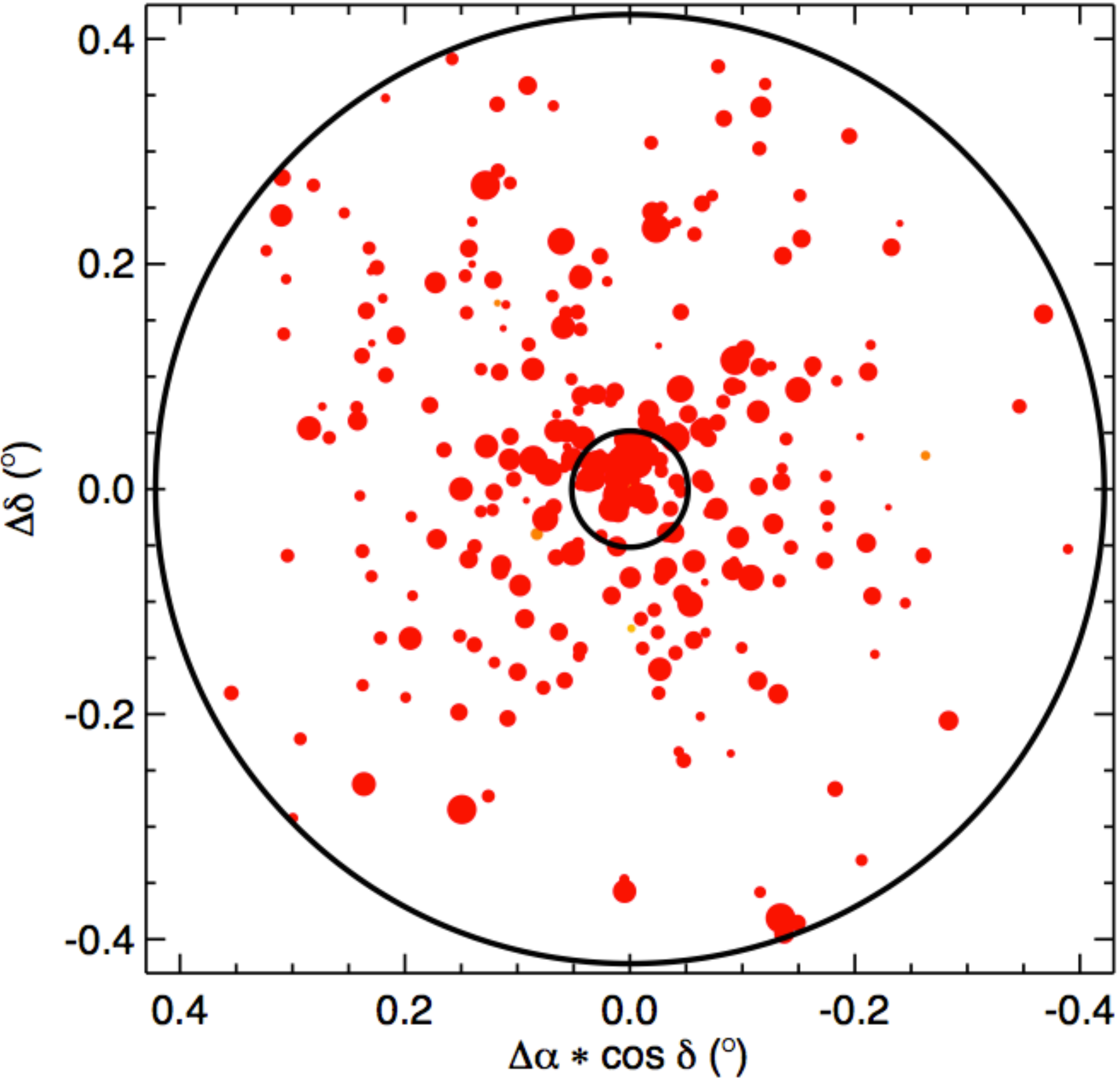}     
    \includegraphics[width=0.333\textwidth]{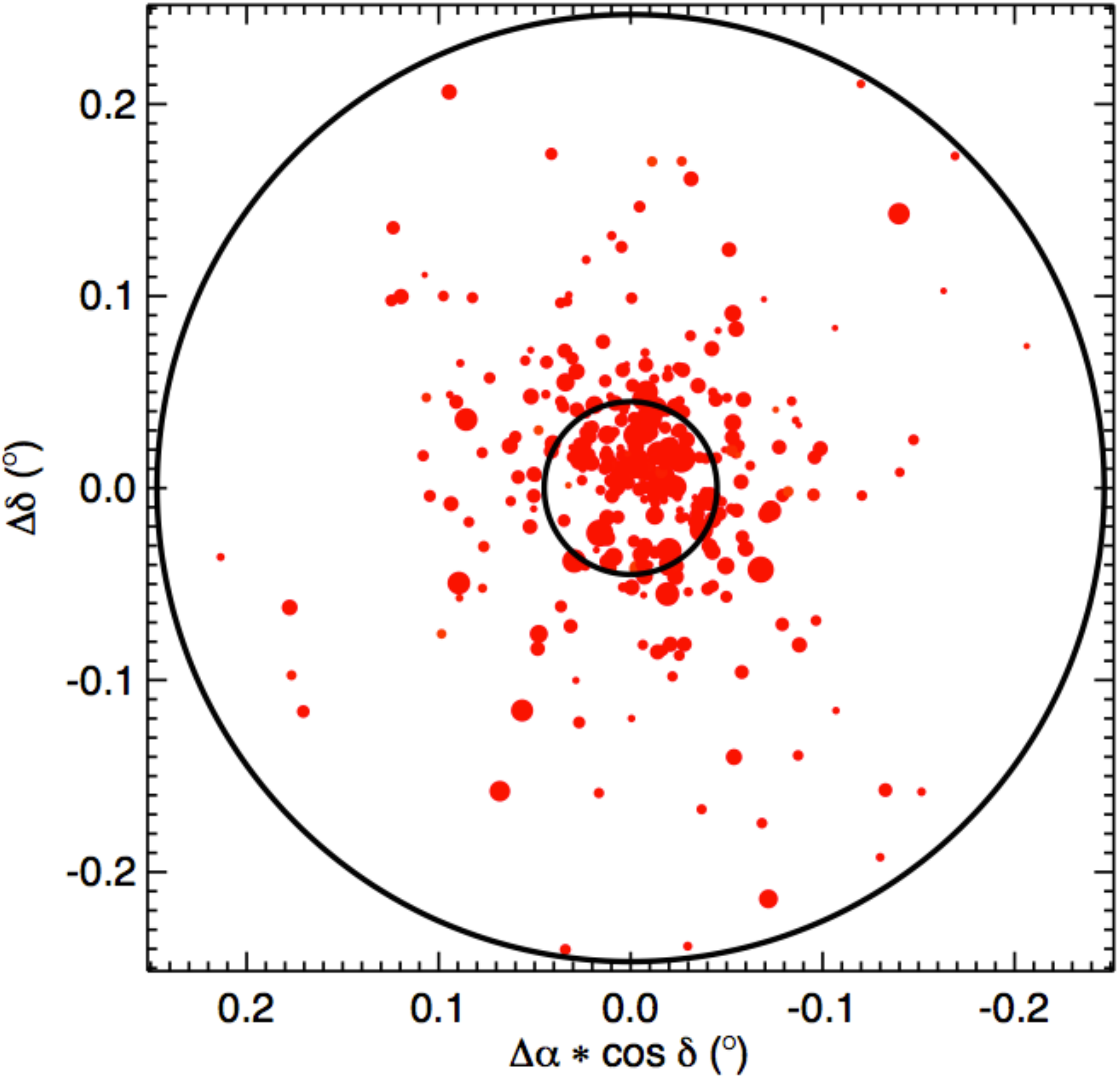}     
    \includegraphics[width=0.333\textwidth]{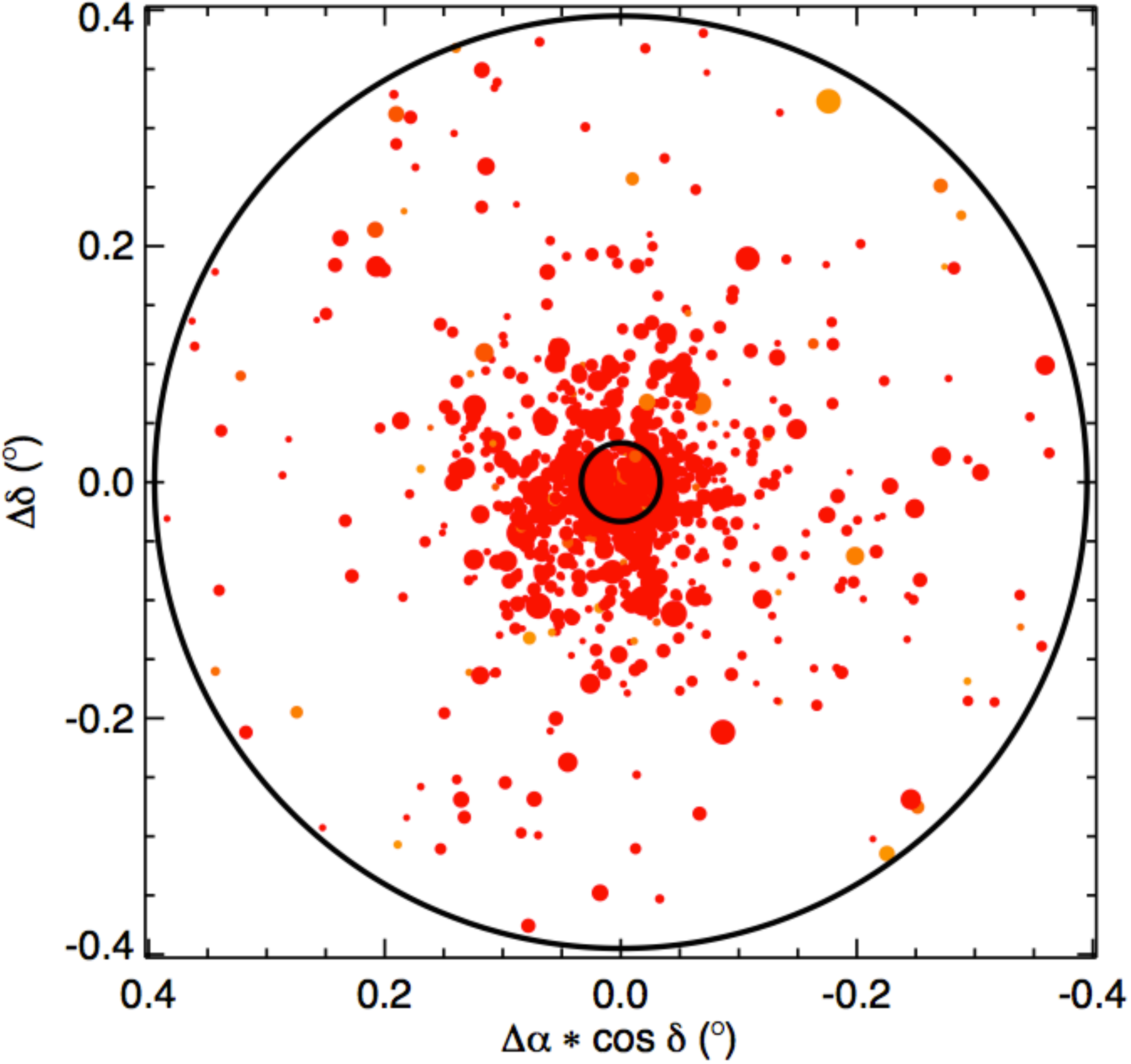}     

    \includegraphics[width=0.333\textwidth]{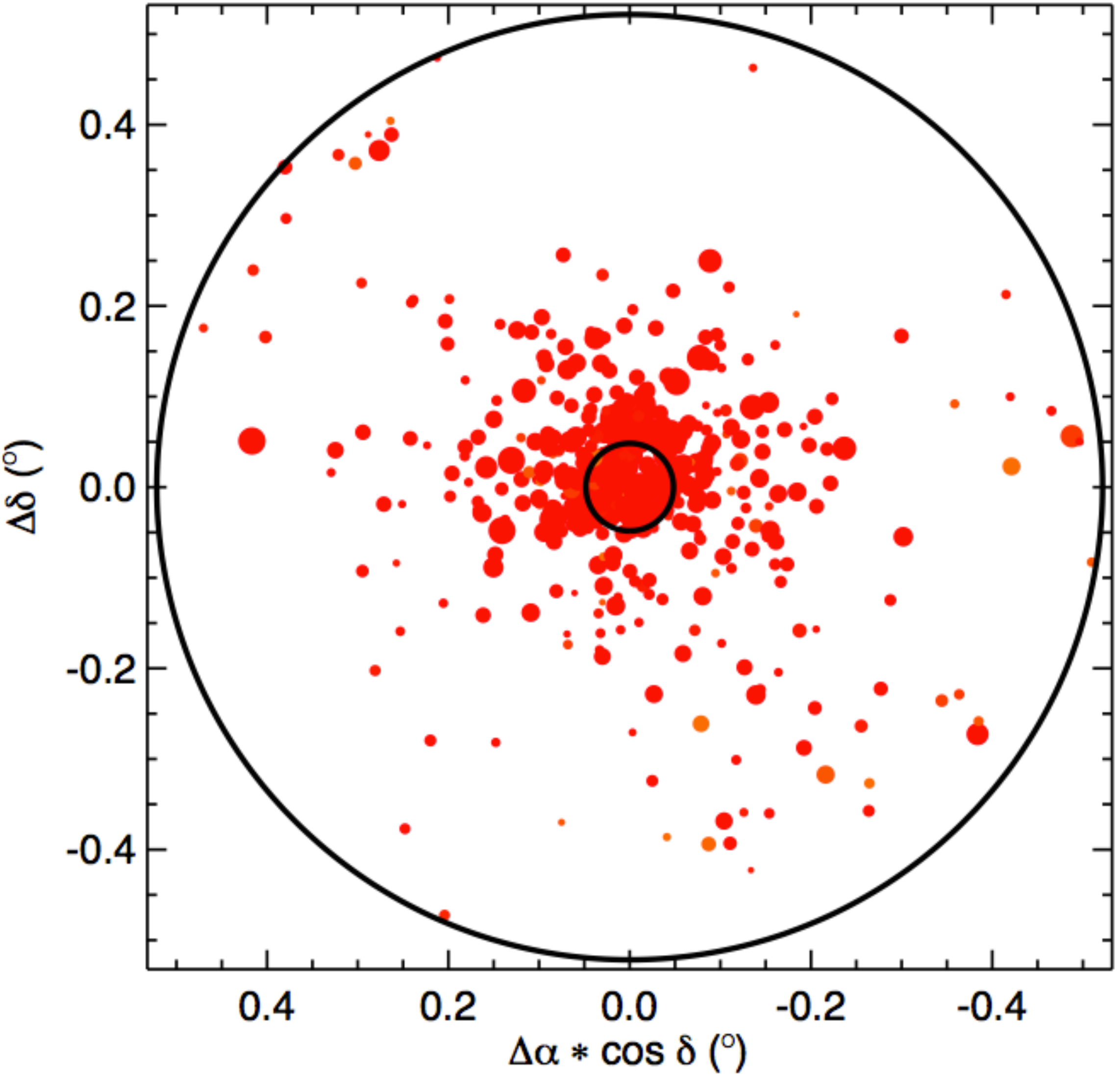}     
    \includegraphics[width=0.333\textwidth]{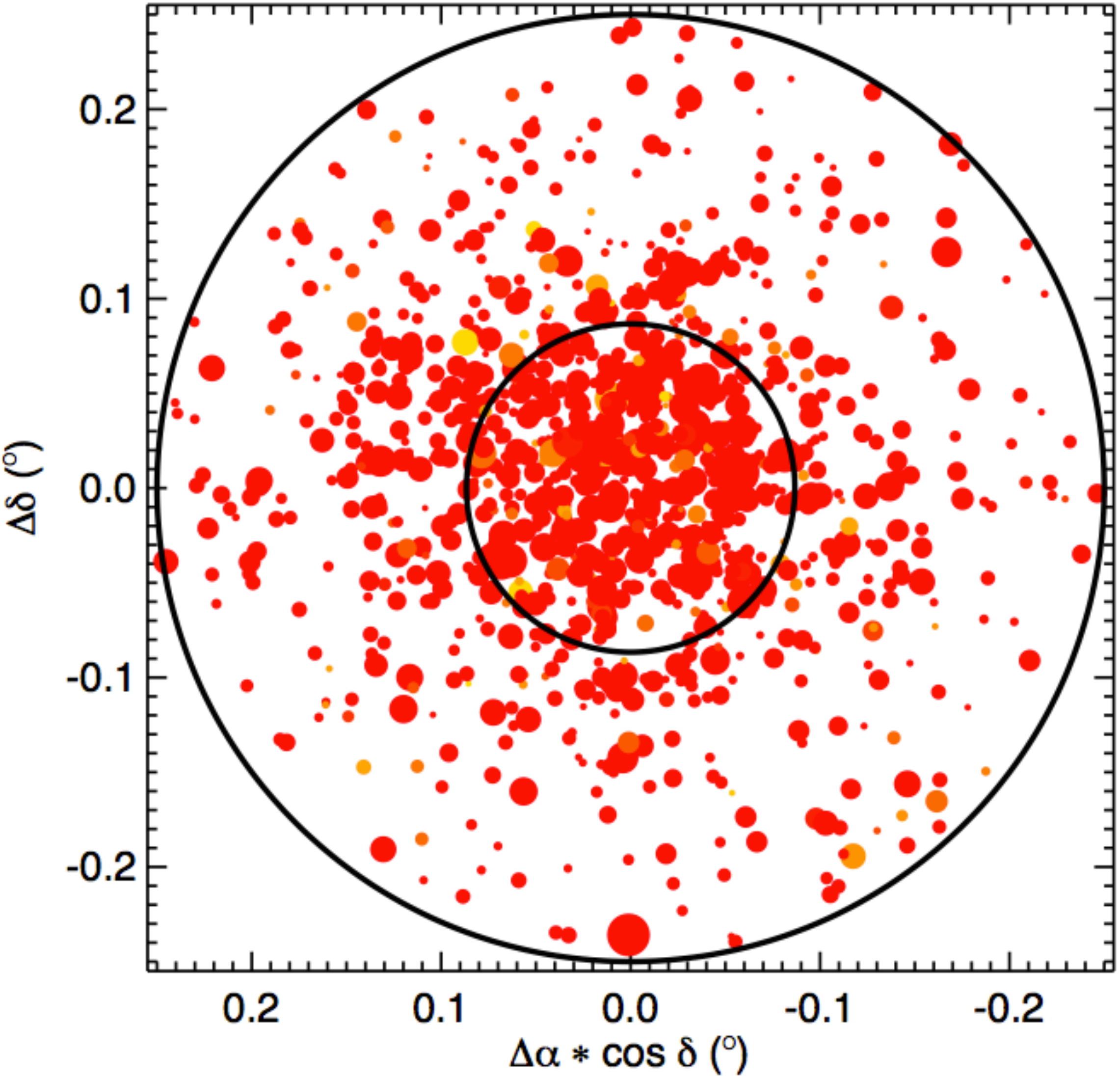}     
    \includegraphics[width=0.333\textwidth]{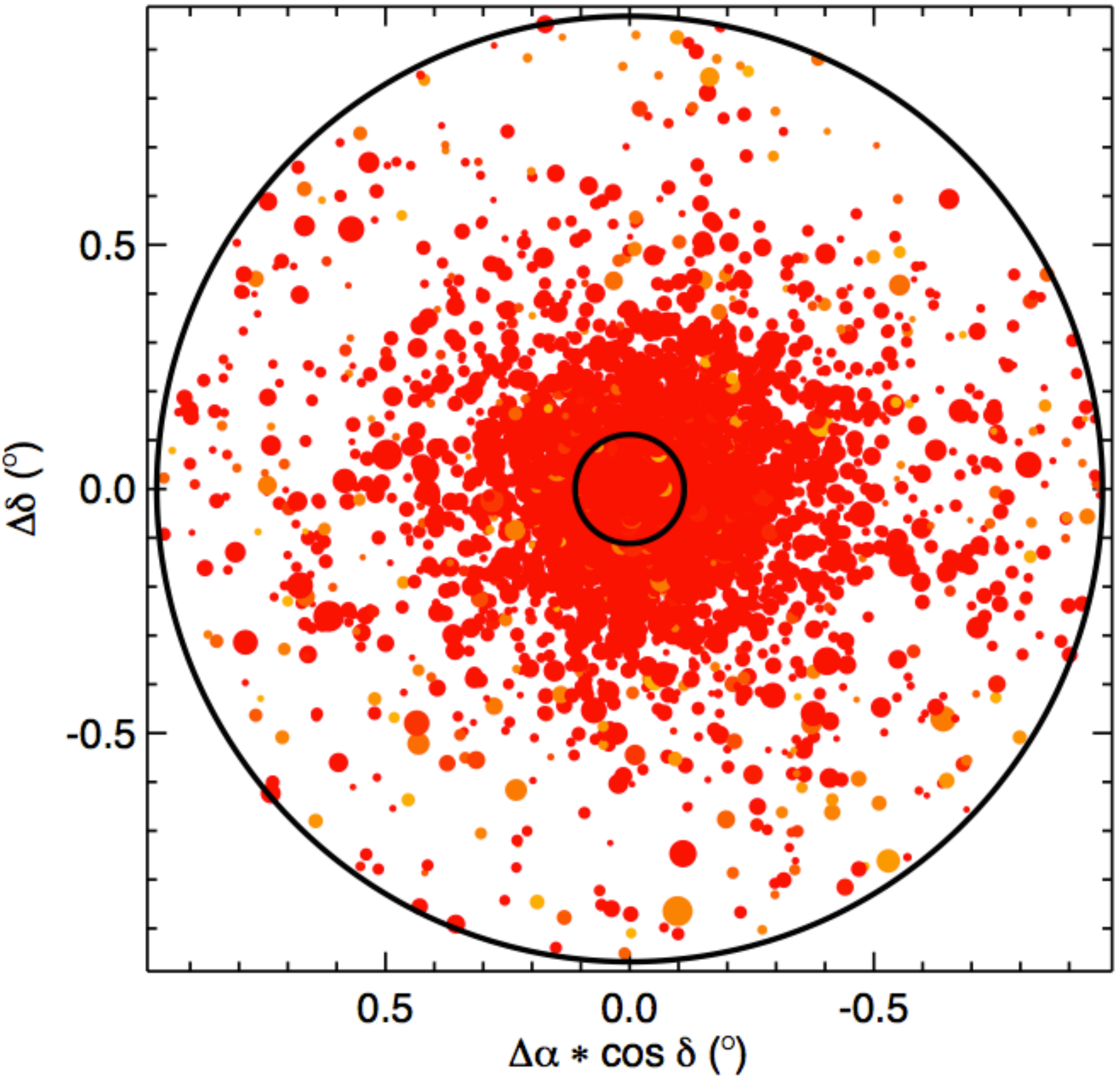}     
     
  }
\caption{ Skymap for the OCs (from top left to bottom right): Czernik\,38, Berkeley\,81, NGC\,6791 (top line), NGC\,6802, NGC\,6811, NGC\,6819 (second line), NGC\,6866, Berkeley89, NGC\,7044 (third line), NGC\,7142, NGC\,7654, NGC\,7789 (bottom line).  }

\label{fig:skymaps_49_60}
\end{center}
\end{figure*}

\bsp

\label{lastpage}

\end{document}